\documentclass[12pt,a4paper,notitlepage]{report}

\usepackage{amssymb}
\usepackage[final]{graphicx}
\graphicspath{{eps/}}
\usepackage{feynmp}
\usepackage{url}
\usepackage{dcolumn}
\usepackage{enumerate}

\newcommand{\dd}{\mathrm{d}}
\newcommand{\df}{\partial}
\newcommand{\dilog}{\mathop{\mathrm{Li}_2}}
\newcommand{\trilog}{\mathop{\mathrm{Li}_3}}
\newcommand{\eps}{\epsilon}

\newcommand{\Dns}{\ensuremath{D^\mathrm{NS}}}
\newcommand{\JB}{\mathrm{JB}}
\newcommand{\Eep}{E_e{\!\!'}\,}

\newcommand{\Angle}{\measuredangle}
\newcommand{\sign}{\mathop\mathrm{sign}}
\newcommand{\GeV}{\ensuremath{\;\mathrm{GeV}}}
\newcommand{\MW}{\ensuremath{M_\mathrm{W}}}
\newcommand{\MZ}{\ensuremath{M_\mathrm{Z}}}
\newcommand{\mrad}{\ensuremath{\;\mathrm{mrad}}}

\newcommand{\overbar}[1]{\ensuremath{\overline{\mbox{#1}}}}
\newcommand{\bigO}[1]{\ensuremath{\mathcal{O}\left(#1\right)}}

\newcolumntype{d}[1]{D{.}{.}{#1}}

\makeatletter

\newcommand{\fmslash}[2][0mu]{%
  \mathchoice
    {\fmsl@sh\displaystyle{#1}{#2}}%
    {\fmsl@sh\textstyle{#1}{#2}}%
    {\fmsl@sh\scriptstyle{#1}{#2}}%
    {\fmsl@sh\scriptscriptstyle{#1}{#2}}}
\newcommand{\fmsl@sh}[3]{%
  \m@th\ooalign{$\hfil#1\mkern#2/\hfil$\crcr$#1#3$}}

\newbox\slashbox \setbox\slashbox=\hbox{$/$}
\newbox\Slashbox \setbox\Slashbox=\hbox{\large$/$}
\def\pFMslash#1{\setbox\@tempboxa=\hbox{$#1$}
  \@tempdima=0.5\wd\slashbox \advance\@tempdima 0.5\wd\@tempboxa
  \copy\slashbox \kern-\@tempdima \box\@tempboxa}
\def\pFMSlash#1{\setbox\@tempboxa=\hbox{$#1$}
  \@tempdima=0.5\wd\Slashbox \advance\@tempdima 0.5\wd\@tempboxa
  \copy\Slashbox \kern-\@tempdima \box\@tempboxa}

\newcommand{\bigggl}{\mathopen\biggg}

\newcommand{\bigggr}{\mathclose\biggg}

\newcommand{\biggg}[1]{{\hbox{$\left#1\vbox to20.5\p@{}\right.\n@space$}}}
\newcommand{\Biggg}[1]{{\hbox{$\left#1\vbox to23.5\p@{}\right.\n@space$}}}

\makeatother

\begin{document}


\begin{titlepage}
  \renewcommand{\thefootnote}{\fnsymbol{footnote}}
  \setcounter{footnote}{0}
  \vspace*{\fill}
  \begin{center}
    \huge\bfseries
    QED Radiative Processes in \\[1mm] Deep Inelastic Scattering \\[1mm]
    at HERA%
    \footnote{Habilitationsschrift,
      Universit\"at Siegen, Germany, February 2002}
  \end{center}
  \vspace*{\fill}
  \begin{center}
    \Large
    Harald Anlauf
  \end{center}
  \vspace*{\fill}
  \begin{abstract}
    High energy scattering processes of charged particles are accompanied
    by radiation of hard photons.  Emission collinear to the incident
    particles, which leads to a reduction of the effective beam energy, and
    the possibility to directly measure these photons at the HERA
    electron-proton collider provides important physics opportunities.  For
    deep inelastic scattering, the measurement of radiative processes
    extends the kinematic range accessible to the HERA experiments to lower
    $Q^2$, as well as helps in separating the proton structure functions
    without the need to run at different collider energies.  QED
    corrections to these radiative processes are discussed, and the
    calculation of the model-independent leptonic corrections is described
    in some detail for the complete one-loop contributions as well as for
    the higher order leading logarithms.
  \end{abstract}
\end{titlepage}







\tableofcontents



\chapter{Introduction}
\label{sec:Intro}

In our quest for the understanding of the structure of matter, scattering
experiments have always played an outstanding r\^ole.  Important landmarks
are the classic experiments by Rutherford \cite{Rutherford:1911} leading to
the discovery of the atomic nucleus, the determination of the size of
nucleons and atomic nuclei from their electromagnetic form factors by
Hofstadter and McAllister \cite{Hofstadter:ae}, and the establishing of the
parton model of the nucleon in deep inelastic lepton-proton scattering at
SLAC \cite{Breidenbach:1969kd}.

Supported by an increasing amount of data from high precision scattering
experiments at high energy electron-positron, proton-antiproton, and
electron-proton colliders, the electroweak Standard Model (SM) including
Quantum Chromodynamics (QCD) is currently accepted as the simplest quantum
field theory describing the observed phenomena of elementary particle
physics.  The fundamental particles of the SM, the leptons, quarks, gauge
fields and the (experimentally not yet established) Higgs are point-like.
The very good agreement of many theoretical predictions and present
experimental results from LEP, Tevatron and HERA indicates that any
substructure of these particles must be smaller than $10^{-19}\,\mbox{m}$,
corresponding to a compositeness scale above several TeV
\cite{LEP+SLD:2001,Grosso-Pilcher:1998vi,compositeness@HERA}.

On the theoretical side, still one of the major challenges is a complete
understanding of hadronic interactions and hadron structure.  The startup
of the lepton-proton collider HERA at DESY with a center of mass energy of
$\sqrt{s} \approx 300 \GeV$ (27.5\GeV\ electrons/positrons on initially
820\GeV, currently 920\GeV\ protons) marked the beginning of a new era of
deep inelastic scattering (DIS) experiments, extending the kinematic domain
for structure function measurements covered by earlier fixed-target
experiments in the Bjorken variable $x$ and momentum transfer $Q^2$ by
several orders of magnitude.  Besides the inclusive (w.r.t.\ the hadronic
final state) electroproduction structure functions, HERA also provides
increasingly precise data for the study of hadronic final states in DIS, in
particular multi-jet events and event shapes.

The large accessible kinematic region at HERA allows the investigation of
different aspects of strong interaction dynamics.  In the perturbative
regime of QCD, much interest is devoted to the study of the transition from
the traditional Altarelli-Parisi (DGLAP) evolution of structure functions
at large $x$ to the Balitsky-Fadin-Kuraev-Lipatov (BFKL)
\cite{BFKL,Schmidt:2001yq} behavior at small $x$, and to a determination
of the gluon density which is expected to be strongly rising for $x \to 0$.
Unitarity requires that the rapid rise predicted by the perturbative
evolution must eventually be damped.  There are indications from the DIS
data at low $x$ that this unitarization manifests itself in a phenomenon
called \emph{geometric scaling} \cite{Stasto:2000er}, where the cross
section is not a function of two independent variables $x$ and $Q^2$ but
rather a function of a single variable $\tau=Q^2/Q_\mathrm{s}^2(x)$.  The
function $Q_\mathrm{s}(x)$ is interpreted as a saturation scale in a
description of non-linear QCD evolution \cite{Gribov:tu,Lublinsky:2001bc}.

Of the structure functions of the proton, $F_2(x,Q^2)$ and $F_L(x,Q^2)$,
the longitudinal one, $F_L$, is much more difficult to access.  The
measurement of $F_L$ at low $x$ is particularly interesting due to its
tight connection to the gluon distribution in the proton.  There exist
several ways to separately extract $F_2$ and $F_L$ from the experimental
data.  One possibility is to run the collider at different ---usually
lower--- center-of-mass energies, which may not be desirable from the point
of view of other parts of the physics program looking for possible hints of
physics beyond the SM at the highest accessible scales \cite{HERA:excess}.

Indirect methods for the determination of $F_L$ can be used at fixed
collision energy but usually require substantial input from theory or other
assumptions, and they depend more or less on the modeling of the hadronic
final state, like extrapolations or QCD fits (see e.g.,
\cite{DIS99:Arkadov,Adloff:2001qk,DIS01:Dubak,EPS01:ZEUS}), or the
measurement of the azimuthal angle distribution of final state hadrons
\cite{Gehrmann:2000vu}.

Krasny et al.~\cite{KPS92} suggested a direct method that utilizes
radiative events with an exclusive hard photon registered (tagged) in a
forward photon detector (PD).  Such a device is actually a central part of
the luminosity monitoring system of the H1 and ZEUS experiments
\cite{H1:rad,Andruszkow:2001jy} that measures the radiative elastic process
$ep \to ep\,+\,\gamma$ \cite{Bethe:1934za}; it was further improved during
the HERA luminosity upgrade \cite{Foster:2001fj}.  The idea of this method
is that emission of photons in a direction close to the incident electron
corresponds to a reduction in the effective beam energy.  The effective
electron energy for each radiative event is determined from the energy of
the photon observed in the PD.

Besides measuring $F_L$, radiative events extend the accessible kinematic
range to lower values of $Q^2$.  This is important for a closing of the gap
between the fixed-target experiments and HERA.  The potential of this
method is supported by preliminary results from the H1 collaboration of an
analysis at low $Q^2$ for $F_2$
\cite{Klein:1998mz,Issever:thesis,Issever:DIS01,H1:2001rad} (for earlier
analyses that did not take into account QED radiative corrections see
\cite{H1:rad,ZEUS96}).  The feasibility of the corresponding determination
of $F_L$ was studied in \cite{FGMZ96}.  However, with currently analyzed
data sets it is not yet possible to compete with $F_L$ from extrapolations
or QCD fits \cite{Issever:thesis}.

A precise analysis of experimental data requires the inclusion of radiative
corrections.  In the case of the exclusive radiative events, even at HERA
energies only the kinematic region of momentum transfers far below the
masses of the W and Z bosons has a significant cross section.  As a
consequence only the QED subset of the electroweak interaction is relevant,
so that we can safely restrict ourselves to the calculation of the QED
corrections.
A minor complication is the well known fact that the full theoretical
control of the radiative corrections in non-radiative deep inelastic
scattering is aggravated by the dependence of the corrections on the
knowledge of the proton structure functions even in kinematic regions which
may be difficult to access.  However, this is almost precisely where a
measurement of the radiative process can help, and why it is important that
this measurement is performed.

From the calculation of radiative corrections to deep inelastic scattering
it is well known that the model independent QED corrections on the lepton
side are most important.  For the tagged photon reaction in the HERA
kinematic regime, the leptonic QED corrections have been discussed at the
leading logarithmic level \cite{Bardin:1997zm,AAKM:ll} and taking into
account next-to-leading logarithms \cite{AAKM:JETP,AAKM:nlo,Anl99:Sigma}.
Since the leading logarithmic corrections lead to compact and transparent
expressions, we use them for a qualitative study of the dependence of the
QED radiation effects on the kinematic reconstruction method.

A central part of the present work deals with the complete leptonic
$\bigO{\alpha}$ corrections%
\footnote{In this work corrections are always counted \emph{relative} to
  the leading contribution for a given final state, which is already
  proportional to $\alpha^3$ for radiative deep inelastic scattering.}
to this process which were outlined in \cite{Anlauf:2001fu}.  We will
present here all essential details of this calculation.  Since it is well
known that also higher-order corrections are very important in a
quantitative description of deep-inelastic scattering, we shall describe in
some length a systematic method to obtain the leading logarithms to the
tagged photon process to all orders.

The organization of this report is as follows.  Section 2 introduces our
notation in the context of deep inelastic scattering to lowest order.  We
recapitulate some basic knowledge of the proton structure functions and
their connection to QCD.  After providing a brief review of the general
features of QED radiative corrections to deep inelastic scattering, section
3 emphasizes the importance of the exclusive radiative processes that are
the main subject of this work.  In section 4 we calculate the QED radiative
corrections to the tagged photon process, beginning with the leading
logarithmic approximation and extending to the complete leptonic
corrections.  Section 5 considers the leading logarithmic contributions at
higher orders.  We conclude with a short summary, recent experimental
results on the structure function $F_2$ obtained with the tagged photon
method and an outlook on the application of photon tagging to $e^+ e^-$
collisions.  Finally, the appendices explain several details of the more
involved parts of the analytic calculations and collect several formulae
that are needed for a numerical implementation.


\begin{fmffile}{figs}
\fmfstraight


\chapter{Deep Inelastic Scattering}

\section{Kinematics}

We start by considering the general deep inelastic electron proton
scattering process
\begin{equation}
\label{eq:dis-kin}
  e(p) + p(P) \to l(p') + X(P_X) \; ,
\end{equation}
where $e$ and $p$ represent the incoming electron and proton, and $l$ and
$X$ the scattered lepton and the final hadronic system, while the
respective four-momenta are given in parentheses.  Although the HERA $ep$
collider is designed to also run with positron beams instead of electrons
and even collected more integrated luminosity with the former, we use the
latter as the generic denotation of the lepton beam below, but we shall
explicitly indicate the dependence on the lepton charge where needed.

\begin{figure}
  \begin{center}
    \begin{picture}(65,45)
\put(0,5){%
\begin{fmfgraph*}(60,35)
\fmfleft{ip,il}
\fmfright{oq1,oq2,oq3,oq4,d1,d2,d3,ol}
\fmf{fermion,tension=4}{il,vl}
\fmf{fermion,tension=2}{vl,ol}
\fmf{photon,tension=4}{vl,ii,vp}
\fmf{dbl_plain_arrow,tension=2}{ip,vp}
\fmf{fermion}{vp,oq4}
\fmf{fermion}{vp,oq3}
\fmf{fermion}{vp,oq2}
\fmf{fermion}{vp,oq1}
\fmfdot{vl}
\fmfblob{0.15w}{vp}
\fmflabel{$p$}{ip}
\fmflabel{$e$}{il}
\fmflabel{$l$}{ol}
\fmfv{label.a=30,label=$\gamma,,Z/W$}{ii}
\end{fmfgraph*}%
}
\put(61,11.5){$\left. \vbox to 10mm{}\right\} X$}
    \end{picture}
    \caption{Deep inelastic lepton-proton scattering}
  \end{center}
\end{figure}
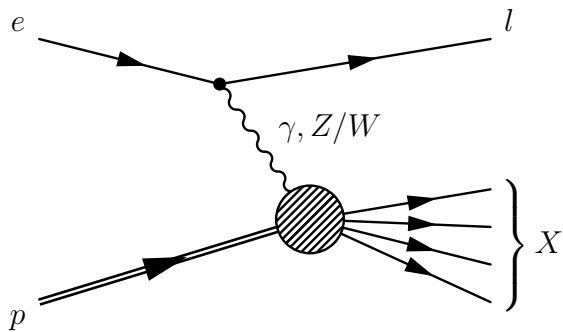

Equation (\ref{eq:dis-kin}) describes the so-called neutral current (NC)
processes, where the scattered lepton equals the incoming one, $l=e$, as
well as charged-current (CC) processes, where the outgoing lepton is an
electron neutrino $\nu_e$ for electron beams and an electron anti-neutrino
$\bar\nu_e$ for positron beams, respectively.

Let us introduce the notation used in this report.  We choose coordinates
in the HERA laboratory system in which the incoming electrons move in the
positive $z$-direction,%
\footnote{As a consequence, one should always keep in mind our definition
  of the \emph{forward direction} being along the positive $z$-axis as the
  \emph{initial lepton direction}, in contrast to the historical decision
  of the HERA experiments to use the direction of the incoming proton beam
  for that purpose.}
whereas the protons move in the negative $z$-direction.  Furthermore, the
direction of the $x$-axis is chosen such that the momentum of the scattered
lepton lies in the $x$-$z$ plane:
\begin{eqnarray}
  p  & = & (E_e, 0, 0,   p_e) \; , \nonumber \\
  P  & = & (E_p, 0, 0, - p_p) \; , \nonumber \\
  p' & = & (E'_l, p'_l \sin\theta, 0, p'_l \cos\theta) \; .
\end{eqnarray}
The HERA $ep$ collider is typically run at an energy of $27.5\GeV$ for the
electron beam, while the nominal energy of the proton ring, initially
$820\GeV$, now reaches up to $920\GeV$.

The kinematics of the scattering process (\ref{eq:dis-kin}) is described by
several Lorentz invariants.  The square of the total center of mass energy
available is given by
\begin{equation}
  s = (p+P)^2 = m^2 + M^2 + S
  \simeq 4 E_e E_p \; ,
\end{equation}
where we introduced the related invariant
\begin{equation}
  S = 2 P \cdot p \; .
\end{equation}
At high energies, the electron mass $m$ and the proton mass $M$ can be
mostly neglected, and $s \simeq S$.  We shall therefore use them almost
synonymously below.  For the HERA beam energies mentioned above, $\sqrt{S}
\simeq 300 \GeV$ and $\sqrt{S} \simeq 318 \GeV$, resp.

The invariant momentum transfer from the lepton to the hadronic system
reads
\begin{equation}
\label{eq:Q2}
  Q^2 \equiv - q^2 = - (p-p')^2
  \simeq 4 E_e E_l' \sin^2 \frac{\theta}{2}
  \; ,
\end{equation}
assuming that the energy and angle of the outgoing lepton can be measured.

Further commonly used invariants are the dimensionless Bjorken variables
\begin{eqnarray}
\label{eq:x}
  x & = &
  \frac{Q^2}{2 P \cdot q}
  \simeq
  \frac{E_e E_l' \sin^2 \frac{\theta}{2}}%
       {E_p \left( E_e - E_l' \cos^2 \frac{\theta}{2} \right)}
  \; ,
  \\
\label{eq:y}
  y & = &
  \frac{P \cdot q}{P \cdot p}
  = \frac{Q^2}{x S}
  \simeq 1 - \frac{E_l'}{E_e} \cos^2 \frac{\theta}{2}
  \; .
\end{eqnarray}
The quantity $y$ corresponds to the relative energy loss of the lepton in
the rest frame of the incoming hadron.  It is easy to show that $x$ and $y$
are restricted to the range
\begin{equation}
  0 \leq x,y \leq 1 \; .
\end{equation}
Finally, the total mass of the hadronic final state $X$ is obtained as
\begin{equation}
  W^2 \equiv
  M_X^2 = (P+q)^2
  = M^2 + \frac{1-x}{x} \, Q^2
  = M^2 + (1-x)yS
  \; .
\end{equation}
Elastic $ep$ scattering corresponds to $x=1$.


\section{Determination of kinematic variables}

Because of four-momentum conservation, the kinematic invariants $x$, $y$
and $Q^2=xyS$ can in principle be determined in different ways from the
measured final state.  Instead of using only the information from the
scattered lepton, one can also take into account the measured hadronic
final state.  In the case of charged current scattering, where the final
state neutrino escapes detection, this is actually necessary.

In the description of deep inelastic scattering in the language of the
naive quark-parton model (QPM), the lepton scatters elastically with a
quasi-free quark within the proton.  The two-body final state of this
subprocess is completely determined by two variables, which can be taken to
be e.g., the energy $\Eep$ and the polar angle $\theta$ of the electron in
the lab frame, or quantities constructed from the hadronic final state.

Assuming local parton-hadron duality, one identifies the energy and
direction of the hadron jet with the energy and direction of the scattered
quark.  Thus, denoting the four-momenta of the produced hadrons by
$p_h=(E_h;\vec{p}_{\perp,h},p_{z,h})$, one can construct the following
quantities:%
\footnote{Remember that the initial proton moves in the negative $z$
  direction.}
\begin{equation}
  \Sigma_h = \sum_h (E_h + p_{z,h})
  \; , \quad
  p_{T,h} =
  \sqrt{
  \left(
  \sum_h \vec{p}_{\perp,h}
  \right)^2 }
  \; ,
\end{equation}
the latter being the total transverse momentum of the hadronic final state.
Neglecting masses, one can define an inclusive angle $\gamma$ of the
hadronic system which then corresponds to the angle of the scattered quark
in the QPM,
\begin{equation}
  \tan \frac{\gamma}{2} = \frac{\Sigma_h}{p_{T,h}}
  \; .
\end{equation}
Obviously one can define similar quantities for the scattered electron,
\begin{equation}
  \Sigma_e = \Eep(1+\cos\theta)
  \; , \quad
  p_{T,e}  = \Eep \sin\theta
  \; , \quad
  \Rightarrow \;
  \tan\frac{\theta}{2} = \frac{\Sigma_e}{p_{T,e}}
  \; .
\end{equation}
Starting from these four variables $[\Eep,\theta,\Sigma_h,\gamma]$, among
which only two are independent, one derives the following ``basic
methods'', that use only two variables:
\begin{itemize}
\item For neutral current events, the electron method $(e)$, which uses
  $\Eep$ and $\theta$, see eqs.~(\ref{eq:Q2})--(\ref{eq:y}), and
\item the double angle method (DA), which uses $\theta$ and $\gamma$
  \cite{Bentvelsen:1992fu}, whereas
\item for charged current events only the ``hadrons only'' method $(h)$,
  also known as Jacquet-Blondel method (JB) \cite{Jacquet-Blondel}, is
  available.
\end{itemize}
The Jacquet-Blondel method defines:
\begin{equation}
\label{eq:JB-vars}
  y_\mathrm{JB} = \frac{\Sigma_h}{2E_e}
  \; , \quad
  Q_\mathrm{JB}^2 = \frac{p_{T,h}^2}{1-y_\mathrm{JB}}
  \; , \quad
  x_\mathrm{JB} = \frac{Q_\mathrm{JB}^2}{y_\mathrm{JB} S}
  \; .
\end{equation}
For neutral current processes, there are possibilities and reasons to also
consider methods that employ more than two of the measured variables to
determine the kinematics, either analytically or using the full event
information for kinematic fits.  Real detectors are not perfect; they
possess finite energy and angle resolution, and they usually do not cover
the full solid angle, so that particles may escape in the beam pipe.
Therefore e.g.\ the energy of the incoming electron may not be well known
as it might emit a collinear photon before the hard collision.

As an example for an analytic method using three variables we mention here
the $\Sigma$ method, which basically uses (\ref{eq:JB-vars}), but replaces
$2E_e$ with the help of four-momentum conservation by the combination
$\Sigma_e+\Sigma_h$, thereby making the determination of the variable $y$
independent of collinear photon emission in the initial state, and replaces
the hadronic transverse momentum $p_{T,h}$ by the leptonic $p_{T,e}$,
\begin{equation}
\label{eq:Sigma-vars}
  y_\Sigma = \frac{\Sigma_h}{\Sigma_e+\Sigma_h}
  \; , \quad
  Q_\Sigma^2 = \frac{p_{T,e}^2}{1-y_\Sigma}
  \; , \quad
  x_\Sigma = \frac{Q_\Sigma^2}{y_\Sigma S}
  \; .
\end{equation}
For a thorough discussion of the essential features of these methods and
some extensions we refer the reader to
\cite{Bassler:1995uq,Bassler:1999tv}, and references cited therein.


\section{The Born cross section for $ep \to eX$}

Let us now turn to the model-independent description of the cross section
for semi-inclusive deep inelastic scattering, where only the final state
electron is measured (electroproduction).  For the time being, we shall
restrict to the case where the momentum transfer $Q^2$ is far below the
electroweak scale, i.\,e., $Q^2 \ll \MZ^2$.  The Born contribution to the
scattering amplitude taking into account only one-photon exchange reads:
\begin{equation}
  S_{fi} =
  i (2\pi)^4 \delta^4(P_X+p'-P-p) \,
  \bar{u}(p')\gamma^\mu u(p) \, \frac{Q_e e^2}{q^2} \,
  \left\langle P_X \left| J_\mu^\mathrm{em}(0) \right| P \right\rangle
  \; .
\end{equation}
where $Q_e$ is the charge of the lepton in units of the proton charge
($Q_e=+1$ for the positron), and $J_\mu^\mathrm{em}(x)$ is the
electromagnetic current operator.

The differential cross section for the scattering of unpolarized particles
reads ($q=p-p'$):
\begin{eqnarray}
\label{eq:dsigma}
  \dd \sigma & = &
  \frac{1}{2 \sqrt{\lambda_S}}
  \left( \frac{e^2}{q^2} \right)^2
  L^{\mu\nu}
  H_{\mu\nu}(P,q) \,
  \frac{\dd^3 \vec{p}\,'}{(2\pi)^3 2\Eep}
\end{eqnarray}
with the lepton tensor:
\begin{eqnarray}
\label{def:lepton-tensor}
  L^{\mu\nu} & = &
  \frac{1}{2} \sum_\mathrm{spins}
  [ \bar{u}(p') \gamma^\mu u(p)]^* \,
    \bar{u}(p') \gamma^\nu u(p)
  \nonumber \\
  & = & \tilde{g}^{\mu\nu} q^2 + 4 \tilde{p}^\mu \tilde{p}^\nu
  \; ,
  \\
  \tilde{g}^{\mu\nu} & = & g^{\mu\nu} - \frac{q^{\mu}q^{\nu}}{q^2}
  \; , \quad
  \tilde{p}^\mu = p^\mu - q^\mu\frac{p \cdot q}{q^2}
  \; .
  \nonumber
\end{eqnarray}
We normalize the hadron tensor $H_{\mu\nu}$ as follows:
\begin{eqnarray}
\label{def:hadron-tensor-H}
  H_{\mu\nu}(P,q) & = &
  \frac{1}{2} \sum_\mathrm{pol} \sum_X
  \left\langle P \left| J_\mu^\mathrm{em}(0) \right| P_X \right\rangle
  \left\langle P_X \left| J_\nu^\mathrm{em}(0) \right| P \right\rangle
  \\
  & = &
  \frac{1}{2} \sum_\mathrm{pol}
  \int \dd^4 x \; e^{i q\cdot x} \,
  \left\langle P \left|
  \left[ J_\mu^\mathrm{em}(x), J_\nu^\mathrm{em}(0) \right]
  \right| P \right\rangle
  \; . \nonumber
\end{eqnarray}
Using current conservation, charge conjugation and parity symmetry, we
decompose this tensor as:
\begin{eqnarray}
\label{eq:hadron-tensor-H}
  H_{\mu\nu}(P,q)
  &=& 4\pi
  \left(
  - \tilde{g}_{\mu \nu} F_1(x,Q^2)
  + \tilde{P}_{\mu} \tilde{P}_{\nu} \, \frac{1}{P\cdot q} \, F_2(x,Q^2)
  \right)
  \nonumber \\
  &=& 4\pi
  \left(
  - \tilde{g}_{\mu \nu} F_1(x,Q^2)
  + \tilde{P}_{\mu} \tilde{P}_{\nu} \, \frac{2x}{Q^2} \, F_2(x,Q^2)
  \right)
  \; ,
  \\
  \tilde P_{\nu} &=& P_{\nu}-q_{\nu}\frac{P\cdot q}{q^2}
  \; , \quad
  x=\frac{Q^2}{2P\cdot q}
  \; , \quad
  Q^2 = -q^2
  \; .
  \nonumber
\end{eqnarray}
The functions $F_1(x,Q^2)$ and $F_2(x,Q^2)$ are the (electromagnetic)
structure functions of the hadron.

The quantity $\lambda_S$ appearing in the flux factor reads:
\begin{equation}
  \lambda_S = S^2 - 4 m^2 M^2 \; .
\end{equation}
Expressing the final electron momentum via the invariants ($y,Q^2$) yields:
\begin{equation}
\label{eq:e-ps-y-Q2}
  \frac{\dd^3 \vec{p}\,'}{(2\pi)^3 2\Eep} =
  \frac{1}{(4\pi)^2} \frac{S}{\sqrt{\lambda_S}} \,
  \dd y \, \dd Q^2
  \; .
\end{equation}
Inserting into (\ref{eq:dsigma}), we obtain for the differential cross
section:
\begin{eqnarray}
  \frac{\dd^2 \sigma}{\dd y \, \dd Q^2}
  & = &
  \frac{\alpha^2}{Q^4}
  \frac{S}{2\lambda_S}
  L^{\mu\nu}
  H_{\mu\nu}(P,q)
  \nonumber \\
  & = &
  \frac{2\pi\alpha^2}{Q^2xy^2}
  \frac{S}{\lambda_S}
  \Biggl[
  \left( 2(1-y) - 2x^2y^2\frac{M^2}{Q^2}
  \right)
  F_2(x,Q^2)
  \nonumber \\ && \qquad \qquad {}
    + 2 \left( 1 - \frac{2m^2}{Q^2}\right) xy^2
  F_1(x,Q^2)
  \Biggr]
\end{eqnarray}
It is convenient to introduce the ratio $R$,
\begin{equation}
  \label{eq:R}
  R(x,Q^2) \equiv
  \frac{F_L(x,Q^2)}{F_T(x,Q^2)} =
  \left( 1+4x^2\frac{M^2}{Q^2} \right)
  \frac{F_2(x,Q^2)}{2xF_1(x,Q^2)} \, - 1
  \; ,
\end{equation}
of the longitudinal and transverse structure functions,
\begin{eqnarray}
  F_L(x,Q^2) & \equiv &
  \left( 1+4x^2\frac{M^2}{Q^2} \right) F_2(x,Q^2) - 2xF_1(x,Q^2)
  \; ,
  \nonumber \\
  F_T(x,Q^2) & \equiv & 2xF_1(x,Q^2) \; .
\end{eqnarray}
Neglecting the small mass of the electron, and changing the kinematic
variables from $(y,Q^2)$ to $(x,y)$, we finally obtain for the Born cross
section:
\begin{eqnarray}
\label{eq:sig-dis-Born}
  \frac{\dd^2 \sigma}{\dd x \, \dd y}
  & = &
  \frac{2\pi\alpha^2}{Q^2xy}
  \left[
    \left( 1+(1-y)^2 + 2x^2y^2\frac{M^2}{Q^2} \right) F_2(x,Q^2)
    - y^2 F_L(x,Q^2)
  \right]
  \nonumber \\
  & = &
  \frac{2\pi\alpha^2}{Q^2xy}
  \Biggl[ 2(1-y) - 2x^2y^2\frac{M^2}{Q^2}
  \nonumber \\ && \qquad \quad {}
    + \left( 1+4x^2\frac{M^2}{Q^2}\right) \frac{y^2}{1+R(x,Q^2)}
  \Biggr]
  F_2(x,Q^2)
  \; .
\end{eqnarray}
The r.h.s.\ of the cross section (\ref{eq:sig-dis-Born}) depends on the two
proton structure functions to be measured, $F_2(x,Q^2)$ and $F_L(x,Q^2)$.
For fixed center of mass energy, only two of the three invariants $x$, $y$
and $Q^2$ can be varied independently.  The separate determination of the
structure functions at fixed $x$ and $Q^2$ requires the variation of $y$
and therefore the variation of the center of mass energy.  For the HERA
collider, this means running with different beam energies
\cite{Bauerdick:1996ey}.


\section{The Born cross section at high energies}

The HERA $ep$ collinear was designed to reach center of mass energies at
the same order of magnitude as the masses of the electroweak gauge bosons,
$\sqrt{S} \simeq 3\cdot 10^2 \GeV \gtrsim \MZ,\MW$, so that both neutral
and charged current reactions can be observed in the same experiments.  For
the neutral current process, $ep \to eX$, this means that we have to take
into account not only photon but also Z-exchange.

The generalization of the cross section (\ref{eq:sig-dis-Born}) that takes
into account both $\gamma$- and Z-exchange and their interference can be
written as (see e.g., \cite{Spiesberger:1992vu}):
\begin{equation}
\label{eq:sig-nc-born}
  \frac{\dd^2\sigma^\mp_\mathrm{NC}}{\dd x \,\dd y}
  =
  \frac{2\pi\alpha^2}{Q^2 xy}
  \left\{
        Y_+ \mathcal{F}_2(x,Q^2)
  \pm Y_- x \mathcal{F}_3(x,Q^2)
      - y^2 \mathcal{F}_L(x,Q^2)
  \right\}
  \; ,
\end{equation}
with
\begin{equation}
  Y_{\pm}(y) = 1 \pm (1-y)^2
  \; .
\end{equation}
In (\ref{eq:sig-nc-born}) we have used the ultrarelativistic limit, $S \gg
m^2,M^2$, which is applicable at HERA energies.  Upper and lower signs
correspond to electron and positron scattering, respectively.

The structure functions $\mathcal{F}_2$ and $\mathcal{F}_L$ are
generalizations of the structure functions $F_2$ and $F_L$ of the previous
section where only photon exchange was taken into account.  The additional
structure function $\mathcal{F}_3$ is due to the parity violating
contributions from Z-exchange to the scattering amplitude.

At HERA energies also the charged current reaction $e p \to \nu_e X$ has an
appreciable cross section, which reads:
\begin{eqnarray}
\label{eq:sig-cc-born}
  \frac{\dd^2\sigma^\mp_\mathrm{CC}}{\dd x \,\dd y}
  & = &
  \frac{G_\mu^2 S}{4\pi}
  \left( \frac{\MW^2}{\MW^2+Q^2} \right)^2
 \\
  & \times &
  \left\{
      Y_+   \mathcal{W}_2^\mp(x,Q^2)
  \pm Y_- x \mathcal{W}_3^\mp(x,Q^2)
    - y^2   \mathcal{W}_L^\mp(x,Q^2)
  \right\}
  \; , \nonumber
\end{eqnarray}
with $G_\mu$ being the muon decay constant, and the charged current
structure functions denoted by $\mathcal{W}_2$, $\mathcal{W}_3$ and
$\mathcal{W}_L$.


\section{Structure functions}

In the naive quark parton model, the deep inelastic scattering cross
section is calculated as an incoherent superposition of the electron
scattering on free partons.  The electroproduction structure functions are
expressed in terms of probability distributions of massless quarks and
antiquarks in the proton,
\begin{equation}
\label{eq:F2-PM}
  F_2^\mathrm{QPM}(x) = 2xF_1^\mathrm{QPM}(x) =
  x \sum_l |Q_l|^2 \, \left[ q_{0l}(x)+\bar{q}_{0l}(x) \right]
  \; .
\end{equation}
Here the sum runs over all quark flavors, $Q_l$ is the charge of the quark
$q_l$ in units of the elementary charge, and $q_{0l}(x)$ and
$\bar{q}_{0l}(x)$ denote the `bare' quark and antiquark distributions.
Bjorken scaling is implemented by assuming $Q^2$-independence of the bare
distributions.  As the Callan-Gross relation, $F_2=2xF_1$, holds in the
naive quark parton model, the longitudinal structure function vanishes,
$F_L^\mathrm{QPM}(x)=0$.


\subsection{Structure functions from perturbative QCD}

The calculation of radiative corrections to scattering processes involving
hadrons in QCD requires the renormalization of the parton densities.  This
results in a logarithmic violation of Bjorken scaling%
\footnote{For an introduction to QCD and hadron structure we refer the
  reader to the many excellent reviews and textbooks, e.g.,
  \cite{Altarelli:1982ax,Buras:1980yt,Dokshitzer:hw,Reya:1979zk} and
  \cite{Ellis:QCD,Muta:QCD,Roberts:1990}.}
and parton densities $q_l(x,Q^2)$ that will weakly depend on the scale
$Q^2$.

In the leading order of QCD, the structure functions $F_{1,2}(x,Q^2)$ are
given in terms of the renormalized parton distributions by
\begin{eqnarray}
\label{eq:F2-lo}
  F_2^\mathrm{LO}(x,Q^2) & = &
  x \sum_l |Q_l|^2 \, \left[ q_l(x,Q^2)+\bar{q}_l(x,Q^2) \right]
  \; , \\
  F_1^\mathrm{LO}(x,Q^2) & = &
  \frac{1}{2} \sum_l |Q_l|^2 \, \left[ q_l(x,Q^2)+\bar{q}_l(x,Q^2) \right]
  \; ,
\end{eqnarray}
where the $q_l(x,Q^2)$ and $\bar{q}_l(x,Q^2)$ now denote the
$Q^2$-dependent quark and antiquark distributions.  At this order there is
no contribution from the gluon distribution $g(x,Q^2)$, and the
longitudinal structure function still vanishes, $F_L^\mathrm{LO}(x,Q^2)=0$.

Beyond the leading order, these simple relations between the quark and
gluon distributions and the physical structure functions are modified.
From the operator product expansion it follows that they may be written as
\cite{Herrod:1980rm}:
\begin{equation}
  F(x,Q^2) =
  \int\limits_x^1 \frac{\dd y}{y}
  \left\{
    \sum_l
    C_l\left( \frac{x}{y}, \alpha_S \right) q_l(y,Q^2) +
    C_g\left( \frac{x}{y}, \alpha_S \right) g(y,Q^2)
  \right\} \; ,
\end{equation}
where the sum runs over all quarks and antiquarks.  The coefficient
functions $C_l$ and $C_g$, the quark and gluon distributions $q_l(x,Q^2)$,
$g(x,Q^2)$, and the QCD coupling $\alpha_S=\alpha_S(Q^2)$ depend on the
chosen renormalization scheme,
and they may also become gauge dependent at higher orders.  However, it is
still useful to write structure functions at next-to-leading order as
\begin{equation}
  F^\mathrm{NLO}(x,Q^2) =
  F^\mathrm{LO}(x,Q^2) + \Delta F(x,Q^2)
  \; ,
\end{equation}
with $\Delta F(x,Q^2)$ formally being $\bigO{\alpha_S}$.

Common renormalization schemes used in deep inelastic scattering are the
\overbar{MS} scheme and the so-called DIS scheme.  In the DIS scheme the
parton densities are redefined such that relation (\ref{eq:F2-lo}) for
$F_2$ also holds beyond the leading order
\cite{Altarelli:1978id,Kubar-Andre:1979uy,Altarelli:1979ub}.  The QCD
corrections to $F_2$ are completely absorbed into the redefinition of the
parton distributions, so that
\begin{eqnarray}
  F_2^\mathrm{DIS}(x,Q^2) & = &
  x \sum_l |Q_l|^2 \,
  \left[ q_l(x,Q^2)+\bar{q}_l(x,Q^2) \right]_\mathrm{DIS}
  \; ,
  \nonumber \\
  \Delta F_2^\mathrm{DIS}(x,Q^2)
  & \equiv & 0 \; .
\end{eqnarray}
The renormalized quark and gluon densities satisfy the Altarelli-Parisi
evolution equations \cite{Altarelli:1977zs}:%
\footnote{Nowadays it has become customary to denote these
  integro-differential equations as
  Dokshitzer\cite{Dokshitzer:sg} -
  Gribov-Lipatov\cite{Gribov:rt,Lipatov:qm} -
  Altarelli-Parisi, or DGLAP equations.}
\begin{eqnarray}
\label{eq:DGLAP}
  \frac{\dd q_l(x,Q^2)}{\dd \ln Q^2}
  & = &
  \frac{\alpha_S(Q^2)}{2\pi}
  \int\limits_x^1 \frac{\dd y}{y} \,
  \Biggl[
    P_{qq} \left( \frac{x}{y} \right) q_l(y,Q^2) +
    P_{qG} \left( \frac{x}{y} \right) g(y,Q^2)
  \Biggr]
  \; , \nonumber \\
  \frac{\dd g(x,Q^2)}{\dd \ln Q^2}
  & = &
  \frac{\alpha_S(Q^2)}{2\pi}
  \int\limits_x^1 \frac{\dd y}{y} \,
  \Biggl[
    \sum_l P_{Gq} \left( \frac{x}{y} \right) q_l(y,Q^2)
  \nonumber \\ && \qquad \qquad \qquad {}
    + P_{GG} \left( \frac{x}{y} \right) g(y,Q^2)
  \Biggr]
  \; ,
\end{eqnarray}
with the splitting functions $P$ being given e.g., in
\cite{Altarelli:1982ax}.

The NLO correction to the structure function $F_1$ in the DIS scheme reads:
\begin{eqnarray}
\label{eq:F1-DIS-nlo}
  \Delta F_1^\mathrm{DIS}(x,Q^2) & = &
  \frac{1}{2}
  \int\limits_x^1 \frac{\dd y}{y}
  \bigggl\{
    \sum_l |Q_l|^2
     \alpha_S(Q^2)
       \Delta f_q\left(\frac{x}{y}\right)
    [q_l(y,Q^2)+\bar{q}_l(y,Q^2)]
  \nonumber
  \\
  && \quad {}
  + 2 \left(\sum_l |Q_l|^2 \right)
    \alpha_S(Q^2)
    \Delta f_g\left(\frac{x}{y}\right)
    g(y,Q^2)
  \bigggr\}
  \; ,
\end{eqnarray}
where the sum runs over all active quark flavors.

At next-to-leading order, the correction terms in (\ref{eq:F1-DIS-nlo}) are
gauge independent, ultraviolet- and infrared-finite; they read
\cite{Altarelli:1978id}:
\begin{eqnarray}
  \alpha_S \Delta f_q(z) & = &
  \alpha_S \left[ f_{q,1}\left(z\right)- f_{q,2}\left(z\right) \right]
  = - \frac{\alpha_S}{2\pi} \, \frac{4}{3} \, 2z
  \; , \nonumber \\
  \alpha_S \Delta f_g(z) & = &
  \alpha_S \left[ f_{g,1}\left(z\right)- f_{g,2}\left(z\right) \right]
  = - \frac{\alpha_S}{2\pi} \, \frac{1}{2} \, 4z(1-z)
  \; .
\end{eqnarray}
The QCD corrections to the structure functions lead to a violation of the
Callan-Gross relation and to a QCD prediction of the leading twist term of
the longitudinal structure at next-to-leading order,
\begin{eqnarray}
\label{eq:FL-nlo}
  F_L(x,Q^2) & = &
  \frac{\alpha_S(Q^2)}{2\pi}
  \int\limits_x^1 \frac{\dd y}{y}
  \left(\frac{x}{y}\right)^2
  \bigggl\{
    \frac{8}{3} F_2(y,Q^2)
  \\
  && \qquad \qquad {}
  + 4 \left(\sum_l |Q_l|^2 \right)
    \left(1-\frac{x}{y}\right)
    y g(y,Q^2)
  \bigggr\}
  \; ,
  \nonumber
\end{eqnarray}
showing that the ratio $R = F_L/(F_2-F_L) \sim \bigO{\alpha_S}$ vanishes
for $Q^2 \to \infty$ because of asymptotic freedom.  Expression
(\ref{eq:FL-nlo}) is also interesting as it predicts a strong variation of
$F_L$ at low $x$ where the gluon distribution is expected to be strongly
rising.  For a discussion of the gluon distribution and further details
regarding the DIS scheme we refer the reader to \cite{Altarelli:1979ub}.

The calculation of the QCD corrections in the \overbar{MS} scheme is
extensively covered in the literature.  The analogous expressions for the
electroproduction structure functions in terms of renormalized parton
densities will not be reproduced here, they can be found e.g., in
\cite{Herrod:1980rm}.

Since the Altarelli-Parisi equations predict only the evolution of the
parton distributions in $Q^2$, the initial conditions have to be fixed
otherwise.  Using certain assumptions about the shape of the parton
distribution at some low initial scale $Q_0^2$, which is typically of the
order $(2\ldots5)\GeV^2$, several groups provide parameterizations of the
evolved distributions that are selected by comparison with experimental
data.  A large collection of parton distributions is available in the
program library PDFLIB \cite{Plothow-Besch:1992qj}.


\subsection{Structure functions in the region of low $Q^2$}
\label{sec:low-Q2}

Exact Bjorken scaling is a built-in property of the naive parton model.
Because of asymptotic freedom it also holds approximately true within QCD
in the region of large momentum transfers, where it receives perturbatively
calculable corrections.  However, one should expect that a naive
extrapolation of perturbative results valid for $Q^2 \gg 1 \GeV^2$ down to
the photoproduction region $Q^2 \approx 0$ will fail sooner or later for
several reasons.

Physically, the spatial resolution power of the virtual photon on the
proton is given by $\sim \hbar/\sqrt{Q^2}$, and therefore a photon of
virtuality $Q^2 \lesssim (\hbar/r_p)^2 \approx (0.2\GeV)^2$, with $r_p \sim
1\,\mathrm{fm}$ being a typical hadron radius, is not able to resolve the
individual partons.  Therefore, the probabilistic, partonic description of
the proton structure must break down at that scale.

On the technical side, in a systematic analysis of the structure functions
in the Bjorken limit using the operator product expansion (see e.g.,
\cite{Roberts:1990}), one obtains an asymptotic expansion of the structure
functions in inverse powers of $Q^2$, e.g.,
\begin{equation}
\label{eq:F2-expansion}
  F_2(x,Q^2) = \sum_{n=0}^\infty \frac{C_n(x,Q^2)}{(Q^2)^n}
  \; .
\end{equation}
The coefficient functions $C_n(x,Q^2)$ depend only weakly, i.e.,
logarithmically on $Q^2$.  The terms with $n=0$ and $n \geq 1$ are usually
referred to as leading and higher twists, respectively.  For approximate
Bjorken scaling, the contribution of the higher twists in
(\ref{eq:F2-expansion}) must be negligible, which is the case only for
$Q^2$ of the order of or above a few$\GeV^2$.  Furthermore, a perturbative
evaluation of the coefficient functions becomes doubtful at small momentum
transfers $Q^2 \lesssim 1\GeV^2$ where the running QCD coupling
$\alpha_S(Q^2)$ approaches unity.

And finally, as we shall see below, electromagnetic gauge invariance
actually requires that the structure functions $F_2$ and $F_1$ vanish in
the limit $Q^2 \to 0$.  This clearly forbids Bjorken scaling at very low
$Q^2$ and calls for a formal resummation of the series
(\ref{eq:F2-expansion}) in that region.

\subsubsection{Constraints for $\mathbf{Q^2 \to 0}$}

To investigate the general behavior and to obtain constraints on the
structure functions for small $Q^2$, let us rewrite the hadronic tensor
(\ref{eq:hadron-tensor-H}) in order to exhibit potential kinematic
singularities as follows \cite{Badelek:1996fe}:
\begin{eqnarray}
  H_{\mu\nu}(P,q) & = &
  4\pi \Biggl[
  - g_{\mu\nu} F_1 + \frac{P_\mu P_\nu}{P \cdot q} F_2
  + \frac{q_\mu q_\nu}{q^2} \left( F_1 + \frac{P \cdot q}{q^2} F_2 \right)
  \nonumber \\ && \quad \;\; {}
  - \frac{P_\mu q_\nu + P_\nu q_\mu}{q^2} F_2
  \Biggr]
  \; .
\end{eqnarray}
As the total cross section for real photons on a proton is finite, these
singularities can only be artifacts of the tensor decomposition of
$H_{\mu\nu}$ into a sum of manifestly gauge-invariant terms.  The limit
$q^2 \to 0$ of this expression clearly must exist.  This allows to impose
the conditions
\begin{eqnarray}
  F_2 & = & \bigO{Q^2}
  \; , \\
  F_1 + \frac{P \cdot q}{q^2} F_2 & = & \bigO{Q^2}
  \; ,
\end{eqnarray}
on the structure functions in the limit $Q^2 \to 0$.

One can go further and relate the structure functions $F_2$ and $F_L$ to
the total photoproduction cross section for transverse and longitudinal
photons,
\begin{eqnarray}
\label{eq:F2-from-sigma}
  F_2 & = & \frac{Q^2}{4\pi^2 \alpha} (\sigma_T + \sigma_L)
  \; , \nonumber \\
  F_L & = & \frac{Q^2}{4\pi^2 \alpha} \sigma_L
  \; .
\end{eqnarray}
The cross sections on the r.h.s.\ are functions
$\sigma_{T,L}=\sigma_{T,L}(Q^2,W^2)$, where $W^2=M^2+2 P\cdot q - Q^2$ is
the invariant mass of the hadronic final state.  The Bjorken variable $x$
appearing in the structure functions on the l.h.s.\ reads:
\begin{equation}
  x = \frac{Q^2}{W^2 - M^2 + Q^2} \; .
\end{equation}
Invoking conservation of the electromagnetic current, which guarantees the
decoupling of the longitudinal degrees of freedom of the photon, the cross
section $\sigma_L \to 0$ for $Q^2 \to 0$ and at fixed $P \cdot q$.  We
therefore expect \cite{Badelek:1996fe,Badelek:1997ap}
\begin{equation}
  F_L = \bigO{Q^4} \; ,
\end{equation}
so that the ratio $R$ should also vanish in this limit,
\begin{equation}
  R(x,Q^2) = \frac{\sigma_L}{\sigma_T}
  = \bigO{Q^2} \; .
\end{equation}
Obviously it is a crucial requirement for reasonable phenomenological
parameterizations of the structure functions in the region of low $Q^2$
that they incorporate the above constraints.

\subsubsection{Models for low $\mathbf{Q^2}$}

There exist several models for the description of high-energy photon-hadron
interactions that automatically provide the proper behavior for $Q^2 \to
0$.  A complete overview is far beyond the scope of this report, so we will
only name a few popular ones, and suggest the interested reader to consult
the reviews \cite{Badelek:1992gs,Cooper-Sarkar:1997jk} for more detailed
descriptions of several phenomenological approaches and dynamical models of
the $F_2$.

In the Vector Meson Dominance (VMD) model (see e.g., \cite{Bauer:1978iq}
and references cited therein), the interaction of a photon with a hadron is
described by assuming that the photon can fluctuate into a vector meson
which then interacts with the hadron.

A qualitative description of the low $x$ behavior of cross sections is
motivated by the parameterization of the high energy behavior of total
cross sections in hadron-hadron interactions in Regge theory
\cite{Collins:1977}.  In this approach, the amplitude for hadronic elastic
scattering processes at center of mass energy $\sqrt{s} = W$ and momentum
transfer $t$ is written as a sum of terms of the form $T(s,t) \sim
\beta_i(t) s^{\alpha_i(t)}$, with $\alpha(t)$ being a so-called Regge
trajectory.  Appealing to the optical theorem and the similarity of
hadron-hadron and photon-hadron total cross sections, the high energy
behavior of the total photoproduction cross section is written as
\begin{equation}
\label{eq:sigma-regge}
  \sigma_\mathrm{tot} = \sum_i \beta_i \, (W^2)^{\alpha_i-1}
  \; ,
\end{equation}
where the $\alpha_i=\alpha_i(0)$ and $\beta_i=\beta_i(0)$ denote the
intercept and the coupling of the corresponding Regge trajectory,
respectively.  Phenomenologically, the energy dependence of the total
hadronic and photoproduction cross sections are fairly well described by
taking into account two terms in (\ref{eq:sigma-regge}): the (soft) pomeron
with intercept $\alpha_\mathrm{I\!P}\simeq 1.08$, which is identified with
an exchange of vacuum quantum numbers, and the Reggeon with intercept
$\alpha_\mathrm{I\!R}\sim 0.5$ \cite{Donnachie:1992ny}, corresponding to an
exchange which is odd under charge conjugation.

Using the relation between the total photoproduction cross section and the
structure functions (\ref{eq:F2-from-sigma}), the Regge parameterization
(\ref{eq:sigma-regge}) then implies the following ansatz for a
parameterization of the structure function $F_2(x,Q^2)$,
\begin{equation}
\label{eq:F2-Regge}
  F_2(x,Q^2) = \sum_i \tilde\beta_i(Q^2) \, (W^2)^{\alpha_i-1}
  \; ,
\end{equation}
which is expected to be valid for high energies and for $W^2 \gg Q^2$.  In
this case, $x=Q^2/(W^2-M^2+Q^2) \simeq Q^2/W^2 \ll 1$.  Therefore, the
parameterization (\ref{eq:F2-Regge}) implies a powerlike small $x$
behavior of $F_2(x,Q^2)$,
\begin{equation}
  F_2(x,Q^2) = \sum_i \tilde\beta_i(Q^2) \, (Q^2)^{\alpha_i-1} \, x^{1-\alpha_i}
  \; .
\end{equation}
Furthermore, from (\ref{eq:F2-Regge}) and (\ref{eq:F2-from-sigma}) we find
that the functions $\tilde\beta_i(Q^2)$ have to satisfy the conditions
$\tilde\beta_i(Q^2)=\bigO{Q^2}$ for $Q^2\to 0$.

The H1 collaboration has extracted the derivative \cite{Adloff:2001rw}
\begin{equation}
  \lambda(x,Q^2) \equiv
  - \left. \frac{\df \ln F_2(x,Q^2)}{\df \ln x} \right|_{Q^2}
  \; ,
\end{equation}
and found $\lambda(x,Q^2)$ to be independent of $x$ for $x < 0.01$ within
the experimental accuracy.  This implies that the leading $x$-dependence of
$F_2$ at low $x$ is consistent with the power-law behavior $F_2 \propto
x^{-\lambda}$ for fixed $Q^2$.

However, experimental data from HERA on $J/\psi$ photoproduction and the
charm structure function of the proton show a steeper rise at low $x$,
providing evidence for a deviation from the simple Pomeron behavior
described above.  A good description of all available data requires to take
into account an additional contribution from a ``hard'' pomeron with
intercept $\approx 1.4$ \cite{Donnachie:1998gm,Donnachie:2001xx}.

A power-like behavior of structure functions for $x \to 0$ compatible with
the hard pomeron can be obtained from perturbative QCD.  Performing an
all-order resummation of large $\alpha_S \ln (1/x)$ terms leads to the
Balitsky-Fadin-Kuraev-Lipatov (BFKL) \cite{BFKL} evolution equation,%
\footnote{See e.g., ref.~\cite{Schmidt:2001yq} for a recent review on BFKL,
  or the textbook by Forshaw and Ross \cite{Forshaw:1997} for a modern
  introduction to QCD and Regge theory.}
whose solution in the leading logarithmic approximation is interpreted as a
pomeron with intercept
\begin{equation}
  \alpha_\mathrm{BFKL} = 1 + \frac{12 \alpha_S}{\pi} \ln 2 \; .
\end{equation}
For this reason the hard pomeron is identified with the perturbative BFKL
pomeron, while the soft pomeron that dominates in the total cross section
is associated with non-perturbative effects.

Recently the color dipole picture has attracted a lot attention.  This
approach considers the fluctuation of the virtual photon into a
quark-antiquark system as an effective color dipole, which then interacts
with the proton through pomeron exchange.  The proper low $Q^2$ behavior is
implemented by suitably parameterized wave functions for transversely and
longitudinally polarized photons.  Some model variants also consider the
proton as a quark-diquark system, i.e., as a color dipole.  For an overview
see e.g., \cite{Forshaw:1999uf,Donnachie:2001wt} and references cited
therein.

Parameterizations of the structure function $F_2$ for phenomenological
applications based on the Regge behavior described above and fit to the
HERA data have been provided by Abramowicz-Levin-Levy-Maor (ALLM)
\cite{Abramowicz:1991xz,ALLM97}, Donnachie and Landshoff
\cite{Donnachie:1998gm,Donnachie:2001xx}, Adel et al.\ \cite{Adel:1996pb},
Capella et al.\ \cite{CKMT}, and Desgrolard et al.\
\cite{Desgrolard:1998kf}.


\subsection{Structure functions at high $Q^2$}

For neutral current scattering processes at high $Q^2$ one has to take into
account both $\gamma$- and Z-exchange and their interference.  The
generalized structure functions $\mathcal{F}_i(x,Q^2)$ describing the
interaction of leptons with charge $Q_l$ and lepton beam polarization $\xi$
may be decomposed (using the notation of \cite{Arbuzov:1996id}) as:
\begin{eqnarray}
\label{eq:F-gen}
  \mathcal{F}_{1,2}(x,Q^2)
  & = &
  F_{1,2}(x,Q^2)
  + 2 |Q_e| (v_e+\lambda a_e) \kappa(Q^2) G_{1,2}(x,Q^2)
  \nonumber \\
  && {}
  + 4 \left(v_e^2 + a_e^2 + 2\lambda v_e a_e \right) \kappa^2(Q^2)
  H_{1,2}(x,Q^2) \; ,
  \nonumber \\
  x \mathcal{F}_3(x,Q^2)
  & = &
  - 2 \sign(Q_l)
  \biggl\{
  |Q_e| (a_e+\lambda v_e) \kappa(Q^2) x G_3(x,Q^2)
  \nonumber \\
  && {} +
  \left[ 2v_e a_e + \lambda(v_e^2+a_e^2) \right] \kappa^2(Q^2) xH_3(x,Q^2)
  \biggr\} \; ,
\end{eqnarray}
with $Q_e=-1$ and
\begin{eqnarray}
  \lambda & = & \xi \sign(Q_l) \; ,
  \nonumber \\
  v_e & = & 1 - 4 \sin^2 \theta_W^\mathrm{eff} \; ,
  \nonumber \\
  a_e & = & 1 \; ,
\end{eqnarray}
and
\begin{equation}
  \kappa(Q^2) =
  \frac{G_\mu \MZ^2}{8\sqrt{2}\pi \alpha(Q^2)} \frac{Q^2}{Q^2+\MZ^2}
  \; .
\end{equation}
The quantity $\sin^2 \theta_W^\mathrm{eff}$ refers to an effective mixing
angle for the renormalized fermion-Z boson couplings.

The leading order expressions of the structure functions read:
\begin{eqnarray}
  F_2(x,Q^2) & = &
  x \sum_q |Q_q|^2 \left[ q(x,Q^2)+\bar{q}(x,Q^2) \right]
  \; , \nonumber \\
  G_2(x,Q^2) & = &
  x \sum_q |Q_q| v_q \left[ q(x,Q^2)+\bar{q}(x,Q^2) \right]
  \; , \nonumber \\
  H_2(x,Q^2) & = &
  x \sum_q \frac{v_q^2 + a_q^2}{4} \left[ q(x,Q^2)+\bar{q}(x,Q^2) \right]
  \; , \\
  x G_3(x,Q^2) & = &
  x \sum_q |Q_q| a_q \left[ q(x,Q^2)-\bar{q}(x,Q^2) \right]
  \; , \nonumber \\
  x H_3(x,Q^2) & = &
  x \sum_q \frac{v_q a_q}{2} \left[ q(x,Q^2)-\bar{q}(x,Q^2) \right]
  \; , \nonumber
\end{eqnarray}
where $q(x,Q^2)$ and $\bar{q}(x,Q^2)$ denote the quark and antiquark
distributions.  The neutral current couplings of the quarks are:
\begin{eqnarray}
  v_q & = & 1 - 4 |Q_q| \sin^2 \theta_W^\mathrm{eff} \; ,
  \nonumber \\
  a_q & = & 1 \; .
\end{eqnarray}

The charged current generalized structure functions,
\begin{eqnarray}
  \mathcal{W}_2(x,Q^2) & = &
  \frac{1+\lambda}{2} \, W_2^{Q_l}(x,Q^2)
  \; ,
  \nonumber \\
  x\mathcal{W}_3(x,Q^2) & = &
  - \sign(Q_l) \,\frac{1+\lambda}{2} \, xW_3^{Q_l}(x,Q^2)
  \; ,
\end{eqnarray}
are obtained from the parton distributions by:
\begin{eqnarray}
\label{eq:W2-W3}
  W_2^+(x,Q^2) & = &
  2x \sum_i \left[ d_i(x,Q^2) + \bar{u}_i(x,Q^2) \right]
  \; ,
  \nonumber \\
  W_2^-(x,Q^2) & = &
  2x \sum_i \left[ u_i(x,Q^2) + \bar{d}_i(x,Q^2) \right]
  \; ,
  \nonumber \\
  xW_3^+(x,Q^2) & = &
  2x \sum_i \left[ d_i(x,Q^2) - \bar{u}_i(x,Q^2) \right]
  \; ,
  \nonumber \\
  xW_3^-(x,Q^2) & = &
  2x \sum_i \left[ u_i(x,Q^2) - \bar{d}_i(x,Q^2) \right]
  \; ,
\end{eqnarray}
with $u_i$ and $d_i$ denoting the densities of the up-type $(u,c,t)$ and
down-type $(d,s,b)$ quarks.

At leading order, the longitudinal structure functions vanish.  We will not
discuss here the QCD corrections to the generalized structure functions
above but refer the reader to the review \cite{vanNeerven:1996ph}.


\chapter{QED Radiative Corrections and Radiative Processes}

The cross sections given in the previous chapter correspond to the Born
approximation in perturbation theory and have to be improved to include
higher order contributions, i.e., radiative corrections.  At HERA energies,
the radiative corrections in general need to be treated within the Standard
Model of the electroweak interaction.

The electroweak radiative corrections to deep inelastic scattering have
already been extensively discussed in the literature (for a review see
\cite{Spiesberger:1992vu,Beyer:pw} and references cited therein).  It is
well known that the numerically most important parts of the corrections
arise from virtual photon corrections, photon emission and light fermion
loops.  Furthermore, in the kinematic region of low to moderate momentum
transfers, $Q^2 \ll \MW^2, \MZ^2$, only the QED part of the electroweak
corrections is relevant.  In this work we will be concerned only with QED
radiative corrections to scattering processes.

This chapter starts by briefly reviewing the most important features of the
$\bigO{\alpha}$ electroweak radiative corrections, but refer the reader who
is interested in details to the literature.  We will then focus on
exclusive radiative processes at this order where the emitted photon is
measured.  As we shall argue, these radiative processes are of particular
interest by themselves.  Not only do they constitute an important part of
the complete radiative corrections, but their measurement gives access to
physical information that may be difficult to obtain otherwise.  The
treatment of QED corrections to these radiative processes, which are a
subset of the higher order corrections to deep inelastic scattering, will
follow in the next chapters.


\section{Classification of radiative corrections}

The basic process in deep inelastic scattering is lepton-quark scattering.
It is well known that the $\bigO{\alpha}$ radiative corrections to the
neutral current cross section can be classified in the following way
\cite{Spiesberger:1992vu}:
\begin{enumerate}[(1)]
\item the \emph{leptonic corrections} are described by Feynman diagrams
  containing an additional photon attached to the lepton line, i.e., the
  virtual corrections from the purely photonic correction to the lepton
  vertex, the photon loop contribution to the self-energies of the external
  leptons, and the photon emission from the lepton line,
\item the \emph{quarkonic corrections} are represented by diagrams with an
  additional photon connected to the quark line, analogous to (1),
\item the \emph{lepton-quark interference} part, consisting of box diagrams
  with at least one photon, and the interference part of the bremsstrahlung
  off leptons and off quarks,
\item and the \emph{purely weak corrections}, that are given by all other
  diagrams that do not contain an additional photon.
\end{enumerate}
Each of the above classes can easily be seen to form a gauge-invariant
(w.r.t.\ QED gauge transformations) subset of the full set of corrections.%
\footnote{For non-abelian gauge theories there exists an algorithm for the
  construction of the minimal gauge-invariant subsets of the tree
  amplitudes at a given order \cite{Boos:1999qc}.%
}$^,$%
\footnote{Class (4) is not irreducible.  For example, one may treat the
  fermion-loop contributions to the vector-boson self energies separately.}
The infrared-singularities of individual contributions therefore cancel
within each of the classes (1--3), while each correction in subset (4) is
infrared-finite.

The contribution of each loop correction diagram $i$ to the cross section
can be expressed by a correction factor $\delta_i$, defined via:
\begin{equation}
\label{eq:delta-i}
  \frac{\dd^2 \sigma_{i}}{\dd x \, \dd y}
  = \frac{\dd^2 \sigma_\mathrm{Born}}{\dd x \, \dd y} \cdot
  \delta_{i}(x,y) \; .
\end{equation}
The total correction from virtual corrections can in principle be obtained
by summing over the individual contributions (\ref{eq:delta-i}).  However,
there are some contributions which are known to be universal and large and
which should be treated in a special way.

The vector boson self energies are dominated by fermion loops. Their
contribution is large because the momentum transfers $Q^2$ attained at HERA
are much larger that the masses of the light fermions, leading to large
logarithms $\sim \ln Q^2/m_f^2$.  The fermion loops of the photon self
energy (the vacuum polarization) can be taken into account by use of the
running QED coupling $\alpha(Q^2)$, thereby summing these leading
logarithms to all orders:
\begin{equation}
  \alpha(Q^2) = \frac{\alpha(0)}{1 - \Pi^\gamma_f(Q^2)} \; .
\end{equation}
Here $\Pi_f^\gamma(Q^2)=\hat\Sigma_f^\gamma(Q^2)/Q^2$ denotes the fermionic
part of the vacuum polarization derived from the renormalized photon
self-energy $\hat\Sigma^\gamma$, and $\alpha(0) \simeq 1/137.036$.  Since
the hadronic part is notoriously difficult to calculate for small $Q^2$, it
is preferably evaluated instead via a dispersion relation and parameterized
for practical applications, see e.g., Burkhardt and Pietrzyk
\cite{Burkhardt:1995tt}.

Another set of contributions to the corrections containing large logarithms
arises from the infrared-divergent virtual corrections and from photon
bremsstrahlung.  This part in general does depend on details of the
experimental setup, i.e., whether the radiated photons can be resolved in
the detector or not.  We shall address this issue later.

According to the above classification of the Feynman diagrams, one
distinguishes the bremsstrahlung contributions into the classes (1--3).
The radiation from the lepton line, which belongs to class (1), is
completely independent of the modeling of the hadron side.%
\footnote{It nevertheless depends on the \emph{shape} of the proton
  structure functions, i.e., their dependence on $x$ and $Q^2$.}
The approach of radiative corrections that restricts to the \emph{leptonic
corrections} and to the vacuum polarization will be referred to as the
\emph{model independent approach}.  The model independent framework has
first been used in deep inelastic scattering in \cite{Mo:1969cg}.  A very
detailed account with application to a large set of experimental
determinations of kinematic variables can be found in
\cite{Akhundov:1996my}.  In this chapter we shall outline the essentials.

In the quark-parton model, there are also contributions from radiation off
the quark line and from lepton-quark interference.  While we shall comment
on the former later on, the latter is in general rather small.  For details
and references we refer the reader again to \cite{Spiesberger:1992vu}.

In the case of charged-current processes, there is no gauge-invariant
separation of the Feynman diagrams as in the NC case.  A similar
decomposition can only be performed in a given physical gauge.  However,
the attribution of the leading logarithmic terms of the corrections to the
lepton and quark lines is still gauge invariant
\cite{Bohm:1987cg,Bardin:1989vz}.  Thus, while the leading logarithms of
the leptonic corrections are model-independent according to the above
definition, the complete set of radiative corrections to CC scattering is
necessarily model-dependent.

The main target of this chapter will be bremsstrahlung in NC processes,
which we shall treat in the model-independent approach.  We will comment
only briefly on radiation in charged-current processes which has a much
smaller cross section, and close with a discussion of emission off the
hadron.

In calculations of cross sections below we always take into account the
large fermionic contributions to the vector boson self energies in the form
of running couplings.  For the neutral current cross section
(\ref{eq:sig-dis-Born}) or (\ref{eq:sig-nc-born}), this implies replacing
the prefactor $\alpha^2$ by $\alpha^2(Q^2)$.  In the charged current case,
however, we shall continue to use eq.~(\ref{eq:sig-cc-born}), which is
accurate enough for our purposes.


\section{Radiative DIS}
\label{sec:rad-dis}

In the following we shall be interested in the model-independent
calculation of the cross section for the radiative deep inelastic
scattering reaction
\begin{equation}
  e(p) + p(P) \to e(p') + \gamma(k) + X(P_X) \, ,
\end{equation}
taking into account only photon exchange%
\footnote{The cross section for this process taking into account both
  photon and Z-exchange is given in detail in
  \cite{Akhundov:1996my,Bardin:1997zm}.}
and emission of the radiated photon off the electron.
%
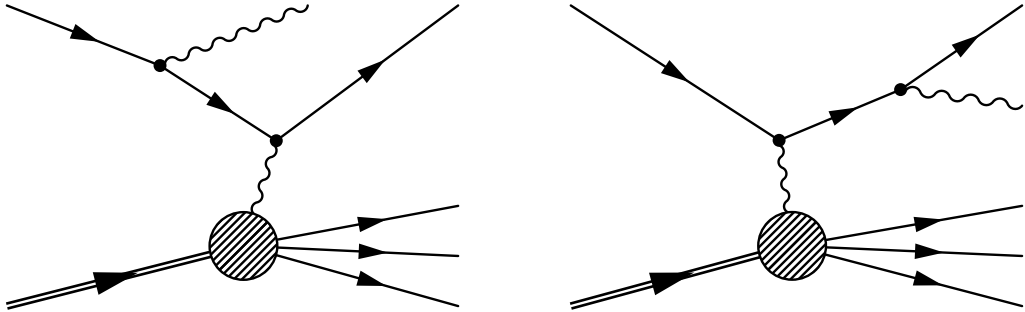
\begin{figure}
  \begin{center}
    \begin{picture}(135,43)
      \put(0,3){%
\begin{fmfgraph}(60,40)
\fmfleft{ip,il}
\fmfright{oq1,oq2,oq3,d1,d2,d3,ol}
\fmftop{a1,a2,og,a3}
\fmf{fermion,tension=1}{il,vg,vl}
\fmf{fermion,tension=1}{vl,ol}
\fmf{photon,tension=0.25}{vg,og}
\fmf{photon,tension=2}{vl,vp}
\fmf{dbl_plain_arrow,tension=3}{ip,vp}
\fmf{fermion}{vp,oq3}
\fmf{fermion}{vp,oq2}
\fmf{fermion}{vp,oq1}
\fmfdot{vl}
\fmfdot{vg}
\fmfblob{0.15w}{vp}
\end{fmfgraph}%
}
      \put(75,3){%
\begin{fmfgraph}(60,40)
\fmfleft{ip,il}
\fmfright{oq1,oq2,oq3,d1,og,d3,ol}
\fmf{fermion,tension=1}{il,vl}
\fmf{fermion,tension=1.5}{vl,vg}
\fmf{fermion,tension=1}{vg,ol}
\fmf{photon,tension=0.5}{vg,og}
\fmf{photon,tension=2}{vl,vp}
\fmf{dbl_plain_arrow,tension=3}{ip,vp}
\fmf{fermion}{vp,oq3}
\fmf{fermion}{vp,oq2}
\fmf{fermion}{vp,oq1}
\fmfdot{vl}
\fmfdot{vg}
\fmfblob{0.15w}{vp}
\end{fmfgraph}%
}
    \end{picture}
    \caption{Model-independent contributions to radiative deep inelastic
      electron-proton scattering}
    \label{fig:rad-dis}
  \end{center}
\end{figure}
%
This bremsstrahlung contribution is often called the Bethe-Heitler (BH)
process.
The contributing Feynman diagrams are shown in figure~\ref{fig:rad-dis}.
The scattering amplitude for this reaction can be factorized into the Born
amplitude for the process of scattering of a virtual photon on the lepton
and the matrix element of the electromagnetic current between the proton
and the hadronic final state,
\begin{equation}
  \mathcal{M}^\mathrm{rad}
  = M_\mu^{e\gamma^* \to e'\gamma} \,
  \frac{g^{\mu\lambda}}{q^2} \,e \,
  \left\langle P_X \left| J_\lambda^\mathrm{em}(0) \right| P \right\rangle
  \; ,
\end{equation}
where $q=p-p'-k$.  The differential cross section for the scattering of
unpolarized particles can then be expressed as a contraction of the hadron
tensor with a tensor describing the QED Compton process:
\begin{eqnarray}
\label{eq:sigma-rad-0}
  \dd \sigma & = &
  \frac{1}{2\sqrt{\lambda_S}} \cdot
  \frac{1}{2} \sum_\mathrm{spins} \,
  \frac{1}{2} \sum_X \,
  \left| \mathcal{M}^\mathrm{rad} \right|^2 \,
  \widetilde{\dd p'} \, \widetilde{\dd k}
  \nonumber \\
  & = &
  \frac{1}{2\sqrt{\lambda_S}} \,
  H^{\mu\nu}(P,q) \cdot
  \left[
  \frac{1}{2} \sum_\mathrm{spins} \,
  M_{\mu}^{e\gamma^*\to e'\gamma}
 (M_{\nu}^{e\gamma^*\to e'\gamma})^*
  \right]
  \widetilde{\dd p'} \, \widetilde{\dd k}
  \; .
\end{eqnarray}
We shall first discuss the expression in square brackets and turn to the
treatment of phase space later.


\subsection{The QED Compton tensor}

Consider the (sub-)process of Compton scattering of a virtual photon on an
electron,
\begin{equation}
\label{eq:virtual-compton-process}
  e(p_1) + \gamma^*(-q) \to e(p_2) + \gamma(k) \; ,
\end{equation}
and let $M_\mu$ be the matrix element of this process with the index $\mu$
describing the polarization state of the virtual photon.  Adopting the
notation and normalization of \cite{KMF87}, we define the unpolarized
Compton tensor
\begin{equation}
\label{def:compton-tensor-Born}
  K_{\mu\nu}(p_1,p_2,k)
  = \frac{1}{(2e^2)^2} \sum_{\mathrm{spins}}
  M_{\mu}^{e\gamma^*\to e'\gamma}
 (M_{\nu}^{e\gamma^*\to e'\gamma})^* \; .
\end{equation}
With the help of current and momentum conservation, this tensor is
conveniently decomposed as follows:
\begin{equation}
\label{eq:compton-tensor-Born}
  K_{\mu\nu}
  = \tilde{g}_{\mu\nu} B_g
  + \sum_{i,j=1,2} \tilde{p}_{i\mu}\tilde{p}_{j\nu} B_{ij}
  \; ,
\end{equation}
where
\begin{equation}
  \tilde{g}_{\mu\nu} = g_{\mu\nu} - \frac{q_{\mu}q_{\nu}}{q^2}
  \; , \quad
  \tilde{p}_{i\mu} = p_{i\mu} - q_{\mu}\frac{p_i \cdot q}{q^2}
  \; , \quad
  i=1,2 \; .
\end{equation}
We introduce the following invariants for this subprocess,
\begin{equation}
  \hat{s} = 2p_2 \cdot k
  \; , \quad
  \hat{t} = - 2p_1 \cdot k
  \; , \quad
  \hat{u} = (p_1 - p_2)^2
  \; ,
\end{equation}
so that, using $p_1^2=p_2^2=m^2$, $k^2 = 0$, we have
\begin{equation}
  \hat{s}+\hat{t}+\hat{u} = q^2 \; .
\end{equation}
The expressions for the quantities $B_{ij}$ in the Born approximation read:
\begin{eqnarray}
\label{eq:compton-born}
  B_g
  &=&
  \frac{1}{\hat{s}\hat{t}}
  \left[(\hat{s}+\hat{u})^2 + (\hat{t}+\hat{u})^2\right]
  - 2m^2q^2\left(\frac{1}{\hat{s}} + \frac{1}{\hat{t}}\right)^2
  \; , \nonumber \\
  B_{11}
  &=& \frac{4q^2}{\hat{s}\hat{t}} - \frac{8m^2}{\hat{s}^2}
  \; , \quad
  B_{22}
  = \frac{4q^2}{\hat{s}\hat{t}} - \frac{8m^2}{\hat{t}^2}
  \; , \\
  B_{12} & = & B_{21} = - \frac{8m^2}{\hat{s}\hat{t}}
  \; . \nonumber
\end{eqnarray}
In the high-energy limit, when at least two of the variables
$\hat{s},\hat{t},\hat{u},q^2$ are large compared to $m^2$, one can drop the
terms proportional to $m^2/\hat{s}\hat{t}$, as their contribution is
suppressed when integrating over any finite part of phase space for the
real photon.  Therefore in the high-energy limit of (\ref{eq:compton-born})
we can set
\begin{eqnarray}
\label{eq:compton-born-HE}
  B_g
  &\to&
  \frac{1}{\hat{s}\hat{t}}
  \left[(\hat{s}+\hat{u})^2 + (\hat{t}+\hat{u})^2\right]
  - 2m^2q^2\left(\frac{1}{\hat{s}^2} + \frac{1}{\hat{t}^2}\right)
  \; , \nonumber \\
  B_{12} & = & B_{21} \to 0
  \; .
\end{eqnarray}


\subsection{The cross section for single photon emission}
\label{sec:single-rad}

With the expression for the lowest order Compton tensor of the previous
section we can now write the unpolarized cross section
(\ref{eq:sigma-rad-0}) for the radiative process as:
\begin{eqnarray}
\label{eq:sigma-rad}
  \dd \sigma & = &
  \left.
  \frac{(4\pi\alpha)^3}{\sqrt{\lambda_S} \, Q_h^4} \,
  K_{\mu\nu}(p,p',k) \, H^{\mu\nu}(P,q_h) \,
  \widetilde{\dd p'} \, \widetilde{\dd k}
  \right|_{q_h=p-p'-k} \; .
\end{eqnarray}
Here we have introduced the hadronic momentum transfer
\begin{equation}
  Q_h^2 \equiv - q_h^2
  \; , \quad
  q_h \equiv P_X - P = p-p'-k \; ,
\end{equation}
to distinguish it from the leptonic momentum transfer
\begin{equation}
  Q_l^2 \equiv - q_l^2
  \; , \quad
  q_l \equiv p-p' \; .
\end{equation}
Note that hadronic and leptonic momentum transfer are related via
\begin{equation}
  Q_h^2 = Q_l^2 + 2p \cdot k - 2p' \cdot k \; .
\end{equation}
In the following we shall need kinematic invariants which are the
``hadronic'' and ``leptonic'' generalizations of the Bjorken variables,
\begin{eqnarray}
\label{eq:x-y-h-l}
  y_h & = &
  \frac{P \cdot q_h}{P \cdot p} = \frac{P \cdot (P_X-P)}{P \cdot p}
  \; ,
  \quad
  x_h = \frac{Q_h^2}{y_h S} = \frac{Q_h^2}{2 P \cdot q_h} \; ,
  \nonumber \\
  y_l & = &
  \frac{P \cdot q_l}{P \cdot p} = \frac{P \cdot (p-p')}{P \cdot p}
  \; , \quad
  x_l = \frac{Q_l^2}{y_l S} = \frac{Q_l^2}{2 P \cdot q_l}
  \; .
\end{eqnarray}
The contraction of the Compton tensor with the hadron tensor can be
decomposed as:
\begin{equation}
\label{eq:KdotH-Born}
  K_{\mu\nu} H^{\mu\nu}
  = 8\pi \left[ S_1 F_1(x_h,Q_h^2) + S_2 F_2(x_h,Q_h^2) \right] \; .
\end{equation}
Retaining only those terms that survive in the limit $S \gg m^2$, but
keeping all terms of order $M^2$, the coefficient functions $S_{1,2}$ can
be expressed as follows:
%
\begin{eqnarray}
\label{eq:S1S2}
  S_1 & = &
  - \frac{1}{\hat{s}\hat{t}}
  \left[(\hat{s}+\hat{u})^2 + (\hat{t}+\hat{u})^2\right]
  + 2m^2q^2\left(\frac{1}{\hat{s}^2} + \frac{1}{\hat{t}^2}\right)
  = - B_g \; , \quad
  \nonumber \\
  S_2 & = &
  - \frac{S}{q^2}
  \Biggl[
    [x_h S + (\hat{t}+\hat{u})] \frac{B_{11}}{4}
  + (1-y_l) [ x_h(1-y_l)S - (\hat{s}+\hat{u}) ] \frac{B_{22}}{4}
  \Biggr]
  \nonumber \\
  &+& \frac{x_h M^2}{q^2} S_1 \; .
  \\
\noalign{\hbox{where}}
  &&
  \hat{s}=2p'\cdot k
  \; , \quad
  \hat{t}=-2p\cdot k
  \; , \quad
  \hat{u}=q_l^2
  \; , \quad
  q^2=q_h^2
  \; .
  \nonumber
\end{eqnarray}
The functions $S_1$ and $S_2$ agree with those given in
\cite{Bardin:1997zm}.

Let us assume that the kinematics of the scattering process is determined
by a measurement of the scattered lepton and the radiated photon.  We can
express the phase space integral for the scattered electron in terms of
$y_l$ and $Q_l^2$, see (\ref{eq:e-ps-y-Q2}):
\begin{equation}
\label{eq:e-ps-yl-Ql2}
  \widetilde{\dd p'} =
  \frac{1}{(4\pi)^2}
  \, \dd y_l \, \dd Q_l^2 \; .
\end{equation}
The differential cross section (\ref{eq:sigma-rad}) then reads:
\begin{eqnarray}
\label{eq:sigma-rad-gen}
  \dd \sigma & = &
  \frac{32\pi^2\alpha^3}{Q_h^4 S} \,
  \left[ S_1 F_1 + S_2 F_2 \right]
  \dd y_l \, \dd Q_l^2 \, \widetilde{\dd k}
  \nonumber \\
  & = &
  \frac{2\alpha^3}{\pi Q_h^4 S} \,
  \left[ S_1 F_1 + S_2 F_2 \right]
  \dd y_l \, \dd Q_l^2 \, E_\gamma \, \dd E_\gamma \; \dd\Omega_\gamma
  \nonumber \\
  & = &
  \frac{2\alpha^3}{\pi} \,
  \frac{y_l}{Q_h^4} \,
  \left[ S_1 F_1 + S_2 F_2 \right]
  \dd x_l \, \dd y_l \, E_\gamma \, \dd E_\gamma \; \dd\Omega_\gamma
  \; .
\end{eqnarray}
In the last line we have change from the set $(y_l,Q_l^2)$ to the set
$(x_l,y_l)$.  Note that the cross section (\ref{eq:sigma-rad-gen}) may be
generalized to also take into account Z-boson exchange by replacing $[S_1
F_1+S_2 F_2]$ by $\sum_{i=1}^3 S_i \mathcal{F}_i$, with the $\mathcal{F}_i$
being the generalized structure functions, and $S_3$ given e.\,g., in
\cite{Bardin:1997zm}.  We will return to it in
section~\ref{sec:dis-rad-general}.


\subsubsection{Parameterizations of the photon phase space}

In the radiative cross section (\ref{eq:sigma-rad-gen}), we have chosen to
express the phase space of the photon in terms of variables
$E_\gamma,\vartheta,\phi$ that are measured in the HERA frame.  Although
Lorentz invariance is not obvious, this representation is useful for
practical applications, as it allows an easy implementation of experimental
cuts and constraints given by the detector.
We will therefore only sketch the relation to the manifestly Lorentz
covariant approach, which is described in great detail in
\cite{Akhundov:1996my}.%
\footnote{For a classic textbook on particle kinematics and the treatment
  of multi-particle phase space see \cite{Byckling:1973}.}

Let us choose a coordinate system for the HERA frame such that the unit
three-vectors in the direction of the incoming and outgoing lepton and the
radiated photon are as follows:
\begin{eqnarray}
  \vec{e}_1 & = & \vec{e}_z = (0,0,1)
  \; , \nonumber \\
  \vec{e}_2 & = & (\sin\theta,0,\cos\theta)
  \; , \nonumber \\
  \vec{e} & = & (\sin\vartheta\cos\phi,\sin\vartheta\sin\phi,\cos\vartheta)
  \; .
\end{eqnarray}
We then have
\begin{eqnarray}
  c & \equiv &
  \vec{e}_1 \cdot \vec{e}_2 = \cos\theta
  \; , \nonumber \\
  c_1 & \equiv &
  \vec{e} \cdot \vec{e}_1 = \cos\vartheta
  \; , \nonumber \\
  c_2 & \equiv &
  \vec{e} \cdot \vec{e}_2 = c \cos\vartheta + s \sin\vartheta\cos\phi
  \; ,
\end{eqnarray}
with $s=\sin\theta=\sqrt{1-c^2}$.  The integral over the solid angle of the
photon may be alternatively expressed in terms of $c_1$, $c_2$ as
\begin{equation}
  \dd\Omega_\gamma
  \equiv \dd(\cos\vartheta)\,\dd\phi
  = J(c_1,c_2) \, \dd c_1 \, \dd c_2
  \; ,
\end{equation}
with the Jacobian
\begin{equation}
  J(c_1,c_2)
  = \frac{2}{|s s_1 \sin\phi|}
  = \frac{2}{\sqrt{1-c^2-c_1^2-c_2^2+2cc_1c_2}}
  =: \frac{1}{\sqrt{\mathcal{D}}}
  \; ,
\end{equation}
where $s_1=\sqrt{1-c_1^2}$.  The allowed region for $(c_1,c_2)$ is given by
the requirement that $\mathcal{D}\geq 0$.  This representation is obviously
symmetric w.r.t.\ incoming and scattered lepton, i.e., $c_1 \leftrightarrow
c_2$.

\begin{figure}[tb]
  \begin{center}
    \begin{picture}(80,90)
      \put(0,5){\includegraphics[width=80mm]{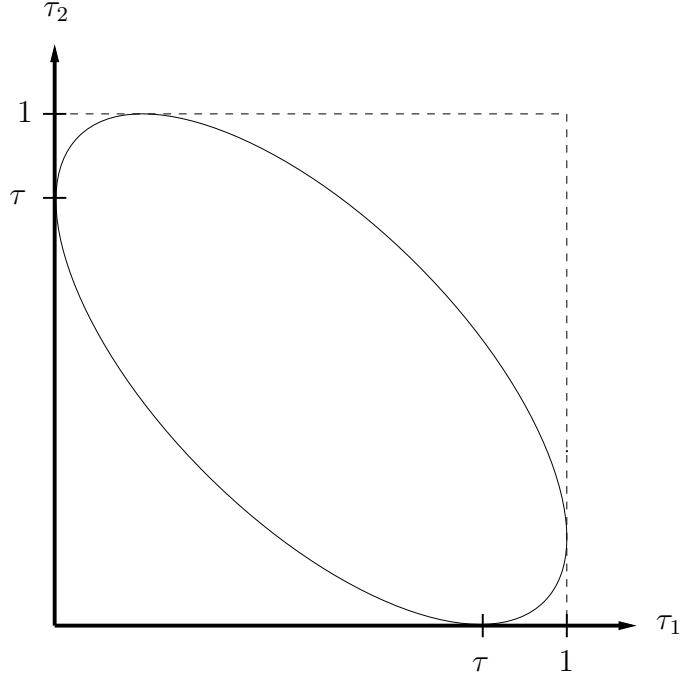}}
      \put(82,6.5){$\tau_1$}
      \put(57.5,1){$\tau$}
      \put(69,1){$1$}
      \put(-4,63){$\tau$}
      \put(-3,74){$1$}
      \put(0.5,88){$\tau_2$}
    \end{picture}
    \caption{Schematic view of the kinematic range for the integration over
      the angles $\tau_1$ and $\tau_2$ for fixed $\tau$.  The allowed range
      corresponding to $\mathcal{D}\geq 0$ is given by the interior of the
      ellipse.}
    \label{fig:ellipse}
  \end{center}
\end{figure}

Furthermore, it is convenient to change variables to
\begin{equation}
  \tau_{1,2} = \frac{1-c_{1,2}}{2}
  \; , \quad
  \tau = \frac{1-c_{1,2}}{2} \; ,
\end{equation}
so that
\begin{equation}
\label{eq:dOmega-t1-t2}
  \dd\Omega_\gamma
  =  \frac{4}{\sqrt{\mathcal{D}}} \, \dd \tau_1 \, \dd \tau_2
  \; .
\end{equation}
Expressing $\mathcal{D}$ in terms of $\tau_{1,2}$, we find:
\begin{equation}
\label{eq:D(tau)}
  \mathcal{D} =
  - 4 \tau \tau_1 \tau_2
  - \left[
     \tau^2 + \tau_1^2 + \tau_2^2
     - 2 \tau \tau_1 - 2\tau \tau_2 - 2\tau_1 \tau_2
    \right] \; .
\end{equation}
The domain $\mathcal{D} \geq 0$ corresponds to the interior of the ellipse
shown in fig.~\ref{fig:ellipse},
which can be parameterized as follows:
\begin{equation}
  0 \leq \tau_1 \leq 1
  \; , \quad
  \tau_2^-(\tau_1) \leq \tau_2 \leq \tau_2^+(\tau_1)
  \; ,
\end{equation}
which is derived from the factorization
\begin{equation}
  \mathcal{D} =
  \left( \tau^+(\tau_1) - \tau_2 \right)
  \left( \tau_2 - \tau^-(\tau_1) \right) \; ,
\end{equation}
with
\begin{equation}
  \tau_2^\pm(\tau_1) =
  \tau (1-\tau_1) + \tau_1(1-\tau)
  \pm 2 \sqrt{\tau (1-\tau) \tau_1 (1-\tau_1)}
  \; .
\end{equation}
Similar expressions are obtained under the exchange $\tau_1 \leftrightarrow
\tau_2$.

We assume now the leptonic determination of the kinematic variables, i.e.,
we keep $x_l,y_l$ fixed.  The upper limit on the energy of the radiated
photon is calculated as follows.  The true invariant mass of the hadronic
final state reads:
\begin{eqnarray}
  W^2
  & = &
  M_X^2 = (P+p-p'-k)^2
  \nonumber \\
  & = & M^2 + (1-x_l)y_l S - 2k \cdot (P+p-p')
  \nonumber \\
  & = & W^2_\mathrm{rec} - 2k \cdot (P+p-p')
  \; ,
\end{eqnarray}
where $W^2_\mathrm{rec}$ is the reconstructed (apparent) hadronic mass if
only the scattered lepton is measured.  The true hadronic mass in inelastic
scattering must be higher than the threshold for pion production,
\begin{equation}
  M_X^2 \geq \bar{M}^2 \equiv (M+m_\pi)^2 \; .
\end{equation}
This leads to
\begin{eqnarray}
\label{eq:Egam-limit}
  2k \cdot (P+p-p')
  & \leq & \left[ M^2 + (1-x_l)y_l S \right] - \bar{M}^2
  \nonumber \\
  & \leq & \left[ M^2 + 2P\cdot(p-p') - Q_l^2 \right] - \bar{M}^2
  \; .
\end{eqnarray}
Expressing this kinematic limit in the HERA frame we obtain, assuming $E_p
\gg M$, $E_e \gg m$:
\begin{equation}
\label{eq:Egam-limit-explicit}
  E_\gamma^\mathrm{max}(\tau_1,\tau_2) =
  \frac{S[1-Y(1-\tau)] - 4E_e^2Y\tau - \left(\bar{M}^2 - M^2 \right)}
       {4\left[E_p(1-\tau_1) + E_e\tau_1 - YE_e\tau_2 \right]}
  \; ,
\end{equation}
with
\begin{eqnarray}
  Y & \equiv & \frac{\Eep}{E_e} = 1 - y_l + x_l y_l \frac{E_p}{E_e}
  \; ,
  \nonumber \\
  \tau & = & \frac{Q_l^2}{4 E_e^2 Y}
  = \frac{x_l y_l E_p}{(1-y_l)E_e + x_l y_l E_p}
  \; .
\end{eqnarray}
Using the above parameterization for the photon angles, we can calculate
the kinematic variables occurring in the Compton tensor
(\ref{eq:compton-born}) and in the coefficient functions $S_{1,2}$ in the
cross section (\ref{eq:sigma-rad-gen}):
\begin{eqnarray}
\label{eq:stu-param}
  \hat{s} & = &
  2 p' \cdot k
  = 2 \Eep E_\gamma (1-\beta_e' c_2)
  \simeq 4 \Eep E_\gamma \left( \tau_2 + \frac{m^2}{4\Eep^2} \right)
  \; ,
  \nonumber \\
  \hat{t} & = &
  - 2 p \cdot k
  = - 2 E_e E_\gamma (1-\beta_e c_1)
  \simeq - 4 E_e E_\gamma \left( \tau_1 + \frac{m^2}{4E_e^2} \right)
  \; ,
  \nonumber \\
  \hat{u} & = & - Q_l^2 \; ,
  \\
  Q_h^2 & = & - q^2
  = - \left( \hat{s} + \hat{t} + \hat{u} \right)
  \simeq Q_l^2 + 4 E_e E_\gamma \tau_1 -  4 \Eep E_\gamma \tau_2
  \; ,
  \nonumber \\
  x_h & = &
  \frac{Q_h^2}{y_l S - 2 P \cdot k}
  \simeq \frac{Q_h^2}{y_l S - 4 E_p E_\gamma (1-\tau_1)}
  \; .
  \nonumber
\end{eqnarray}
Note that we explicitly kept the leading lepton mass terms in $\hat{s}$ and
$\hat{t}$.  It is necessary to retain them when they occur as denominator
factors and we want to integrate over the regions $\tau_1 \to 0$ or $\tau_2
\to 0$, where the emitted photon is collinear to either the incoming or the
outgoing lepton.

Inspecting (\ref{eq:stu-param}) and remembering (\ref{eq:dOmega-t1-t2}) one
easily sees with a little algebra that we can express the integration over
photon phase space in terms of invariants,
\begin{equation}
\label{eq:dk-by-inv}
  \widetilde{\dd k}
  = \mathcal{J}' \, \dd \hat{s} \; \dd |\hat{t}| \; \dd x_h
  \; ,
\end{equation}
with the Jacobian $\mathcal{J}'$ being given in \cite{Akhundov:1996my}.
With the help of relations like $Q_h^2=Q_l^2-\hat{s}-\hat{t}$ one may
switch to other equivalent sets of kinematic invariants as integration
variables.

For a fixed energy of the radiated photon, the cross section
(\ref{eq:sigma-rad-gen}) obviously exhibits a strong dependence on the
integration variables in three domains of phase space: (1) $\tau_1 \to 0$,
(2) $\tau_2 \to 0$, (3) $Q_h^2 \to 0$.  One can decompose the cross section
so that each contribution peaks in only one of these regions (the peaks
were called ``$s$-peak'', ``$p$-peak'' and ``$t$-peak'' in
\cite{Mo:1969cg}).  We shall now investigate this peaking behavior in a
simplified fashion.


\subsubsection{Emission collinear to the incoming electron}

In the region $\tau_1 \to 0$, where the photon is almost collinear to the
incoming electron (``initial state radiation'', ISR), the kinematic
variable $\hat{t}$ of the Compton subprocess becomes very small; it
approaches $|\hat{t}| \simeq \bigO{m^2}$ rather than being of order
$Q_l^2$.  As a consequence, the expression for the differential cross
section simplifies significantly.  Defining the variable
\begin{equation}
  z = \frac{E_e - E_\gamma}{E_e} \; ,
\end{equation}
which describes the relative energy fraction of the incoming electron after
emission of the photon, we obtain the relations
\begin{equation}
\label{eq:isr-scal}
  Q_h^2 = zQ_l^2
  \; , \quad
  y_h = y_l-(1-z)
  \; , \quad
  x_h = \frac{Q_h^2}{y_hS}
  = \frac{x_l y_l z}{y_l+z-1}
  \; .
\end{equation}
This implies $0 \leq y_h \leq z$ and  $(1-z) \leq y_l \leq 1$.

The coefficient functions $S_{1,2}$ reduce to:
\begin{eqnarray}
\label{eq:S1S2-ISR}
  S_1 & \simeq &
  \frac{Q_h^2}{z}
  \left(
    \frac{1+z^2}{1-z} \, \frac{1}{(-\hat{t})} - 2z \, \frac{m^2}{\hat{t}^2}
  \right)
  \; ,
  \\
  S_2 & \simeq &
  \frac{S}{Q_h^2}
  \, \left[ x_h z S \left( 1 - \frac{M^2 Q_h^2}{(zS)^2} \right) - Q_h^2 \right]
  \left(
    \frac{1+z^2}{1-z} \, \frac{1}{(-\hat{t})} - 2z \, \frac{m^2}{\hat{t}^2}
  \right)
  \; .
  \nonumber
\end{eqnarray}
Obviously, these functions factorize into the respective Born expressions
$S_1^B$, $S_2^B$ (see \cite{Akhundov:1996my,Bardin:1997zm}) and a universal
function given in parentheses.

For initial state radiation and at high electron energy it is convenient to
write the photon phase space as
\begin{equation}
  \widetilde{\dd k} = \frac{1}{(4\pi)^2} \,
  \dd z \; \dd |\hat{t}| \; \frac{\dd\phi}{2\pi}
  \; .
\end{equation}
The radiative cross section (\ref{eq:sigma-rad-gen}) schematically
simplifies to:
\begin{equation}
\label{eq:sig-isr}
  \dd\sigma
  \sim
  \dd\sigma^\mathrm{Born}(x_h,y_h/z,Q_h^2)
  \cdot
  \frac{\alpha}{2\pi}
  \left(
    \frac{1+z^2}{1-z} \, \frac{1}{(-\hat{t})} - 2z \, \frac{m^2}{\hat{t}^2}
  \right)
  \dd z \, \dd|\hat{t}| \; \frac{\dd\phi}{2\pi}
  \; .
\end{equation}
When the radiated photon is not observed, we have to integrate over all
available phase space as discussed above.  The integral over $\hat{t}$
leads to large logarithm in the electron mass, of the order
$\ln(E_e^2/m^2)$ resp.\ $\ln(Q_l^2/m^2)$:
\begin{equation}
  \dd\sigma^\mathrm{ISR}
  \approx
  \left[
    \frac{\alpha}{2\pi}
    \left(
      \frac{1+z^2}{1-z} \, \ln \frac{Q_l^2}{m^2}
      - \frac{2z}{1-z}
    \right)
  \right]
  \cdot
  \dd\sigma^\mathrm{Born}(x_h,y_h/z,Q_h^2) \;
  \dd z
  \; .
\end{equation}
The soft photon singularity encountered in the integration over $z$ for $z
\to 1$ is canceled by matching infrared singular terms in the virtual
corrections.


\subsubsection{Emission collinear to the outgoing electron}

For $\tau_2 \to 0$, the photon becomes collinear to the scattered electron
(``final state radiation'', FSR).  In this case we have $\hat{s} \simeq
\bigO{m^2}$.  Defining analogously
\begin{equation}
\label{eq:zf}
  z_f = \frac{\Eep+E_\gamma}{\Eep} \; ,
\end{equation}
we obtain
\begin{equation}
\label{eq:fsr-scal}
  Q_h^2 = z_fQ_l^2
  \; , \quad
  y_h = 1-z_f(1-y_l)
  \; , \quad
  x_h
  = \frac{x_l y_l z_f}{1-z_f(1-y_l)}
  \; .
\end{equation}
The coefficient functions again exhibit factorization,
\begin{eqnarray}
  S_1 & \simeq &
  \frac{Q_h^2}{z_f}
  \left(
    \frac{1+z_f^2}{z_f-1} \, \frac{1}{\hat{s}} - 2z_f \, \frac{m^2}{\hat{s}^2}
  \right)
  \; ,
  \\
  S_2 & \simeq &
  \frac{S}{z_f Q_h^2} \,
  \left[
    x_h S \left( 1 - \frac{M^2 Q_h^2}{S^2} \right) - Q_h^2
  \right]
  \left(
    \frac{1+z_f^2}{z_f-1} \, \frac{1}{\hat{s}} - 2z_f \, \frac{m^2}{\hat{s}^2}
  \right)
  \; .
  \nonumber
\end{eqnarray}
We write the photon phase for final state radiation as
\begin{equation}
  \widetilde{\dd k} = \frac{1}{(4\pi)^2} \,
  \dd z_f \; \dd \hat{s} \; \frac{\dd\phi'}{2\pi}
  \; ,
\end{equation}
with the azimuthal angle $\phi'$ measured w.r.t.\ the outgoing electron.
The radiative cross section (\ref{eq:sigma-rad-gen}) now reads:
\begin{equation}
  \dd\sigma
  \sim
  \dd\sigma^\mathrm{Born}(x_h,y_h,Q_h^2)
  \cdot \frac{\alpha}{2\pi}
  \left(
    \frac{1+z_f^2}{z_f-1} \, \frac{1}{\hat{s}} - 2z_f \, \frac{m^2}{\hat{s}^2}
  \right)
  \frac{\dd z_f}{z_f^3} \, \dd \hat{s}
  \; \frac{\dd\phi'}{2\pi}
  \; .
\end{equation}
The integration over the photon solid angle again leads to a large
logarithm in the electron mass, $\ln (\Eep/m)^2$.  The soft photon
singularity for $z_f \to 1$ cancels against the remaining infrared
divergent parts of the virtual correction.  The sum of virtual corrections
at relative order $\bigO{\alpha}$ and the contributions from initial and
final state radiation at the same order is infrared finite, as required by
the Bloch-Nordsieck theorem.

The above discussion assumes that the detector is able to distinguish the
scattered electron from an almost collinear photon.  This is the case if
the detector determines the momentum of the electron directly, e.g., by the
bending of its trajectory in a magnetic field.  We will refer to such a
measurement as an \emph{exclusive} one.

On the other hand, electromagnetic calorimeters essentially provide
information about the deposited electromagnetic energy and the location and
profile of the generated showers.  When the showers of nearby hits overlap,
it may no longer be possible to distinguish the particles that initiated
these showers, and only the sum of their energies can be determined
accurately.  For the present work we will assume for simplicity that a
minimum angle exists, below which electrons and photons cannot be separated
anymore, and denote this type of measurement as a \emph{calorimetric} one.

In the calorimetric case, collinear final state radiation therefore does
not change the measured kinematic variables if the momenta of final
electron and photon are combined, $p'_\mathrm{cal}=p'+k$, since
\begin{equation}
  Q^2_\mathrm{cal} = -(p-p'_\mathrm{cal})^2 = Q_h^2
  \; , \quad
  x_\mathrm{cal}
  = \frac{Q^2_\mathrm{cal}}{2 P \cdot (p-p'_\mathrm{cal})}
  = x_h
  \; .
\end{equation}
and similarly $y_\mathrm{cal} = y_h$.  Using $p' \simeq
p'_\mathrm{cal}/z_f$, we can rewrite eq.~(\ref{eq:e-ps-yl-Ql2})
as
\begin{equation}
  \widetilde{\dd p'} \to
  \frac{1}{z_f^2} \, \widetilde{\dd p'}_\mathrm{cal}
  =
  \frac{1}{z_f^2}
  \cdot
  \frac{1}{(4\pi)^2}
  \, \dd y_\mathrm{cal} \, \dd Q_\mathrm{cal}^2
  \; ,
\end{equation}
and we identify the leptonic kinematic variables with the measured ones.
After the appropriate variable transformation in the Born cross section,
the radiative cross section becomes proportional to the non-radiative one:
\begin{equation}
\label{eq:sig-fsr}
  \dd\sigma
  \sim
  \dd\sigma^\mathrm{Born}(x_l,y_l,Q_l^2)
  \cdot \frac{\alpha}{2\pi}
  \left(
    \frac{1+z_f^2}{z_f-1} \, \frac{1}{\hat{s}} - 2z_f \, \frac{m^2}{\hat{s}^2}
  \right)
  \frac{\dd z_f}{z_f^3} \, \dd \hat{s}
  \; \frac{\dd\phi'}{2\pi}
  \; .
\end{equation}
Performing the integration over the photon solid angle, integrating over
$z_f$ and combining with the virtual corrections, one can show that --
besides the cancellation of infrared singularities as in the exclusive case
-- the large logarithm in the electron mass cancels for final state
radiation, and only a logarithm in the resolution parameter remains, in
accordance with the Kinoshita-Lee-Nauenberg theorem \cite{KLN}.


\subsubsection{The QED Compton peak}

A moderately strong peaking behavior of the cross section is also found
for $Q_h^2 \to 0$.  In this case the exchanged photon between the hadron
and the electron becomes almost real; its momentum is almost collinear to
the incoming proton.  The contributing region in phase space lies in the
vicinity of the upper boundary in figure~\ref{fig:ellipse}, $\tau_2 \to
\tau_2^+(\tau_1)$, corresponding to the outgoing electron and photon having
opposite azimuthal angles and compensating transverse momenta in the HERA
frame.  The hard scattering process with large momentum transfer occurs
between this photon and the electron, leading to its denomination as the
QED Compton process in deep inelastic scattering.  Approximating the
four-momentum of the exchanged photon as
\begin{equation}
  q_h \simeq - \xi P + q_{\perp,h} \; ,
\end{equation}
such that $\xi$ represents the momentum fraction of the proton taken by
this photon, and $q_{\perp,h}$ being a small transverse momentum, we find
for the invariants of the electron-photon scattering process:
\begin{equation}
  \hat{s} \simeq \xi S
  \; , \quad
  \hat{t} \simeq - \xi (1-y_l) S
  \; , \quad
  \hat{u} \simeq - \xi y_l S
  \; , \quad
  q^2 \approx 0
  \; .
\end{equation}
Assuming that we determine the kinematic variables from a measurement of
the scattered electron, $x_l \simeq \xi$.  The coefficient functions
factorize into one piece proportional to the differential cross section for
Compton scattering,
\begin{eqnarray}
  S_1 & \simeq &
  \frac{1+(1-y_l)^2}{1-y_l}
  \; ,
  \\
  S_2 & \simeq &
  \frac{1+(1-y_l)^2}{1-y_l}
  \cdot
  \left[
    \frac{x_h - x_l}{x_l^2} - \frac{x_h M^2}{Q_h^2}
  \right]
  \; ,
  \nonumber
\end{eqnarray}
and one can express the differential cross section in the form
\cite{Bluemlein:kz,Anlauf:vi}
\begin{equation}
  \dd \sigma \simeq
  \frac{2\pi\alpha^2}{S} \,
  \frac{1+(1-y_l)^2}{x_l(1-y_l)} \cdot \gamma(x_l,Q_l^2) \,
  \dd x_l \, \dd y_l \; ,
\end{equation}
with $\gamma(x_l,Q_l^2)$ being the inelastic contribution to the photon
distribution within the proton \cite{Anlauf:1992wr,Arbuzov:1996id},
\begin{eqnarray}
\label{eq:photon-dist}
  \gamma(x_l,Q_l^2) & = &
  \frac{\alpha}{2\pi}
  \int\limits_{x_l}^1 \frac{\dd z}{x_l z}
  \int\limits_{(Q_h^2)^\mathrm{min}}^{(Q_h^2)^\mathrm{max}}
  \frac{\dd Q_h^2}{Q_h^2}
  \Biggl[
  \left(
    1+(1-z)^2 + 2x_l^2 \frac{M^2}{Q_h^2}
  \right)
  F_2 \left(\frac{x_l}{z},Q_h^2 \right)
  \nonumber \\ && \qquad \qquad {}
    - z^2 F_L \left(\frac{x_l}{z},Q_h^2 \right)
  \Biggr] \; .
\end{eqnarray}
With the integration limits $(Q_h^2)^\mathrm{min,max}$ given below in
(\ref{eq:limits-Qh2-approx}), it is evident that the logarithm obtained
from the $Q_h^2$-integration, $\ln (Q_l^2/M^2)$, does not depend on a small
mass.  It therefore is much smaller than the corresponding logarithms in
the electron mass for ISR and FSR.  Nevertheless, the QED Compton
contribution is significant in the region of large $y_l$.

An accurate calculation of the photon distribution, $\gamma(x,Q^2)$,
requires a detailed modeling of the contributions to the proton structure
functions from low mass hadronic final states.  However, it should be noted
that also elastic proton scattering contributes to the QED Compton process
at HERA, and it even dominates the inelastic contribution
\cite{Courau:1992ht}.  A separation of elastic and inelastic contributions
may be difficult but possible via the residual transverse momentum of the
hadronic system.  On the other hand, the elastic contribution is
essentially independent of $Q^2$.  Provided sufficient statistics, the QED
Compton process could also be used as a test of QCD, as the $Q^2$ evolution
of the photon distribution within the proton differs from that of the
colored partons \cite{DeRujula:1998yq}.  No separation of elastic and
inelastic contributions would be required for that purpose.


\subsection{Kinematic effects of photon radiation}
\label{sec:kin-effects}

It was already mentioned in the introduction to this chapter that the
radiative corrections are sensitive to the shape of the structure
functions.  Following Krasny \cite{Krasny:HERA1991} we shall illuminate the
connection of this shape dependence to the contributions with emission of
hard photons.  We assume that only the scattered electron is measured.

\subsubsection{Kinematic limits}

If we are only interested in the differential cross section in the lepton
variables, $\dd^2\sigma/\dd x_l \, \dd y_l$, we may integrate over the
phase space of the photon.  When no experimental cuts are to be applied it
is convenient to express this phase space integral in terms of kinematic
invariants, e.\,g., $x_h$, $Q_h^2$, and $\hat{s}$ or $\hat{t}$, see
(\ref{eq:dk-by-inv}).  Then the integration over $\hat{s}$ (resp.\
$\hat{t}$) may be performed analytically.  The domain of integration for
the other two variables is constrained by the inequality
(ref.~\cite{Akhundov:1996my}, eq.~B.4):
\begin{equation}
\label{eq:kin-domain-ineq}
  \left[ y_l^2 Q_h^2 + y_h^2 Q_l^2 - y_l y_h (Q_l^2+Q_h^2) \right] S^2
  - M^2 (Q_l^2-Q_h^2)^2
  \leq 0 \; .
\end{equation}
Taking $(x_h,Q_h^2)$ as remaining integration variables, the boundary of
the domain described by (\ref{eq:kin-domain-ineq}) reads:
\begin{eqnarray}
\label{eq:limits-Qh2}
  x_h^\mathrm{min} & = & x_l \; ,
  \nonumber \\
  (Q^2_h)^\mathrm{max(min)} & = &
  x_h \,
  \frac{(x_h y_l S - Q_l^2)\left(y_l S \pm \sqrt{\lambda_q}\right)
        +2x_h M^2 Q_l^2}%
       {2[x_h y_l S - Q_l^2+x_h^2 M^2]}
  \; ,
  \\
\noalign{\hbox{where}}
  \lambda_q & = & y_l^2 S^2 + 4 M^2 Q_l^2
  \; .
  \nonumber
\end{eqnarray}
In the high-energy limit, $S \gg M^2$, the boundaries for $Q_h^2$ following
from (\ref{eq:limits-Qh2}) are
\begin{eqnarray}
\label{eq:limits-Qh2-approx}
  (Q^2_h)^\mathrm{min}
  & \simeq &
  \frac{x_h x_l^2 M^2}{x_h-x_l(1-x_h^2M^2/Q_l^2)}
  \; ,
  \nonumber \\
  (Q^2_h)^\mathrm{max}
  & \simeq &
  x_h y_l S = \frac{x_h Q_l^2}{x_l} \; .
\end{eqnarray}
The upper limit on $x_h$ and a further restriction on $Q_h^2$ is obtained
from the inelastic threshold:
\begin{equation}
  M_X^2 = M^2 + \frac{1-x_h}{x_h} Q_h^2 \geq \bar{M}^2 \; .
\end{equation}
Obviously, $x_h=1$ corresponds to elastic scattering of the proton.  For
$x_h \to 1$, the upper limit (\ref{eq:limits-Qh2-approx}) then implies
\begin{equation}
  x_h^\mathrm{max}
  \simeq 1 - \frac{x_l \left(\bar{M}^2 - M^2\right)}{Q_l^2} \; ,
\end{equation}
and the actual lower limit on $Q_h^2$ reads
\begin{equation}
\label{eq:Qh2min}
  (Q^2_h)^\mathrm{min} \simeq
  \max \left(
  \frac{x_h x_l^2 M^2}{x_h-x_l(1-x_h^2M^2/Q_l^2)}
  ,
  \frac{x_h}{1-x_h} \left(\bar{M}^2 - M^2 \right)
  \right) \; .
\end{equation}
Alternatively, one can switch to $(x_h,y_h)$ as integration variables.  In
that case,
\begin{eqnarray}
\label{eq:limits-yh}
  y_h^\mathrm{min} & \simeq &
  \frac{x_l^2 y_l M^2 / S}{(x_h - x_l)y_l + x_h^2 M^2/S}
  \; ,
  \nonumber \\
  y_h^\mathrm{max} & = & y_l
  \; .
\end{eqnarray}
Neglecting the proton mass, the above conditions simplify to:
\begin{equation}
  x_l \leq x_h \leq 1
  \; , \quad
  0 \leq y_h \leq y_l
  \; .
\end{equation}


\subsubsection{The radiatively corrected cross section}

As discussed above, the measured differential cross section depends on the
structure functions in the whole domain $x_h \geq x_l$ and $y_h \leq y_l$.
At leading order this dependence can be written in the form of a
convolution,
\begin{equation}
\label{eq:Krasny-sig-meas}
  \frac{\dd^2 \sigma^\mathrm{meas}}{\dd x_l \, \dd y_l}
  \simeq
  \int\limits_{x_l}^{x_h^\mathrm{max}} \dd x_h
  \int\limits_{y_h^\mathrm{min}}^{y_l} \dd y_h
  \; K(x_l,y_l;x_h,y_h) \,
  \frac{\dd \sigma^\mathrm{Born}(x_h,y_h)}{\dd x_h \, \dd y_h}
  \; .
\end{equation}
The analytic form of the kernel $K(x_l,y_l;x_h,y_h)$ can be found e.g., in
\cite{Bohm:1986na}.  We do not need its precise form here, but the relevant
approximations have been given in section~\ref{sec:single-rad}.

\begin{figure}
  \begin{center}
    \begin{picture}(90,180)
      \put(0,100){\includegraphics[scale=1.2]{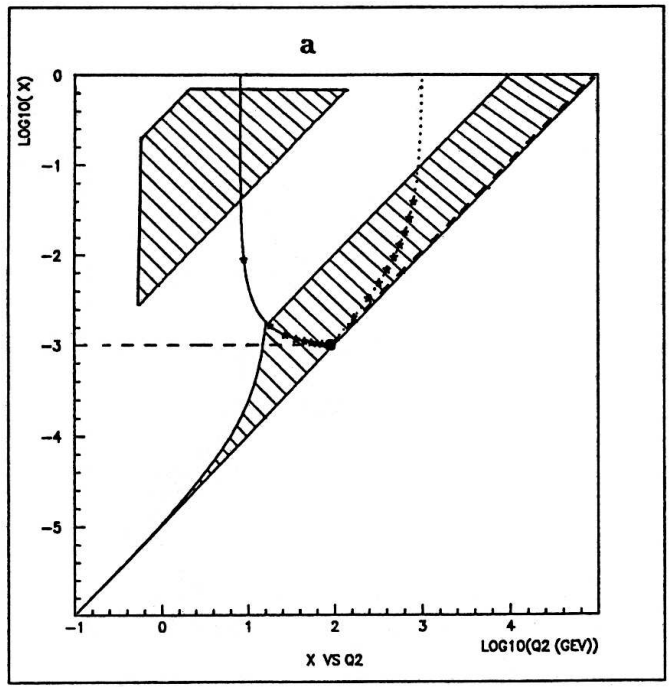}}
      \put(0,  0){\includegraphics[scale=1.2]{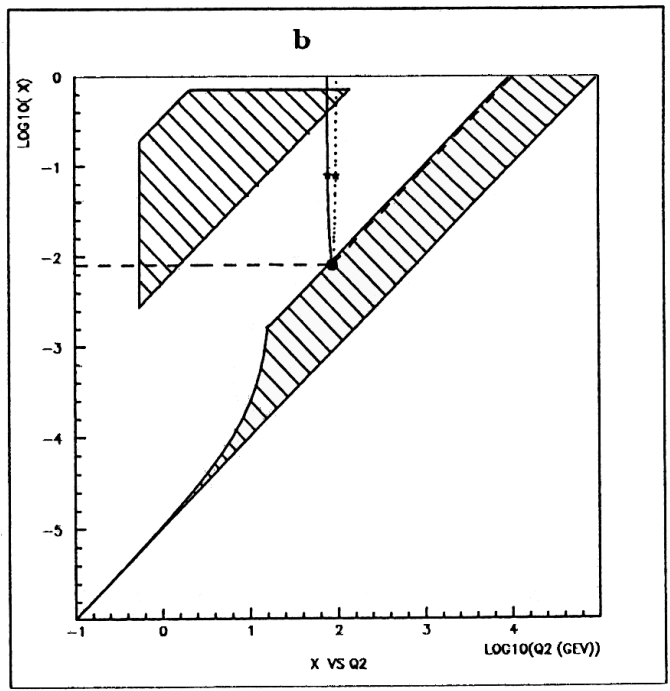}}
    \end{picture}
    \caption{The topology of the $(x,Q^2)$ domains contributing to the
      radiative cross section at HERA. (a) $x_l=10^{-3}$, $Q_l^2=95\GeV$;
      (b) $x_l=9 \cdot 10^{-3}$, $Q_l^2=95\GeV$.  (Taken from
      \cite{Krasny:HERA1991})}
    \label{fig:DIS-rad-topology}
  \end{center}
\end{figure}

The implications of hard photon radiation are most easily understood by
considering the following simple examples \cite{Krasny:HERA1991}.  The
boundaries of the integration region contributing to the inclusive cross
section at $x_l=10^{-3}$, $Q_l^2=95\GeV$ and at $x_l=9 \cdot 10^{-3}$,
$Q_l^2=95\GeV$ are shown as dashed lines in
fig.~\ref{fig:DIS-rad-topology}a and fig.~\ref{fig:DIS-rad-topology}b,
respectively.
The shaded areas roughly correspond to the kinematic domains accessible to
fixed-target experiments (upper left corner) and the HERA experiments.  The
HERA band is essentially limited by the measurement of the scattered
electron energy and the geometrical acceptance of the detectors
\cite{Klein:wm}.%
\footnote{The kinematic domain covered by the HERA experiments has been
  extended over time, mostly towards lower values of $x$ and $Q^2$, by
  successive upgrades of the detectors (e.g., small angle taggers) or using
  a longitudinally shifted position of the interaction vertex.  However, an
  `unexplored gap' between the shaded domains still remains.}
The solid curve on the left side of each point $(x_l,y_l)$ corresponds to
the integration region contributing in the case when the photon is
collinear to the incoming electron; it is described by the parameterization
(see eq.~(\ref{eq:isr-scal})):
\begin{equation}
  x_h = \frac{x_l y_l z}{y_l + z - 1}
  \; , \quad
  Q_h^2 = z Q_l^2
  \; ,
\end{equation}
with $y_l = Q_l^2/x_l S$ and $z=(E_e-E_\gamma)/E_e$.  Each asterisk on the
curve represents a $z(E_\gamma)$, with $E_{\gamma,j}=3j\GeV$.  The asterisk
closest to the full circle corresponds to $j=1$, i.e., $E_\gamma=3\GeV$.
The upper limit on the photon energy in this case follows from the
condition $(1-y_l)/(1-x_l y_l) \leq z \leq 1$.

Similarly, the dotted curve on the right side of the $(x_l,y_l)$ point
corresponds to the case where the photon is emitted collinear to the
scattered electron.  Here,
\begin{equation}
  x_h = \frac{x_l y_l z_f}{1 - z_f(1-y_l)}
  \; , \quad
  Q_h^2 = z_f Q_l^2
  \; ,
\end{equation}
with $y_l$ as above and $z_f=(\Eep+E_\gamma)/\Eep$.  The kinematically
allowed range is $1 \leq z_f \leq 1/(1-y_l(1-x_l))$.  The symbols on the
curve correspond to the same choice of photon energies as above,
$E_{\gamma,j}=3j\GeV$.

The dominant contributions to the measured cross section
(\ref{eq:Krasny-sig-meas}) originate from small bands centered around the
curves discussed above; they correspond to the domain where the kernel $K$
is large.  The scale of width of these bands can be estimated using the
approximations (\ref{eq:sig-isr}) and (\ref{eq:sig-fsr}); for the photon
solid angle they correspond to cones of half opening angle $\sim m/E_e$ and
$\sim m/\Eep$ around initial and final electron, respectively.  Another
significant contribution to the cross section is the region where the Born
differential cross section in the integrand is large, i.e., where $Q_h^2$
is close to its lower limit (\ref{eq:Qh2min}).  In this region the photon
exchanged between the electron and the hadronic system is almost real, and
the hard scattering process is interpreted as QED Compton scattering,
$e\gamma\to e\gamma$, between the electron and the photon emitted from a
quark or proton.

As can be seen from fig.~\ref{fig:DIS-rad-topology}, moving the point
$(x_l,y_l)$ around within the HERA kinematic domain, one can easily find
regions where the dominant contributions to the convolution integral
(\ref{eq:Krasny-sig-meas}) involve mostly ``uncharted territory''.  This is
especially the case for small $x \lesssim 10^{-4}$ and $Q^2 \lesssim
10\GeV$.

At this point uncertainties enter the radiatively corrected cross section.
While one might expect reliable predictions for the structure functions
from perturbative QCD already for moderate momentum transfers, say, above
$4 \GeV^2$, this appears questionable when the non-perturbative regime $Q^2
\lesssim 1 \GeV^2$ is approached.  Therefore, extrapolations of the
structure functions such as those mentioned in section~\ref{sec:low-Q2}
will be needed, leading to a model-dependence and uncertainty in the
predictions of the radiative corrections.

We would like to point out that in the case of HERA this strong
model-dependence concerns mainly initial state radiation and the Compton
process.  The Compton contribution to the structure function measurement
can however be suppressed by imposing some minimum experimental cut on the
transverse momentum or invariant mass of the outgoing hadronic system.  For
final state radiation, the calorimetric measurement of electrons and
photons leads to a cancellation of the large logarithmic contribution as
discussed above.  It also does not shift the kinematic variables.

The uncertainties related to the model-dependence can be brought under
control if one can experimentally detect events with emitted hard photons
and either rejects them (see e.g., the discussion in \cite{Jadach:et} for
using tagged ISR at HERA), or if one turns necessity into a virtue and
actually accepts these events and recalculates the kinematic variables of
the hard subprocess taking into account the four-momentum of the emitted
photon.  This way one can actually extend the kinematic region in which the
structure functions of the proton can be measured.  Furthermore, we shall
see another virtue of the measurement of radiative events in the next
section in that it helps in the determination of the longitudinal structure
function $F_L$.

The above reasoning in principle also applies to other methods of
determination of the kinematic variables, although the relations between
the measured kinematic variables and those of the hard scattering process
differ.

Finally we would like to mention that the neglect of photon emission off
the hadrons in the above discussion is not a problem.  We shall argue in
section~\ref{sec:rc-from-hadron} that initial state radiation can be
absorbed into the evolution of the parton distributions, while final state
radiation off hadrons does not shift the kinematic variables at all.


\section[Collinear radiation and tagged photon processes]%
	{Collinear radiation and tagged photon \\ processes}
\label{sec:coll-intro}


The measurement of deep inelastic scattering with an exclusive photon can
be expected to be a challenging task, as the radiative cross section
(\ref{eq:sigma-rad-gen}) is formally suppressed by a factor $\alpha/\pi$
relative to the non-radiative cross section.  We mentioned three regions of
photon phase space where the radiative cross section is enhanced and can
partially compensate the large suppression.  However, only the emission
collinear to the incoming electron appears to have been exploited yet.


\begin{figure}[tb]
  \begin{center}
    \includegraphics[scale=0.5,angle=90]{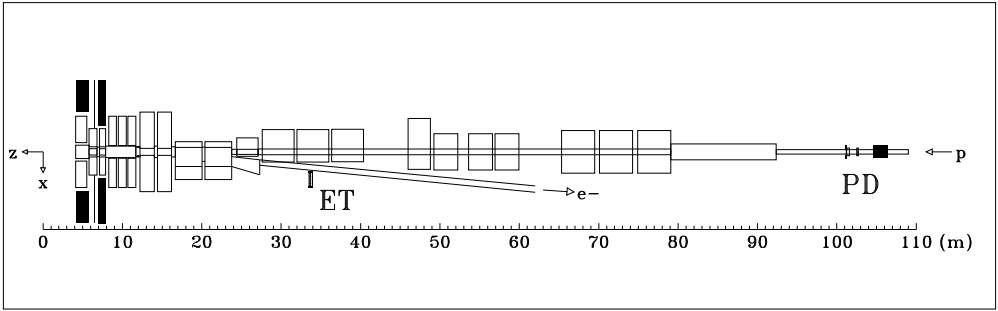}
    \caption{Schematic view of the H1 luminosity system. (Taken from
      \cite{Klein:1998mz})}
    \label{fig:H1-lumi}
  \end{center}
\end{figure}



\begin{figure}[tb]
  \begin{center}
    \includegraphics[scale=0.75]{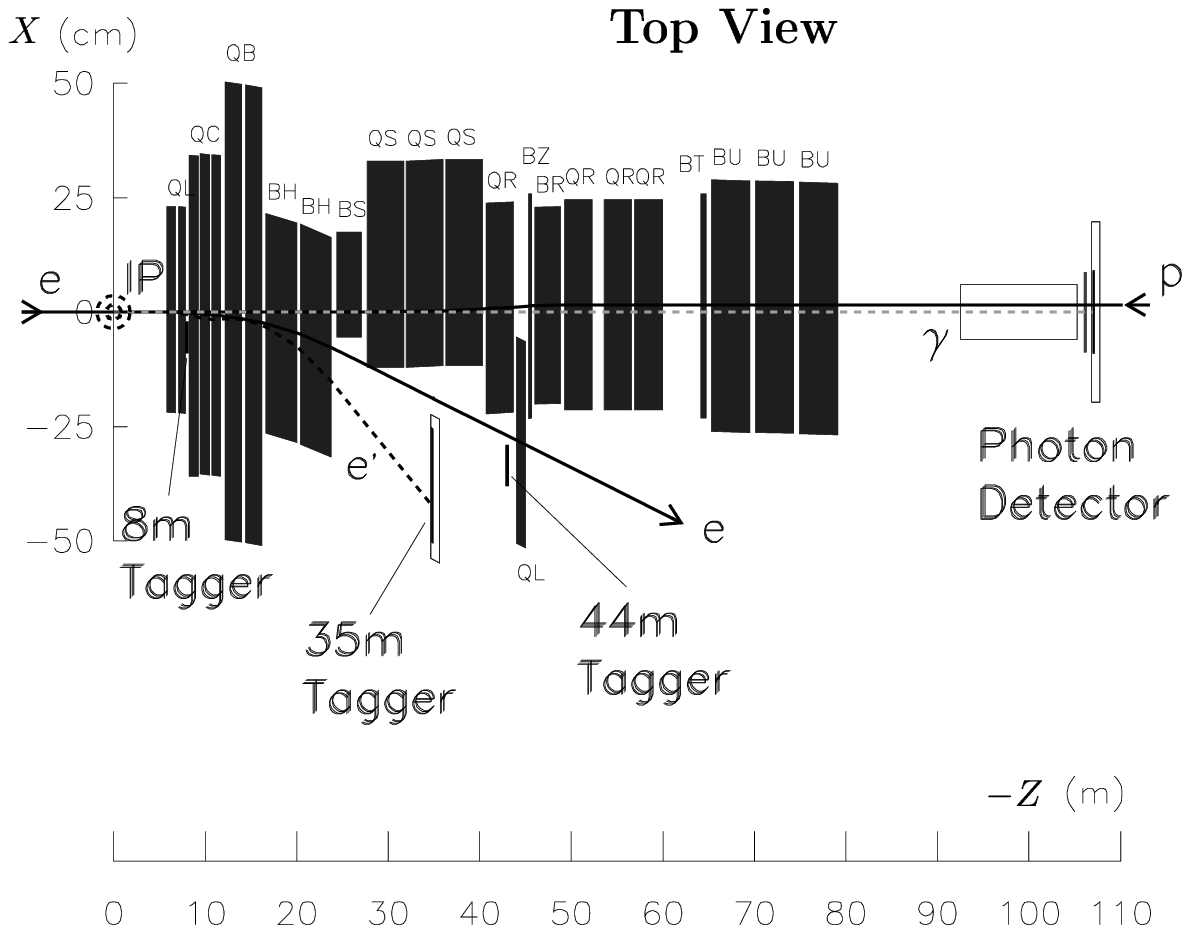}
    \caption{General layout of the ZEUS luminosity monitor. (From
      \cite{Andruszkow:2001jy})}
    \label{fig:ZEUS-lumi}
  \end{center}
\end{figure}


The main reaction used for the determination of the luminosity at the H1
and ZEUS experiments at the HERA collider is the radiative elastic
scattering process $ep \to ep + \gamma$.  For the identification of the
emitted photon, both experiments possess a small photon detector (PD) in
the very forward direction, see fig.~\ref{fig:H1-lumi} for H1 and see
fig.~\ref{fig:ZEUS-lumi} for ZEUS.
The solid angle covered by these photon detectors is roughly described by
$0 \leq \vartheta_\gamma \leq \vartheta_0$ about the direction of the
initial electron beam, with the maximum angle $\vartheta_0$ being of the
order of $(0.45 \ldots 0.5)\mrad$ for both experiments.

This forward photon detector is an ideal device for tagging radiative deep
inelastic scattering events.  Since the maximum angle $\vartheta_0$ is
small compared to the scattering angle of the electron $\theta$ that is
measured in the main detector, the differential cross section for these
radiative events factorizes exactly.  Assuming azimuthal symmetry,
integrating the Compton tensor over the solid angle covered by the photon
detector yields:
\begin{eqnarray}
\label{eq:K-isr-int}
  \frac{E_e^2}{\pi}
  \int\limits_\mathrm{PD} \dd\Omega_\gamma \; K_{\mu\nu}
  & = &
  \left[
    \frac{1+z^2}{1-z} \, \ln \left( 1 + \frac{E_e^2 \vartheta_0^2}{m^2} \right)
    - \frac{2z}{1-z} \,
    \left( 1 + \frac{m^2}{E_e^2 \vartheta_0^2} \right)^{-1}
  \right]
  \nonumber \\
  & \times &
  \frac{1}{z} \,
  \left( - Q_l^2 \tilde{g}_{\mu\nu}
         + 4z \tilde{p}_{\mu} \tilde{p}_{\nu} \right)
  \; + \; \bigO{\vartheta_0^2} \; .
\end{eqnarray}
The tensor structure in the last line is proportional to the non-radiative
lepton tensor (\ref{def:lepton-tensor}):
\begin{equation}
  \frac{1}{z(1-z)} \,
  L_{\mu\nu}(zp,p') \; .
\end{equation}
We therefore find:
\begin{eqnarray}
\label{eq:sig-isr-int}
  \frac{\dd^3\sigma}{\dd x_l \, \dd y_l \, \dd z}
  & = &
  \frac{1}{z} \,
  \dd\sigma^\mathrm{Born}(x_h,y_h,Q_h^2)
  \\
  & \times &
  \frac{\alpha}{2\pi}
  \left[
    \frac{1+z^2}{1-z} \, \ln \left( 1 + \frac{E_e^2 \vartheta_0^2}{m^2} \right)
    - \frac{2z}{1-z} \,
    \left( 1 + \frac{m^2}{E_e^2 \vartheta_0^2} \right)^{-1}
  \right]
  \; . \nonumber
\end{eqnarray}
Inserting the conditions at HERA, we find for $E_e = 27.5 \GeV$ and
$\vartheta_0=(0.45 \ldots 0.5)\mrad$:
\begin{equation}
  \zeta_0 :=
  \frac{E_e^2 \vartheta_0^2}{m^2} \simeq (6-7) \cdot 10^2 \gg 1 \; ,
\end{equation}
thus $\zeta_0 \gg 1$ even if $\vartheta_0 \ll 1$.  Neglecting terms of
order $\bigO{\zeta_0^{-1}}$ and $\bigO{\vartheta_0}$, the cross section
(\ref{eq:sig-isr-int}) simplifies to:
\begin{equation}
\label{eq:sig-isr-int-simp}
  \frac{\dd^3\sigma}{\dd x_l \, \dd y_l \, \dd z}
  =
  \frac{\alpha}{2\pi} P(z,L_0)
  \cdot
  \frac{\dd^2\sigma^\mathrm{Born}(x_h,y_h/z,Q_h^2)}{\dd x_h\,\dd(y_h/z)}
  \; ,
\end{equation}
where we have introduced
\begin{equation}
  P(z,L_0) = \frac{1+z^2}{1-z} L_0 - \frac{2z}{1-z}
  \; , \quad 
  L_0 = \ln \zeta_0
  \; .
\end{equation}
Evaluating the ``large logarithm'' $L_0$ for the HERA forward photon
detectors we obtain:
\begin{equation}
\label{eq:L0-PD}
  L_0 \approx 6.5 \; .
\end{equation}
The cross section (\ref{eq:sig-isr-int-simp}) becomes even simpler when we
interpret the emission of the photon as a reduction of the electron energy,
$E_e^\mathrm{eff} = z E_e$.  The energy fraction $z$ can be invariantly
written as:
\begin{equation}
\label{def:z}
  z = \frac{2P \cdot (p - k)}{S} \; .
\end{equation}
Furthermore we choose the following set of invariant kinematic variables
that use the scattered electron and take into account the energy loss due
to photon emission \cite{KPS92}:
\begin{equation}
\label{eq:kin-vars}
  \hat{Q}^2 = -(p-p'-k)^2
  \; , \quad
  \hat{x}   = \frac{\hat{Q}^2}{2P \cdot (p-p'-k)}
  \; , \quad
  \hat{y}   = \frac{P \cdot (p-p'-k)}{P \cdot (p-k)}
  \; .
\end{equation}
Obviously,
\begin{equation}
  \hat{Q}^2 = \hat{x}\hat{y} zS \; .
\end{equation}
The physical range of the shifted Bjorken variables reads:
\begin{equation}
  0 \leq \hat{x} , \hat{y} \leq 1 \; .
\end{equation}
The shifted variables (\ref{eq:kin-vars}) are related to the standard
Bjorken variables via:
\begin{equation}
\label{eq:xyQ-shifted}
  \hat{Q}^2 = z Q_l^2
  \; , \qquad
  \hat{x} = \frac{x_l y_l z}{y_l+z-1}
  \; , \qquad
  \hat{y} = \frac{y_l+z-1}{z}
  \; ,
\end{equation}
and with the Jacobian for the variable transformation:
\begin{equation}
\label{eq:jacobian-l-shifted}
  \left| \frac{\df (\hat{x},\hat{y})}{\df (x_l,y_l)} \right|
  = \frac{y_l}{y_l + z - 1}
  = \frac{y_l}{z \hat{y}}
  \; .
\end{equation}
In terms of shifted variables, the radiative cross section
(\ref{eq:sig-isr-int-simp}) assumes a very simple, suggestive form,
\begin{equation}
\label{eq:Born}
  \frac{\dd^3\sigma}{\dd\hat{x}\,\dd\hat{y}\,\dd z}
  =
  \frac{\alpha}{2\pi} \, P(z,L_0) \,
  \frac{\dd^2 \sigma_\mathrm{Born}}{\dd\hat{x}\,\dd\hat{y}}
  \; .
\end{equation}
The differential Born cross section on the r.h.s.\ is a function of the
shifted variables (\ref{eq:xyQ-shifted}) only.
For $S$, $\hat{x}$ and $\hat{Q}^2$ fixed, the variable
$\hat{y}=\hat{Q}^2/(\hat{x}zS)$ can still be varied through its dependence
(\ref{def:z}) on the tagged photon energy.  Thus a measurement of cross
section (\ref{eq:Born}) for different $z$ can be used to separate $F_2$ and
$F_L$ at fixed center of mass energy \cite{KPS92}, see also
(\ref{eq:sig-dis-Born}).  Furthermore, from (\ref{eq:xyQ-shifted}) it is
obvious that smaller values of $\hat{Q^2}$ can be attained for fixed energy
and angle of the scattered electron than for the non-radiative measurement.


\begin{figure}[tb]
  \begin{center}
    \begin{picture}(100,100)
     \put(0,0){\includegraphics[scale=0.5]{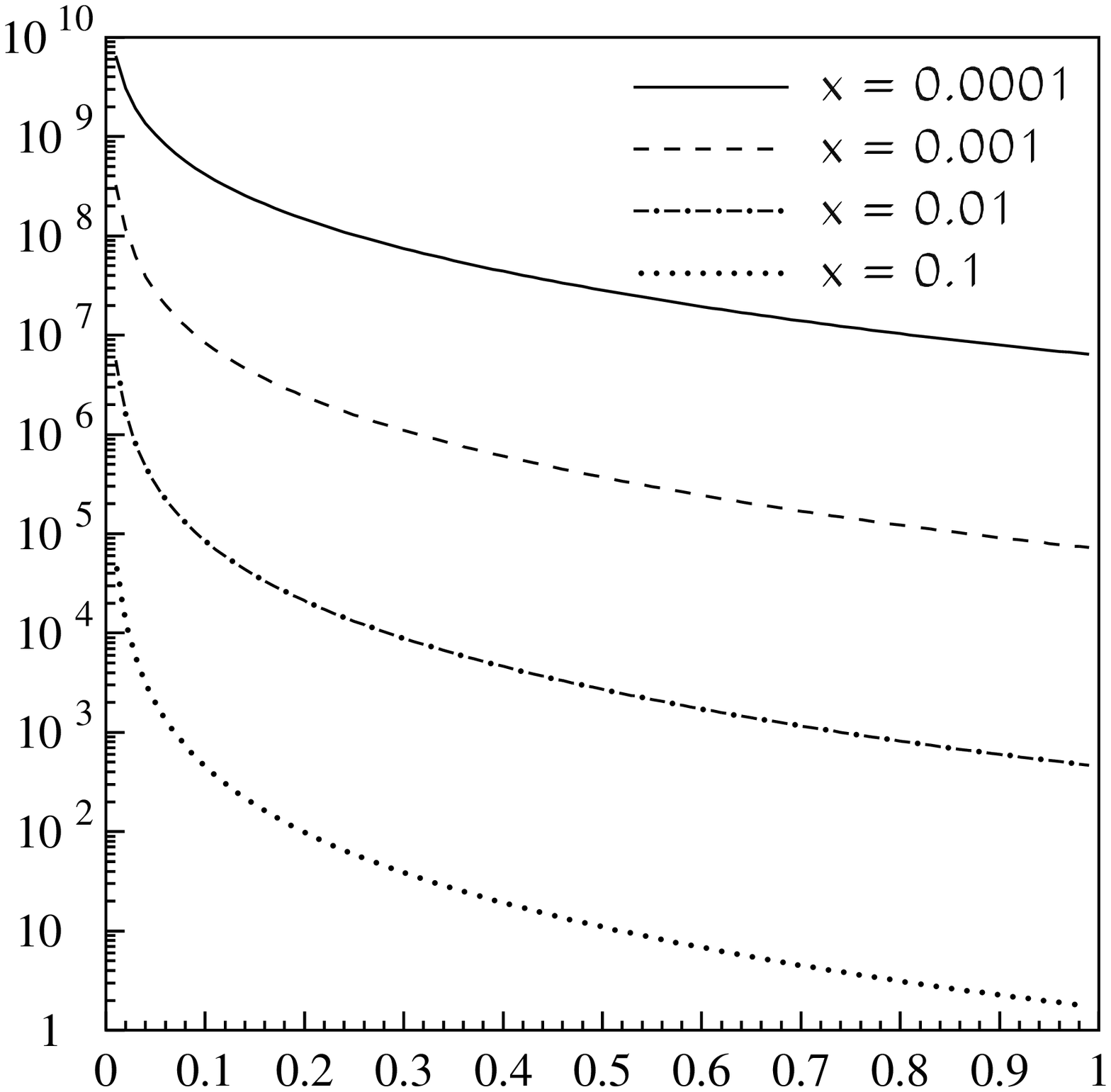}}
     \put(49,0){$\hat{y}$}
     \put(-5,35){\rotatebox{90}{$\dd^3\sigma/\dd\hat{x}\,\dd\hat{y}\,\dd z$ [pb]}}
    \end{picture}
    \caption{The lowest order radiative cross section (\ref{eq:Born}) for
      $E_\gamma=10\GeV$.}
    \label{fig:sigma-born}
  \end{center}
\end{figure}


To give an impression of the radiative cross section, we show in
figure~\ref{fig:sigma-born} the differential cross section (\ref{eq:Born})
for a tagged photon energy of $E_\gamma=10\GeV$.  As parameters we used
\begin{equation}
\label{eq:HERA-tag-param}
  E_e = 27.5 \GeV
  \; , \quad
  E_p = 820 \GeV
  \; , \quad
  \vartheta_0 = 0.5 \mrad
  \; ,
\end{equation}
the ALLM97 \cite{ALLM97} parameterization of the proton structure function
$F_2$, and for simplicity a fixed value $R=0.3$.  Similar results are
obtained using modern structure functions from \texttt{PDFLIB}
\cite{Plothow-Besch:1992qj}.  However, varying $R$ in the range allowed by
experimental data (see e.g., \cite{Aid:1996au}) or suggested by models (see
e.g., \cite{Badelek:1997ap}) can lead to a sizable variation of the cross
section for large $\hat{y}$.

For later convenience we shall also introduce here the quantity
$\tilde\Sigma$ which is related to the lowest order cross section as
follows:
\begin{equation}
\label{def:Sigma}
  \tilde\Sigma(\hat{x},\hat{y},\hat{Q}^2)
  =
  \frac{1}{\hat{y}} \,
  \frac{\dd^2 \sigma_\mathrm{Born}}{\dd\hat{x}\,\dd\hat{y}}
  \; .
\end{equation}
so that, together with (\ref{eq:jacobian-l-shifted}):
\begin{equation}
  \frac{z}{y_l} \,
  \frac{\dd^3\sigma}{\dd x_l\,\dd y_l\,\dd z}
  =
  \frac{1}{\hat{y}} \,
  \frac{\dd^3\sigma}{\dd\hat{x}\,\dd\hat{y}\,\dd z}
  =
  \frac{\alpha}{2\pi} \, P(z,L_0) \,
  \tilde\Sigma(\hat{x},\hat{y},\hat{Q}^2)
  \; .
\end{equation}


\section{The general radiative DIS process}
\label{sec:dis-rad-general}

In the kinematic region of high momentum transfers $Q^2 \gtrsim \MZ^2$,
which are attainable at HERA, not only photon but also Z exchange have to
be taken into account.  The calculations of section~\ref{sec:rad-dis} can
be generalized by reconsidering the Compton tensor
(\ref{def:compton-tensor-Born}) and replacing the virtual photon by a
left-handed vector current and contracting it with the generalized hadron
tensor.  The result is simply a replacement of (\ref{eq:KdotH-Born}):
\begin{equation}
\label{eq:KdotH-NC}
  K_{\mu\nu} H^{\mu\nu}
  \to
  8\pi \left[
    S_1 \mathcal{F}_1(x_h,Q_h^2) + S_2 \mathcal{F}_2(x_h,Q_h^2)
    + S_3 \mathcal{F}_3(x_h,Q_h^2)
  \right] \; .
\end{equation}
The $\mathcal{F}_i$ are the generalized structure functions
(\ref{eq:F-gen}), $S_{1,2}$ are identical with (\ref{eq:S1S2}), and
\cite{Bardin:1997zm}:
\begin{eqnarray}
  S_3 & = &
  \frac{x_h S}{\hat{s}\hat{t}}
  \left( q^2+\hat{u} - y_l (q^2-\hat{t}) \right)
  +
  \frac{q^2+\hat{u}}{2} \left( \frac{1}{\hat{t}}-\frac{1}{\hat{s}} \right)
  \nonumber \\
  & + &
  \frac{m^2}{\hat{t}^2} \left( q^2 - 2(1-y_l) x_h S \right)
  - \frac{m^2}{\hat{s}^2} \left( q^2 + 2 x_h S \right) \; .
\end{eqnarray}
Taking over the derivation of the differential cross section
(\ref{eq:sigma-rad}), we obtain:
\begin{eqnarray}
\label{eq:sigma-rad-NC}
  \dd^5 \sigma & = &
  \frac{2\alpha^3}{\pi} \,
  \frac{y_l}{Q_h^4} \,
  \left( \sum_i S_i \mathcal{F}_i \right)
  \dd x_l \, \dd y_l \, E_\gamma \, \dd E_\gamma \; \dd\Omega_\gamma
  \; .
\end{eqnarray}
This cross section exhibits the same peaking behavior as discussed above.
For example, in the region of collinear initial state radiation we use the
approximations (\ref{eq:S1S2-ISR}) for $S_{1,2}$, along with
\begin{eqnarray}
  S_3 & \simeq &
  \left[ x_h S - \frac{Q_h^2}{2z} \right]
  \left(
    \frac{1+z^2}{1-z} \, \frac{1}{(-\hat{t})} - 2z \, \frac{m^2}{\hat{t}^2}
  \right)
  \; .
  \nonumber
\end{eqnarray}
Consequently, the tagged photon cross section at high momentum transfers
factorizes:
\begin{eqnarray}
\label{eq:sig-nc-rad}
  \frac{\dd^3 \sigma_\mathrm{NC}}{\dd \hat{x} \, \dd \hat{y} \, \dd z}
  =
  \frac{\alpha}{2\pi} P(z,L_0) \cdot
  \frac{\dd^2 \sigma_\mathrm{NC}^\mathrm{Born}}{\dd \hat{x} \, \dd \hat{y}}
  \left( \hat{x}, \hat{y} , \hat{Q}^2 \right)
  \; ,
\end{eqnarray}
with the neutral current Born cross section (\ref{eq:sig-nc-born})
appearing on the r.h.s.

For $Q^2 \ll \MZ^2$, eq.~(\ref{eq:sig-nc-rad}) agrees well with
(\ref{eq:Born}).  At large momentum transfers $Q^2 \gtrsim \MZ^2$, where
also Z exchange plays a significant r\^ole, the cross section is rather
small, as it is roughly suppressed by a factor $\alpha/\pi$ relative to the
non-radiative cross section, leading to a rather small number of events.
Besides, in this region the QCD prediction of $\mathcal{F}_L$ should be
reliable, and it is more important to measure $\mathcal{F}_3$.  This,
however, is better achieved by using different lepton charges rather than
photon tagging.

The radiative neutral current process can in principle also be used to test
the validity of the Standard Model.  In \cite{Cornet:1996sf} the
possibility of a measurement of possible anomalous triple gauge boson
couplings (Z$\gamma\gamma$, ZZ$\gamma$) was studied.  However it was found
that the sensitivity achieved at HERA is too small.


\section{Tagged photons in charged current reactions}

Radiative processes with hard photon emission do also occur in charged
current reactions.  However, the cross section for the non-radiative
process $ep \to \nu X$ is already quite small, and ``paying'' another
factor $\alpha/\pi$ further reduces it, making it difficult if not
impossible to gather sufficient statistics even after the HERA luminosity
upgrade.  Nevertheless, we shall briefly discuss it here for the sake of
completeness.  It is worth to mention that in contrast to neutral current
processes the doubly differential cross section $\dd^2
\sigma_\mathrm{CC}/\dd x\,\dd y$ is still a rather flat distribution in $y$
at HERA energies, as the denominator of the W propagator is:
\[
  Q^2+\MW^2=xyS+\MW^2 \sim \bigO{S} \sim \bigO{\MW^2} \; .
\]
%

In the charged current case there are no leptonic variables at our
disposal.  The only option to determine the kinematics is thus the
measurement of the hadron variables $\Sigma_h$ and $p_{T,h}$, and the
energy of the tagged photon.  In analogy to the non-radiative
Jacquet-Blondel method (\ref{eq:JB-vars}) we define shifted hadron
variables:
\begin{equation}
\label{eq:JB-vars-shifted}
  \hat{y}_\JB = \frac{\Sigma_h}{2zE_e}
  \; , \quad
  \hat{Q}_\JB^2 = \frac{p_{T,h}^2}{1-\hat{y}_\JB}
  \; , \quad
  \hat{x}_\JB = \frac{\hat{Q}_\JB^2}{\hat{y}_\JB z S}
  \; .
\end{equation}
Their relation to the unshifted variables is:
\begin{equation}
\label{eq:JB-vars-shifted-rel}
  \hat{y}_\JB = \frac{y_\JB}{z}
  \; , \quad
  \hat{Q}_\JB^2 = Q_\JB^2 \, \frac{1-y_\JB}{1-y_\JB/z}
  \; , \quad
  \hat{x}_\JB = x_\JB \, \frac{1-y_\JB}{1-y_\JB/z}
  \; .
\end{equation}
This apparently implies $\hat{Q}_\JB^2 \geq Q_\JB^2$, in contrast to
$\hat{Q}^2=zQ_l^2 \leq Q_l^2$ for the lepton method.  However, it should be
mentioned that this does \emph{not} mean that one can really access events
at higher $Q^2$ than in the non-radiative case.  It only tells us that the
kinematic effect of collinear radiation makes radiative events (with
untagged photons) \emph{appear} at even lower measured values of $Q^2$.

As we have argued at the beginning of the chapter, there is no gauge
invariant subset of the Feynman diagrams for single photon emission in the
case of charged current processes, and one cannot define leptonic or
hadronic corrections.  Nevertheless, the leading contribution to the
radiative cross section for emission in the forward direction has the same
universal structure as in the neutral current case.  This may easily be
verified by performing an explicit calculation, using a physical gauge for
the emitted photon.  As we intend to drop all terms at
$\bigO{\vartheta_0^2}$ from the integrated cross section, the only relevant
Feynman diagram that needs to be taken into account is the one with
emission from the incoming lepton.

Let us choose an axial gauge, defined with the help of an arbitrary
light-like four vector $c$ ($c \neq k$, $c^2=0$).  The sum over photon
polarizations then reads
\begin{equation}
  \sum_{\lambda=\pm} \eps_\mu^*(k,\lambda) \eps_\nu(k,\lambda)
  = - g_{\mu\nu} + \frac{ k_\mu c_\nu + k_\nu c_\mu}{k \cdot c}
  \; .
\end{equation}
The choice of the vector $c$ is not important as long as we make sure that
its direction is far outside the cone around the incoming electron.  In the
present case it is convenient to take it proportional to the momentum of
the outgoing (massless) neutrino, $c=p_\nu$.

An explicit calculation confirms the tagged photon cross section,
integrated over the photon polar angle in the range $0 \leq \vartheta \leq
\vartheta_0$ and expressed in terms of the shifted variables
(\ref{eq:JB-vars-shifted}):
\begin{eqnarray}
  \frac{\dd^3 \sigma_\mathrm{CC}}{\dd \hat{x}_\JB \, \dd \hat{y}_\JB \, \dd z}
  =
  \frac{\alpha}{2\pi} P(z,L_0) \cdot
  \frac{\dd^2 \sigma_\mathrm{CC}^\mathrm{Born}}{\dd \hat{x}_\JB \, \dd \hat{y}_\JB}
  \left( \hat{x}_\JB, \hat{y}_\JB , \hat{Q}_\JB^2 \right)
  \; .
\end{eqnarray}
with $\hat{Q}_\JB^2 = \hat{x}_\JB \hat{y}_\JB z S$.  At HERA energies, we
may assume that the longitudinal structure function $\mathcal{W}_L$ is
small and calculable, being proportional to $\alpha_S(\hat{Q}_\JB^2)$.  A
possible benefit of using radiative charged current events, more important
than extending the accessible kinematic region, may be an improvement in
the separation of the structure functions $\mathcal{W}_2$ and
$x\mathcal{W}_3$, which also depend on the lepton charge, see
(\ref{eq:W2-W3}).  However, considering the small variation of the relative
size of the respective coefficients $Y_+$ and $Y_-$ and the projected
integrated luminosity of HERA this appears very difficult.

Non-collinear emission of photons in charged current reactions can also be
used to check the WW$\gamma$ vertex of the Standard Model.  Helbig and
Spiesberger \cite{Helbig:1991iw} have compared the sensitivity of the
reaction $ep \to \nu \gamma X$ to the W production process at HERA and
concluded that it is too small to be competitive to other determinations.
(For constraints from $e^+e^-$ colliders see e.g., \cite{Heister:2001qt}
but also the comment in \cite{LEP+SLD:2001}).


\section{Radiation from the hadron}
\label{sec:rc-from-hadron}

In the above discussions we have put aside the QED corrections from the
hadron side.  These correction are in practice quite small, despite the
small masses of the light quarks, and may be treated perturbatively.  We
will follow here the reasoning of Kripfganz and Perlt
\cite{Kripfganz:1988bd}.

In calculating the radiative corrections to high energy processes we
generally encounter mass singular terms of the type
\[
  \left[ \frac{\alpha}{2\pi} \ln \frac{Q^2}{m^2} \right]^n \; ,
\]
where $Q^2$ is a typical large invariant and $m$ a mass of any of the
external particles.  In the case of QCD it has been shown that the mass
singularities of this type can be factorized and absorbed into the hadronic
structure or fragmentation functions \cite{QCD:factorization}.  Performing
the factorization at some reference scale $Q_0^2$ then shifts all
logarithms in the small masses into the distributions, and one may expect
that the resulting net correction will be typically of order $\alpha/(2\pi)
\cdot \ln(Q^2/Q_0^2)$.  Therefore, the final result of a perturbative QCD
calculation does not involve any large logarithm of a small quark mass.

This procedure can be extended to take into account the inclusive QED
corrections to structure functions.  Starting from the evolution equations
for the parton distributions (\ref{eq:DGLAP}), we replace the quark
splitting function
\begin{eqnarray}
\label{eq:DGLAP+QED}
  \frac{\alpha_S}{2\pi} P_{qq}(x)
  & \to &
  \frac{\alpha_S}{2\pi} P_{qq}^{(1,0)}(x) +
  \left( \frac{\alpha_S}{2\pi} \right)^2 P_{qq}^{(2,0)}(x) + \ldots
  \nonumber \\
  & + &
  \frac{\alpha}{2\pi} P_{qq}^{(0,1)}(x) +
  \frac{\alpha\alpha_S}{(2\pi)^2} P_{qq}^{(1,1)}(x) + \ldots
  \nonumber \\
  & + & \bigO{\alpha^2}
  \; .
\end{eqnarray}
The first line in (\ref{eq:DGLAP+QED}) represents the usual expansion of
the splitting function in QCD.  The results for the $P_{qq}^{(0,m)}$ can be
obtained from the $P_{qq}^{(n,0)}$ by replacing the gluons by photons,
i.e., substitution of the group theoretical factors of $SU(3)$ by the
corresponding ones of the $U(1)$,
\[
  C_F \to 1
  \; , \quad
  C_A \to 0 \; ,
\]
and taking into account the quark electric charges.  Note that the leading
logarithmic contributions $P_{qq}^{(1,0)}$ and $P_{qq}^{(0,1)}$ are
proportional,
\[
  P_{qq}^{(0,1)} = \frac{3 Q_q^2}{4} \, P_{qq}^{(1,0)} \; .
\]
As long as we are only interested in the leading order QED correction to
the quark distributions, we expand them in powers of of $\alpha$,
\begin{eqnarray}
  q(x,Q^2) =
  q^{(0)}(x,Q^2) + \frac{\alpha}{2\pi} q^{(1)}(x,Q^2) + \ldots \; ,
  \\
  g(x,Q^2) =
  g^{(0)}(x,Q^2) + \frac{\alpha}{2\pi} g^{(1)}(x,Q^2) + \ldots \; ,
\end{eqnarray}
where the upper index (0) denotes the pure QCD evolved parton
distributions, and (1) the $\bigO{\alpha}$ correction.  To this order we
may take $\alpha$ fixed; any $Q^2$-dependence is then shifted to order
$\alpha^2$.  Kripfganz and Perlt explicitly demonstrate in
\cite{Kripfganz:1988bd} that the $\bigO{\alpha}$ correction
$q^{(1)},g^{(1)}$ obtained from the solution to the generalized DGLAP
equations with (\ref{eq:DGLAP+QED}) do not contain logarithms in the
fermion mass.

\begin{figure}
  \begin{center}
    \includegraphics[scale=0.75]{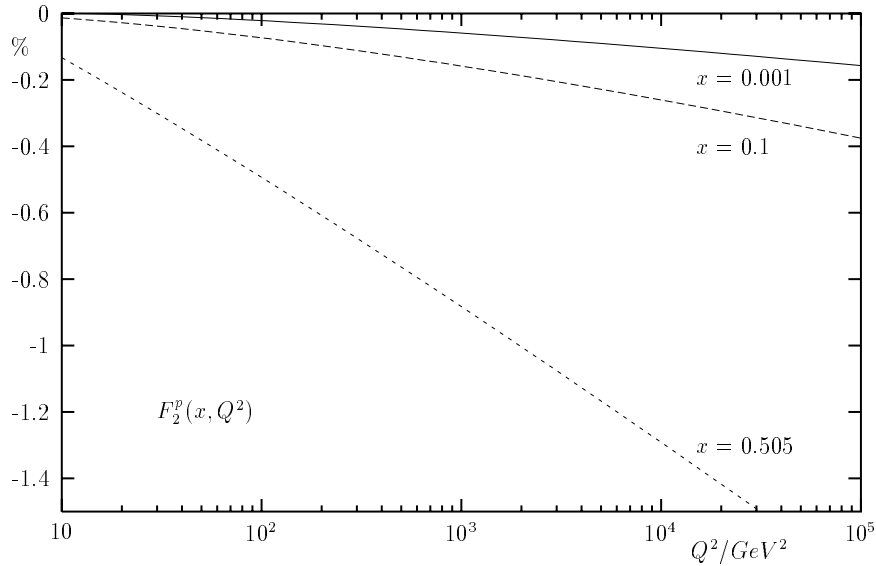}
    \caption{$Q^2$ dependence of the QED corrections to the structure
      function $F_2^p$ for deep inelastic lepton-proton scattering at
      $x=0.001$, $x=0.1$ and $x=0.505$ in percent.  (Taken from
      \cite{Spiesberger:1995dm}).}
    \label{fig:F2p-QED-HS}
  \end{center}
\end{figure}

The QED corrections to parton distributions have also been studied in
\cite{Spiesberger:1995dm}.  Although depending on the input parton
distributions, the corrections to structure functions in general stay at
the per mille level, except for $x\to 1$ where they may reach of the order
of 1~percent.  A typical result for the structure function $F_2$ in shown
in fig.~\ref{fig:F2p-QED-HS}

Photon radiation from hadrons at large angles w.r.t.\ the hadron is small
and calculable within perturbative QCD.  For the tagged photon process
discussed above, its contribution is $\bigO{\vartheta_0^2}$ and thus
negligible.


\chapter{QED Corrections to Radiative Scattering}
\label{sec:qed-cor-rad}

From the viewpoint of perturbation theory, the radiative cross sections
discussed in the previous chapter are interesting hybrids.  On one hand,
these radiative processes are already contained in the radiative
corrections to the deep inelastic scattering process as part of the hard
photon corrections.  On the other hand, they correspond to the Born
approximation for the scattering process with the given number of detected
particles which is ``DIS plus one exclusive photon''.  For a precise
description of the cross sections, radiative corrections to these processes
need to be considered.  Due to the abovementioned relation the corrections
to radiative processes are necessarily part of the higher order corrections
to non-radiative processes.

This chapter is devoted to a calculation of the QED corrections to the
tagged photon process at the next order in perturbation theory.  Note that
they are of order $\bigO{\alpha}$ relative to the radiative process, but of
order $\bigO{\alpha^2}$ relative to the non-radiative DIS process.  Since
the QED corrections are dominated by large logarithms in the electron mass,
we start with a discussion of these corrections in a simplified framework
by focusing just on the leading logarithmic terms \cite{AAKM:ll}.  We then
illustrate the influence of the choice of reconstruction method of the
kinematic variables on the size of the corrections.  The qualitative
features of the corrections turn out to be similar to those in the case of
non-radiative DIS.

We will not repeat here the calculation of the subset of next-to-leading
logarithms which were presented in \cite{AAKM:JETP,AAKM:nlo} for leptonic
variables and in \cite{Anl99:Sigma} for the $\Sigma$ variables, but explain
in some detail the full calculation of the model-independent
$\bigO{\alpha}$ corrections by the present author \cite{Anlauf:2001fu}.  A
systematic approach to QED corrections taking into account the leading
logarithms at all orders follows in the next chapter.  Our restriction to
the corrections to the tagged photon process excludes the consideration of
$e^+ e^-$ pair production corrections in DIS (\cite{Merenkov:rr}, but see
also \cite{Hoffmann:1999jg} and references) which at this order do not
contribute.
Furthermore, we will not discuss here the QED corrections to the QED
Compton process but refer the reader to
\cite{Anlauf:vi,Anlauf:1992wr,Courau:1992ht}.


\section{Radiative corrections in the leading logarithmic approximation}
\label{sec:tagged-ll}

As discussed in the previous chapter, there are two potential sources of
large corrections to the cross section that contribute terms of the order
of $\alpha/\pi \cdot \ln(Q^2/m_f^2)$, with $m_f$ being the mass of a light
fermion.  The first one are the vacuum polarization corrections to the
propagator of the exchanged boson between the electron and the hadronic
system, which we assume to be taken into account and resummed using
renormalized (running) couplings.  The other source of large corrections is
the radiation of photons off light fermions.

Following \cite{AAKM:ll}, we shall suppose a calorimetric experimental
setup, where a hard photon radiated collinearly to the outgoing electron
line cannot be distinguished from a bare outgoing electron, so that final
state radiation can be neglected to the desired accuracy.  We also assume
some minimum lower cut on the transverse momentum of the outgoing hadronic
system, in order to suppress the contribution from QED Compton events for a
leptonic measurement of the kinematic variables.  Therefore, the only large
corrections we explicitly need to consider below originate from the
radiation of photons off the incoming electron.  We will treat these in the
so-called structure function formalism.


\subsection{Structure function formalism}

In the structure function formalism for QED
\cite{Kuraev:1985hb,Altarelli:1986kq,Beenakker:1989km}, a radiatively
corrected inclusive cross section is written as a convolution of the
electron non-singlet structure function $D^\mathrm{NS}(z,Q^2)$ with the
Born cross section $\sigma_0$ for the hard scattering process taken at
reduced center of mass energy.  Schematically,
\begin{equation}
\label{eq:sigma-tot-inclusive}
  \sigma_\mathrm{RC}(s) =
  \int \dd z \; D^\mathrm{NS}(z,Q^2) \, \sigma_0(zs;Q^2)
  \; ,
\end{equation}
with $Q^2$ being some typical large scale of the hard process.  The
electron non-singlet structure function,
\begin{equation}
\label{eq:DNS}
  D^\mathrm{NS}(z,Q^2) =
  \delta(1-z) +
  \frac{\alpha L}{2\pi} \, P^{(1)}(z)
  +
  \frac{1}{2!} \left( \frac{\alpha L}{2\pi} \right)^2 P^{(2)}(z) +
  \bigO{ (\alpha L)^3 } \; ,
\end{equation}
which is a QED analog of the parton distributions in QCD, depends on the
large scale $Q^2$ only via the large logarithm $L = \ln(Q^2/m_e^2)$.  It is
known to properly sum the leading logarithmic contributions $(\alpha L)^n$
to all orders in perturbation theory \cite{Beenakker:1989km}.

For our purposes we need here the first two coefficients of the power
series expansion (\ref{eq:DNS}) of $\Dns$.  Introducing a small auxiliary
parameter $\eps$ that serves as an infrared (IR) regulator to separate
virtual and soft from hard photon contributions, these two coefficients
read:
\begin{eqnarray}
\label{eq:P1}
        P^{(1,2)}(z) & = & P^{(1,2)}_\delta \cdot \delta(1-z) +
        P^{(1,2)}_\Theta(z) \cdot \Theta(1-\eps-z) \; , \nonumber \\
\noalign{\hbox{with}}
        P^{(1)}_\Theta(z) & = & \frac{1+z^2}{1-z} \; , \qquad
        P^{(1)}_\delta = 2 \ln \eps + \frac{3}{2} \; ,
\end{eqnarray}
and similarly (see e.g.\ \cite{Beenakker:1989km}),
\begin{eqnarray}
  \label{eq:P2}
  P^{(2)}_\Theta(z) & = &
  \int_{z/(1-\eps)}^{1-\eps} \frac{\dd t}{t}
  P^{(1)}_\Theta(t) P^{(1)}_\Theta\left(\frac{z}{t}\right)
  + 2 P^{(1)}_\delta P^{(1)}_\Theta(z)
  \\
  & = & 2 \left[
  \frac{1+z^2}{1-z} \left(2\ln(1-z) - \ln z + \frac{3}{2} \right)
  + \frac{1+z}{2} \ln z - 1 + z \right] \;.
  \nonumber
\end{eqnarray}
Inspecting the lowest order tagged photon cross section (\ref{eq:Born}),
\[
  \frac{\dd^3\sigma_\mathrm{lo}}{\dd\hat{x}\,\dd\hat{y}\,\dd z}
  =
  \frac{\alpha}{2\pi} \, P(z,L_0) \,
  \frac{\dd^2 \sigma_\mathrm{Born}}{\dd\hat{x}\,\dd\hat{y}}
  \; ,
\]
and comparing with the $\bigO{\alpha}$ contribution in the integrand to the
inclusive cross section (\ref{eq:sigma-tot-inclusive}), we conclude that
the logarithmic piece of $P(z,L_0)$ is indeed contained in the first order
correction%
; the difference in the logarithms
($L-L_0$) is accounted for by the cross section for photons emitted at
angles larger than $\vartheta_0$.

\subsection{Leading logarithms at the next order}

Let us now turn to the contributions to the cross section at order
$\bigO{\alpha^2}$ \cite{AAKM:ll}.  The typical maximum emission angle
$\vartheta_0$ is about $(0.45 \ldots 0.5)\mrad$ for the HERA photon
detectors.  The logarithm $L_0$ appearing in (\ref{eq:Born}) is moderately
large, $L_0 \simeq 6.5$, see (\ref{eq:L0-PD}).  The complement
\[
  L_1 \equiv L - L_0 \approx 6 \ldots 16 \gg 1
  \quad
  \mbox{(for e.g., $Q^2$ = 0.1 \ldots 1000~GeV$^2$)}
\]
is of similar magnitude or even larger than $L_0$.  Starting at this order
we thus have two competing large logarithms $L_0,L_1$.  As a consequence,
in the leading logarithmic approach at $\bigO{\alpha^2}$ we need to
consider all double logarithms of the type $\alpha^2 L_0^2$, $\alpha^2 L_0
L_1$.  There cannot be terms $\alpha^2 L_1^2$ since we require at least one
tagged photon.

The corrections receive contributions from virtual corrections to collinear
emission, one collinear plus one soft photon, two collinear photons, and
one collinear plus one non-collinear photon.  Since the logarithmic
contribution from the non-collinear photon arises mainly from emission at
small angles outside the photon detector, we shall also denote it as a
semi-collinear one.

The sum of the contributions of virtual and soft corrections to collinear
photon emission, with the emission angle of the soft photon being
integrated over the full solid angle, but the hard photon only over the
solid angle of the PD, can be obtained from the expression for the one-loop
Compton tensor.  The logarithmic part reads \cite{KMF87}:
\begin{equation}
\label{eq:V+S-ll}
  \left(\frac{\alpha}{2\pi}\right)^2 L_0
  \left[ (L_0 + L_1) P^{(1)}_\delta -  (L_0 + 2 L_1) \ln z \right]
  P^{(1)}_\Theta(z) \cdot \sigma_0(\hat{x},\hat{y},z) \; .
\end{equation}
For the double collinear contribution we assume that only the sum of the
energies of photons in the PD is measured.  In this case $z = 1 - (\sum_i
E_\gamma^{(i)})/E_e$.  The contribution from two hard photons in the PD,
with the energy fraction of each being larger than $\eps$, then yields
\cite{Mer88}:
\begin{equation}
\label{eq:cc}
  \left(\frac{\alpha}{2\pi}\right)^2 \frac{L_0^2}{2}
  \left[ P^{(2)}_\Theta(z) - 2 P^{(1)}_\Theta(z)
  	\left( P^{(1)}_\delta - \ln z \right) \right]
  \cdot \sigma_0(\hat{x},\hat{y},z) \; ,
\end{equation}
see also (\ref{eq:P2}).

The final contribution is associated with one collinear photon hitting the
PD and the other one being emitted at an angle larger than $\vartheta_0$.
Let us try to understand the origin of the large logarithms in this case.
For the hard photon that hits the PD, the major contribution comes from the
region of polar angles $\vartheta_\gamma^{(1)} \approx m_e/E_e \ll
\vartheta_0$.  Similarly, for the other photon the biggest contribution
comes from polar angles close to the lower limit, $\vartheta_\gamma^{(2)}
\gtrsim \vartheta_0$.  Thus, the leading contributions come essentially
from the region where we have strong ordering in angles,
$(\vartheta_\gamma^{(2)}/\vartheta_\gamma^{(1)})^2 \approx \zeta_0 \gg 1$.
One can easily see by an explicit calculation that one obtains the double
logarithms entirely from the contribution where the photon hitting the PD
is emitted first and the ``lost'' photon is emitted second, while the
reversed case is suppressed as long as the following condition is
satisfied:
\begin{equation}
\label{eq:simple-ordering}
  x_2 \cdot \zeta_0 \gg 1
  \; , \qquad
  \mbox{ where } x_2 = \frac{E_\gamma^{(2)}}{E_e} \; .
\end{equation}
The contributions from regions where condition (\ref{eq:simple-ordering})
is not met are subleading.  Thus we shall assume in the following $\eps \ll
1$, but $\eps \cdot \zeta_0 \gg 1$.

Taking the ordering of photon emissions into account, we calculate the
contribution from one tagged plus one undetected photon using the quasireal
electron method \cite{Baier:1973ms},
\begin{equation}
\label{eq:sc}
  \left(\frac{\alpha}{2\pi}\right)^2 L_0 L_1 \cdot
  P^{(1)}_\Theta(z)
  \int^{x_2^{\mathrm{max}}}_{\eps}
  \frac{\dd x_2}{z}
  P^{(1)}_\Theta\left(1-\frac{x_2}{z}\right)
  \tilde{\sigma}_0(\hat{x},\hat{y};z-x_2) \; ,
\end{equation}
where $\tilde{\sigma}_0$ is understood to be expressed in terms of the
``true'' kinematic variables $x_t$, $y_t$ of the hard subprocess,
\begin{equation}
\label{eq:J}
  \tilde{\sigma}_0(\hat{x},\hat{y};z-x_2) \equiv
  \sigma_0(x_t, y_t, z-x_2) \cdot
  \mathcal{J}(\hat{x},\hat{y};1-x_2/z) \; .
\end{equation}
Note that this contribution explicitly depends on the experimental
determination of the kinematic variables $\hat{x}$ and $\hat{y}$, since the
almost collinear emission of the second photon shifts the ``true''
kinematic variables ($x_t,y_t$) with respect to the measured ones
($\hat{x},\hat{y}$).
The Jacobian
%
\begin{equation}
  \mathcal{J} =
  \det \left( \frac{\df(x_t,y_t)}{\df(\hat{x},\hat{y})} \right)
\end{equation}
accounts for scaling properties of the chosen kinematic variables under
radiation of the second photon.  The upper limit of the $x_2$-integration
in (\ref{eq:sc}) is given by either some experimental cut on the maximum
energy of the second photon, or by the kinematic limit, which also depends
on the choice of the reconstruction method.

After a change of variables, $x_2 = zu$, $u_0 = x_2^{\mathrm{max}}/z$, the
integral in (\ref{eq:sc}) may be conveniently decomposed into IR divergent
($\eps$ dependent) and IR convergent contributions as (suppressing the
arguments $\hat{x}$, $\hat{y}$):
\begin{eqnarray}
\label{eq:sc-rewrite}
\lefteqn{
  \int^{x_2^{\mathrm{max}}}_{\eps}
  \frac{\dd x_2}{z} \,
  P^{(1)}_\Theta\left(1-\frac{x_2}{z}\right)
  \tilde{\sigma}_0(z-x_2)  =  } \nonumber \\
  & = &
  \int^{u_0}_{\eps/z} \dd u \;
  P^{(1)}_\Theta(1-u)
  \left[ \left(\tilde{\sigma}_0(z(1-u)) - \tilde{\sigma}_0(z) \right)
  + \tilde{\sigma}_0(z) \right]
  \nonumber \\
  & = &
  \sigma_0(z) \cdot \left[
  \int_0^{u_0} \dd u \;
  P^{(1)}_\Theta(1-u)
  \left( \frac{\tilde{\sigma}_0(z(1-u))}{\tilde{\sigma}_0(z)} - 1 \right)
  \right. \nonumber \\ & & \left.
  \qquad \qquad
	+ 2 \ln z
  - P^{(1)}_\delta
  - \int_{u_0}^1 \dd u \; P^{(1)}_\Theta(1-u) \right],
\end{eqnarray}
where in the last step we have extended the $u$-integration of the IR
convergent piece to 0 due to the smallness of $\eps$, and we have used
the property
\[
  \int_0^1 \dd u \; P^{(1)}(u) = 0
\]
in the simplification of the IR divergent piece.

Adding up the contributions (\ref{eq:V+S-ll}), (\ref{eq:cc}), and
(\ref{eq:sc}), we see that the dependence on the auxiliary parameter $\eps$
cancels, as it should.


\subsection{The radiatively corrected cross section}

We are now able to write down the result for the leading logarithmic
radiative corrections.  The correction $\delta_{\mathrm{RC}}$ is defined
via:
\begin{equation}
  \frac{\dd^3\sigma_{\mathrm{RC}}}{\dd\hat{x} \, \dd\hat{y} \, \dd z} =
  \frac{\dd^3\sigma_{\mathrm{lo}}}{\dd\hat{x} \, \dd\hat{y} \, \dd z} \cdot
  \left( 1 + \delta_{\mathrm{RC}}(\hat{x},\hat{y},z) \right) \, .
\end{equation}
Since we restricted ourselves to the leading logarithms, we must retain in
the correction factor $(1+\delta_{\mathrm{RC}})$ only the logarithmic terms
$L_0,L_1$ from (\ref{eq:V+S-ll}), (\ref{eq:cc}), and
(\ref{eq:sc}).
We thus obtain:%
\footnote{In contrast to \cite{AAKM:ll} we do not count here the
  contribution from vacuum polarization as part of the radiative
  corrections, as it is contained in the leading order tagged photon cross
  section $\dd^3\sigma_{\mathrm{lo}}/\dd\hat{x} \, \dd\hat{y} \, \dd z$.}
\begin{eqnarray}
\label{eq:delta-ll}
  \delta_{\mathrm{RC}} & = &
  \frac{\alpha L_0}{4\pi}
  \frac{1-z}{1+z^2} P^{(2)}_\Theta(z)
  \nonumber \\
  &+& \frac{\alpha L_1}{2\pi} \Biggl[ \int_0^{u_0} \dd u \;
  P^{(1)}_\Theta(1-u)
  \left( \frac{\tilde{\sigma}_0(z(1-u))}{\tilde{\sigma}_0(z)} - 1 \right)
  \nonumber \\
  &-&
  \int_{u_0}^1 \dd u \; P^{(1)}(1-u) \Biggr] \; .
\end{eqnarray}
Remember that the contribution from the undetected hard photon depends on
the choice of kinematic variables and on the upper limit $u_0$.


\begin{table}
\begin{center}
\begin{tabular}{|c|c|c|c|c|c|}
\hline
Method	& $Q^2_t$ & $x_t$ & $y_t$
	& $z'_\mathrm{min}$ & $\mathcal{J}(\hat{x},\hat{y};z')$ \\
\hline
lepton	& $z'\hat{Q}^2$ & $\frac{\hat{x}\hat{y}z'}{\hat{y}+z'-1}$
	& $\frac{\hat{y}+z'-1}{z'}$
	& $\frac{1-\hat{y}}{1-\hat{x}\hat{y}}$
	& $\frac{\hat{y}}{\hat{y}+z'-1}$ \\
\hline
    JB	& $\frac{1-\hat{y}_\mathrm{JB}}{1-\hat{y}_\mathrm{JB}/z'}\,\hat{Q}^2_\mathrm{JB}$
	& $\frac{1-\hat{y}_\mathrm{JB}}{1-\hat{y}_\mathrm{JB}/z'}\,\hat{x}_\mathrm{JB}$
	& $\frac{\hat{y}_\mathrm{JB}}{z'}$
	& $\frac{\hat{y}_\mathrm{JB}}{1-\hat{x}_\mathrm{JB}(1-\hat{y}_\mathrm{JB})}$
	& $\frac{1-\hat{y}_\mathrm{JB}}{z'-\hat{y}_\mathrm{JB}}$ \\
\hline
$\Sigma$& $\hat{Q}^2_\Sigma$ & $\frac{\hat{x}_\Sigma}{z'}$ & $\hat{y}_\Sigma$
	& $\hat{x}_\Sigma$ & $\frac{1}{z'}$ \\
\hline		
\end{tabular}
\end{center}
\caption{Scaling properties of the kinematic variables under initial state
  radiation for different experimental methods of their determination.}
\label{tab:jacobian}
\end{table}


\subsection{Numerical results}

To give an impression of the significance of the radiative corrections at
leading logarithmic accuracy as given by eq.~(\ref{eq:delta-ll}), we shall
present some typical numerical results.  The relations between the measured
and the true kinematic variables in (\ref{eq:J}) for some reconstruction
methods are given in table~\ref{tab:jacobian}; they agree with those for
the calculation of the leading logarithms from initial state radiation to
non-radiative DIS (see e.g., \cite{Akhundov:1996my,Arbuzov:1996id}).  The
variable $z'$ corresponds to $1-u$ in (\ref{eq:delta-ll}), so that $u_0 = 1
- z'_\mathrm{min}$.


We use the same set of HERA parameters as in section~\ref{sec:coll-intro},
eq.~(\ref{eq:HERA-tag-param}), the ALLM97 parameterization and a fixed
ratio $R=F_L/F_T=0.3$, unless stated otherwise.  As a representative
intermediate value for the tagged energy we take $E_\gamma=10\GeV$.  No cut
will be applied to the energy of the second (``lost'') photon.


\begin{figure}[tb]
  \begin{center}
    \begin{picture}(100,90)
     \put(0,0){\includegraphics[scale=0.5]{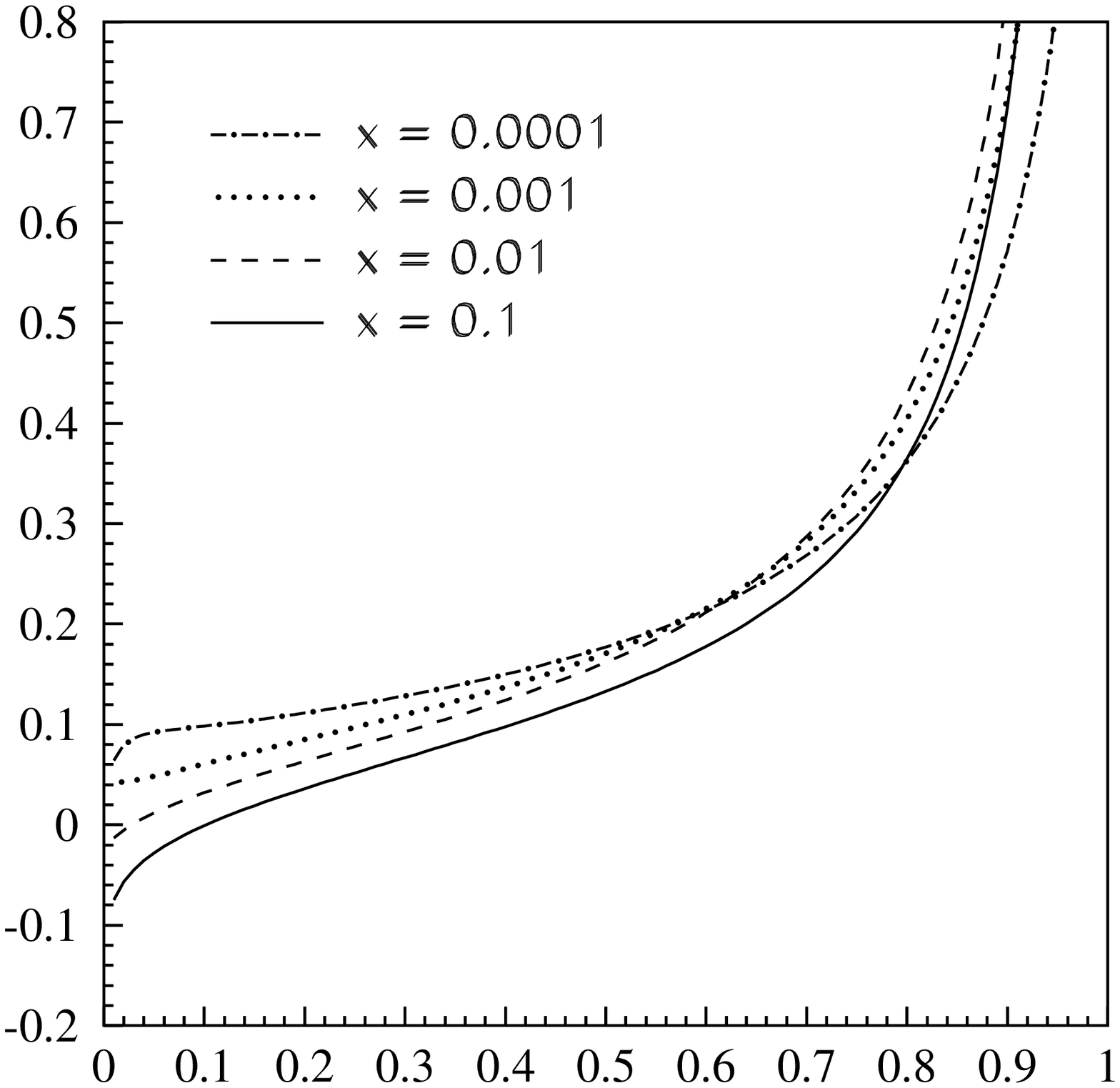}}
     \put(49,0){$\hat{y}$}
     \put(-6,50){$\delta_\mathrm{RC}$}
    \end{picture}
    \caption{The leading logarithmic radiative corrections to the tagged
    photon cross section for $E_\gamma=10\GeV$.  Lepton method.}
    \label{fig:corr-ll-e}
  \end{center}
\end{figure}


\begin{figure}[tb]
  \begin{center}
    \begin{picture}(80,70)
     \put(0,0){\includegraphics[scale=0.4]{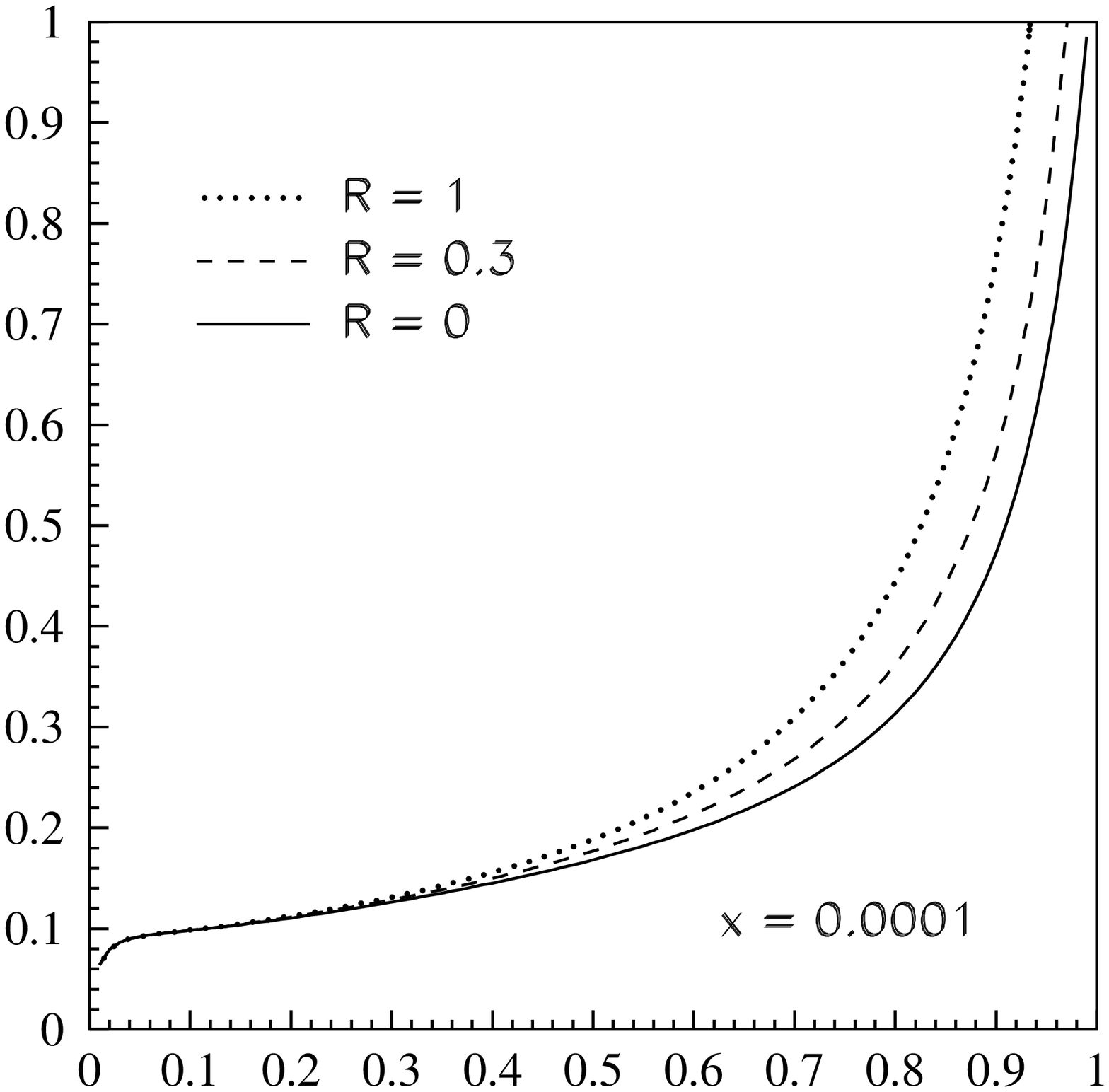}}
     \put(39,-1){$\hat{y}$}
     \put(-6,40){$\delta_\mathrm{RC}$}
    \end{picture}
    \caption{$R$-dependence of the leading logarithmic radiative
      corrections to the tagged photon cross section for $E_\gamma=10\GeV$.
      Lepton method.}
    \label{fig:corr-ll-r}
  \end{center}
\end{figure}


Figure~\ref{fig:corr-ll-e} shows the radiative correction
$\delta_\mathrm{RC}$ (\ref{eq:delta-ll}) for the reconstruction of the
kinematic variables using the electron method.  In the region of very small
$\hat{y}_l$, the radiative corrections may become strongly negative.  The
contributions from virtual and soft photon corrections dominate, because
the kinematic limit on the energy of the undetected photon tends to zero as
\begin{equation}
\label{eq:u0-l}
  u_0 = 1-z'_\mathrm{min} =
  \frac{\hat{y}_l (1-\hat{x}_l)}{1 - \hat{x}_l \hat{y}_l}
  \; .
\end{equation}
For $\hat{y}_l \to 1$, the phase space for photon emission increases, and
we find the typical strong rise of the corrections, which is due to the
large shift between the ``true'' and the measured the kinematic variables
under hard photon radiation as discussed in section~\ref{sec:kin-effects},
and due to the steep dependence of the lowest order neutral current cross
section for small $\hat{y}$, resp.\ $\hat{Q}^2$.


The radiative corrections for the electron method exhibit a very strong
dependence on the ratio $R=F_L/F_T$, see fig.~\ref{fig:corr-ll-r}.  Again,
this effect is largest for large $\hat{y}$.  Increasing $R$ increases the
correction, mainly because it reduces the hard cross section $\sigma_0$ for
$\hat{y} \to 1$, while hard collinear radiation probes $x_t>\hat{x}$,
$y_t<\hat{y}$.  This strong dependence of the radiative corrections on a
quantity (here: $R$) poorly known in other regions of phase space is a nice
example of the kinematic effects we discussed in
section~\ref{sec:kin-effects}.


\begin{figure}
  \begin{center}
    \begin{picture}(80,155)
     \put(0,80){\includegraphics[scale=0.4]{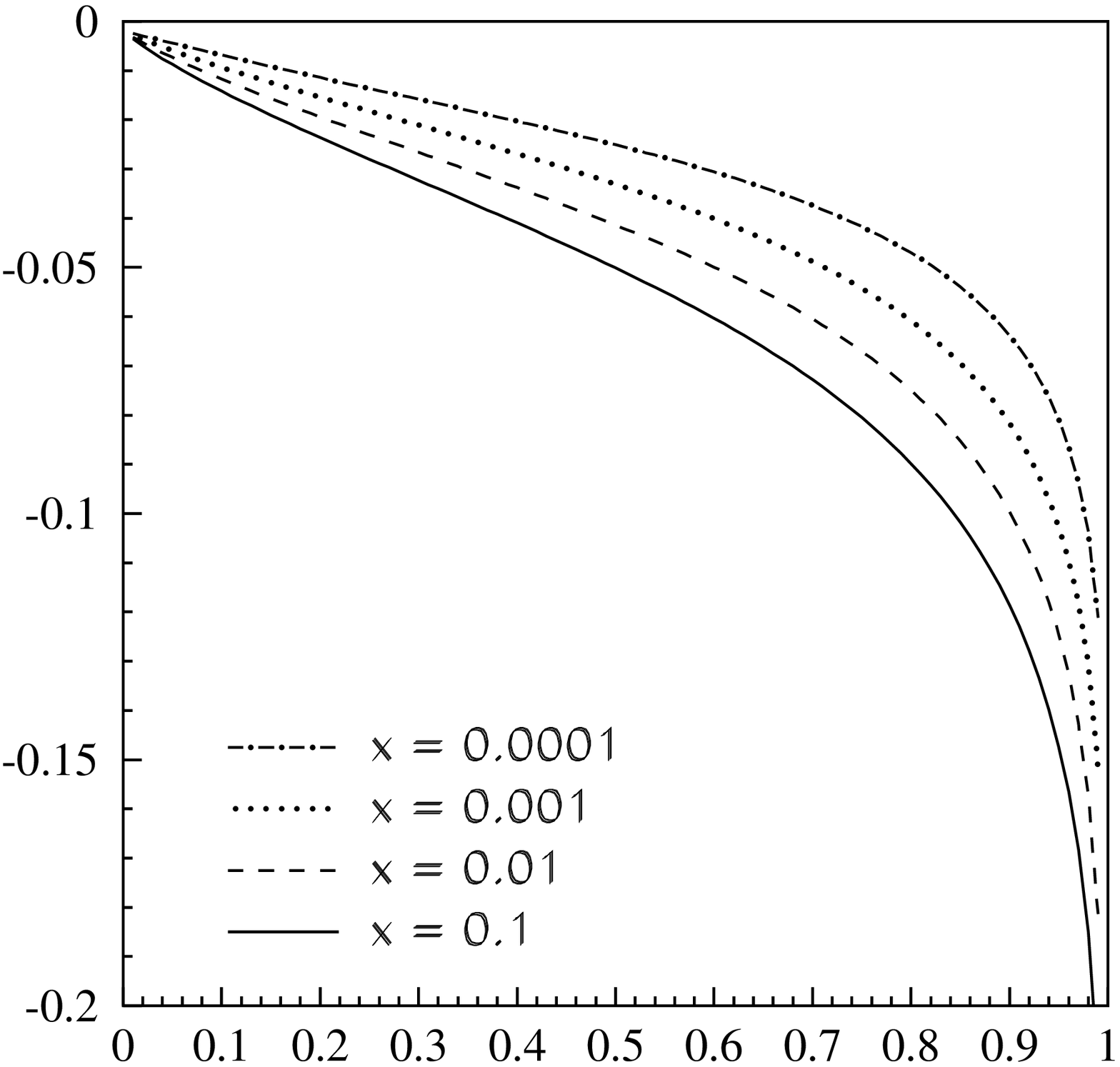}}
     \put(0,0){\includegraphics[scale=0.4]{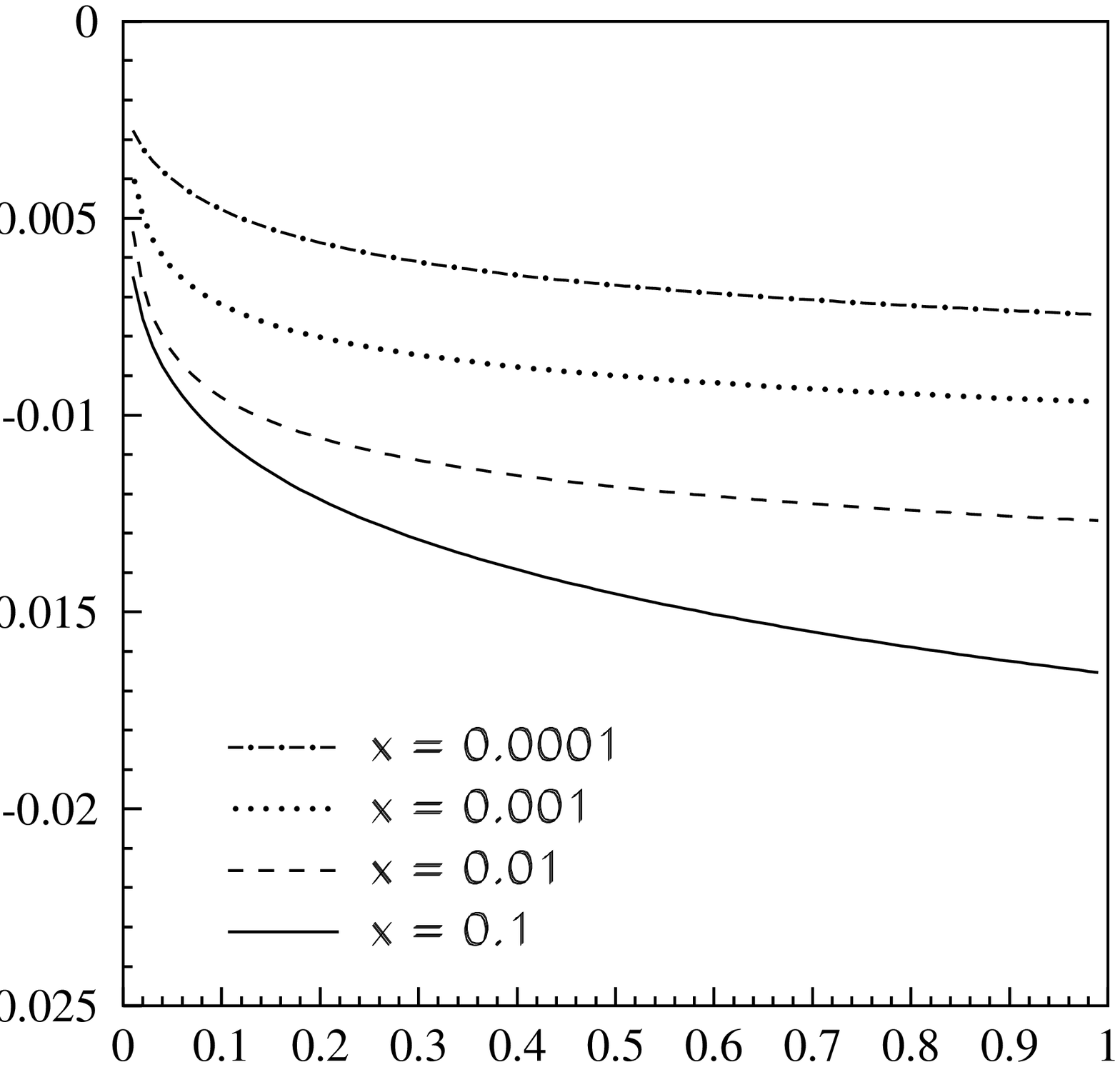}}
     \put(39,-1){$\hat{y}$}
     \put(-7,40){$\delta_\mathrm{RC}$}
     \put(39,79){$\hat{y}$}
     \put(-7,120){$\delta_\mathrm{RC}$}
     \put(-20,145){(a)}
     \put(-20,65){(b)}
    \end{picture}
    \caption{Leading logarithmic radiative
      corrections to the tagged photon cross section for $E_\gamma=10\GeV$:
      (a) Jacquet-Blondel method, (b) $\Sigma$ method.}
    \label{fig:corr-ll-jb+sig}
  \end{center}
\end{figure}


Figure~\ref{fig:corr-ll-jb+sig}a shows the leading logarithmic radiative
corrections for the Jacquet-Blondel method, again for a tagged photon
energy of 10\GeV.  In this case the corrections are negative for $\hat{y}
\to 1$, because in this limit the phase space for the undetected photon
tends to zero,
\begin{equation}
  u_0 = 1 - z'_\mathrm{min} =
  \frac{(1-\hat{x}_\mathrm{JB})(1-\hat{y}_\mathrm{JB})}%
       {1-\hat{x}_\mathrm{JB}(1-\hat{y}_\mathrm{JB})}
  \; ,
\end{equation}
whereas for small $\hat{y}_\mathrm{JB}$ the corrections remain moderate.
Since the Jacquet-Blondel variables correspond to a ``more inclusive''
measurement than the leptonic variables, the corrections due to radiation
of an additional photon are generally smaller.

Finally, figure~\ref{fig:corr-ll-jb+sig}b displays the corresponding
results for the $\Sigma$ method.  Here the corrections are surprisingly
small.  This apparent suppression is easily traced back to the weak
dependence on initial state radiation of the reconstruction of the
kinematic variables using the $\Sigma$ method, see
table~\ref{tab:jacobian}.  At the same time the $R$-dependence is
negligible \cite{Anl99:Sigma}.


\section{Complete leptonic corrections}
\label{sec:complete}

As explained in the previous chapter, the subset of leptonic QED
corrections to deep inelastic scattering is gauge invariant.  Assuming
one-photon exchange it also factorizes, thus allowing a discussion isolated
from the hadronic part.  In the present section, we shall therefore
consider the complete set of QED corrections to the Compton subprocess
\begin{equation}
\label{eq:compton-process}
  e(p_1) + \gamma^*(-q) \to e(p_2) + \gamma(k) \; ,
\end{equation}
with the emission angle of the photon being integrated over the PD, while
taking into account the corrections from virtual and real QED corrections.
While performing this integration, we require that the remaining part of
the amplitude of the full process, $\mathcal{M}(\gamma^*+p \to X)$, depends
only weakly on the transverse recoil momentum of the forward photon.  This
implies the condition $p_1 \cdot k \ll p_1 \cdot p_2$, i.e., $\vartheta_0
\ll \theta$.

The presentation below follows the outline given in \cite{Anlauf:2001fu},
but provides more details.


\subsection{Compton tensor}

We start our calculation with the Compton tensor $K_{\mu\nu}$ as defined in
(\ref{def:compton-tensor-Born}).  For the discussion of the radiative
corrections to this tensor we write its decomposition up to and including
1-loop contributions as follows \cite{KMF87}:
\begin{eqnarray}
\label{eq:compton-tensor}
  \nonumber
  K_{\mu\nu}^\mathrm{1-loop}
  &=& \frac{1}{2} \left( P_{\mu\nu} + P_{\nu\mu}^{*} \right)
  \; , \\
  P_{\mu\nu}
  &=& \tilde{g}_{\mu\nu}\left(B_g + \frac{\alpha}{2\pi}T_g\right)
  + \sum_{i,j=1,2} \tilde{p}_{i\mu}\tilde{p}_{j\nu}
    \left(B_{ij} + \frac{\alpha}{2\pi}T_{ij}\right)
  \; ,
  \\ \nonumber
  \tilde{g}_{\mu\nu} &=& g_{\mu\nu} - \frac{q_{\mu}q_{\nu}}{q^2}
  \; , \quad
  \tilde{p}_{i\mu} = p_{i\mu} - q_{\mu}\frac{p_i \cdot q}{q^2}
  \; , \quad
  i=1,2 \; .
\end{eqnarray}
The expressions for the quantities $B_{ij}$ corresponding to the Born
approximation are given in (\ref{eq:compton-born}).  The general results
for the one-loop QED contributions $T_g$, $T_{ij}$ in the high energy limit
can be found in \cite{KMF87}; they are lengthy and will not be reproduced
here.

The next step will be the integration over the solid angle of the photon
that is tagged in the forward detector.  Since we are interested in the
region of almost collinear emission, $k \simeq (1-z) p_1$, we neglect the
transverse momentum of the emitted photon in the tensor decomposition
(\ref{eq:compton-tensor}), as it is of order $\bigO{E_e \vartheta_0}$, and
use momentum conservation to set $\tilde{p}_2 = z \tilde{p}_1$.  To the
same accuracy, the kinematic variables of the Compton subprocess are
related to those of the radiative DIS process via:
\[
  \hat{u} = - \frac{\hat{Q}^2}{z}
  \; , \quad
  q^2 = (p_1-k-p_2)^2 \simeq - \hat{Q}^2
  \; , \quad
  \hat{s} \simeq \frac{1-z}{z} \, \hat{Q}^2
  \; .
\]


\subsection{Virtual and soft corrections}

The virtual corrections to the Compton tensor are furthermore conveniently
decomposed into a piece containing the universal infrared singular
contributions, which are proportional to the Born contributions, and an
infrared finite remainder:
\begin{equation}
\label{eq:T-decomp}
  T_g = \rho B_g + T_g'
  \; , \quad
  T_{ij} = \rho B_{ij} + T_{ij}'
  \; , \quad
  i,j=1,2
  \; .
\end{equation}
The precise form of the infrared singular expression $\rho$ depends on the
regularization procedure.  The calculation of the one-loop contributions in
\cite{KMF87} uses a fictitious photon mass $\lambda$ in the calculation of
the virtual corrections and obtains:
\begin{equation}
  \rho = 4(L_Q-1)\ln\frac{\lambda}{m} - L^2_Q + 3L_Q + 3\ln z
  +\frac{\pi^2}{3} -\frac{9}{2}
  \; ,
\end{equation}
with
\begin{equation}
  L_Q = \ln \frac{-\hat{u}}{m^2} = \ln\frac{Q^2}{m^2}
  \; .
\end{equation}
The integration of the Compton tensor over the solid angle of the collinear
photon is described in some detail in appendix~\ref{sec:int-virt}.  We
quote here only the final result:
\begin{eqnarray}
  \label{eq:virt-int}
  \frac{E_e^2}{\pi}
  \int \dd\Omega_k \;
  K_{\mu\nu}^\mathrm{1-loop}
  & = &
  \left( - Q_l^2 \tilde{g}_{\mu\nu}
         + 4z \tilde{p}_{1\mu} \tilde{p}_{1\nu} \right) \times
  \\
  && {} \frac{1}{1-z}
  \left[
    \left( 1 + \frac{\alpha}{2\pi} \rho \right) P(z,L_0)
	     - \frac{\alpha}{2\pi} T
  \right]
  + \mathcal{O}\left(\vartheta_0^2,\zeta_0^{-1}\right)
  \; ,
  \nonumber
\end{eqnarray}
where
\begin{eqnarray}
\label{eq:virt-int-T}
  T &=&
  (A\ln z + B) P(z,L_0)
  + C L_0 + D \; ,
  \nonumber \\
  A &=& 2 L_Q - L_0 - 2\ln(1-z) \; ,
  \nonumber \\
  B &=& \ln^2 z - 2 \dilog (1-z) - \frac{1}{2} \; ,
  \nonumber \\
  C &=&
  - \frac{2z}{1-z}\ln z
  - z \; ,
  \\
  D &=&
  - \frac{1-6z+4z^2}{1-z} \left( \dilog(1-z) + \ln z \ln(1-z) \right)
  \nonumber \\
  &-& 2z \ln^2(1-z) + \frac{8z}{1-z} \ln z
  - \frac{4 \pi^2}{3} z + 1 \; .
  \nonumber
\end{eqnarray}
Here
\begin{equation}
  \dilog(x) = -\int\limits_{0}^{x}\frac{\dd y}{y}\ln(1-y)
\end{equation}
is the Spence function or dilogarithm \cite{Lewin:1981}.
The single and double logarithmic terms in $L_0$ and $L_Q$ of expression
(\ref{eq:virt-int-T}) agree with \cite{AAKM:nlo}.

The dependence of the virtual corrections on the unphysical parameter
$\lambda$ is canceled by the contribution from emission of an additional
soft photon, calculated using the same regularization method.  Since soft
photon emission factorizes, its contribution is proportional to the Born
piece $B_{\mu\nu}$ and can be combined with the $\rho$-piece of the virtual
corrections.  Requiring that the energy fraction of the second (soft)
photon in units of the energy of the incoming electron does not exceed
$\eps$, with $\eps \ll 1$, and adding the contribution from soft photon
emission to the virtual correction then amounts to the replacement of the
quantity $\rho$ in (\ref{eq:virt-int}) by $\tilde\rho$, see \cite{KMF87}:
\begin{equation}
\label{eq:rho-v+s}
  \tilde{\rho} = 2(L_Q - 1)\ln\frac{\eps^2}{Y}
  + 3L_Q + 3\ln z - \ln^2Y - \frac{\pi^2}{3} - \frac{9}{2}
  + 2\dilog\left(\frac{1+c}{2}\right)
  \, , \quad
\end{equation}
with
\begin{equation} \label{eq:Y-c}
  Y = \frac{\Eep}{E_e}
  \quad \mbox{and} \quad
  c \equiv \cos\theta = \cos\Angle(\vec{p},\vec{p}\,{}')
\end{equation}
being the relative energy of the scattered electron and the cosine of the
scattering angle in the lab system, respectively.


\subsection{Double hard bremsstrahlung: ISR}
\label{sec:double-hard}

\subsubsection{Kinematics of the double Compton process}

\begin{figure}
  \begin{center}
\begin{picture}(137,28)
\put(0,4){%
\begin{fmfgraph*}(40,20)
\fmfbottomn{b}{5}
\fmftopn{t}{5}
\fmfleftn{l}{5}
\fmfrightn{r}{5}
\fmf{fermion}{l5,v1,v2,v,r5}
\fmf{phantom}{v1,b2}
\fmf{phantom}{v,b4}
\fmffreeze
\fmf{photon}{v1,t2}
\fmf{photon}{v2,t3}
\fmf{photon}{v,b4}
\fmfv{l.a=90,label=$k_1$}{t2}
\fmfv{l.a=90,label=$k_2$}{t3}
\fmfv{l.a=-90,label=$\mu$}{b4}
\fmfv{l.a=90,label=$p_1$}{l5}
\fmfv{l.a=90,label=$p_2$}{r5}
\fmfdot{v,v1,v2}
\end{fmfgraph*}%
}
\put(48.5,4){%
\begin{fmfgraph*}(40,20)
\fmfbottomn{b}{5}
\fmftopn{t}{5}
\fmfleftn{l}{5}
\fmfrightn{r}{5}
\fmf{fermion}{l5,v1,v2,v,r5}
\fmf{phantom}{v1,b2}
\fmf{phantom}{v,b4}
\fmffreeze
\fmf{photon}{v1,t2}
\fmf{photon}{v2,b3}
\fmf{photon}{v,t4}
\fmfv{l.a=90,label=$k_1$}{t2}
\fmfv{l.a=-90,label=$\mu$}{b3}
\fmfv{l.a=90,label=$k_2$}{t4}
\fmfv{l.a=90,label=$p_1$}{l5}
\fmfv{l.a=90,label=$p_2$}{r5}
\fmfdot{v,v1,v2}
\end{fmfgraph*}%
}
\put(97,4){%
\begin{fmfgraph*}(40,20)
\fmfbottomn{b}{5}
\fmftopn{t}{5}
\fmfleftn{l}{5}
\fmfrightn{r}{5}
\fmf{fermion}{l5,v1,v2,v,r5}
\fmf{phantom}{v1,b2}
\fmf{phantom}{v,b4}
\fmffreeze
\fmf{photon}{v1,b2}
\fmf{photon}{v2,t3}
\fmf{photon}{v,t4}
\fmfv{l.a=-90,label=$\mu$}{b2}
\fmfv{l.a=90,label=$k_1$}{t3}
\fmfv{l.a=90,label=$k_2$}{t4}
\fmfv{l.a=90,label=$p_1$}{l5}
\fmfv{l.a=90,label=$p_2$}{r5}
\fmfdot{v,v1,v2}
\end{fmfgraph*}%
}
\end{picture}
    \caption{Three of the six Feynman diagrams contributing to the double
      Compton process (\ref{eq:dbl-compt}).  The remaining three diagrams
      are obtained by exchanging $k_1 \leftrightarrow k_2$.}
    \label{fig:Feynman-double-Compton}
  \end{center}
\end{figure}
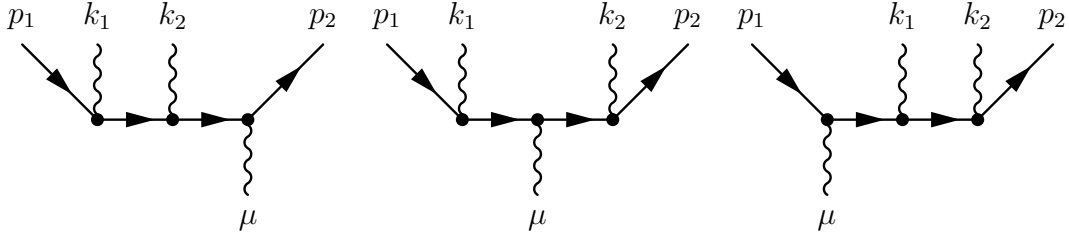

The other contributions to the radiative corrections involve the emission
of two hard photons.  This part is quite elaborate, as we need to keep the
electron mass different from zero even at high energies.  Techniques for
the efficient calculation of processes with double photon emission have
been developed by the CALKUL collaboration (see \cite{Berends:1984ez} and
references cited therein).  It is however not necessary to deal with the
full expressions for our purposes, since we are only interested in the hard
photon corrections to the tagged photon cross section.  We shall therefore
consider the double Compton process of
fig.~\ref{fig:Feynman-double-Compton},
\begin{equation}
\label{eq:dbl-compt}
  e(p_1) + \gamma^*(-q) \to e(p_2) + \gamma(k_1) + \gamma(k_2)
  \; .
\end{equation}
with the implicit assumption that one of the final photons is emitted
almost collinearly to the incoming electron.  Furthermore, we will keep
only those terms that contribute to integrated cross sections at high
energy, i.e., for $|(p_1-p_2)^2|, |q^2| \gg m^2$.

For explicit calculations in the HERA frame the following parameterization
of the momenta is convenient:
\begin{eqnarray}
  p_1 & = & (E_e; p_e \vec{e}_z) = E_e \cdot (1;\beta_e \vec{e}_z) \; ,
  \\
  p_2 & = & (\Eep; p_e' \vec{n}) = \Eep \cdot(1;\beta_e' \vec{n}) \; ,
  \quad
  \vec{n} = (\sin\theta,0,\cos\theta) \; ,
  \nonumber \\
  k_i & = &
  x_i E_e \cdot (1; \vec{n}_i)
  \; , \quad
  \vec{n}_i =
  (\sin\vartheta_i\cos\phi_i,\sin\vartheta_i\sin\phi_i,\cos\vartheta_i)
  \; , \quad
  i=1,2
  \; .  \nonumber
\end{eqnarray}
The $x_i \equiv E_{\gamma,i}/E_e$ are the energy fractions of the photons
in units of the electron beam energy.

Instead of the polar angles the photons, $\vartheta_i$, we shall make
frequent use of the substitution
\begin{equation}
  \tau_i = \frac{1-\cos\vartheta_i}{2}
  \; , \quad
  (0 \leq \tau_i \leq 1) \; .
\end{equation}
Collinear emission of photon $i$ obviously corresponds to $\tau_i \to 0$.

For the kinematic invariants of the Compton subprocess we introduce the
notation:
\begin{eqnarray*}
  z_i  & = & 2 p_1 \cdot k_i
  \; , \quad
  z_i' = 2 p_2 \cdot k_i
  \; , \\
  \sigma & = & 2 k_1 \cdot k_2 = (k_1+k_2)^2 \; , \\
  \Delta  & = & - [(p_1-k_1-k_2)^2 - m^2] = z_1 + z_2 - \sigma \; , \\
  \Delta' & = & [ (p_2+k_1+k_2)^2 - m^2 ] = z_1' + z_2' + \sigma \; , \\
  Q_l^2 & = & - (p_1-p_2)^2
  \; , \\
  Q_h^2 & = & - q^2 = Q_l^2 + z_1 + z_2 - z_1' - z_2' - \sigma \; .
\end{eqnarray*}
%
%
In the kinematic region of interest at least one of the invariants
$z_{1,2}$ will be small, i.e., $\bigO{m^2}$:
\begin{eqnarray}
\label{eq:zi-general}
  z_i & = &
  2 p \cdot k_i
  = 2 x_i E_e^2 (1 - \beta_e \cos\vartheta_i)
  \nonumber \\
  & = &
  2 x_i E_e^2 (1 - \beta_e + 2\beta_e \tau_i)
  \simeq x_i \left( m^2 + 4E_e^2 \tau_i \right)
  \; .
\end{eqnarray}
In the last line we neglected terms of relative order $\bigO{m^2/E_e^2}$.
Note that also the invariant $\Delta$ may become small in the double
collinear region, as
\begin{eqnarray}
  \sigma
  & = & 2 k_1 \cdot k_2
  = 2 x_1 x_2 E_e^2 (1-\vec{n}_1 \cdot \vec{n}_2)
  \nonumber \\
  & = & 2 x_1 x_2 E_e^2
  [1-\cos\vartheta_1\cos\vartheta_2
    -\sin\vartheta_1\sin\vartheta_2\cos(\phi_1-\phi_2)]
  \\
  & = & 4 x_1 x_2 E_e^2
  \left[\tau_1+\tau_2-2\tau_1\tau_2
  -2\sqrt{\tau_1\tau_2(1-\tau_1)(1-\tau_2)}\cos(\phi_1-\phi_2)\right]
  \nonumber \; .
\end{eqnarray}
The kinematic restriction to at least one almost collinear photon allows
certain simplifications of the expressions we encounter.  To exploit these
systematically we used the symbolic manipulation program \texttt{FORM}
\cite{FORM}.


\subsubsection{Double Compton tensor}

For the treatment of two photon emission off unpolarized electrons, we
normalize the double Compton tensor analogously to
(\ref{def:compton-tensor-Born}) as \cite{Anlauf:2001fu}%
\footnote{Note that the normalization chosen here differs from
  \cite{Anlauf:2001fu} by a factor of 1/4 in order to conveniently absorb
  factors of 4 in other places below.}
\begin{equation}
\label{def:double-compton-tensor}
  K_{\mu\nu}^{\gamma\gamma}
  = \frac{1}{(2e^2)^3} \sum_{\mathrm{spins}}
  M_{\mu}^{e\gamma^*\to e'\gamma\gamma}
 (M_{\nu}^{e\gamma^*\to e'\gamma\gamma})^* \; ,
\end{equation}
where now $M_{\mu}^{e\gamma^*\to e'\gamma\gamma}$ is the matrix element of
the double Compton process (\ref{eq:dbl-compt}), with the index $\mu$
describing the polarization of the virtual photon.


\subsubsection{Double collinear emission}

When both photons are emitted almost collinearly to the incoming electron,
it is convenient to express the photon four-momenta occurring in the
numerator of the expression for the double Compton tensor schematically as:
\begin{equation}
\label{eq:ki-approx}
  k_i = x_i p_1 + k_{i,\perp} \; .
\end{equation}
The components of the transverse parts $k_{i,\perp}$ are typically of the
order $\bigO{m}$ and limited to be smaller than $x_i E_e \vartheta_0$; they
can be safely neglected.  Similar simplifications can be applied for
$z_i'$, $\Delta'$ and $Q_h^2$.  We will therefore set:
\begin{eqnarray}
  z_i' & \simeq & 2 x_i p_1 \cdot p_2 \simeq x_i Q_l^2 \; ,
  \nonumber \\
  \Delta' & \simeq & (x_1+x_2) Q_l^2 \; ,
  \nonumber \\
  Q_h^2 & \simeq & (1-x_1-x_2) Q_l^2 \; ,
  \nonumber \\
  \tilde{k}_i & \simeq & x_i \tilde{p}_1 \; ,
\end{eqnarray}
and by momentum conservation:
\[
  \tilde{p}_2 \simeq
  \tilde{p}_1 - \tilde{k}_1 - \tilde{k}_2 = (1-x_1-x_2) \tilde{p}_1 \; .
\]
The expression for the double Compton tensor simplifies tremendously when
we note that in the double collinear region:
\[
   \frac{z_i}{Q_l^2} \, , \; \frac{\Delta}{Q_l^2} \;
   \lesssim \; \bigO{\vartheta_0^2} \; .
\]
Introducing the abbreviations
\begin{equation}
  r_1=1-x_1 \; , \quad r_2=1-x_2 \; , \quad z=1-x_1-x_2 \; ,
\end{equation}
the double Compton tensor takes a rather compact form:
\begin{eqnarray}
\label{eq:K-2-coll}
  K_{\mu\nu}^\mathrm{2-coll} & = &
  \left[
  - \tilde{g}_{\mu\nu} Q_l^2
  + 4 z \left( \tilde{p}_{1\mu} \tilde{p}_{1\nu} \right)
  \right]
  \nonumber \\ & \times &
  \Biggl[
    \frac{1+z^2}{x_1 x_2} \frac{1}{z_1 z_2}
  - \frac{z}{\Delta^2} \left( \frac{z_1}{z_2} + \frac{z_2}{z_1} \right)
  + \frac{1}{x_1 x_2}
    \left( \frac{r_1^3+z r_2}{z_1 \Delta}
         + \frac{r_2^3+z r_1}{z_2 \Delta} \right)
  \nonumber \\ &&
  \; {}
  - 2 \frac{m^2}{\Delta}
    \left( \frac{r_1^2+z^2}{x_2 z_1^2}
         + \frac{r_2^2+z^2}{x_1 z_2^2}
         + \frac{(1-z)(r_1 r_2 + z)}{x_1 x_2 z_1 z_2}
    \right)
  \\ &&
  \; {}
  - 4 z\, \frac{m^2}{\Delta^2} \left( \frac{1}{z_1} + \frac{1}{z_2} \right)
  + 4 z\, \frac{m^4}{\Delta^2} \left( \frac{1}{z_1} + \frac{1}{z_2} \right)^2
  \Biggr] \; .
  \nonumber
\end{eqnarray}
This expression is consistent with \cite{Mer88}, where leading and
next-to-leading logarithms were calculated.


\subsubsection{The integration over the photon phase space}

We assume that the photon detector measures only the sum of the energies
even if the two photons hit in different positions.  Taking into account
the symmetry factor $1/2!$ for the emitted photons, we need to calculate:
\begin{equation}
\label{eq:int-2-coll}
  \frac{e^4}{4} \cdot
  \frac{1}{2!}
  \int\limits_\mathrm{PD}
  \widetilde{\dd k}_1 \, \widetilde{\dd k}_2 \;
  \Theta(x_1 - \eps) \, \Theta(x_2 - \eps) \;
  \delta(x_1 + x_2 - (1-z)) \;
  K_{\mu\nu}^\mathrm{2-coll} \; .
\end{equation}
The infrared cutoff $\eps$ is necessary to regularize the singularity when
either photon becomes soft.

For a general shape of the PD the integrations in (\ref{eq:int-2-coll}) can
only be performed numerically.  To proceed analytically we require, in
addition to $\vartheta_0 \ll 1$, azimuthal symmetry of the PD.  The result
obtained then also serves as a useful cross check of a numerical
implementation.  The integrations over the photons in (\ref{eq:int-2-coll})
are therefore understood as
\begin{eqnarray}
  \int\limits_\mathrm{PD} \widetilde{\dd k}_i
  & = &
  \frac{1}{(4\pi)^2}
  \int x_i \, \dd x_i \;
  \frac{E_e^2}{\pi}
  \int \dd \Omega_i \; \Theta(\vartheta_0-\vartheta_i)
  \; ,
\end{eqnarray}
with no restriction on the azimuthal integration.

Under these conditions the integrals over the solid angles of the photons
can be performed completely.  A list of the relevant integrals is given in
appendix~\ref{sec:double-coll-int}, along with a description of their
calculation.

The remaining integration over the relative photon energies however is very
tedious.  Nevertheless, one can decompose the result of
(\ref{eq:int-2-coll}) in the following form:
\begin{equation}
\label{eq:2-coll-int}
  \left[
  - \tilde{g}_{\mu\nu} Q_l^2
  + 4 z \left( \tilde{p}_{1\mu} \tilde{p}_{1\nu} \right)
  \right] \times
  \frac{\alpha^2}{8 \pi^2}
  \left[
    P^{\mathrm{(2)}}_\mathrm{log}(z)
    + P^{\mathrm{(2),IR-div.}}_\mathrm{nonlog}(z)
    + P^{\mathrm{(2),IR-fin.}}_\mathrm{nonlog}(z)
  \right] \; .
\end{equation}
The previously known leading terms containing double and single logarithms
$L_0$ are contained in $P^{\mathrm{(2)}}_\mathrm{log}(z)$:
\begin{eqnarray}
\label{eq:P2-log}
  P^{\mathrm{(2)}}_\mathrm{log}(z)
  & = &
  \left[
    - 4 \frac{1+z^2}{1-z} \ln \frac{\eps}{1-z}
    + (1+z) \ln z - 2(1-z)
  \right] L_0^2
  \nonumber \\
  &+&
  \left[
    6(1-z)
    + \frac{3+z^2}{1-z} \ln^2 z
    + 4 \frac{(1+z)^2}{1-z} \ln \frac{\eps}{1-z}
  \right] L_0
  \nonumber \\
  & = &
  \left[
    P_\Theta^{(2)}(z)
    + 2 \, \frac{1+z^2}{1-z}
    \left( \ln z - \frac{3}{2} - 2 \ln \eps \right)
  \right] L_0^2
  \\
  &+&
  \left[
    6(1-z)
    + \frac{3+z^2}{1-z} \ln^2 z
    + 4 \frac{(1+z)^2}{1-z} \ln \frac{\eps}{1-z}
  \right] L_0
  \; , \nonumber
\end{eqnarray}
with the second-order leading-log radiator
\begin{eqnarray*}
  P_\Theta^{(2)}(z)
  & = &
  2 \left[
    \frac{1+z^2}{1-z} \left( 2 \ln(1-z) - \ln z + \frac{3}{2} \right)
    + \frac{1}{2} (1+z) \ln z - 1 + z
  \right] \; .
\end{eqnarray*}
The nonleading terms are split into an infrared divergent piece that
depends on $\ln \eps$,
\begin{eqnarray}
\label{eq:P-nonlog-IR-div}
  P_\mathrm{nonlog}^\mathrm{IR-div}(z)
  = \frac{8z}{1-z} \ln \frac{\eps}{1-z}
  \; ,
\end{eqnarray}
and an infrared finite piece $P^{\mathrm{(2),IR-fin.}}_\mathrm{nonlog}(z)$.
An outline of the calculation and more information on the last piece are
given in appendix~\ref{sec:double-coll-int}.

Inspecting (\ref{eq:P2-log}), (\ref{eq:P-nonlog-IR-div}) we find that the
terms depending on the soft-photon cutoff parameter $\eps$ in
(\ref{eq:2-coll-int}) do factor nicely, as expected from the usual
soft-photon factorization:
\begin{equation}
  \left[
  - \tilde{g}_{\mu\nu} Q_l^2
  + 4 z \left( \tilde{p}_{1\mu} \tilde{p}_{1\nu} \right)
  \right] \times
  \left(\frac{\alpha}{2 \pi}\right)^2 P(z,L_0)
  \cdot 2(L_0-1) \ln \frac{1}{\eps}
  \; .
\end{equation}


\subsection{Final state collinear radiation}

Consider now the kinematic region where one photon is emitted almost
collinearly to the incoming electron and the other one close to the
outgoing electron.  We suppose that the angle between the scattered
electron and the second photon is smaller than some angular separation
parameter $\vartheta_0'$, i.e.,
\begin{equation}
\label{eq:cluster-cond}
  \Angle(\vec{k}_2,\vec{p}{\,'}) \leq \vartheta'_0
  \; , \quad \mbox{with} \quad \vartheta'_0 \ll \theta
  \; .
\end{equation}
Without loss of generality we therefore set:
\begin{equation}
\label{eq:k1k2-isr+fsr}
  k_1 = x_1 p_1 + k_{1,\perp}
  \; , \quad
  k_2 = \frac{\xi}{1-\xi} \, p_2 + k_{2,\perp}
  \; ,
\end{equation}
where we again assume that the indicated transverse parts of the photon
momenta are small, being typically of the order $\bigO{m}$, c.f.\ the
discussion of (\ref{eq:ki-approx}).  These simplifications lead to:
\begin{eqnarray}
  z_1' & \simeq & x_1 Q_l^2
  \; , \quad
  z_2 \simeq \frac{\xi}{1-\xi} Q_l^2 \; ,
  \nonumber \\
  \Delta & \simeq & (1-x_1) \frac{\xi}{1-\xi} Q_l^2 \; ,
  \nonumber \\
  \Delta' & \simeq & \frac{x_1}{1-\xi} Q_l^2 \; ,
  \nonumber \\
  Q_h^2 & \simeq & \frac{1-x_1}{1-\xi} Q_l^2
  \nonumber \\
  \tilde{k}_1 & \simeq & x_1 \tilde{p}_1 \; , \quad
  \tilde{k}_2 \simeq \frac{\xi}{1-\xi} \tilde{p}_2 \; ,
\end{eqnarray}
and by momentum conservation:
\[
  \tilde{p}_2 + \tilde{k}_2 =
  \tilde{p}_1 - \tilde{k}_1
  \quad \Rightarrow \quad
  \tilde{p}_2 \simeq (1-x_1)(1-\xi) \tilde{p}_1 \; .
\]
Furthermore:
\[
  \frac{z_1}{Q_l^2} \; \lesssim \; \bigO{\vartheta_0^2}
  \; , \quad
  \frac{z_2'}{Q_l^2} \; \lesssim \; \bigO{\vartheta_0'{}^2} \; ,
\]
and the double Compton tensor considerably simplifies to
\begin{eqnarray}
\label{eq:K-fsr}
  K_{\mu\nu}^{2\gamma,\mathrm{FSR}}
  & \simeq &
  \left[
    - \tilde{g}_{\mu\nu} \, \frac{Q_l^2}{1-\xi}
    + 4 (1-x_1) \cdot \left( \tilde{p}_{1\mu} \tilde{p}_{1\nu} \right)
  \right]
  \nonumber \\ & \times &
  \left[
    \frac{1+(1-x_1)^2}{x_1}\frac{1}{z_1} - 2(1-x_1) \frac{m^2}{z_1^2}
  \right]
  \nonumber \\ & \times &
  \left[
    \frac{1+(1-\xi)^2}{\xi}\frac{1}{z_2'} - 2 \frac{m^2}{z_2'{}^2}
  \right]
  \; ,
\end{eqnarray}
exhibiting a complete factorization of collinear initial and final state
radiation, respectively.  For the application to the tagged photon process,
we have to identify $x_1=1-z$ in the above expressions.

\subsubsection{The integration over the photon phase space}

Integrating the expression (\ref{eq:K-fsr}) for the double Compton tensor
over the solid angle of photon 1 at fixed energy we obtain:
\begin{eqnarray}
\label{eq:int-fsr-1}
  \frac{E_e^2}{\pi} \int\limits_\mathrm{PD} \dd\Omega_1 \;
  K_{\mu\nu}^{2\gamma,\mathrm{FSR}}
  & = &
  \left[
    - \tilde{g}_{\mu\nu} \, \frac{Q_l^2}{1-\xi}
    + 4 (1-x_1) \cdot \left( \tilde{p}_{1\mu} \tilde{p}_{1\nu} \right)
  \right] \times
  \nonumber \\ &&
  \frac{1}{x_1} \, P(1-x_1,L_0) \times
  \nonumber \\ &&
  \left[
    \frac{1+(1-\xi)^2}{\xi}\frac{1}{z_2'} - 2 \frac{m^2}{z_2'{}^2}
  \right] \; .
\end{eqnarray}
The corresponding integration over the final state photon depends on the
details of the experimental situation, i.e., whether the detector is able
to resolve a photon collinear to the electron, or whether it just measures
the sum of their energies.  We denominated these cases in the discussion in
section~\ref{sec:single-rad} as \emph{exclusive} and \emph{calorimetric}
measurement, respectively.

Let us consider first the exclusive case, when only $p_2$ of the scattered
electron is measured but the momentum $k_2$ of the almost collinear photon
is missed.  The phase space for the second photon can be rewritten as:
\begin{equation}
  \int \widetilde{\dd k}_2 =
  \frac{1}{(4\pi)^2} \,
  \frac{\Eep^2}{\pi}
  \int \frac{\xi \, \dd \xi}{(1-\xi)^3}
  \int\limits_{\vartheta_2' \leq \vartheta'_0} \dd \Omega_2'
  \; ,
\end{equation}
with $\Eep$ being the measured energy of the scattered electron and the
polar angle $\vartheta_2'$ of the photon being measured with respect to the
scattered electron.  Performing the integration over the solid angle leads
to:
\begin{eqnarray}
\label{eq:fsr-int-excl}
  \lefteqn{
  \int \widetilde{\dd k}_2
  \left[
    \frac{1+(1-\xi)^2}{\xi}\frac{1}{z_2'} - 2 \frac{m^2}{z_2'{}^2}
  \right]
  } \nonumber \\
  & = &
  \frac{1}{(4\pi)^2}
  \int \frac{\dd \xi}{(1-\xi)^2} \,
  \left[
    \frac{1+(1-\xi)^2}{\xi}\,(L'-1) + \xi
  \right]
  \; ,
\end{eqnarray}
with
\begin{equation}
  L' = \ln \frac{E_e^2 \vartheta_0'{}^2}{m^2} + 2 \ln Y
  \; , \quad
  Y = \frac{\Eep}{E_e}
  \; .
\end{equation}
The lower limit of the $\xi$ integration in (\ref{eq:fsr-int-excl}) is
given by
\begin{equation}
  \xi_\mathrm{min} = \frac{\eps}{Y} \; ,
\end{equation}
while the upper limit depends on details of the full process and on the
kinematic reconstruction method.  It needs to be discussed in the concrete
application.

In the calorimetric case we assume that only the four-momentum of a cluster
(or electromagnetic jet) can be measured when condition
(\ref{eq:cluster-cond}) is met:
\begin{equation}
  p'_\mathrm{cal} \equiv p'_\mathrm{cluster} = p_2 + k_2 \; .
\end{equation}
Expressing the occurrences of $p_2$ in the numerator of (\ref{eq:int-fsr-1})
in terms of $p'_\mathrm{cal}$, the tensor structure in the first line
becomes independent of $\xi$,
\begin{equation}
  \left[
    - \tilde{g}_{\mu\nu} \, \left(Q^2_l\right)_\mathrm{cal}
    + 4 (1-x_1) \cdot \left( \tilde{p}_{1\mu} \tilde{p}_{1\nu} \right)
  \right]
  \; ,
\end{equation}
because the calorimetrically measured invariant momentum transfer is:
\begin{equation}
  \left(Q^2_l\right)_\mathrm{cal}
  = - (p_1 - p'_\mathrm{cal})^2
  = \frac{Q_l^2}{1-\xi} \simeq \frac{Q_h^2}{1-x_1} \; .
\end{equation}
In the definition of measured cross sections only $p'_\mathrm{cal}$ can be
used.  We therefore replace the phase space of the outgoing electron by the
cluster using the substitution:
\begin{equation}
  \int \widetilde{\dd p}_2
  \int \widetilde{\dd k}_2
  \quad \rightarrow \quad
  \int \widetilde{\dd p}'_\mathrm{cal}
  \int \widetilde{\dd k}_2
  \;
  (1-\xi)^2
  \; .
\end{equation}
Combining the Jacobian $(1-\xi)^2$ with the integration over the solid
angle of the photon yields:
\begin{eqnarray}
\label{eq:fsr-int-incl}
  \lefteqn{
  \int \widetilde{\dd k}_2
  \;
  (1-\xi)^2
  \left[
    \frac{1+(1-\xi)^2}{\xi}\frac{1}{z_2'} - 2 \frac{m^2}{z_2'{}^2}
  \right]
  } \nonumber \\
  & = &
  \frac{1}{(4\pi)^2}
  \int \dd \xi \,
  \left[
    \frac{1+(1-\xi)^2}{\xi}\,(L''-1) + \xi
  \right]
  \; ,
\end{eqnarray}
with
\begin{equation}
  L'' = \ln \frac{E_e^2 \vartheta_0'{}^2}{m^2} + 2 \ln Y
  + 2 \ln \frac{\Eep}{E'_\mathrm{cluster}}
  = L' + 2 \ln (1-\xi)
  \; ,
\end{equation}
since in the inclusive case the ratio $Y$ is given by:
\begin{equation}
  Y = \frac{E'_\mathrm{cluster}}{E_e} \; .
\end{equation}
Inserting the limits for the $\xi$-integration which are now independent of
the full process,
\begin{equation}
  \frac{\eps}{Y} \leq \xi \leq 1 \; ,
\end{equation}
and evaluating the r.h.s.\ of (\ref{eq:fsr-int-incl}) we finally obtain:
\begin{equation}
\label{eq:fsr-incl}
  \frac{1}{(4\pi)^2}
  \left[
    \left( 2 \ln \frac{Y}{\eps} - \frac{3}{2} \right)(L'-1)
    + 3 - \frac{2\pi^2}{3}
  \right]
  \; .
\end{equation}


\subsection{Semi-collinear emission}
\label{sec:sc}

The final situation covers the configuration where one photon is emitted
almost collinearly to the incoming electron, while the other photon is
emitted at a large angle with respect to both the incoming and outgoing
electron directions, i.e., for $\vartheta_2 > \vartheta_0$ and
$\vartheta_2' > \vartheta_0'$.  We denote this kinematic domain as the
semi-collinear one.

To be consistent with the above treatment of the double collinear emission,
we need to perform the solid angle integration over the collinear photon
and drop all contributions of the order $\mathcal{O}(\vartheta_0)$ and
$\mathcal{O}(\zeta_0^{-1})$.
In doing so, a subtlety arises from the propagator denominators
$\Delta=z_1+z_2-\sigma$.  For strongly ordered angles $\vartheta_1 \ll
\vartheta_2$, corresponding to $z_1 \ll z_2$, one may approximate:
\begin{equation}
  \label{eq:sc-approx}
  k_1 \simeq x_1 p_1
  \; , \quad
  \Delta \simeq (1-x_1) z_2 + \bigO{z_1}
  \qquad (\mbox{for} \; \vartheta_1 \ll \vartheta_2)
  \; .
\end{equation}
This leads to a factorized expression not only for the differential cross
section, but also after integration over the collinear photon, as long as
$\vartheta_2 \gg \vartheta_0$.
However, for $\vartheta_1 \approx \vartheta_2$, which can happen if photon
2 is emitted close to the forward cone that is specified by the solid angle
of the PD, approximation (\ref{eq:sc-approx}) no longer works well.  One
obtains contributions from further terms of an expansion around the limit
$\vartheta_1 \to 0$, which lead to a more complicated, steeper
$\vartheta_2$-dependence that spoils the factorization.
Fortunately, these terms fall off rapidly and essentially contribute only
in the small region $\vartheta_0 \lesssim \vartheta_2 < \mbox{few} \cdot
\vartheta_0$.  Since these extra-terms are important only for this
particular kinematic configuration of almost double collinear emission we
shall denote them the \emph{quasi-collinear contribution}.
The details of this calculation are given in appendix~\ref{sec:sc-calc}.

Assuming that photons in this narrow region outside the PD will not be
measured, we explicitly integrate these additional terms over the photon
solid angle and split their contribution schematically as follows:
\begin{eqnarray}
  \label{eq:leptonic-semi-coll}
  \lefteqn{
  \frac{E_e^4}{\pi^2}
  \int\limits_{\vartheta_1<\vartheta_0} \dd\Omega_1
  \int\limits_{\vartheta_2>\vartheta_0} \dd\Omega_2 \;
  K_{\mu\nu}^{\mathrm{semi-coll}}
  }
  \nonumber \\
  & \simeq &
  \frac{E_e^2}{\pi} \int\limits_{\vartheta_2>\vartheta_0} \dd\Omega_2 \;
  \Biggl\{
  \Biggl[
  - \tilde{g}^{\mu\nu}
    \frac{(r_1(Q_l^2+z_2))^2+(r_1Q_l^2-z_2')^2}{r_1 z_2 z_2'}
  \nonumber
  \\
  && \qquad \qquad \qquad \quad {}
  + 4 r_1^2 \tilde{p}_1^\mu \tilde{p}_1^\nu
    \frac{Q_h^2}{r_1 z_2 z_2'}
  + 4 \tilde{p}_2^\mu \tilde{p}_2^\nu
    \frac{Q_h^2}{r_1 z_2 z_2'}
  \Biggr]
  \cdot
  \frac{1}{x_1 r_1} \, P(r_1,L_0)
  \Biggr\}
  \nonumber \\
  & + &
  \left[
  - \tilde{g}_{\mu\nu} Q_l^2
  + 4 (r_1-x_2) \left( \tilde{p}_{1\mu} \tilde{p}_{1\nu} \right)
  \right] \cdot
  \frac{1}{x_1 x_2} \, H(x_1,x_2)
  \; ,
\end{eqnarray}
with $r_1=1-x_1$.  The expression for the function $H(x_1,x_2)$, which
collects the non-factorizing, quasi-collinear terms, is given in
appendix~\ref{sec:sc-calc}, eq.~(\ref{eq:function-H}).  It is
infrared-finite, $\vartheta_0$-independent for small $\vartheta_0$, and
does not contain any logarithm of a large scale.

The integrand in the first part on the r.h.s.\ of
(\ref{eq:leptonic-semi-coll}) can be rewritten as
\begin{equation}
\label{eq:leptonic-sc-fact}
  \Biggl\{ \ldots \Biggr\} \simeq
  \frac{1}{x_1} \, P(r_1,L_0)
  \cdot
  \frac{1}{r_1} K_{\mu\nu}^\mathrm{Born}(r_1 p_1,p_2,k_2)
  \; ,
\end{equation}
where for the sake of consistency one should drop terms of order $m^2$ on
the r.h.s.\ in the expression for the lowest order Compton tensor
(\ref{def:compton-tensor-Born}).

Since in our decomposition of phase space only photon 1 reaches the PD, we
have to identify $r_1$ by $z$ and $x_1$ by $1-z$ in the above expressions.
However, we still need to integrate over the phase space of the other
photon that is emitted at large angles.  This calculation depends on the
complete scattering process and in general requires a numerical
integration.


\section{Corrections to radiative DIS}

Let us turn to the description of the radiative deep inelastic scattering
process with photon tagging,
\[
  e(p) + p(P) \to
  e(p') + X(P_X) + \gamma(k_1) \; \left( {} + \gamma(k_2) \right)
  \; ,
\]
including the complete leptonic QED radiative corrections calculated above.
It is straightforward to contract the expressions for the radiatively
corrected Compton tensor with the hadron tensor (\ref{eq:hadron-tensor-H}).
We assume here the kinematic reconstruction by measurement of the scattered
lepton and later outline the changes for other methods.

Applying the results from the previous section, we find for the
contribution from virtual and soft corrections to the cross section:
\begin{equation}
\label{eq:V+S}
  \frac{1}{\hat{y}} \,
  \frac{\dd^3\sigma_\mathrm{V+S}}{\dd\hat{x}\,\dd\hat{y}\,\dd z} =
  \frac{\alpha^2}{4\pi^2} \left[ P(z,L_0) \tilde{\rho} - T \right]
  \tilde\Sigma(\hat{x},\hat{y},\hat{Q}^2) \; ,
\end{equation}
where $\tilde\rho$ is taken from (\ref{eq:rho-v+s}) with
\begin{eqnarray}
\label{eq:Y-c-elastic}
  Y & \equiv & \frac{\Eep}{E_e}
  = z(1 - \hat{y}) + \hat{x}\hat{y} \, \frac{E_p}{E_e}
  \; ,
  \nonumber \\
  c & \equiv & \cos\theta =
  \frac{z(1 - \hat{y}) E_e - \hat{x}\hat{y} E_p}
       {z(1 - \hat{y}) E_e + \hat{x}\hat{y} E_p}
  \; .
\end{eqnarray}

In the calculation of the contributions from the emission of two hard
photons, we decompose the phase space into three regions discussed in the
previous section (see also \cite{AAKM:nlo}): (i) both hard photons hit the
forward photon detector, i.e., both are emitted within a narrow cone around
the electron beam $(\vartheta_{1,2} \leq \vartheta_0, \, \vartheta_0 \ll
1)$; (ii) one photon is tagged in the PD, while the other is collinear to
the outgoing electron $(\vartheta_2' \equiv \Angle(\vec{k}_2,\vec{p}{\,'})
\leq \vartheta'_0)$; and finally (iii) the second photon is emitted at
large angles (i.e., outside the defined narrow cones) with respect to both
incoming and outgoing electron momenta.  For the sake of simplicity, we
assume that $m/E_e \ll \vartheta'_0 \ll \theta$.

The contribution from the kinematic region (i), with both hard photons
being tagged in the PD, but only the sum of their energies measured, reads:
\begin{eqnarray}
\label{eq:sig-i}
  \frac{1}{\hat{y}} \,
  \frac{\dd^3 \sigma^{\gamma\gamma}_\mathrm{(i)}}{\dd\hat{x}\,\dd\hat{y}\, \dd z}
  & = &
  \frac{\alpha^2}{8\pi^2}
  \left[
    P^{\mathrm{(2)}}_\mathrm{log}
    + P^{\mathrm{(2),IR-div.}}_\mathrm{nonlog}
    + P^{\mathrm{(2),IR-fin.}}_\mathrm{nonlog}
  \right]
  \tilde\Sigma
  \; ,
\end{eqnarray}
see eq.~(\ref{eq:2-coll-int}).

In region (ii) we need to distinguish between the cases of whether the
outgoing electron in the presence of a collinear photon can be measured
separately (exclusively), or whether its energy and momentum are detected
together with the electron (inclusively), as this affects the reconstructed
kinematic variables.

For the exclusive event selection, when only the scattered electron is
detected, we obtain from (\ref{eq:int-fsr-1}) and (\ref{eq:fsr-int-excl}):
\begin{equation}
\label{eq:sig-ii-excl}
  \frac{1}{\hat{y}} \,
  \frac{\dd^3 \sigma^{\gamma\gamma}_{\mathrm{(ii),excl}}}%
       {\dd\hat{x}\,\dd\hat{y}\, \dd z} =
  \frac{\alpha^2}{4\pi^2} P(z,L_0)
  \int\limits_{\xi_\mathrm{min}}^{\xi_\mathrm{max}}
  \frac{\dd \xi}{(1-\xi)^2}
  \left[ \frac{1+(1-\xi)^2}{\xi} \left(L'-1\right) + \xi \right]
  \tilde\Sigma_f \; ,
\end{equation}
where
\begin{eqnarray}
  \tilde\Sigma_f & = & \tilde\Sigma(x_f,y_f,Q_f^2)
  \; , \quad
  L' = \ln \frac{E_e^2\vartheta_0'{}^2}{m^2} + 2 \ln Y
  \; , \nonumber \\
  x_f  & = & \frac{\hat{x} \hat{y}}{\hat{y} - \xi}
  \; , \quad
  y_f = 1 - \frac{1-\hat{y}}{1-\xi}
  \; , \quad
  Q_f^2 = \frac{\hat{Q}^2}{1-\xi}
  \; , \\
  \xi_\mathrm{min} & = & \frac{\eps}{Y}
  \; , \quad
  \xi_\mathrm{max}
  = \frac{(1-\hat{x})\hat{y} - \left(\bar{M}^2 - M^2 \right)/(zS)}%
         {1 - \left(\bar{M}^2 - M^2 \right)/(zS)}
  \simeq (1-\hat{x})\hat{y}
  \; . \nonumber
\end{eqnarray}

In the case of a calorimetric event selection, where only the sum of the
energies of the outgoing electron and collinear photon is measured and
taken into account in the determination of the kinematic variables, the
corresponding contribution from (\ref{eq:int-fsr-1}) and
(\ref{eq:fsr-incl}) is proportional to the lowest order contribution and
reads
\begin{eqnarray}
\label{eq:sig-ii-cal}
  \frac{1}{\hat{y}} \,
  \frac{\dd^3 \sigma^{\gamma\gamma}_{\mathrm{(ii),cal}}}{\dd\hat{x}\,\dd\hat{y}\, \dd z}
  & = &
  \frac{\alpha^2}{4\pi^2} P(z,L_0)
  \!\!
  \int\limits_{\xi_\mathrm{min}}^1
  \!\! \dd \xi
  \left[ \frac{1+(1-\xi)^2}{\xi}
    \left(L' - 1 + 2\ln(1-\xi) \right) + \xi \right]
  \tilde\Sigma
  \nonumber \\
  & = &
  \frac{\alpha^2}{4\pi^2} P(z,L_0)
  \left[
    \left( 2 \ln \frac{Y}{\eps} - \frac{3}{2} \right)(L'-1)
    + 3 - \frac{2\pi^2}{3}
  \right]
  \tilde\Sigma
  \, , \qquad \;
\end{eqnarray}
see also \cite{AAKM:JETP,AAKM:nlo,Anl99:Sigma}.

For the calculation of the contribution from the semi-collinear region
(iii) we apply the decomposition (\ref{eq:leptonic-semi-coll}).  Due to the
factorization property (\ref{eq:leptonic-sc-fact}) of the leading term, it
is useful to introduce the ``radiation kernel'' \cite{AAKM:nlo}:
\begin{eqnarray}
\label{eq:Igamma}
  I^\gamma(zp,p',k_2)
  & \equiv & \frac{1}{8\pi} \,
  K_{\rho\sigma}(zp,p',k_2)H^{\rho\sigma}(P,zp-p'-k_2)
  \\
  & = &
  \frac{1}{z z_2 z_2'}
  \Biggl[
  G \cdot F_1(x_h,Q_h^2) +
  \Biggl(
    x_h (zS)^2[1+(1-\hat{y})^2]
    - \frac{x_h M^2}{Q_h^2} G
  \nonumber \\ && \qquad \; {}
    + (zS) [ (1-\hat{y})(\hat{Q}_l^2-z_2')-(\hat{Q}_l^2+z z_2)]
  \Biggr) F_2(x_h,Q_h^2)
  \Biggr]
  \, , \nonumber
\end{eqnarray}
which can be obtained from eqs.\ (\ref{eq:KdotH-Born}), (\ref{eq:S1S2}).
Here:
\begin{eqnarray}
  Q_h^2 & = & \hat{Q}_l^2 + z z_2 -z_2'
  \; , \quad
  z_2  = 2p \cdot k_2
  \; , \quad
  z_2' = 2p' \cdot k_2
  \; ,
  \nonumber \\
  x_h & = &
  \frac{Q_h^2}{2P \cdot (zp-p'-k_2)}
  = \frac{Q_h^2}{\hat{y}zS - 2P \cdot k_2}
  \; ,
  \nonumber \\
  G & = &
  \left(\hat{Q}_l^2+z z_2\right)^2+\left(\hat{Q}_l^2-z_2'\right)^2
  \; .
\end{eqnarray}
The semi-collinear contribution then reads:
\begin{eqnarray}
\label{eq:sig-iii}
  \frac{1}{\hat{y}} \,
  \frac{\dd^3 \sigma^{\gamma\gamma}_\mathrm{(iii)}}{\dd\hat{x}\,\dd\hat{y}\, \dd z}
  & = &
  \frac{\alpha^2}{\pi^2} \, P(z,L_0)
  \int\frac{\dd^3 k_2}{|\vec{k}_2|}
  \, \frac{\alpha^2(Q_h^2)}{Q_h^4}
  \, I^{\gamma}(zp,p',k_2)
  \nonumber \\
  & + &
  \frac{\alpha^2}{4\pi^2}
  \int\limits_{\eps}^{x_2^t}
  \dd x_2 \,
  \frac{z}{z-x_2} \, H(1-z,x_2) \, \tilde\Sigma(x_t,y_t,Q_t^2)
  \; .
\end{eqnarray}
In the integration over the solid angle the cones corresponding to double
collinear ISR (region (i), half opening angle $\vartheta_0$) and FSR
(region (ii), half opening angle $\vartheta_0'$) have to be excepted.
The upper limit on the photon energy is obtained from
(\ref{eq:Egam-limit-explicit}) by substituting $E_e \to zE_e$, $S \to zS$,
$Y \to Y/z$:
\begin{equation}
  E_2^\mathrm{max} =
  \frac{S[z-Y(1-\tau)] - 4E_e^2zY\tau - \left(\bar{M}^2 - M^2 \right)}
       {4\left[E_p(1-\tau_1) + zE_e\tau_1 - YE_e\tau_2 \right]}
  \, .
\end{equation}
For the second, quasi-collinear part, we have used the abbreviations
\begin{equation}
  x_t = \frac{(z-x_2) \hat{x}\hat{y}}{z\hat{y}-x_2}
  \; , \quad
  y_t = \frac{z\hat{y}-x_2}{z-x_2}
  \; , \quad
  Q_t^2 = \hat{Q}^2 \frac{z-x_2}{z}
  \; ,
\end{equation}
and under HERA conditions and for small $\hat{x}$ the upper integration
limit simplifies to
\begin{equation}
  \label{eq:x2t}
  x_2^t
  = z \frac{(1-\hat{x})\hat{y} - \left(\bar{M}^2 - M^2 \right)/(zS)}%
	   {1 - \hat{x}\hat{y}}
  \simeq z \hat{y} \; .
\end{equation}

The total contribution from QED radiative corrections is finally obtained
by adding up (\ref{eq:V+S}), (\ref{eq:sig-i}), (\ref{eq:sig-iii}), and,
depending on the chosen event selection, (\ref{eq:sig-ii-excl}) or
(\ref{eq:sig-ii-cal}).  As has been demonstrated explicitly in
\cite{AAKM:nlo} the unphysical IR regularization parameter $\eps$ cancels
in the sum.

In the case of a bare electron measurement, the angle $\vartheta_0'$ plays
the r\^ole of an (unphysical) regularization parameter that serves to
define the phase space region of collinear final state radiation.  It can
be shown to cancel between the contributions from regions (ii) and (iii)
and to drop out of the final result if chosen small enough, see
\cite{AAKM:nlo}.

For a calorimetric measurement of the scattered electron, $\vartheta_0'$
corresponds to a resolution of the detector and is actually physical.
Therefore the cross section will depend on it.  However, in this case the
mass singularity that is connected with final state collinear emission
cancels in the sum of the contributions from virtual and soft corrections
and from regions (ii) and (iii), in accordance with the
Kinoshita-Lee-Nauenberg theorem \cite{KLN}.  For sufficiently small
$\vartheta_0'$ the resulting cross section depends logarithmically on
$\vartheta_0'$, while for a coarse detector, i.e., for $\vartheta_0' \sim
\bigO{1}$, the result agrees at the leading logarithmic level with the
correction obtained in section~\ref{sec:tagged-ll}.




\begin{figure}[tb]
  \begin{center}
    \begin{picture}(100,90)
     \put(0,0){\includegraphics[scale=0.5]{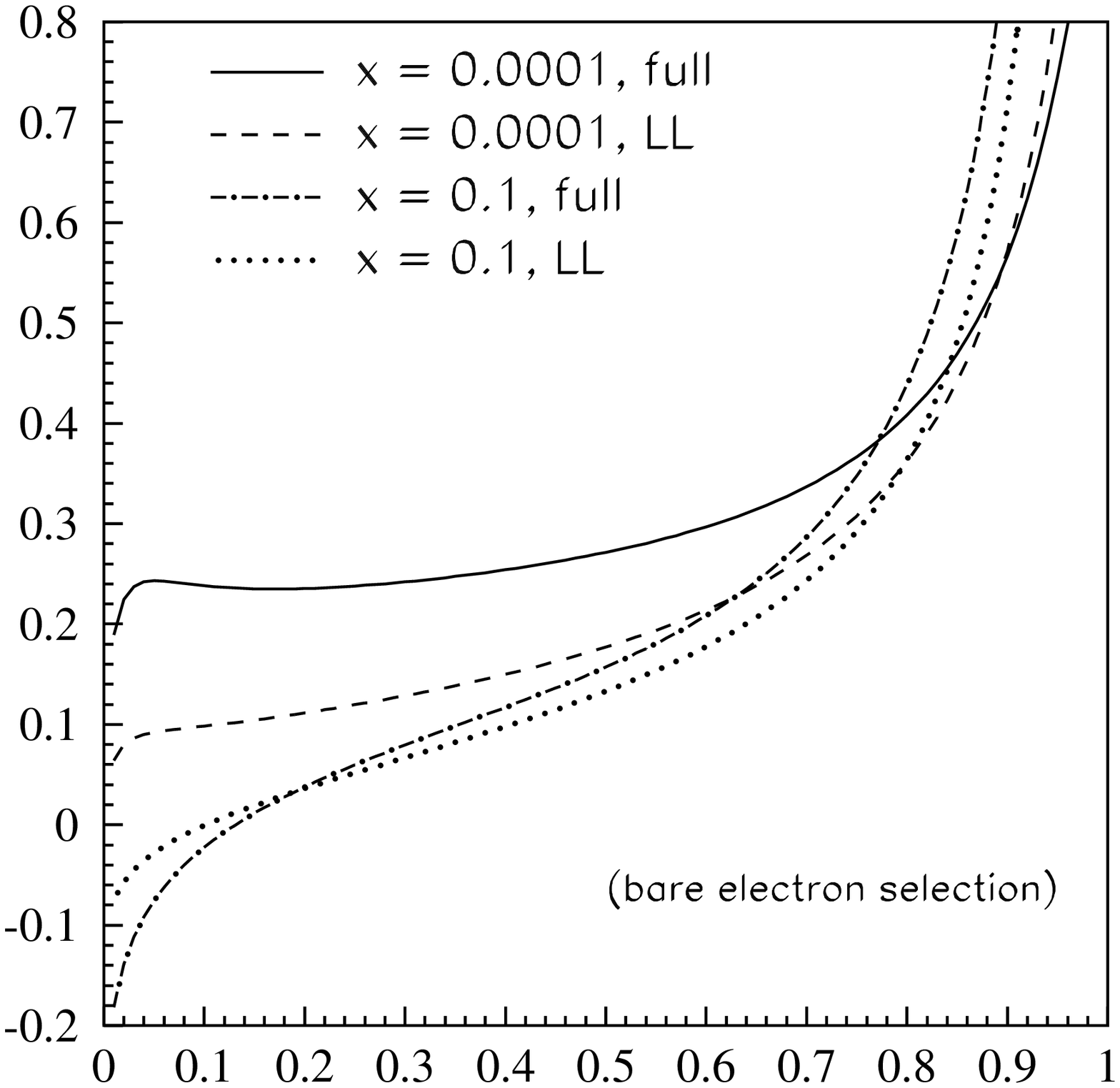}}
     \put(49,0){$\hat{y}$}
     \put(-6,50){$\delta_\mathrm{RC}$}
    \end{picture}
    \caption{Radiative corrections to the tagged photon cross section with
      leading logarithmic and with full leptonic accuracy for the bare
      electron selection and for $E_\gamma=10\GeV$.}
    \label{fig:corr-e-bare}
  \end{center}
\end{figure}


Let us now compare the radiative corrections calculated with full leptonic
accuracy to those obtained by considering only the leading logarithms.  For
the same set of parameters as in section~\ref{sec:tagged-ll},
figure~\ref{fig:corr-e-bare} displays the radiative correction
\begin{equation}
\label{eq:delta-rc}
	\delta_\mathrm{RC} =
	\frac{\dd^3\sigma}{\dd^3\sigma_\mathrm{Born}} - 1
\end{equation}
at leading logarithmic and with full leptonic accuracy for the bare
electron measurement for $\hat{x}=0.1$ and $\hat{x}=10^{-4}$ and for a
tagged energy $E_\gamma=10\GeV$.  We find a large difference of the order
of 5 to 10 percent between both calculations, especially in the region of
small $\hat{x}$ and for large $\hat{y}$.


\begin{figure}[tb]
  \begin{center}
    \begin{picture}(100,90)
     \put(0,0){\includegraphics[scale=0.5]{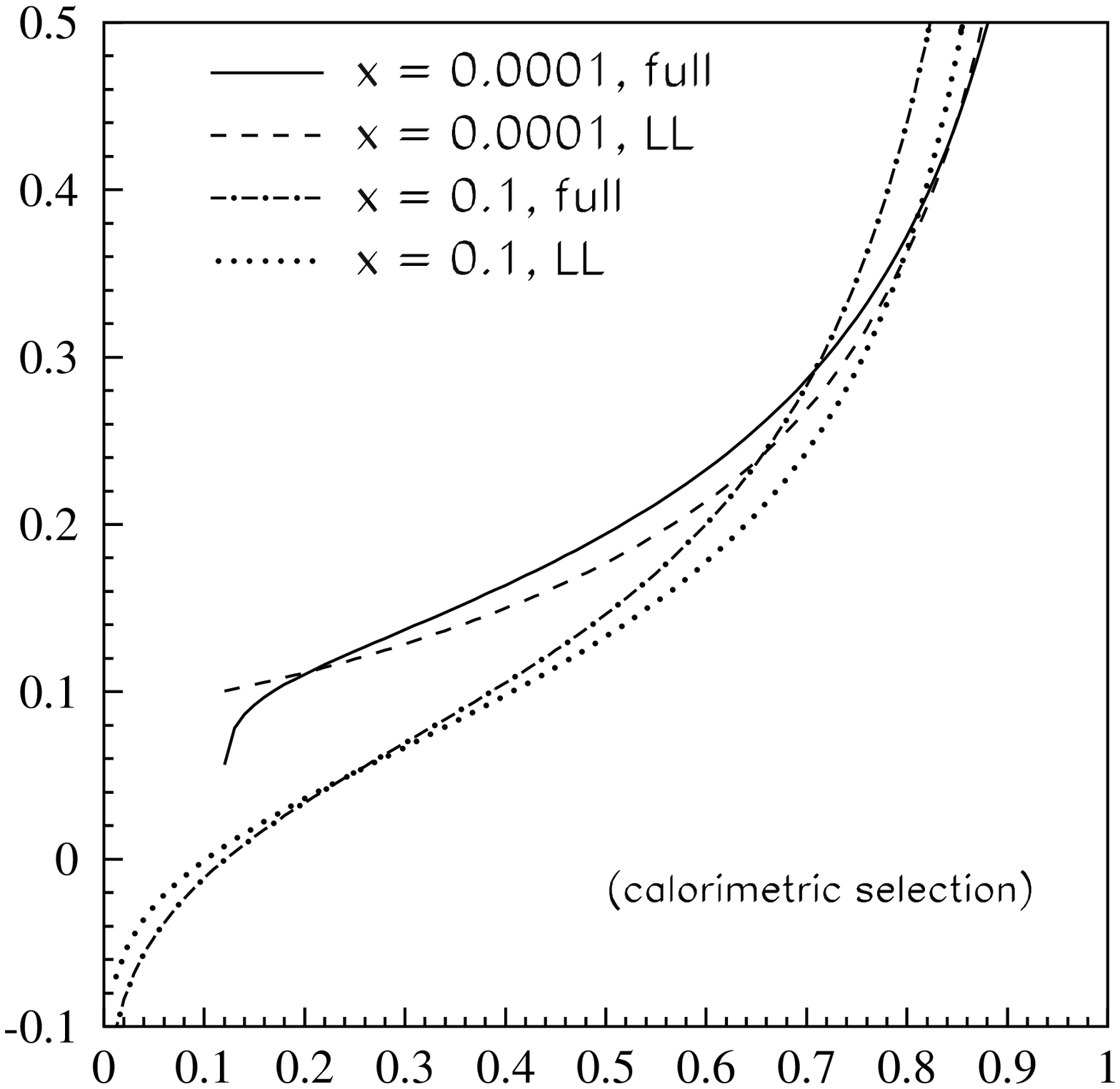}}
     \put(49,0){$\hat{y}$}
     \put(-6,50){$\delta_\mathrm{RC}$}
    \end{picture}
    \caption{Radiative corrections to the tagged photon cross section with
      leading logarithmic and with full leptonic accuracy for calorimetric
      event selection with $\vartheta_0' = 50 \mrad$ and for
      $E_\gamma=10\GeV$.} 
    \label{fig:corr-e-calo}
  \end{center}
\end{figure}


In the case of the HERA main detectors, the measurement of the energy of
the scattered electron in the electromagnetic calorimeter is performed with
a typical separation from an almost collinear photon of the order of
(several cm)/(1 meter) $\sim \bigO{50\mrad}$.  Figure~\ref{fig:corr-e-calo}
displays the corresponding comparison of the radiative corrections for the
calorimetric selection, assuming
\begin{equation}
  \vartheta_0' = 50 \mrad \; .
\end{equation}
Apparently the difference between the leading logarithmic and the full
leptonic result becomes significantly smaller, especially at smaller
$\hat{x}$ and for small $\hat{y}$, reaching a level of the order of 5
percent.  This suggests that the leading logarithmic approximation gives a
reasonable estimate for the radiative corrections for a sufficiently
inclusive measurement, at least for the reconstruction of the kinematic
variables using the lepton method.


\begin{figure}[tb]
  \begin{center}
    \begin{picture}(100,90)
     \put(0,0){\includegraphics[scale=0.5]{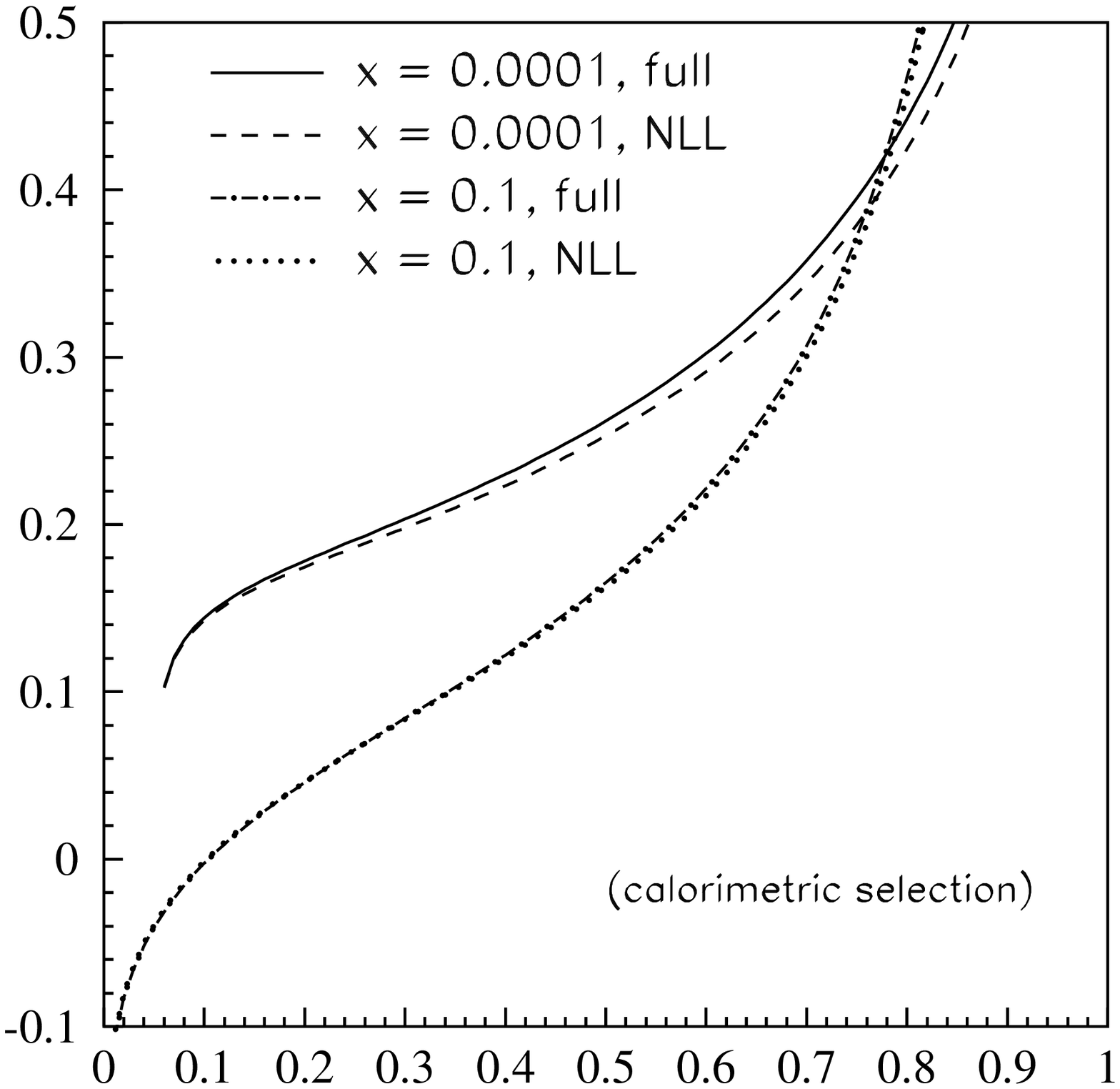}}
     \put(49,0){$\hat{y}$}
     \put(-6,50){$\delta_\mathrm{RC}$}
    \end{picture}
    \caption{Comparison of the radiative corrections with next-to-leading
      logarithmic and with full leptonic accuracy for a calorimetric event
      selection with $\vartheta_0' = 50 \mrad$ and for $E_\gamma=20\GeV$.}
    \label{fig:corr-e-acc}
  \end{center}
\end{figure}


In the references \cite{AAKM:JETP,AAKM:nlo,Anl99:Sigma} a calculation of
the radiative corrections was performed that takes into account the leading
and next-to-leading logarithms of the full corrections.  We can now compare
this approximation with the full result.  It turns out that it is rather
close to the full calculation \cite{Anlauf:2001fu}, and that the
non-logarithmic terms, typically contributing at the per mille level, are
potentially significant only at small $\hat{x}$ and for a large tagged
photon energy.  To demonstrate this we compare in
figure~\ref{fig:corr-e-acc} the corrections in the calorimetric case taking
into account the next-to-leading logarithms vs.\ the full leptonic
corrections for a higher tagged energy of $E_\gamma = 20\GeV$.  Similar
results are obtained for the bare electron measurement as the
non-logarithmic terms do not change the contribution from region (ii) and
the large-angle part of region (iii).


\begin{figure}[tb]
  \begin{center}
    \begin{picture}(100,90)
     \put(0,0){\includegraphics[scale=0.5]{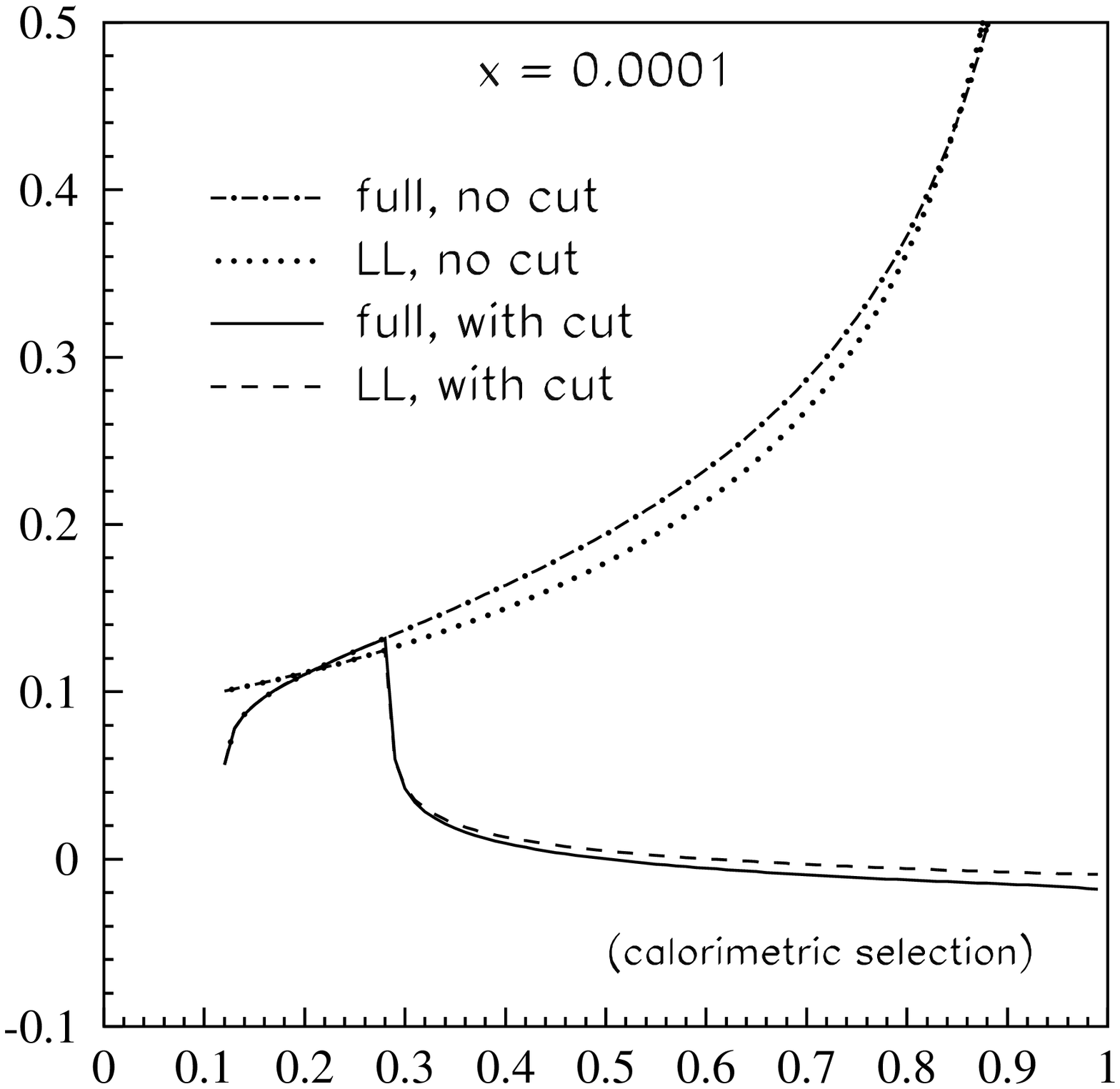}}
     \put(49,0){$\hat{y}$}
     \put(-6,50){$\delta_\mathrm{RC}$}
    \end{picture}
    \caption{Comparison of the radiative corrections with leading
      logarithmic and with full leptonic accuracy for a calorimetric event
      selection without cuts and with cut (\ref{eq:Emiss-cut}) for
      $E_\gamma=10\GeV$.}
    \label{fig:corr-e-cut}
  \end{center}
\end{figure}


The size of the contribution from hard photon emission to the radiative
corrections can be reduced if one applies experimental cuts.  As an example
which is applicable at HERA we define a ``missing longitudinal energy''
using longitudinal momentum conservation:
\begin{equation}
\label{eq:Emiss}
  E_\mathrm{miss} :=
  E_e - E_\mathrm{PD} - \frac{\Sigma_h + \Eep(1+\cos\theta)}{2}
  = \frac{P \cdot k_2}{2 E_p}
  = \frac{1}{2} \left( E_\gamma^{(2)} + p_{\gamma,z}^{(2)} \right)
  \; ,
\end{equation}
where $E_\mathrm{PD}$ represents the actually measured energy in the photon
detector.  Assuming 100\% efficiency of the PD and the second photon being
lost in the forward beam pipe outside the PD, the quantity
$E_\mathrm{miss}$ is roughly equal to the energy of this lost photon.  We
then apply the following cut:
\begin{equation}
\label{eq:Emiss-cut}
  \Delta_\mathrm{miss} := \frac{E_\mathrm{miss}}{E_\mathrm{PD}}
  < 0.5 \; ,
\end{equation}
and compare the radiative corrections with and without this cut for the
calorimetric selection.  For the same set of parameters and for
$\hat{x}=10^{-4}$ and $E_\gamma = 10\GeV$ the result is displayed in
figure~\ref{fig:corr-e-cut}.

We observe that the size of the corrections is much reduced, and that the
difference between the leading logarithmic and the full result is also
lessened slightly.  A main reason is that this cut is mostly efficient for
the contribution from very hard lost photons which would lead to the
largest shifts between reconstructed and ``true'' kinematic variables.
Therefore this cut is effective only for sufficiently large $\hat{y}$, see
also (\ref{eq:u0-l}).  It also affects the QED Compton contribution which
plays a r\^ole only at large $\hat{y}$.


\begin{figure}[tb]
  \begin{center}
    \begin{picture}(100,90)
     \put(0,0){\includegraphics[scale=0.5]{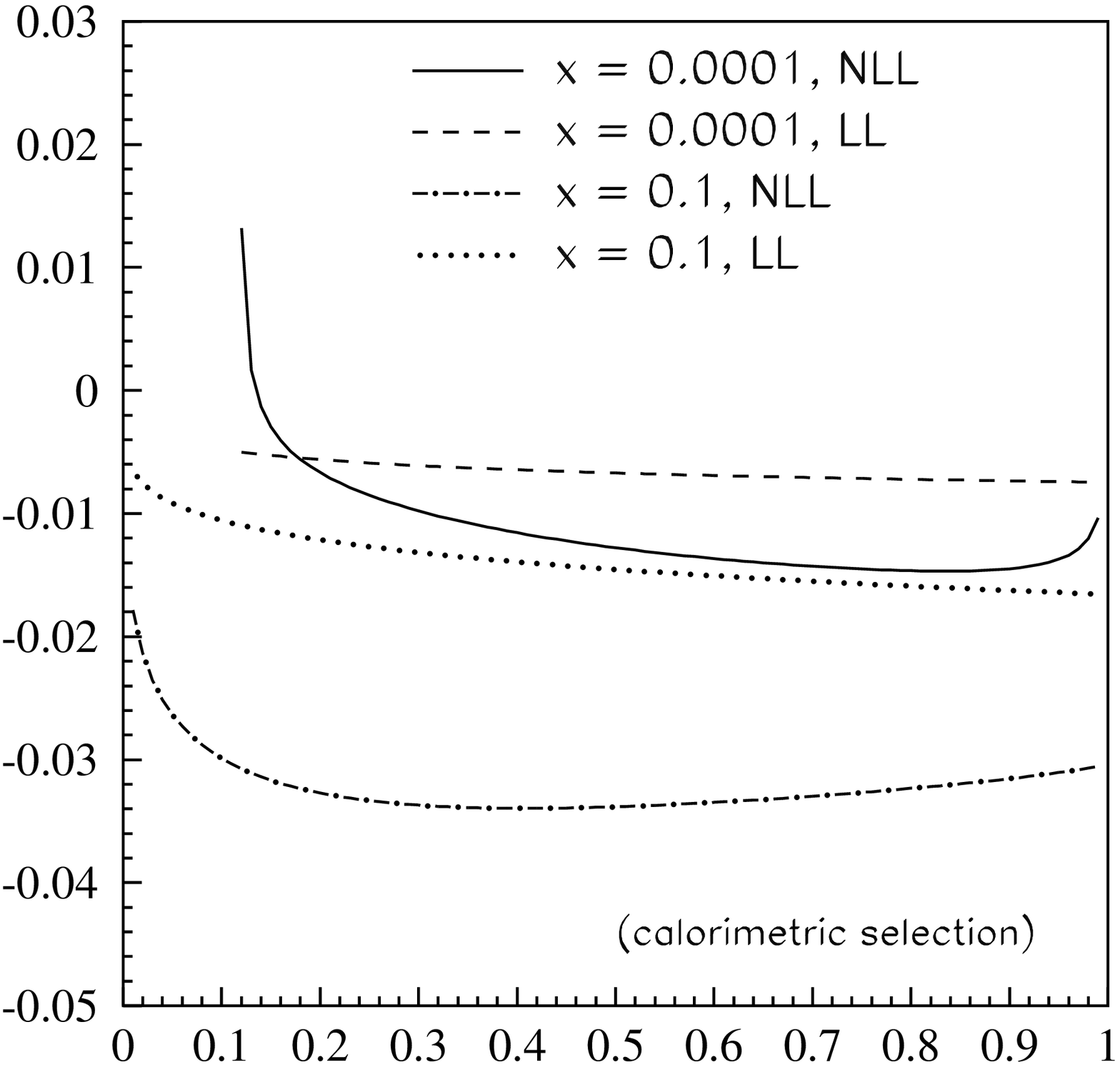}}
     \put(49,0){$\hat{y}$}
     \put(-9,50){$\delta_\mathrm{RC}$}
    \end{picture}
    \caption{Radiative corrections with leading and with next-to-leading
      logarithmic accuracy for the $\Sigma$ method for a calorimetric event
      selection with $\vartheta_0' = 50 \mrad$ and for $E_\gamma=10\GeV$.}
    \label{fig:corr-sig-calo}
  \end{center}
\end{figure}


For other reconstructions of the kinematic variables the above formulae
have to be adapted in the contributions from final state radiation (ii) and
for the semi-collinear contribution (iii).  Ref.~\cite{Anl99:Sigma}
discusses the modification of the calculation of the next-to-leading
logarithms for the $\Sigma$ method.  The derivation of the boundaries of
the photon phase space is more elaborate than for the electron method, so
that we will not treat it here and only quote the results.

We found in section~\ref{sec:tagged-ll} that the leading logarithmic
corrections for the $\Sigma$ method were quite small, see
fig.~\ref{fig:corr-ll-jb+sig}b.  The reason for this behavior can be
attributed to the scaling properties of the kinematic variables under
collinear photon emission, see table~\ref{tab:jacobian}.  However, as can
be seen from fig.~\ref{fig:corr-sig-calo} the more accurate next-to-leading
logarithmic calculation adds terms that are the order of a few percent and
thus of similar size as the leading logarithmic estimate, even for the
calorimetric selection.  As can be expected the corrections become
significantly larger for a bare electron measurement but still stay well
below those for the kinematic reconstruction using the lepton method.
Furthermore, the $R$ dependence of the corrections remains small even after
taking into account the next-to-leading logarithmic terms.  For more
details we refer the reader to \cite{Anl99:Sigma}.


\chapter{Estimates of Higher Order QED Corrections}
\label{sec:higher-orders}

In this chapter we shall attempt to obtain an all orders estimate of the
leading logarithmic contributions to the radiative corrections to the deep
inelastic scattering process with tagged initial state radiation.  To this
end, we will first rederive the final result of section \ref{sec:tagged-ll}
in a simpler framework.  We then generalize the arguments to all orders to
find a closed expression after resummation.

Our starting point will be the factorization theorem for the QED
radiatively corrected, photon inclusive cross section for high energy
scattering processes,
\begin{equation}
\label{eq:QED-ft}
  \sigma_\mathrm{RC}(s) =
  \int\limits_0^1 \dd z \; D(z,\mu_F^2) \, \sigma_h(zs;\mu_F^2)
  \; .
\end{equation}
Here $s$ stands for a typical large scale of the process.  All large
logarithms in the small electron mass $m$ are contained in the radiator
function $D(z,\mu_F^2)$, while the ``hard cross section'' $\sigma_h$ is
finite for $m \to 0$.  Also, the cancellation of infrared singularities
from QED virtual corrections and from soft photon emission takes place
within this radiator function at all orders.  The factorization theorem
(\ref{eq:QED-ft}) is a corollary of the corresponding theorems for the
factorization of infrared and mass singularities in QCD
\cite{QCD:factorization}, but has also been proven explicitly for QED in
\cite{Beenakker:1989km}.  Note that we have suppressed in (\ref{eq:QED-ft})
the dependence of the hard cross section on the renormalization scale
$\mu_R^2$, which is formally unrelated to $\mu_F^2$.
The purpose of factorization in QCD in the above example is to extract all
soft (long distance) contributions into the distribution $D$, so that the
hard process receives only perturbatively calculable short distance
corrections.

Eq.\ (\ref{eq:QED-ft}) introduces on the r.h.s.\ an arbitrary unphysical
scale, $\mu_F^2$.  Physical results, and thus the l.h.s.\ should in
principle be independent of the choice of this so-called factorization
scale.  This formal independence can be used to derive renormalization
group type equations.  In practical applications, however, things are more
complicated.  Calculating the radiator function using a partial resummation
of large terms at all orders in the coupling constant while treating the
hard part at fixed order in perturbation theory leads to an unavoidable
residual dependence of the r.h.s.\ on $\mu_F^2$ in higher orders that are
not completely calculated.  On the other hand, one can show that the
leading dependence of the uncalculated corrections to $\sigma_h$ is
logarithmic in $\mu_F^2$.  If all small scales can be extracted into the
radiator function $D$, one may hope that the missing logarithmic terms in
$\sigma_h$ will be small if this factorization scale is chosen close to a
typical large scale.

In the simplest case, the leading logarithmic approximation, the hard cross
section is approximated by the so-called ``improved Born cross section''
for the hard scattering process by replacing coupling constants by the
running couplings taken at the scale $\mu_F^2$.  For the photon inclusive
cross section with no explicit registration of the radiated photons there
essentially is a single large scale, e.g., the total energy $s$ or a
momentum transfer $Q^2$.  The ambiguity in selecting one of these scales
reflects the uncertainty of this approximation; it can only be resolved by
a calculation with higher (next-to-leading) accuracy, i.e., taking into
account terms beyond the Born cross section.  Nevertheless, the leading
logarithmic result often gives a reasonable estimate.  The structure
function method in the form as applied in section~\ref{sec:tagged-ll}
corresponds to this simple approximation.


\section{Exclusive photons in the leading logarithmic approximation}

The radiator function $D(z,\mu^2)$ appearing in (\ref{eq:QED-ft}) satisfies
an Altarelli-Parisi type evolution equation.  As we shall be interested in
the leading logarithmic contribution only, we restrict ourselves to the
electron non-singlet structure function $D^\mathrm{NS}(z,\mu^2)$, which is
the solution of the integro-differential equation,%
\footnote{For the sake of simplicity we shall ignore the running of the QED
  coupling.}
\begin{equation}
\label{eq:Dns-evol}
  \mu^2 \, \frac{\dd}{\dd \mu^2} \, D^\mathrm{NS}(z,\mu^2)
  = \frac{\alpha}{2\pi}
  \int\limits_z^1 \frac{\dd y}{y} \,
  P_{ee}(y)
  D^\mathrm{NS}\left( \frac{z}{y},\mu^2 \right) \; ,
\end{equation}
with the electron splitting function
\begin{equation}
\label{eq:Pee-reg}
  P_{ee}(x) =
  \lim_{\eps \to 0}
  \left(
    \Theta(1-\eps-x) \, \frac{1+x^2}{1-x}
    - \delta(1-x) \int\limits_{0}^{1-\eps-x} \frac{1+y^2}{1-y} \;  \dd y
  \right)
\end{equation}
and with initial condition
\begin{equation}
\label{eq:Dns-0}
  D^\mathrm{NS}(z,m^2) = \delta(1-z) \; .
\end{equation}
The normalization of the non-singlet structure function,
\begin{equation}
\label{eq:Dns-norm}
  \int\limits_0^1 \dd z \; D^\mathrm{NS}(z,\mu^2) = 1
  \; ,
\end{equation}
valid for any $\mu^2$, is a consequence of
\begin{equation}
\label{eq:Pee-norm}
  \int\limits_0^1 \dd x \; P_{ee}(x) = 0
  \; ,
\end{equation}
which expresses the order-by-order cancellation of the infrared
singularities between virtual corrections ($x=1$) and soft photon emission
($x \to 1$).

The formal solution of the integro-differential equation
(\ref{eq:Dns-evol}) with initial condition (\ref{eq:Dns-0}) reads:
\begin{equation}
\label{eq:Dns-formal}
  D^\mathrm{NS}(z,\mu^2) =
  \delta(1-z) +
  \frac{\alpha}{2\pi}
  \int\limits_{m^2}^{\mu^2} \frac{\dd Q^2}{Q^2}
  \int\limits_{z}^1 \frac{\dd y}{y}
  P_{ee}(y)
  D^\mathrm{NS}\left( \frac{z}{y},Q^2 \right)
  \; .
\end{equation}
A perturbative solution is obtained from (\ref{eq:Dns-formal}) by
iteration.  It may be written in different equivalent forms, e.g.,
\begin{eqnarray}
  D^\mathrm{NS}(z,\mu^2)
  & = &
\label{eq:Dns-pert-i}
  \delta(1-z)
  + \frac{\alpha}{2\pi}
    \int\limits_{m^2}^{\mu^2} \frac{\dd Q_1^2}{Q_1^2} \, P_{ee}(z)
  \\
  & + &
    \left( \frac{\alpha}{2\pi} \right)^2
    \int\limits_{m^2}^{\mu^2} \frac{\dd Q_1^2}{Q_1^2}
    \int\limits_{Q_1^2}^{\mu^2} \frac{\dd Q_2^2}{Q_2^2}
    \int\limits_z^1 \frac{\dd y}{y} \,
    P_{ee}(y) P_{ee}\left( \frac{z}{y} \right)
  + \ldots
  \nonumber \\
  & = &
\label{eq:Dns-pert-ii}
  \delta(1-z)
  + \frac{\alpha}{2\pi}
    \int\limits_{m^2}^{\mu^2} \frac{\dd Q_1^2}{Q_1^2} \, P_{ee}(z)
  \\
  & + &
  \frac{1}{2!}
  \left( \frac{\alpha}{2\pi}
    \int\limits_{m^2}^{\mu^2} \frac{\dd Q^2}{Q^2}
  \right)^2
  \int\limits_z^1 \frac{\dd y}{y} \,
  P_{ee}(y) P_{ee}\left( \frac{z}{y} \right)
  + \ldots
  \; .
  \nonumber
\end{eqnarray}
As long as we are interested in photon inclusive radiative corrections
only, both forms (\ref{eq:Dns-pert-i}) and (\ref{eq:Dns-pert-ii}) are
equally useful.  However, the first variant (\ref{eq:Dns-pert-i}) turns out
to be powerful for the treatment of exclusive photon emission in QED.  This
can be seen as follows.

For QED radiative corrections, the origin of the leading large logarithms,
$L\equiv\ln(\mu^2/m^2)$, is identified by analyzing the peaking behavior
of the squared matrix elements for photon radiation in the region of phase
space where the photons are widely separated in polar angle (measured
w.r.t.\ the radiating fermion).

Let us start by considering a typical expression corresponding to single
photon emission.  In that case, the large logarithm results from the
integration over the photon polar angle, which is of the type ($\tau \equiv
(1-\cos\vartheta)/2$):%
\footnote{In the approximation of small angles, one may equally well work
  with the photon rapidity
  $y=\frac{1}{2}\ln[(1+\cos\vartheta)/(1-\cos\vartheta)] \simeq
  -\frac{1}{2}\ln\tau$.}
\begin{equation}
\label{eq:LL-ang-dis}
  \int \frac{\dd (p \cdot k)}{p \cdot k}
  \simeq \int \frac{\dd (\cos\vartheta)}{1 - \beta\cos\vartheta}
  \simeq \int \frac{\dd \tau}{\tau + (1-\beta)/2}
  \simeq \int \frac{\dd \tau}{\tau + m^2/(4E_e^2)}
  \; .
\end{equation}
In the high-energy limit, one may drop the term $m^2/(4E_e^2)$ in the
denominator if one cuts off the $\tau$-integration at
\begin{equation}
  \tau^\mathrm{min} = \frac{m^2}{4E_e^2} \; ,
\end{equation}
corresponding to a fictitious minimum angle $\vartheta_\mathrm{min} \simeq
m/E_e$.  The large logarithm $L$ of the non-singlet structure function
$\Dns$ is then mimicked by assuming a fictitious maximum%
\footnote{For a general discussion of inclusive radiative corrections, the
  choice of $\tau^\mathrm{min}$ and $\tau^\mathrm{max}$ is completely
  arbitrary, as long as $\tau^\mathrm{max}/\tau^\mathrm{min}=\mu^2/m^2$.
  We do know, however, the full $\bigO{\alpha}$ result, and can match the
  leading logarithmic result at this order to the full answer, confirming
  the above choice for $\tau^\mathrm{min}$.}
\begin{equation}
\label{eq:tau-max}
  \tau^\mathrm{max} = \frac{\mu^2}{4E_e^2} \; .
\end{equation}
It must be emphasized that this clearly is an effective but \emph{not} a
physical cutoff.  First, at large angles the approximations leading to
(\ref{eq:LL-ang-dis}) will no longer be valid.  Second, $\tau^\mathrm{max}$
in (\ref{eq:tau-max}) obviously is not restricted to be smaller than unity
for arbitrary $\mu^2$.

The $n$ photon contributions ($n \geq 2$) in eq.\ (\ref{eq:Dns-pert-i}) can
be interpreted analogously.  However, we need to take into account the fact
that the leading contributions are obtained from the integration over
configurations in phase space where the photon polar angles are strongly
ordered,
\begin{equation}
  \vartheta_1 \ll \vartheta_2 \ll \cdots \ll \vartheta_n
  \; ,
\end{equation}
which is equivalent to saying that the (negative) virtualities of the
electron after radiation, roughly given by $2 p\cdot k_i$, are strongly
ordered for comparable photon energies.  This suggests to replace in
(\ref{eq:Dns-pert-i}) the nested integrations
\begin{equation}
\label{eq:LL-Q2-dis-n}
  \int\limits_{m^2}^{\mu^2} \frac{\dd Q_1^2}{Q_1^2}
  \int\limits_{Q_1^2}^{\mu^2} \frac{\dd Q_2^2}{Q_2^2}
  \; \ldots
  \int\limits_{Q_{n-1}^2}^{\mu^2} \frac{\dd Q_n^2}{Q_n^2}
\end{equation}
by
\begin{equation}
\label{eq:LL-ang-dis-n}
  \int\limits_{\tau^\mathrm{max}}^{\tau^\mathrm{max}}
  \frac{\dd \tau_1}{\tau_1}
  \int\limits_{\tau_1}^{\tau^\mathrm{max}}
  \frac{\dd \tau_2}{\tau_2}
  \; \ldots
  \int\limits_{\tau_{n-1}}^{\tau^\mathrm{max}}
  \frac{\dd \tau_n}{\tau_n}
  \; .
\end{equation}
The method described above corresponds to the introduction of
``unintegrated splitting functions'' for the generation of exclusive
photons from QED photon showers in Monte Carlo event generators like
KRONOS~\cite{Anlauf:1992wr}, which are chosen in such a way that they
reproduce the angular distribution for single photon emission in the region
of small angles as given by the peaking behavior of the electron
propagator,%
\footnote{Although the Monte Carlo event generator KRONOS generates the
  proper large inclusive logarithms, the actual implementation of the
  minimum and maximum values for the polar angles generally leads to
  emission of exclusive photons at too large polar angles.}
\begin{equation}
  e^2 \int \widetilde{\dd k}_i \,
  \frac{1}{p^{(i)} \cdot k_i} \;
  \frac{1+(1-x_i)^2}{x_i}
  = \frac{\alpha}{2\pi}
  \int\limits_{\tau_i^\mathrm{min}}^{\tau^\mathrm{max}}
  \frac{\dd \tau_i}{\tau_i}
  \int \dd x_i \; P_{e\gamma}(x_i) \; ,
\end{equation}
with
\begin{equation}
  P_{e\gamma}(x)
  = \frac{1+(1-x)^2}{x} \, \Theta(x-\eps)
  + \left( 2 \ln \eps + \frac{3}{2} \right) \delta(x) \; ,
\end{equation}
and the small parameter $\eps$ serving as infrared-regulator.

In the following we shall assume the photon angle distributions
(\ref{eq:LL-ang-dis-n}) for an estimate of the radiative corrections to the
tagged photon process.

Although the rewriting of the scale-ordered integrations
(\ref{eq:LL-Q2-dis-n}) in the perturbative expansion (\ref{eq:Dns-pert-i})
by the angular-ordered integrations (\ref{eq:LL-ang-dis-n}) looks very
suggestive, a remark on the proper cancellation of infrared singularities
is in order.

In fixed order perturbation theory, the Bloch-Nordsieck theorem guarantees
that the emission of soft photons is a Poisson process.  The probability
for emitting $n$ photons in the energy interval $\omega \in
[\omega_\mathrm{min}, \omega_\mathrm{max}]$ is proportional to
\begin{equation}
\label{eq:p-n-soft}
  \frac{1}{n!}
  \left[
    \frac{\alpha}{\pi} \ln \left( \frac{\mu^2}{m^2} \right)
    \ln \left( \frac{\omega_\mathrm{max}}{\omega_\mathrm{min}} \right)
  \right]^n \; .
\end{equation}
The obvious divergence for $\omega_\mathrm{min} \to 0$ is known to be
canceled by taking into account all appropriate virtual corrections at the
same order of perturbation theory.

From the reasoning further above, we know that in the regime of hard photon
radiation the emission of photons is ordered in virtualities resp.\ angles.
Applying this ordering condition to the phase space of the photon leads to
a cancellation of the combinatorial factor $1/n!$; the same will happen in
eq.\ (\ref{eq:p-n-soft}) with the combination $1/n! \cdot\ln^n(\mu^2/m^2)$.

We shall interpret this outcome in the following way: emission of a single
(hard) photon results in an off-shell electron.  The finite virtuality
effectively restricts the phase space accessible to further photon
emissions to the range of larger angles.  At the same time, the virtual
corrections to the amplitude after the $i$-th photon emission are
regularized or ``cut off'' by the virtuality of the electron, $Q_i^2$,
instead of $m^2$.  The ordered perturbative expansion (\ref{eq:Dns-pert-i})
then expresses this order-by-order cancellation due to the normalization
property (\ref{eq:Pee-norm}) of the electron splitting function even for
our interpretation of ordered photon emission.


\section{Leading radiative corrections to DIS with tagged ISR}

Let us now turn to an alternative derivation of the leading radiative
corrections to DIS with tagged initial state radiation using the above
reasoning.  We shall find that not only can the result of section
\ref{sec:tagged-ll} be reproduced with little effort but also generalized.

As a start, the lowest order cross section for the tagged photon process
(\ref{eq:Born}) follows from the first term of the expansion of the
electron non-singlet structure function.  Integrating over the solid angle
of the forward photon detector, $\vartheta \leq \vartheta_0$, we obtain:
\begin{equation}
\label{eq:sig-ll-lo}
  \frac{\dd^3 \sigma^{(1)}}{\dd x \, \dd y \, \dd z}
  =
  \frac{\alpha}{2\pi}
  \int\limits_{\tau^\mathrm{min}}^{\tau_0} \frac{\dd \tau}{\tau} \;
  P_{ee}(z) \,
  \frac{\dd^2 \sigma_\mathrm{Born}(x,y;z)}{\dd x \, \dd y}
  =
  \frac{\alpha L_0}{2\pi} \, P^{(1)}(z) \,
  \sigma_0(x,y;z)
  \; ,
\end{equation}
with $\tau_0=(1-\cos\vartheta_0)/2$, $L_0=\ln(\tau_0/\tau^\mathrm{min})$,
and the abbreviation (c.f.\ section \ref{sec:tagged-ll}):
\begin{equation}
  \sigma_0(x,y;z) =
  \frac{\dd^2 \sigma_\mathrm{Born}(x,y;z)}{\dd x \, \dd y}
  \; .
\end{equation}

At order $\bigO{\alpha^2}$, corresponding to double photon emission, we
decompose the contributions into two cases.  Case (a) corresponds to both
photons hitting the photon detector.  Assuming that only the sum of their
energies is measured, we find:
\begin{eqnarray}
\label{eq:LL-2a}
  \frac{\dd^3 \sigma^\mathrm{(2,a)}}{\dd x \, \dd y \, \dd z}
  & = &
  \left( \frac{\alpha}{2\pi} \right)^2
  \int\limits_{\tau^\mathrm{min}}^{\tau_0} \frac{\dd \tau_1}{\tau_1}
  \int\limits_{\tau_1}^{\tau_0} \frac{\dd \tau_2}{\tau_2} \;
  \int\limits_{z}^1 \frac{\dd \xi}{\xi} \;
  P_{ee}(\xi) \,
  P_{ee}\left(\frac{z}{\xi} \right) \,
  \sigma_0(x,y;z)
  \nonumber \\
  & = &
  \frac{1}{2!} \left( \frac{\alpha L_0}{2\pi} \right)^2 P^{(2)}(z) \,
  \sigma_0(x,y;z)
  \; .
\end{eqnarray}
In case (b), one photon enters the photon detector, while the other is
emitted at a larger angle.  Again, to logarithmic accuracy we may neglect
the transverse momentum of both photons, as we did in section
\ref{sec:tagged-ll}.  Taking the angle ordering into account yields:
\begin{eqnarray}
\label{eq:sig-ll-b}
  \frac{\dd^3 \sigma^\mathrm{(2,b)}}{\dd x \, \dd y \, \dd z}
  & = &
  \left( \frac{\alpha}{2\pi} \right)^2
  \int\limits_{\tau^\mathrm{min}}^{\tau_0} \frac{\dd \tau_1}{\tau_1} \;
  P_{ee}(z)
  \int\limits_{\tau_0}^{\tau^\mathrm{max}} \frac{\dd \tau_2}{\tau_2} \;
  \int\limits \dd z' \;
  P_{ee}\left( z' \right) \,
  \tilde\sigma_0(z,y;z,z')
  \nonumber \\
  & = &
  \left( \frac{\alpha}{2\pi} \right)^2 L_0 L_1 \, P^{(1)}(z) \;
  \int\limits_{z'_\mathrm{min}}^1 \dd z' \;
  P_{ee}\left( z' \right) \,
  \tilde\sigma_0(x,y;z,z')
  \; ,
\end{eqnarray}
with $L_1=\ln(\tau^\mathrm{max}/\tau_0)=\ln(Q^2/m^2)-L_0$.
Remember that the semi-collinear contribution does depend on the method of
reconstruction of the kinematic variables.  Therefore, $\tilde\sigma_0$ in
the above expression is understood to be expressed in terms of the ``true''
kinematic variables of the hard subprocess,
\begin{equation}
  \tilde\sigma_0(x,y;z,z') \equiv
  \sigma_0(x_\mathrm{true},y_\mathrm{true};z z')
  \cdot \mathcal{J}(x,y;x_\mathrm{true},y_\mathrm{true};z') \; ,
\end{equation}
with the Jacobian $\mathcal{J}$ depending on the chosen set of kinematic
variables, see table~\ref{tab:jacobian}.  The integration limit
$z'_\mathrm{min}$ also depends on kinematic cuts.

Inserting the regularized form of the electron splitting function
(\ref{eq:Pee-reg}) in the last line of (\ref{eq:sig-ll-b}) leads to:
\begin{eqnarray}
\label{eq:LL-2b}
  \frac{\dd^3 \sigma^\mathrm{(2,b)}}{\dd x \, \dd y \, \dd z}
  & = &
  \left( \frac{\alpha}{2\pi} \right)^2 L_0 L_1 \, P^{(1)}(z) \;
  \nonumber \\
  & \times &
  \left[ \;
  \int\limits_{z'_\mathrm{min}}^1 \dd z' \;
  P^{(1)}(z') \,
  \left(
  \tilde\sigma_0(x,y;z,z')
  - \tilde\sigma_0(x,y;z,z'=1)
  \right)
  \right.
  \nonumber \\
  && \; \left. {}
  - \tilde\sigma_0(x,y;z,z'=1)
  \int\limits_{0}^{z'_\mathrm{min}}\dd z' \; P^{(1)}(z')
  \right]
  \; .
\end{eqnarray}
Adding contributions (\ref{eq:LL-2a}) and (\ref{eq:LL-2b}), identifying
$z'=1-u$, and noting that
\[
  \tilde\sigma_0(x,y;z,z'=1)=\sigma_0(x,y;z)
  \; ,
\]
we recover result (\ref{eq:delta-ll}).

The above procedure may be generalized to higher orders in $\alpha$; we
only need to pay attention to the angle ordering to find the proper
combinatorial factors.  As a further explicit example we sketch the
application at order $\alpha^3$.

The contribution from three forward (collinear) photons reads:
\begin{equation}
\label{eq:sigma-3a}
  \frac{1}{3!} \left( \frac{\alpha L_0}{2\pi} \right)^3
  P_\Theta^{(3)}(z)
  \int \dd z' \; \delta(1-z') \,
  \tilde\sigma_0(x,y;z,z') \; .
\end{equation}
The semi-collinear part receives contributions from two collinear and one
large-angle photon,
\begin{equation}
  \frac{1}{2!} \left( \frac{\alpha L_0}{2\pi} \right)^2
  P_\Theta^{(2)}(z)
  \int\limits \dd z' \;
  \left( \frac{\alpha L_1}{2\pi} \right) P^{(1)}(z') \,
  \tilde\sigma_0(x,y;z,z') \; ,
\end{equation}
and from one collinear and two large-angle photons:
\begin{equation}
  \left( \frac{\alpha L_0}{2\pi} \right) P_\Theta^{(1)}(z)
  \int\limits \dd z' \;
  \frac{1}{2!} \left( \frac{\alpha L_1}{2\pi} \right)^2
  P^{(2)}(z') \,
  \tilde\sigma_0(x,y;z,z') \; .
\end{equation}
It is now straightforward to perform a summation of the leading
contributions to all orders.  Those parts corresponding to collinear
emission with no photon at large angles obviously add up to
\begin{equation}
\label{eq:sigma-ao-coll}
  D^\mathrm{NS}(z;L_0) \cdot
  \int \dd z' \; \delta(1-z') \,
  \tilde\sigma_0(x,y;z,z')
  \; ,
\end{equation}
with the suggestive notation $D^\mathrm{NS}(z;L_0)$ indicating that it
coincides with the inclusive non-singlet structure function,
\begin{equation}
\label{eq:Dns-alt}
  D^\mathrm{NS}(z;L) =
  \sum_{n=0}^\infty \frac{1}{n!}
  \left( \frac{\alpha L}{2\pi} \right)^n P^{(n)}(z)
  \; ,
\end{equation}
and thus satisfies an evolution equation,
\begin{equation}
  \frac{\df}{\df L} \, D^\mathrm{NS}(z; L)
  = \frac{\alpha}{2\pi}
  \int\limits_z^1 \frac{\dd y}{y} \,
  P_{ee}(y)
  D^\mathrm{NS}\left( \frac{z}{y}; L \right)
  \; , \quad
  D^\mathrm{NS}(z;0) = \delta(1-z)
  \; .
\end{equation}
The slight change in notation here is intended in order to emphasize that
the present version of $\Dns$ has no exclusive meaning in the sense of the
discussion in the previous subsection; it just represents the formal sum on
the r.h.s.\ of (\ref{eq:Dns-alt}).  Furthermore, and even more importantly,
while $D^\mathrm{NS}(z;L)$ appears to be well defined, the partial cross
section (\ref{eq:sigma-ao-coll}) with only forward photons is by itself
\emph{not} a meaningful, physical expression, as it lacks contributions
from large angle photons.  This is tacitly expressed by the
$\delta$-function in the variable $z'$ that is related to the energy loss
outside the solid angle covered by the forward photon detector, indicating
that it is a virtual correction part.  Strictly speaking,
(\ref{eq:Dns-alt}) has to be multiplied by the probability for \emph{no
emission at large angles}.

The full result for the cross section is therefore obtained by summing the
contributions from \emph{any} number of collinear and \emph{any} number of
large-angle photon emissions while taking into account the angle ordering.
It is easily seen that this procedure finally leads to the following
expression for the all-order leading logarithmic result:
\begin{eqnarray}
\label{eq:sig-ll-all-order}
  \frac{d^3\sigma_\mathrm{RC}^{\mathrm{LL}(\infty)}}{\dd x \, \dd y \, \dd z}
  & = &
  D^\mathrm{NS}(z;L_0) \cdot
  \int\limits_{z'_\mathrm{min}}^1 \dd z' \;
  D^\mathrm{NS}(z';L_1) \,
  \tilde\sigma_0(x,y;z,z')
  \nonumber \\
  & = &
  D^\mathrm{NS}(z;L_0) \cdot
  \left[
  \sigma_0(x,y;z)
  \cdot
  \left( 1 -
    \int\limits_0^{z'_\mathrm{min}} D^\mathrm{NS}(z';L_1) \; \dd z'
  \right)
  \right.
  \\
  && \qquad {} +
  \left.
  \int\limits^1_{z'_\mathrm{min}} D^\mathrm{NS}(z';L_1)
  \left(
    \tilde\sigma_0(x,y;z,z')
  - \tilde\sigma_0(x,y;z,1)
  \right) \dd z'
  \right]
  \; . \nonumber
\end{eqnarray}
In the last step we have used the normalization (\ref{eq:Dns-norm}) of the
non-singlet structure function and exhibited the cancellation of the
singular behavior for $z' \to 1$.

Expanding the all-order result (\ref{eq:sig-ll-all-order}) in powers of
$\alpha$ confirms that the terms up to order $\bigO{\alpha^2}$ terms agree
with the result of section~\ref{sec:tagged-ll}.


\section{The non-singlet structure function}

In the previous section we found a very compact expression for the tagged
photon cross section including the leading QED radiative corrections
resummed to all orders.  It involves the electron non-singlet structure
function $\Dns(z;L)$, where $L$ stands for any of the large logarithms $L_0
= \ln (E_e^2 \vartheta_0^2/m^2)$, $L_1=\ln(Q^2/m^2)-L_0$.  During the
calculation of the one-loop radiative corrections to this process in
chapter~\ref{sec:qed-cor-rad}, we already encountered the first two orders
of a perturbative expansion in $\alpha$ of $\Dns$.  We shall now sketch
methods aimed at practical determinations of this function for numerical
applications, mostly following Je{\.z}abek \cite{Jezabek:1991bx},
Przybycie\'n \cite{Przybycien:1993qe}, and references quoted in these
papers.

Introducing the parameter%
\footnote{In the usual treatment of the Gribov-Lipatov equation, the
  running of the QED coupling $\alpha$ is incorporated by replacing $\beta$
  by
\[
  \beta(\mu^2) =
  \frac{2}{\pi} \int\limits_{m^2}^{\mu^2}
  \frac{\alpha(Q^2) \, \dd Q^2}{Q^2} \; .
\]
This modification serves to take into account the additional corrections
from pair radiation.  However, the forward taggers at HERA clearly detect
only real (on-shell) photons.  We shall therefore disregard these
contributions at least for the tagged collinear initial state radiation and
work with constant $\alpha$.}
\begin{equation}
  \beta = \frac{2 \alpha}{\pi} \, L
  \; ,
\end{equation}
the Gribov-Lipatov evolution equation for the electron non-singlet
structure function reads:
\begin{equation}
\label{eq:GL-eq}
  \frac{\df \Dns(z,\beta)}{\df \beta}
  = \frac{1}{4}
  \int\limits_z^1 \frac{\dd y}{y} \;
  P_{ee}(y) \, \Dns\left(\frac{z}{y},\beta \right)
  \; ,
\end{equation}
with boundary condition
\begin{equation}
  \Dns(z,0) = \delta(1-z) \; .
\end{equation}
The integro-differential equation (\ref{eq:GL-eq}) can be solved in various
ways.

A perturbative solution is obtained by a power series ansatz,
\begin{equation}
\label{eq:Dns-ser}
  \Dns(z,\beta) =
  \sum_{n=0}^N \frac{1}{n!} \left( \frac{\beta}{4} \right)^n
  P^{(n)}(z)
  + \bigO{\beta^{N+1}} \; ,
\end{equation}
yielding
\begin{equation}
  P^{(0)}(z) = \delta(1-z) \; ,
\end{equation}
and the recurrence relation
\begin{equation}
\label{eq:P-recurs}
  P^{(n)}(z) =
  \int\limits_z^1 \frac{\dd y}{y} \;
  P_{ee}(y) \, P^{(n-1)}\left(\frac{z}{y} \right)
  \; , \quad n \geq 1
  \; .
\end{equation}
From (\ref{eq:Pee-norm}) we find:
\begin{equation}
  \int\limits_0^1 \dd z \; P^{(n)}(z) = 0
  \; , \quad n \geq 1
  \; .
\end{equation}
The coefficient functions $P^{(n)}$ are clearly divergent for $z \to 1$ and
need regularization.  Using a small auxiliary parameter $\eps \to 0$ as in
(\ref{eq:Pee-reg}), the first few terms obtained from recurrence relation
(\ref{eq:P-recurs}) read (see also
\cite{Skrzypek:1992vk,Przybycien:1993qe,Arbuzov:1999cq} and references):
\begin{eqnarray}
\label{eq:P-n-pert}
  P^{(i)}(z) &=&
  \lim_{\eps \to 0} \left[
    P^{(i)}_{\delta} \delta(1-z) +
    P^{(i)}_{\Theta}(z)\Theta(1-z-\eps)
  \right]
  \; , \\ \nonumber
  P^{(1)}_{\Theta}(z) &=& \frac{1+z^2}{1-z}\ , \quad
  P^{(1)}_{\delta} = 2\ln\eps + \frac{3}{2} \; ,
  \\ \nonumber
  P^{(2)}_{\Theta}(z) &=&
  2\Biggl[\frac{1+z^2}{1-z}\left(2\ln(1-z)-\ln z +\frac{3}{2}\right)
  +\frac{1}{2}(1+z)\ln z-1+z\Biggr] \, ,
  \\ \nonumber
  P^{(2)}_{\delta}
  &=& \left( 2\ln\eps + \frac{3}{2} \right)^2 - \frac{2 \pi^2}{3}
  \; ,
  \nonumber \\
  P^{(3)}_\Theta(z)
  & = &
  24 \, \frac{1+z^2}{1-z}\Biggl( \frac{1}{2}\ln^2(1-z)
  + \frac{3}{4}\ln(1-z) - \frac{1}{2}\ln{z}\ln(1-z)
  \nonumber \\
  && \qquad \qquad {} + \frac{1}{12}\ln^2z
  - \frac{3}{8}\ln{z} + \frac{9}{32} - \frac{\pi^2}{12} \Biggr)  
  \\
  & + & 6(1+z)\ln{z}\ln(1-z) - 12(1-z)\ln(1-z)
  + \frac{3}{2}(5-3z)\ln{z}
  \nonumber \\
  & - & 3(1-z) - \frac{3}{2}(1+z)\ln^2z + 6(1+z)\dilog(1-z) \; ,
  \nonumber \\
  P^{(3)}_{\delta}
  & = &
  \left(2 \ln\eps + \frac{3}{2} \right)^3
  - 2\pi^2 \left(2 \ln\eps + \frac{3}{2} \right) + 16\zeta(3)
  \; . \nonumber
\end{eqnarray}
As can already be seen from these expressions, the power series expansion
(\ref{eq:Dns-ser}) does not converge uniformly.  As $z \to 1$, the
coefficients behave as
\begin{equation}
  P_\Theta^{(n)}(z) \sim n \cdot 2^n \cdot \frac{\ln^{n-1}(1-z)}{1-z}
  \; .
\end{equation}
This is an obvious hint towards a closed form of the solution to the
evolution equation in the limit $z \to 1$ which was found by Gribov:
\begin{equation}
\label{eq:D-Gribov}
  D_G(z,\beta) =
  \frac{\exp\left[ \beta/2 \cdot (3/4 - \gamma) \right]}{\Gamma(1+\beta/2)}
  \, \frac{\beta}{2} \, (1-z)^{\beta/2-1} \; .
\end{equation}
Here $\gamma=0.57721\ldots$ is Euler's constant.
The softening of the perturbative behavior of the non-integrable
singularity, $(1-z)^{-1}$, to an integrable one is a consequence of the
resummation of multiple soft photon emission to all orders
(``exponentiation'').

The knowledge of the solution (\ref{eq:D-Gribov}) has lead to several
proposals to improve the accuracy of finite order perturbative
approximations to $\Dns$.  This was pioneered by Kuraev and Fadin
\cite{Kuraev:1985hb} who suggested the prescription
\begin{equation}
\label{eq:Dns-KF}
  \Dns(z,\beta) =
  D_G(z,\beta) +
  \sum_{n=1}^N \beta^n \xi_n(z)
  + \bigO{\beta^{N+1}} \; ,
\end{equation}
where the functions $\xi_n(z)$ are derived from the requirement that
(\ref{eq:Dns-ser}) and (\ref{eq:Dns-KF}) are identical up to order
$\beta^N$.

An alternative prescription was formulated by Jadach and Ward
\cite{Jadach:1988gb}.  They suggested a factorized ansatz,
\begin{eqnarray}
\label{eq:Dns-JW}
  \Dns(z,\beta)
  & = & D_G(z,\beta) \, \Phi(z,\beta)
  \nonumber \\
  & = & D_G(z,\beta)
  \sum_{n=0}^{N-1} \left(\frac{\beta}{2}\right)^n \phi_n(z)
  + \bigO{\beta^{N+1}}
  \; .
\end{eqnarray}
Again, the coefficient functions $\phi_n(z)$ can in principle be determined
by comparing the expansions (\ref{eq:Dns-ser}) and (\ref{eq:Dns-JW}).  It
is however advantageous to first derive an evolution equation for the
function $\Phi(z,\beta)$, which can in turn be used to obtain a recurrence
formula for the $\phi_n(z)$ \cite{Jezabek:1991bx}.  The first three
coefficients read:
\begin{eqnarray}
\label{eq:phi-n-JW}
  \phi_0(z) & = &
  \frac{1+z^2}{2}
  \; , \nonumber \\
  \phi_1(z) & = &
  - \frac{1}{8} \left[ 2(1-z)^2 + (1+3z^3) \ln z \right]
  \; , \\
  \phi_2(z) & = &
  \frac{1}{8} \Biggl[ (1-z)^2 + \frac{1-4z+3z^2}{2} \ln z
  + \frac{1+7z^2}{12} \ln^2 z
  \nonumber \\
  && \quad {}
  + (1-z^2) \dilog(1-z)
  \Biggr] \; . \nonumber
\end{eqnarray}
Further coefficients of the Jadach-Ward series were calculated analytically
and numerically by Przybycie\'n \cite{Przybycien:1993qe}, where it was also
found that the higher order terms approach a very regular pattern.  It is
worthwhile to mention that the known coefficients of this series are much
more compact than those of the perturbative series (\ref{eq:P-n-pert}).

Numerical studies of the solutions to the Gribov-Lipatov equation obtained
by Monte Carlo calculation and by inverse Mellin transforms show that, if
truncated at the same order, the Jadach-Ward series converges significantly
better than the Kuraev-Fadin series and provides a particularly good
approximation to the full result already with the first three terms for
typical LEP and HERA energies
\cite{Jadach:1990vz,Skrzypek:1990qs,Skrzypek:1992vk} except for very small
$z$.  However, even this can be remedied.  Inspecting the coefficient
functions (\ref{eq:phi-n-JW}) for $z \to 0$ it is evident that the poor
convergence in this limit is due to $\phi_n(z) \sim \ln^n z$.  Studying the
$z \to 0$ behavior of the Gribov-Lipatov equation, this has lead
Je{\.z}abek to suggest a further improvement of the Jadach-Ward series that
we will not pursue here.  For details we refer the reader to
\cite{Jezabek:1991bx}.


\section{Numerical results for higher order corrections}

In order to assess the importance of the higher order QED corrections to
the tagged photon process, we have calculated the radiative correction
factor
\begin{equation}
  \delta_\mathrm{h.o.} =
  \left(
  \left.
  \frac{\dd^3\sigma_\mathrm{h.o.}}{\dd x \, \dd y \, \dd z}
  \right/ \!\!
  \frac{\dd^3\sigma_\mathrm{l.o.}}{\dd x \, \dd y \, \dd z}
  \right) - 1 \; ,
\end{equation}
with $\sigma_\mathrm{l.o.}$ being the purely leading order, logarithmic
part of the lowest order radiative cross section (\ref{eq:sig-ll-lo}), and
$\sigma_\mathrm{h.o.}$ being the radiatively corrected cross section taking
into account corrections at the level of leading logarithms as discussed
above.

\begin{table}
\begin{center}
\begin{tabular}{|c|d{2.4}|d{2.4}|d{2.4}|d{2.4}|}
\hline
 $\delta_\mathrm{h.o.}$
 & \multicolumn{2}{c|}{$x=10^{-4}$} & \multicolumn{2}{c|}{$x=10^{-1}$} \\
\hline
 & \multicolumn{1}{c|}{$\bigO{\alpha}$ LL}
 & \multicolumn{1}{c|}{all-order LL}
 & \multicolumn{1}{c|}{$\bigO{\alpha}$ LL}
 & \multicolumn{1}{c|}{all-order LL} \\
\hline
 $y=0.05$
 &  0.074 &  0.068
 & -0.040 & -0.043 \\
\hline
 $y=0.2$
 & 0.093 & 0.089
 & 0.024 & 0.019 \\
\hline
 $y=0.5$
 & 0.159 & 0.156
 & 0.120 & 0.116 \\
\hline
 $y=0.8$
 & 0.344 & 0.344
 & 0.351 & 0.350 \\
\hline
 $y=0.95$
 & 0.842 & 0.849
 & 1.412 & 1.438 \\
\hline
\end{tabular}
\end{center}
\caption{Comparison of higher order leading logarithmic radiative
  corrections at $\bigO{\alpha}$ versus the all-order LL resummed
  calculation for the electron method for a tagged energy of $5\GeV$}
\label{tab:delta-ho-ll-5}
\end{table}

\begin{table}
\begin{center}
\begin{tabular}{|c|d{2.4}|d{2.4}|d{2.4}|d{2.4}|}
\hline
 $\delta_\mathrm{h.o.}$
 & \multicolumn{2}{c|}{$x=10^{-4}$} & \multicolumn{2}{c|}{$x=10^{-1}$} \\
\hline
 & \multicolumn{1}{c|}{$\bigO{\alpha}$ LL}
 & \multicolumn{1}{c|}{all-order LL}
 & \multicolumn{1}{c|}{$\bigO{\alpha}$ LL}
 & \multicolumn{1}{c|}{all-order LL} \\
\hline
 $y=0.05$
 &  0.118 &  0.113
 & -0.016 & -0.019 \\
\hline
 $y=0.2$
 & 0.148 & 0.146
 & 0.049 & 0.045 \\
\hline
 $y=0.5$
 & 0.212 & 0.213
 & 0.147 & 0.145 \\
\hline
 $y=0.8$
 & 0.385 & 0.393
 & 0.379 & 0.387 \\
\hline
 $y=0.95$
 & 0.681 & 0.702
 & 1.428 & 1.486 \\
\hline
\end{tabular}
\end{center}
\caption{Comparison of higher order leading logarithmic radiative
  corrections at $\bigO{\alpha}$ versus the all-order LL resummed
  calculation for the electron method for a tagged energy of $20\GeV$}
\label{tab:delta-ho-ll}
\end{table}

We have used the same sets of parameters as in the previous cases and the
ALLM97 parameterization of the structure function of the proton, with fixed
$R=0.3$.  As an example, we have performed the numerical calculation for
the electron method.  Tables \ref{tab:delta-ho-ll-5} and
\ref{tab:delta-ho-ll} compare the numerical results for the radiative
correction factor $\delta_\mathrm{h.o.}$ for the leading logarithmic
correction as obtained in section~\ref{sec:tagged-ll} with the all-order
resummed result (\ref{eq:sig-ll-all-order}), evaluated using the
Jadach-Ward approximation (\ref{eq:Dns-JW}) of the electron non-singlet
structure function, taking into account the first three coefficients
(\ref{eq:phi-n-JW}) of the expansion.

As can be seen from these tables, the difference between the first-order
corrections and the all-order resummed result is typically at the level of
a few per mille.  The difference slightly increases for small $x$ and $y
\to 0$, where the phase space for photon emission is strongly constrained
and effects of multiple soft photon emission become important.

For $y \to 1$, there is also a significant difference between the
$\bigO{\alpha}$ and the all-order corrections.  To understand and to check
this finding, we have also calculated the $\bigO{\alpha^2}$ leading
logarithmic corrections using eqs.~(\ref{eq:sigma-3a}ff).  It turns out
that the difference between the $\bigO{\alpha^2}$ and the all-order result
is at the level of only a few$\times 10^{-4}$ for $0.01 < y < 0.99$,
establishing that the differences in these tables also for large $y$ are
truly an indication of the size of higher order corrections beyond the
first order, being essentially saturated by the contributions at
$\bigO{\alpha^2}$.  Note, however, that the effects from higher-order
corrections are small compared to the uncertainties of the radiative
corrections due the present errors in the ratio $R$ even for rather small
values of $y$.

Finally, there is still the question of the appropriate scale to be used in
the leading logarithmic approximation.  For the lowest order contribution
to the radiative cross section, we can determine the ``optimal'' scale by
``matching'' the cross section (\ref{eq:sig-ll-lo}) to the cross section
obtained by a fixed order calculation, by comparing
\begin{equation}
  \frac{1+z^2}{1-z} \, L_0^\mathrm{LLA}
  \quad \mbox{vs.} \quad
  \frac{1+z^2}{1-z} \, L_0 - \frac{2z}{1-z}
  \equiv
  \frac{1+z^2}{1-z} \, (L_0-1) + 1-z
  \; .
\end{equation}
Obviously, the complete lowest order cross section is better approximated
by $L_0^\mathrm{LLA} \simeq L_0$ for $z \to 0$, corresponding to emission
of a very hard collinear photon, while it appears more reasonable to set
$L_0^\mathrm{LLA} \simeq L_0-1$ for $z \to 1$.  Taking this variation as an
indication of the intrinsic uncertainty of the leading logarithmic
approximation, we have checked the contributions from higher order
corrections by varying the collinear logarithm from $L_0$ to $L_0-1$.  We
found a change in the contributions from higher order corrections being
typically of the order of a few$\times 10^{-4}$ except in the region of
large $x$ and $y \to 1$, where it can reach the per mille level.  However,
this remaining scale ambiguity, which is resolvable by a calculation of the
next-to-leading logarithmic contributions at order $\bigO{\alpha^2}$,
appears to be well below the anticipated statistical accuracy of the
corresponding experiments even after the upgrade of the HERA collider.


\chapter{Concluding Remarks and Outlook}
\label{sec:conclusion}


In this report we have considered QED radiative processes at HERA, focusing
on deep inelastic scattering with an exclusive photon being tagged in a
forward photon detector.  These processes can be used to extend the
effective kinematic range for structure function measurements accessible
with the HERA experiments down to lower values of the invariant momentum
transfer $Q^2$.  The information obtained this way is important for
accurate calculations of the radiative corrections to non-radiative deep
inelastic scattering.  The extended range allows testing the domain of
applicability of perturbative QCD, as well as studying the transition into
the non-perturbative regime which is described by phenomenological models.
Furthermore, a measurement of the triple differential cross section
$\dd^3\sigma/\dd\hat{x}\,\dd\hat{y}\,\dd z$, eq.~(\ref{eq:Born}), enables
the separation of the structure functions of the proton $F_2(x,Q^2)$ and
$F_L(x,Q^2)$ and thus a direct measurement of the longitudinal structure
function without the need to run the HERA collider at different c.m.s.\
energies.

The understanding and control of the radiative corrections is crucial for
precise theoretical predictions of cross sections.  We discussed the
calculation of the most important QED corrections to the tagged photon
cross section in various approximations.  The leading logarithmic
approximation provides a rough estimate of the radiative corrections.  It
is quite useful for inclusive measurements, when final state radiation off
the electron is unimportant (``calorimetric selection''), and describes the
full leptonic corrections with an accuracy of several percent.  Its compact
expressions make it suitable for a qualitative understanding of the leading
contributions to the corrections for different kinematic reconstruction
methods.

The most important result presented in this work is the calculation of the
full gauge-invariant set of leptonic QED corrections.  We find that the
difference between the full corrections and the leading logarithmic
approximation is very significant.  Depending on the reconstruction method
and whether the measurement is calorimetric or exclusive, this difference
can easily reach 5 or even 10 percent.  Nevertheless, an approximation that
keeps only the leading and next-to-leading logarithms appears to work well
within an accuracy of typically a few per mille.

Finally, we estimated the size of the QED corrections at higher orders.
Motivating an exclusive interpretation of the Gribov-Lipatov equation
similar to QED photon shower algorithms in some Monte Carlo event
generators, we obtained a compact, closed expression for the all-order
leading logarithmic contributions to the tagged photon cross section.  For
this particular process, these higher order corrections turn out to be less
important than the full corrections at relative order $\bigO{\alpha}$.
Furthermore, they are typically smaller than those induced by the present
uncertainty in the longitudinal structure function, $F_L$, at low $Q^2$.

The apparently large dependence of the radiative corrections on $F_L$ for
the lepton method is not a real problem.  Provided sufficient data is
available, it can be solved similar to the case of non-radiative DIS by
iteration, where the result of the analysis is fed back into the
calculation of the radiative corrections.
We thus conclude that the QED radiative corrections to the tagged photon
process at HERA are well under control with the calculations described in
this report.
What remains to be done is an implementation of the QED corrections in a
Monte Carlo event generator that is able to perform the calculations also
for less symmetric and more complicated setups than assumed here.


\begin{figure}
  \begin{center}
    \begin{picture}(137,120)
      \put(-3,0){\includegraphics[scale=0.75]{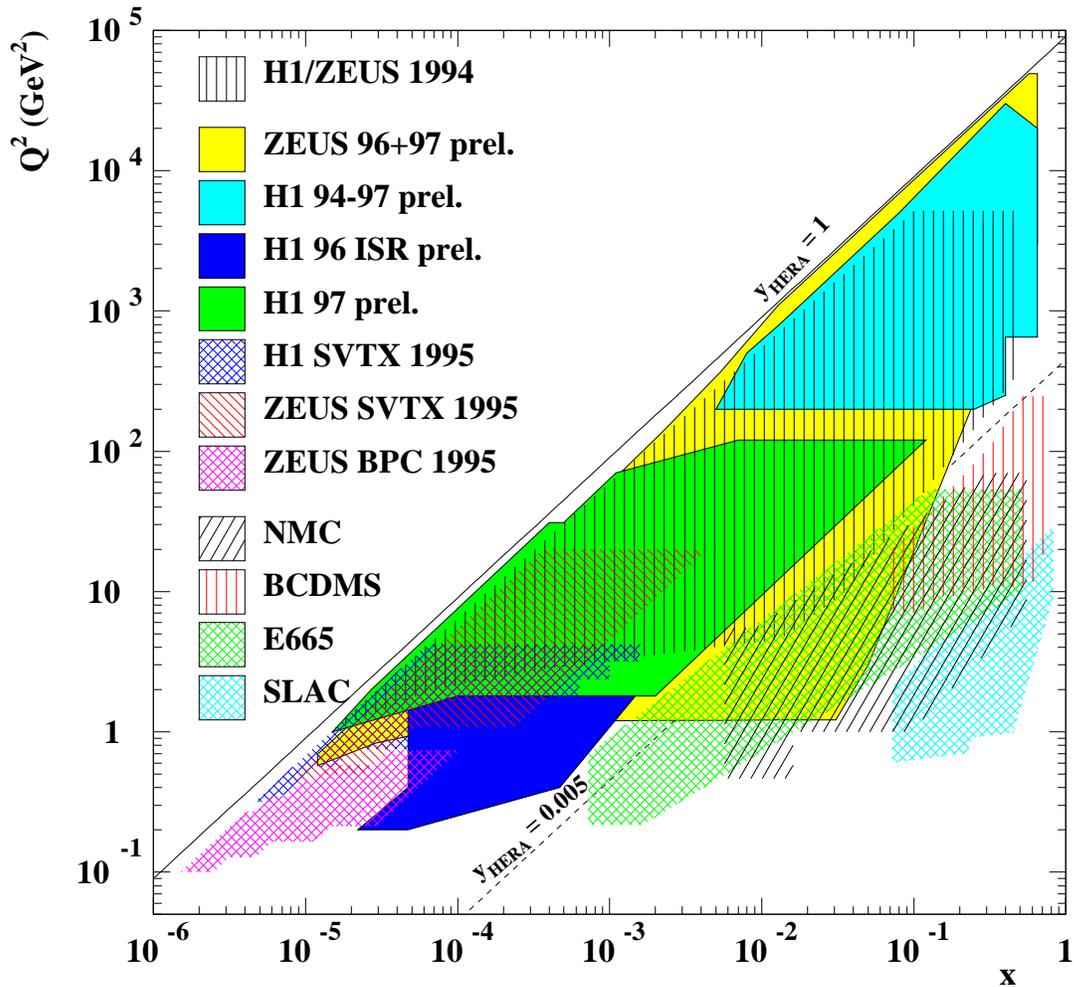}}
    \end{picture}
    \caption{Measured regions of $F_2$ in the $(x,Q^2)$ kinematic plane, as
      shown at the ICHEP98 conference.  The nominal acceptance region of
      the HERA measurements corresponds to $y_\mathrm{HERA} > 0.005$.  The
      fixed-target experimental data occupy the region of high $x$ at low
      $Q^2$.  (Taken from \cite{Doyle:1998qg}).}
    \label{fig:HERA-dis-kinem}
  \end{center}
\end{figure}

For an overview of the experimental situation,
figure~\ref{fig:HERA-dis-kinem} displays the kinematic coverage of the
($x,Q^2$) plane as presented at the ICHEP98 conference in Vancouver
\cite{Doyle:1998qg}.  The contributions from HERA to the determination of
the structure function $F_2$ essentially lie to the left of and above the
dashed line that roughly describes the acceptance limit of the HERA
experiments, while those from the fixed-target experiments are to the right
and below this line.
The HERA experiments cover a wide range in $y=Q^2/(xS)$ at sufficiently
large $x$ and $Q^2$ by combining several measurements, but in the
interesting region of low $x$ and for $Q^2 \lesssim 1 \GeV^2$ the
kinematically domain accessible to non-radiative DIS becomes very narrow.
The large, dark blue area in that region, indicated ``H1 96 ISR prel.''
corresponds to the kinematic domain that was added by analyzing events with
tagged initial state radiation \cite{Klein:1998mz}; it overlaps with other
measurements only at the left and upper boundaries.
The analyzed radiative events obviously populate only the region of rather
low $Q^2$, justifying a posteriori our calculation that took into account
only photon exchange and neglected contributions from the Z boson.


\begin{figure}
  \begin{center}
    \begin{picture}(135,175)
      \put(-1,0){\includegraphics[scale=0.65]{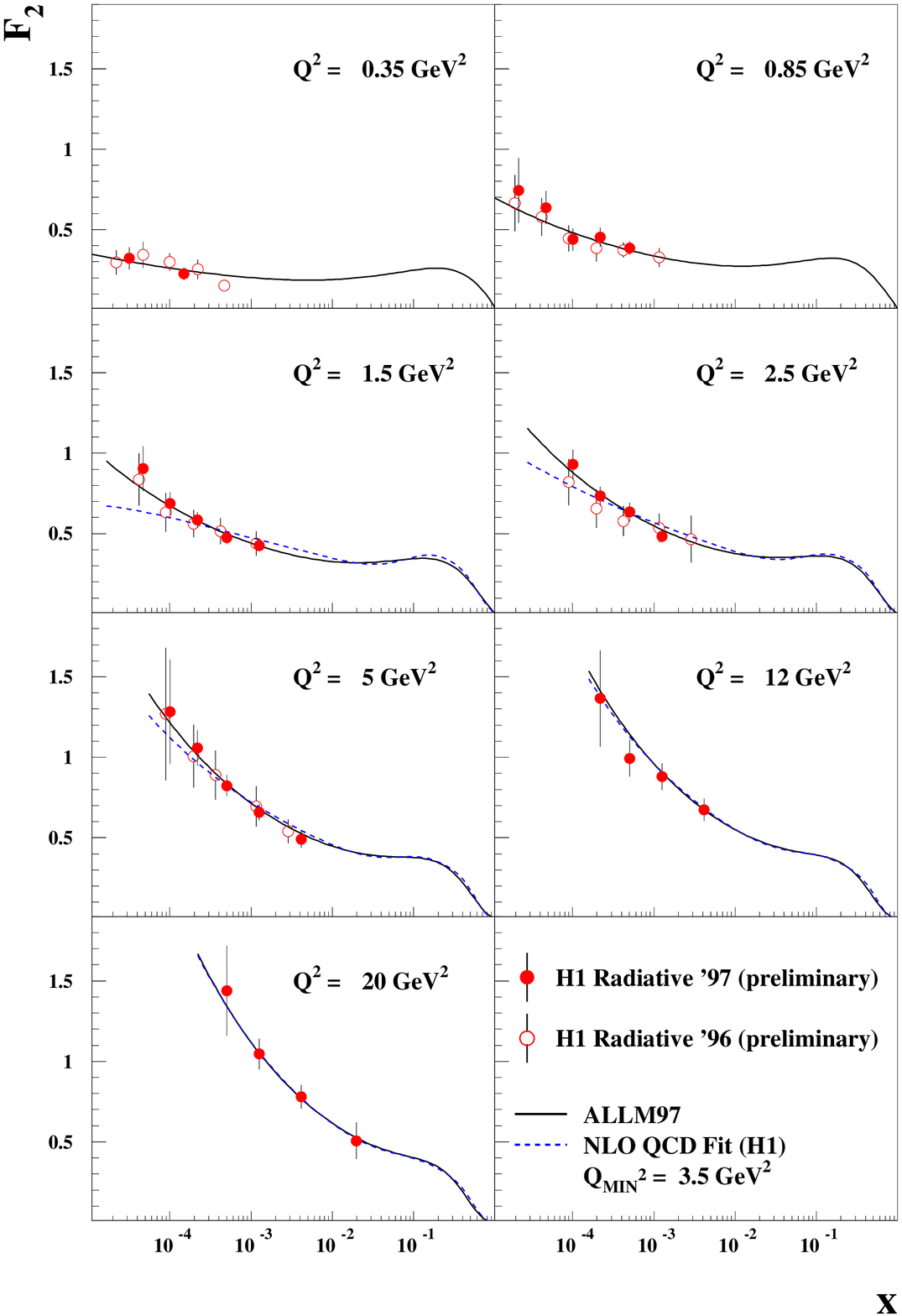}}
    \end{picture}
    \caption{The structure function $F_2$ as determined from tagged ISR
      events as a function of $x$ for fixed values of $Q^2$.  The data are
      compared to the ALLM97 parameterization (solid curve) and a NLO QCD
      fit to non-radiative $F_2$ data for $Q^2 \geq 3.5\GeV^2$ (dashed
      curve).  (Taken from \cite{Issever:DIS01}).}
    \label{fig:H1-F2-ISR}
  \end{center}
\end{figure}


An improved determination of the structure function $F_2$ from tagged ISR
events including the H1 97 ISR data has been presented at the DIS01
conference by the H1 collaboration \cite{Issever:DIS01} and is shown in
fig.~\ref{fig:H1-F2-ISR}.  The results of this analysis look very
promising.  There is good agreement with results on $F_2$ from
non-radiative events in the kinematic regions where both methods overlap.
Furthermore, it should be emphasized that the ISR data on $F_2$ describe
the important domain where the transition from perturbative QCD ($Q^2 \gg 1
\GeV^2$) to photoproduction ($Q^2 \approx 0 \GeV^2$) occurs, within a
single, coherent data set.  This eliminates systematic errors stemming from
uncertainties in the normalization when combining different data sets.

In principle it should be possible to simultaneously obtain the ratio
$R=F_L/F_T$ during this analysis from the $\hat{y}$-dependence of the
tagged photon cross section (\ref{eq:Born}) at fixed $\hat{x}$ and
$\hat{Q}^2$.  The extraction method described in \cite{FGMZ96} has been
applied to the H1 1996-97 data set \cite{Issever:thesis}, resulting so far
only in upper limits on $F_L$ and $R$ which are not yet competitive with
more model dependent indirect determinations.  Improvements in the
detectors and more data may change the situation.

In the present paper we have concentrated on DIS with unpolarized beams.
For polarized protons there are additional spin-dependent structure
functions $g_1$, $g_2$.  These are being measured at the HERMES experiment
at HERA and at low $Q^2$, using the polarization of the lepton beam and a
polarized fixed target.

At the upgraded HERA machine, large longitudinal polarization of the lepton
beams can also be expected for the H1 and ZEUS experiments.  A primary
motivation for lepton polarization is the spin dependence of weak
interactions which can be exploited in searches for physics beyond the
Standard Model, e.g., extensions to the electroweak sector with heavy
vector bosons which couple to right-handed currents, or supersymmetry.

If acceleration of polarized protons were possible in the HERA ring, the
kinematic domain accessible to spin-dependent lepton-proton scattering
could be dramatically extended.  Gakh et al.\ \cite{Gakh:2000mz} have
considered DIS with tagged photons for polarized beams and performed
calculations for the corresponding leptonic QED radiative corrections with
logarithmic accuracy.  The leading logarithmic corrections coincide with
with those discussed in this work, but the next-to-leading logarithms are
spin-dependent.  The model-dependent lepton-quark interference terms which
contribute to asymmetries such as spin-dependence have not yet been
calculated.  However, the initial plans for proton polarization in
the HERA ring were set back with the recent HERA upgrade.

In the description of radiative processes in this report we have neglected
the contributions which are due to emission off the hadron side; we argued
that they are not relevant for the determination of the leptoproduction
structure functions.  On the other hand, the case of radiative
(quasi-)elastic scattering,
\[
  e + p \to e + \gamma + p \; ,
\]
with the photon being well separated from the incident and scattered
lepton, has recently received much attention.  This process receives
contributions from bremsstrahlung off from the lepton (Bethe-Heitler
process, BH) as discussed in this report, but also from Deeply Virtual
Compton Scattering (DVCS) on the proton,
\[
  \gamma^* + p \to \gamma + p \; ,
\]
which gives access to the new class of so-called \emph{generalized} or
\emph{non-forward} or \emph{skewed} parton distributions (GPDs)
\cite{DVCS}.  Briefly put, GPDs are probability amplitudes to knock out a
parton from a hadron and to put it back with a different longitudinal
momentum.  DVCS has a reliable theoretical basis within perturbative QCD.
Its measurement provides a further opportunity to gain more insight into
nucleon properties such as spin content \cite{DVCS-review}.  The extraction
of DVCS from the HERA data also requires to properly take into account the
irreducible background from the BH process including the QED radiative
corrections.

Radiative processes also play an important r\^ole at $e^+ e^-$ colliders.
Most prominent are the ``radiative return events'' at LEP when running at
energies far beyond the Z resonance: an incoming electron or positron may
lose just as much energy as needed so that the effective collision energy
matches the Z mass.  Similar radiative return processes are also visible at
other colliders, e.g., $B$ and $\phi$ factories, and can be used to measure
hadron production cross sections and resonances below the nominal machine
c.m.s.\ energy \cite{tagging:hadronic}.  The high luminosity of these
factories can easily compensate the large suppression of the radiative
process by a factor $\alpha/\pi$ despite the absence of a dedicated photon
detector at small angles.

Another important application of radiative return events is the improvement
of measurements of the ratio
\[
  R_\mathrm{had}(s) \equiv
  \frac{\sigma(e^+ e^- \to \mbox{hadrons})}%
       {\sigma(e^+ e^- \to \mu^+ \mu^-)} \; ,
\]
which is needed as input in the determination of the hadronic contribution
to the running QED coupling $\alpha(Q^2)$.  Measurements performed at the
DA$\Phi$NE collider have provided encouraging results \cite{Denig:2001ra}
and have also been used in the determination of the hadronic contributions
to the muon $g-2$ \cite{Spagnolo:1998mt}.


\chapter*{Acknowledgments}


I would like to take this opportunity to thank Andrej B. Arbuzov and Eduard
A. Kuraev for a fruitful collaboration on the subject.  Edward E. Boos,
Hans D. Dahmen, Panagiotis Manakos and Thorsten Ohl deserve special thanks
for providing continuous support and helping in various ways to ``keep me
going''.
%
%
I enjoyed discussions with colleagues, especially the organizers of the
HERA 1998 Monte Carlo workshop \cite{HERAMC:1998} and in particular the
convenors of the working group WG70 on QED radiative effects.
Furthermore, I thank Manfred Fleischer for drawing my attention to this
interesting subject, and finally \c{C}i\u{g}dem \.I\c{s}sever for actually
using some of the results presented here in the difficult radiative $F_2$
analysis.

Financial support from the Bundesministerium f\"ur Bildung und Forschung
(BMBF), Germany, is gratefully acknowledged.


\appendix


\chapter{Auxiliary Calculations}


In this appendix we present some details on how to efficiently perform the
integration over the solid angle of the radiated photons in the forward
region for the virtual corrections and the double collinear contributions.

Denoting the energy fraction of the $i$-th photon in units of the incident
electron energy, $x_i = E_{\gamma,i}/E_e$, we write:
\begin{eqnarray}
\label{eq:dOmega-rewrite}
  \int \widetilde{\dd k}_i
  & = &
  \frac{1}{(4\pi)^2}
  \int x_i \, \dd x_i \;
  \frac{E_e^2}{\pi}
  \int \dd \Omega_i
  \; .
\end{eqnarray}
When a photon is almost collinear to a fermion it is convenient to rewrite
the scalar product between the fermion and photon momenta as follows:
\begin{eqnarray}
\label{eq:zi-coll}
  z_i & = &
  2 p \cdot k_i
  = 2 E_e E_{\gamma,i} (1 - \beta_e \cos\vartheta_i)
  = 2 E_e E_{\gamma,i} (1 - \beta_e + \beta_e \tau_i)
  \nonumber \\
  & = &
  \frac{2}{1+\beta_e} x_i
  \left(m^2 + \frac{2\beta_e(1+\beta_e)}{4} \, 4E_e^2 \tau_i \right)
  \nonumber \\
  & \simeq & x_i \left( m^2 + 4E_e^2 \tau_i \right)
  =: x_i m^2 (1+\zeta_i)
  \; .
\end{eqnarray}
In the last line we neglected terms of order $\bigO{m^2/E_e^2}$, and we
introduced the variable
\begin{equation}
\label{def:zeta-i}
  \zeta_i = \frac{4E_e^2}{m^2} \, \tau_i \; ,
\end{equation}
which varies between 0 and $4E_e^2/m^2$.

Inserting this relation between in the expression for the photon phase
space we obtain the parameterization:
\begin{eqnarray}
\label{eq:int-dk-as-dzeta}
  \int \widetilde{\dd k_i}
  & = &
  \frac{1}{(4\pi)^2}
  \int x_i \, \dd x_i \; m^2
  \int \frac{\dd \phi_i}{2\pi} \, \dd \zeta_i
  \; .
\end{eqnarray}


\section{Integrals for virtual corrections}
\label{sec:int-virt}

For the calculation of the virtual corrections to the tagged photon process
we need to integrate the virtually corrected Compton tensor
(\ref{eq:compton-tensor}) over the phase space of the (single) radiated
photon, which is the solid angle of the photon detector PD:
\begin{equation}
  \int\limits_\mathrm{PD} \dd\Omega_\gamma \; K_{\mu\nu}
  \; .
\end{equation}
We shall assume throughout that the photon detector be azimuthally
symmetric with respect to the direction of the incoming electron direction
and covering the range of polar angles $0 \leq \vartheta_\gamma \leq
\vartheta_0$, with $\vartheta_0 \ll \theta$, where $\theta \sim \bigO{1}$
represents the scattering angle of the outgoing electron.

Introducing
\begin{equation}
  \zeta_0 := \frac{E_e^2 \vartheta_0^2}{m^2} \gg 1
  \; , \quad
  L_0 := \ln \zeta_0 \gg 1 \; ,
\end{equation}
we can treat $L_0$ as a \emph{large logarithm} for the conditions of the
HERA photon detectors.  It is large in the same sense as
$L_Q=\ln(Q_l^2/m^2)\gg 1$.  We therefore denote as large logarithmic terms
those ones that contain at least one factor of $L_0$ or $L_Q$.

Consistently neglecting of terms of order $\bigO{\vartheta_0^2}$, we may
drop terms proportional to the transverse momentum of the radiated photon
with respect to the incoming lepton.  Therefore we take $k \simeq (1-z)p_1$
and set
\begin{equation}
  \tilde{p}_2 \simeq z \tilde{p}_1
\end{equation}
in the integrand.  Among the kinematic invariants of the Compton
subprocess, $\hat{s}$, $\hat{t}$, $\hat{u}$ and $q^2$, only $\hat{t}$ is
small and may become of order $m^2$.  We parameterize it as
\begin{equation}
  \hat{t} \equiv -2 p_1 \cdot k = -m^2(1-z)(1+\zeta)
  \; , \quad
  0 \leq \zeta \leq \zeta_0 \; .
\end{equation}
Furthermore, we set
\begin{equation}
  \hat{s} = q^2 - \hat{t} - \hat{u}
  \; , \quad
  \hat{u} = - Q_l^2
  \; , \quad
  q^2 = -(1-z) Q_l^2
  \; .
\end{equation}
Inserting the decomposition (\ref{eq:T-decomp}) into
(\ref{eq:compton-tensor}), we obtain:
\begin{eqnarray}
\label{eq:K-int}
  \frac{E_e^2}{\pi}
  \int\limits_\mathrm{PD} \dd\Omega_\gamma \; K_{\mu\nu}
  & = &
  \frac{1}{1-z}
  \left( - Q_l^2 \tilde{g}_{\mu\nu}
         + 4z \tilde{p}_{1\mu} \tilde{p}_{1\nu} \right)
  \left( 1 + \frac{\alpha}{2\pi} \rho \right) P(z,L_0)
  \nonumber \\
  & + &
  \frac{E_e^2}{\pi}
  \int\limits_\mathrm{PD} \dd\Omega_\gamma \,
  \frac{\alpha}{2\pi}
  \Bigl[
    \tilde{g}_{\mu\nu} T_g'
  \\ && \qquad \qquad \qquad {}
    + \tilde{p}_{1\mu} \tilde{p}_{1\nu}
    \left( T'_{11} + z^2 T'_{22} + z \left( T'_{12} + T'_{21} \right) \right)
  \Bigr]
  \nonumber
  \; .
\end{eqnarray}
The straightforward integration of $T_g'$ is essentially elementary and
leads to transcendent functions including logarithms and dilogarithms.  In
the convention of Lewin \cite{Lewin:1981}, the dilogarithm (Spence
function) reads:
\begin{equation}
  \dilog(x) = -\int\limits_{0}^{x}\frac{\dd y}{y}\ln(1-y)
  \; .
\end{equation}
Neglecting terms of $\bigO{\zeta_0^{-1}}$ we find:
\begin{eqnarray}
\label{eq:int-Tg}
  \frac{E_e^2}{\pi} \int \dd\Omega_\gamma \; T_g'
  & = &
  \frac{Q_l^2}{1-z}
  \left[ (A\ln z + B) P(z,L_0) + C L_0 + D \right]
  =: \frac{Q_l^2}{1-z} \, T \; ,
  \nonumber \\
  A &=& 2 L_Q - L_0 - 2\ln(1-z) \; ,
  \nonumber \\
  B &=& \ln^2 z - 2 \dilog (1-z) - \frac{1}{2} \; ,
  \nonumber \\
  C &=&
  - \frac{2z}{1-z}\ln z
  - z \; ,
  \\
  D &=&
  - \frac{1-6z+4z^2}{1-z} \left( \dilog(1-z) + \ln z \ln(1-z) \right)
  \nonumber \\
  &-& 2z \left( \ln^2(1-z) + \frac{2 \pi^2}{3} \right)
  + \frac{8z}{1-z} \ln z
  + 1
  \nonumber \; .
\end{eqnarray}
The integration over solid angle for the other coefficients of the tensor
decomposition, $T'_{ij}, \; i,j=1,2$, is more involved.  In that case the
evaluation of the non-logarithmic terms (in the sense of being free of
$L_Q$ and $L_0$ in the final result) unavoidably leads to trilogarithms,
defined as (see \cite{Lewin:1981}):
\begin{equation}
  \trilog(t) = \int\limits_0^t \frac{\dd t'}{t'} \, \dilog(t')
  \; .
\end{equation}
The trilogarithms originate from a common integral that can be expressed in
the following way:
\begin{eqnarray}
  \int\limits_{t_0+\delta}^{\infty}
  \dd t \, \left( \frac{1}{t} - \frac{1}{t-t_0} \right) \dilog(1-t)
  & = &
  \ln (\delta) \dilog(1-t_0)
  + \trilog(t_0) - \trilog(1-t_0)
  \nonumber \\ &+&
    \trilog(1)
  - \frac{\pi^2}{6} \ln t_0 + \frac{1}{2} \ln^2 t_0 \ln(1-t_0)
  \nonumber \\ &+&
  \bigO{\delta} \; .
\end{eqnarray}
This expression, regularized by a small parameter, $\delta \ll 1$, is valid
for $0 < t_0 < 1$.  As the integral converges at the upper limit, we have
extended the upper limit to infinity.  We shall not reproduce here the
somewhat lengthy calculation which can be done with the help of
\cite{Lewin:1981} and only requires some care in finding the proper
analytic continuations of intermediate expressions.

The results for the integrations over the photon solid angle read:
\begin{eqnarray}
\label{eq:int-Tij}
  \frac{E_e^2}{\pi}
  \int \dd\Omega_\gamma \; T_{11}'
  & = &
  \frac{z}{(1-z)^3}
  \Biggl[
    -2 \left( A\ln z + B \right) \frac{1+(1-z)^2}{1-z} L_0
    - (3-z) A L_0
  \nonumber \\
  && {}
    - \frac{4z(2-z)}{1-z} E
    + \left( \frac{3-8z}{z} C + \frac{5-11z+5z^2}{1-z} \right) L_0
    + 2 C
  \nonumber \\
  && {}
    +2z - 4
    - \frac{2(1-4z+z^2)}{1-z} \left( \dilog(1-z) + \ln z \ln(1-z) \right)
  \nonumber \\
  && {}
    + (3-z) \left( \ln^2(1-z) + \frac{2\pi^2}{3} \right)
    - \frac{2(1-z)}{z} \ln(1-z)
  \Biggr] \, ,
  \nonumber \\
  \frac{E_e^2}{\pi}
  \int \dd\Omega_\gamma \; T_{22}'
  & = &
  \frac{1}{z(1-z)^3}
  \Biggl[
   -2z^2 \left( A\ln z + B \right) \frac{1+2(1-z)^2}{1-z} L_0
  \nonumber \\
  && {}
    +  (1-3z) A L_0
    + 8z(1-z)(A\ln z+B)
    - 4(1-z)^2 D
  \nonumber \\
  && {}
    - \left( \frac{1-4z^2+8z^3}{z} C + \frac{(1-2z)(3-2z)(1-z-z^2)}{1-z} \right) L_0
  \nonumber \\
  && {}
    - \frac{2(1+4z-15z^2+8z^3)}{1-z}
      \left( \dilog(1-z) + \ln z \ln(1-z) \right)
  \nonumber \\
  && {}
    - (1-11z+8z^2) \left( \ln^2(1-z) + \frac{2\pi^2}{3} \right)
    - \frac{4z(4-3z)}{1-z} E
  \nonumber \\
  && {}
    - \frac{2(1-z)(1+2z)}{z} \ln(1-z)
    + 2 C
    + 4 - 6z
  \Biggr] \, ,
  \\
  \frac{E_e^2}{\pi}
  \int \dd\Omega_\gamma \; T_{12}'
  & = &
  \frac{1}{(1-z)^3}
  \Biggl[
    \frac{2z(2-z)}{1-z}
    \left[
    \left( A\ln z + B \right) L_0
      + 2E
    \right]
  \nonumber \\
  && {}
    + (3-z) A L_0
    - \left( \frac{3-8z}{z} C + \frac{5-11z+5z^2}{1-z} \right) L_0
    - 2 C
  \nonumber \\
  && {}
    + 2z
    + \frac{2(3-6z+z^2)}{1-z} \left( \dilog(1-z) + \ln z \ln(1-z) \right)
  \nonumber \\
  && {}
    - (3-z) \left( \ln^2(1-z) + \frac{2\pi^2}{3} \right)
    + \frac{2(1-z)}{z} \ln(1-z)
  \Biggr] \, ,
  \nonumber \\
  \frac{E_e^2}{\pi}
  \int \dd\Omega_\gamma \; T_{21}'
  & = &
  \frac{1}{(1-z)^3}
  \Biggl[
    \left( A\ln z + B \right) \frac{2z^2}{1-z} L_0
    - (1-3z) A L_0
    - 2 C + 2z
  \nonumber \\
  && {}
    + \left( \frac{1-4z^2+8z^3}{z} C + \frac{(1-2z)(3-2z)(1-z-z^2)}{1-z} \right) L_0
  \nonumber \\
  && {}
    + \frac{2(1-6z+15z^2-8z^3)}{1-z}
      \left( \dilog(1-z) + \ln z \ln(1-z) \right)
  \nonumber \\
  && {}
    + (1-11z+8z^2) \left( \ln^2(1-z) + \frac{2\pi^2}{3} \right)
    + \frac{4z(4-3z)}{1-z} E
  \nonumber \\
  && {}
    + \frac{2(1-z)(1+2z)}{z} \ln(1-z)
    - 4 (1-z)^2 C L_0
  \Biggr] \, ,
  \nonumber
\end{eqnarray}
with the abbreviations $A$, $B$, $C$ and $D$ as given above and
\begin{eqnarray}
  E & = &
  \trilog(1-z) - \trilog(z) + \trilog(1) - \frac{\pi^2}{6} \ln(z)
  \nonumber \\
  & - & \left( \dilog(1-z) + \frac{1}{2} \ln z \ln (1-z) \right) \ln (1-z)
  \; .
\end{eqnarray}
The single and double logarithmic terms in $L_0$ and $L_Q$ of the above
expressions (\ref{eq:int-Tg}) and (\ref{eq:int-Tij}) agree with
ref.~\cite{AAKM:nlo}.

It is easy to verify that the above integrals satisfy the relation
\begin{equation}
  \int\limits_\mathrm{PD} \dd \Omega_\gamma \,
  \left\{ 4zT_g'
    + Q_l^2\left[ T_{11}'+z^2T_{22}'+z\left(T_{12}'+T_{21}'\right)\right]
  \right\} = 0 \; ,
\end{equation}
which, after insertion into (\ref{eq:K-int}), leads to the factorization of
the virtual corrections as in eq.~(\ref{eq:virt-int-T}).

As a further check
we test the proper behavior in the soft photon limit, i.e., for $z \to 1$.
The coefficients $A$, $B$, $C$ and $D$ are at most logarithmically
divergent, and we find up to terms of order $\bigO{(1-z)^0}$:
\begin{eqnarray}
  \left.
  \frac{E_e^2}{\pi}
  \int\limits_\mathrm{PD} \dd\Omega_\gamma \; K_{\mu\nu}
  \right|_{z\to 1}
  & \simeq &
  \left( - Q_l^2 \tilde{g}_{\mu\nu}
         + 4\tilde{p}_{1\mu} \tilde{p}_{1\nu} \right)
  \frac{L_0-1}{1-z}
  \left[ 1 + \frac{\alpha}{2\pi} \left( \rho + \frac{1}{2} \right) \right]
  \nonumber \\
  & = &
  \left( - Q_l^2 \tilde{g}_{\mu\nu}
         + 4\tilde{p}_{1\mu} \tilde{p}_{1\nu} \right)
  \frac{ L_0 - 1 }{1-z} \, \left(F_1^\mathrm{(e)}(-Q_l^2)\right)^2
  . \qquad \quad {}
\end{eqnarray}
Here $F_1^\mathrm{(e)}(-Q_l^2)$ is the Dirac form factor of the electron in
the one-loop approximation.  The result (\ref{eq:virt-int-T}) thus fulfills
soft photon factorization as required, see also \cite{KMF87}, eq.~(36).


\section{Integrals for double collinear emission}
\label{sec:double-coll-int}

The calculation of the contribution from double collinear emission in
section~\ref{sec:double-hard} assumes that only the sum of the photon
energies, $(1-z)E_e$, will be measured in the forward photon detector.

With the help of (\ref{eq:dOmega-rewrite}) we split the integration of
expression (\ref{eq:2-coll-int}) over the restricted two-photon phase space
in the following way:
\begin{eqnarray}
  \label{eq:2-coll-ps}
  \lefteqn{
  \int \widetilde{\dd k}_1 \; \widetilde{\dd k}_2 \;
  \Theta(\vartheta_0 - \vartheta_1) \,
  \Theta(\vartheta_0 - \vartheta_2) \;
  \delta\left( (1-z) - (x_1+x_2) \right)
  \Bigl[ \ldots \Bigr]
  } \nonumber \\
  & = &
  \frac{1}{(4\pi)^4}
  \int x_1 \, \dd x_1 \int x_2 \, \dd x_2 \;
  \delta\left( (1-z) - (x_1+x_2) \right)
  \overline{\Bigl[ \ldots \Bigr]}
  \; .
\end{eqnarray}
In the last line we adopted the notation of Arbuzov et
al.~\cite{Arbuzov:1997qb},
\begin{eqnarray}
\label{def:overline()}
  \overline{\Bigl[ \ldots \Bigr]}
  & := &
  \frac{E_e^4}{\pi^2}
  \int \dd\Omega_1 \, \dd\Omega_2 \;
  \Theta(\vartheta_0 - \vartheta_1) \,
  \Theta(\vartheta_0 - \vartheta_2) \;
  \Bigl[ \ldots \Bigr]
  \; ,
\end{eqnarray}
for the angular part of the integrals.

Equation (\ref{eq:zi-coll}) shows that in the collinear region the $z_i$
are typically of the order of or larger than $m^2$, even in the limit of
high energies.  This knowledge allows us to perform the proper counting of
powers in selecting those terms in the differential cross section for
single and double photon emission that contribute in the collinear region.
For example,
\begin{eqnarray}
  \sigma
  & = & 2 k_1 \cdot k_2
  = 2 E_e^2 x_1 x_2 (1-\vec{n}_1 \cdot \vec{n}_2)
  \nonumber \\
  & = & 2 x_1 x_2 E_e^2
  [1-\cos\vartheta_1\cos\vartheta_2
    -\sin\vartheta_1\sin\vartheta_2\cos(\phi_1-\phi_2)]
  \nonumber \\
  & = & 4 x_1 x_2 E_e^2
  \left[\tau_1+\tau_2-2\tau_1\tau_2
  -2\sqrt{\tau_1\tau_2(1-\tau_1)(1-\tau_2)}\cos(\phi_1-\phi_2)\right]
  \nonumber \\
  & = &
  x_1 x_2 m^2
  \biggl[\zeta_1(1-\tau_2)+\zeta_2(1-\tau_1)
  \nonumber \\
  && \qquad \quad {}
  -2\sqrt{\zeta_1\zeta_2} \sqrt{(1-\tau_1)(1-\tau_2)}
    \cos(\phi_1-\phi_2)\biggr]
  \nonumber \\
  & \simeq &
  x_1 x_2 m^2
  \biggl[\zeta_1+\zeta_2
  -2 \sqrt{\zeta_1\zeta_2} \cos(\phi_1-\phi_2)\biggr]
  \; .
\end{eqnarray}
In the last line it is assumed that both angles $\tau_{1,2} \ll 1$ in the
collinear region.  We then find:
\begin{eqnarray}
  \Delta & = & z_1 + z_2 - \sigma
  \nonumber \\
  & \simeq &
  m^2 \left[
  x_1+x_2+x_1\zeta_1r_2+x_2\zeta_2r_1
  + 2x_1x_2 \sqrt{\zeta_1\zeta_2} \cos(\delta\phi)
  \right]
  \, , \quad
\end{eqnarray}
with $r_{1,2} = 1-x_{1,2}$ and $\delta\phi \equiv \phi_2-\phi_1$.

In the double collinear limit, the only non-trivial dependence on the
azimuthal angles of the photons appears in the expression $\Delta$.
Performing the integration over the relative azimuthal angles of the two
photons yields:
\begin{eqnarray}
  \int\limits_{-\pi}^{\pi} \frac{\dd (\delta\phi)}{2\pi} \, \frac{1}{\Delta}
  & \simeq &
  \frac{1}{m^2} \Bigl[
   \left( x_1 \zeta_1 r_2 + x_2 \zeta_2 r_1 + x_1 + x_2 \right)^2
  - 4 x_1^2 x_2^2 \zeta_1 \zeta_2
  \Bigr]^{-1/2}
  \nonumber \\ & =: & \frac{1}{m^2} D^{-1/2} \; ,
  \nonumber \\
  \int\limits_{-\pi}^{\pi} \frac{\dd (\delta\phi)}{2\pi} \, \frac{1}{\Delta^2}
  & \simeq &
  \frac{1}{m^4} \cdot
  \frac{x_1 \zeta_1 r_2 + x_2 \zeta_2 r_1 + x_1 + x_2}{D^{3/2}}
  \; .
\end{eqnarray}
At this point it turns out to be useful to introduce yet another
substitution for the polar angle variables of the photons:
\begin{equation}
\label{def:eta-zeta}
  \eta_1 = x_1 r_2 (1+\zeta_1)
  \; , \quad
  \eta_2 = x_2 r_1 (1+\zeta_2)
  \; .
\end{equation}
Inserting these definitions into the above expression for $D$, we obtain:
\begin{eqnarray}
\label{def:D}
  D(\eta_1,\eta_2) & = &
  \eta_1^2 + \eta_2^2 + 2 \eta_1 \eta_2 \frac{1-x_1-x_2-x_1x_2}{r_1 r_2}
  + \frac{4x_1x_2}{r_1 r_2} (r_1 \eta_1 + r_2 \eta_2)
  \nonumber \\
  & = &
  \eta_1^2 + \eta_2^2 + 2 \eta_1 \eta_2 \cos\psi
  + 2 (r_1 \eta_1 + r_2 \eta_2)(1-\cos\psi)
  \; ,
\end{eqnarray}
where
\begin{equation}
  \label{def:cospsi}
  \cos\psi := \frac{1-x_1-x_2-x_1x_2}{r_1 r_2}
  = 1 - \frac{2 x_1 x_2}{r_1 r_2}
  \; .
\end{equation}
Obviously, $-1 < \cos\psi \leq 1$ in the physically allowed range $0 \leq
x_i < 1$, $0 \leq x_1+x_2 < 1$.
Therefore we find:
\begin{eqnarray*}
  \int\limits_{-\pi}^{\pi} \frac{\dd (\delta\phi)}{2\pi} \, \frac{1}{\Delta}
  & \simeq &
  \frac{1}{m^2} \cdot \frac{1}{\sqrt{D(\eta_1,\eta_2)}} \; ,
  \\
  \int\limits_{-\pi}^{\pi} \frac{\dd (\delta\phi)}{2\pi} \, \frac{1}{\Delta^2}
  & \simeq &
  \frac{1}{m^4} \cdot
  \frac{\eta_1 + \eta_2 + 2 x_1 x_2}{[D(\eta_1,\eta_2)]^{3/2}} \; .
\end{eqnarray*}


\subsection{Integrals over photon angles}

We shall now provide the relevant angular integrals needed for the
contribution of two photons emitted almost collinearly to the incoming
electron.  With the replacement (\ref{def:eta-zeta}) definition
(\ref{def:overline()}) reads:
\begin{eqnarray}
  \overline{\Bigl[ \ldots \Bigr]}
  & = &
  m^4
  \int\limits_0^{\zeta_0} \dd\zeta_1 \, \dd\zeta_2 \;
  \int\limits_{-\pi}^{\pi}
  \frac{\dd\phi_1}{2\pi} \, \frac{\dd\phi_2}{2\pi} \,
  \Bigl[ \ldots \Bigr]
  \nonumber \\
  & = &
  \frac{m^4}{x_1 x_2 r_1 r_2}
  \int\limits_{x_1 r_2}^{x_1 r_2(1+\zeta_0)} \dd\eta_1
  \int\limits_{x_2 r_1}^{x_2 r_1(1+\zeta_0)} \dd\eta_2
  \int\limits_{-\pi}^{\pi}
  \frac{\dd\phi_1}{2\pi} \, \frac{\dd\phi_2}{2\pi} \,
  \Bigl[ \ldots \Bigr]
  \; , \qquad
\end{eqnarray}
and we have:
\begin{equation}
\label{eq:zi-etai}
  z_1 = \frac{m^2 \eta_1}{r_2}
  \; , \quad
  z_2 = \frac{m^2 \eta_2}{r_1}
  \; .
\end{equation}
The calculation is performed under the assumption that $\zeta_0 \gg 1$ and
$\vartheta_0^2 \ll 1$.

We begin with the integrals that give rise to double and single large
logarithms $L_0$.


\subsubsection{$\mathbf{1/(z_1 z_2)}$:}

This one is really trivial:
\begin{equation}
\label{eq:intlead1}
  \overline{ \left[ \frac{1}{z_1 z_2} \right] }
  = \frac{1}{x_1 x_2} \, L_0^2 \; .
\end{equation}


\subsubsection{$\mathbf{1/(z_1 \Delta)}$:}

We start with the integration over azimuthal angles:
\begin{equation}
  m^4 \int \dd\zeta_1 \, \dd\zeta_2
  \int\limits_{-\pi}^{\pi}
  \frac{\dd\phi_1}{2\pi} \, \frac{\dd\phi_2}{2\pi} \;
  \frac{1}{z_1 \Delta}
  =
  \frac{1}{x_1 x_2 r_1}
  \int \frac{\dd \eta_1}{\eta_1}
  \int \frac{\dd \eta_2}{\sqrt{D(\eta_1,\eta_2)} }
  \; .
\end{equation}
Next, the integral over $\eta_2$ is easily evaluated:
\begin{equation}
\label{eq:int-over-root-D}
  \int \frac{ \dd\zeta_2}{ \sqrt{D(\eta_1,\eta_2)} }
  = \ln E(\eta_1,\eta_2)
  \; ,
\end{equation}
with
\begin{equation}
  E(\eta_1,\eta_2)
  = \sqrt{D} + \frac{1}{2} \frac{\df D}{\df \eta_2}
  = \eta_1 \cos\psi + \eta_2 + r_2(1-\cos\psi) + \sqrt{D}
  \; .
\end{equation}
Let us also define
\begin{eqnarray}
  \tilde{D}(\eta_1,\eta_2)
  & := & \eta_1^2 + \eta_2^2 + 2\eta_1 \eta_2 \cos\psi
  \; ,
  \nonumber \\
  \tilde{E}(\eta_1,\eta_2)
  & := &
    \sqrt{\tilde{D}} + \frac{1}{2} \frac{\df \tilde{D}}{\df \eta_2}
  = \eta_1 \cos\psi + \eta_2 + \sqrt{\tilde{D}}
  \; ,
\end{eqnarray}
which reproduce the leading asymptotic behavior of $D$ and $E$ for
$\eta_{1,2} \gg 1$.
Some special values that will be needed below are:
\begin{eqnarray}
  \sqrt{D(\eta_1,\eta_2=x_2r_1)} & = &
  \eta_1 + x_2(1+x_1)
  \; ,
  \nonumber \\
  E(\eta_1,\eta_2=x_2r_1) & = &
  \frac{2}{r_1 r_2} (\chi \eta_1 + x_2 r_2)
  \; ,
  \\
  \sqrt{\tilde{D}(\eta_1,\eta_2=0)} & = &
  \eta_1
  \; ,
  \nonumber \\
  \tilde{E}(\eta_1,\eta_2=0) & = &
  (1+\cos\psi) \eta_1 = \frac{2 \chi}{r_1 r_2} \eta_1
  \; ,
  \nonumber \\
  \sqrt{\tilde{D}(\eta_1=0,\eta_2)} & = &
  \eta_2
  \; ,
  \nonumber \\
  \tilde{E}(\eta_1=0,\eta_2) & = & 2 \eta_2
  \; ,
  \nonumber \\
\noalign{\hbox{where}}
\label{def:chi}
  \chi & = & 1-x_1-x_2 \; .
\end{eqnarray}

Before proceeding with the $\eta_1$ integration, let us note that we are
finally interested in definite integrals over $\eta_2$, with lower limit
$\eta_2^\mathrm{min}=x_2r_1$.  Choosing the integration constant in
(\ref{eq:int-over-root-D}) appropriately so that the antiderivative
vanishes if $\eta_2$ is taken at the lower limit, we can decompose the
logarithm in the integrand in the following way:
\begin{equation}
\label{eq:decomp-E}
  \ln \frac{E(\eta_1,\eta_2)}{E(\eta_1,x_2r_1)}
  = \ln \frac{\tilde{E}(0,\eta_2)}{\tilde{E}(\eta_1,0)}
  - \ln \frac{E(\eta_1,x_2 r_1)}{\tilde{E}(\eta_1,0)}
  + \ln \frac{\tilde{E}(\eta_1,\eta_2)}{\tilde{E}(0,\eta_2)}
  + \ln \frac{E(\eta_1,\eta_2)}{\tilde{E}(\eta_1,\eta_2)}
  \; .
\end{equation}
Since we shall be interested in the case $\eta_2^\mathrm{max} \gg 1$, there
will be certain simplifications possible when considering the integral over
$\eta_1$.

The first integral is readily evaluated,
\begin{eqnarray}
  \int \frac{\dd\eta_1}{\eta_1} \;
  \ln \frac{\tilde{E}(0,\eta_2)}{\tilde{E}(\eta_1,0)}
  & = &
  \int \frac{\dd\eta_1}{\eta_1} \;
  \ln \frac{2\eta_2}{(1+\cos\psi)\eta_1}
  \nonumber \\
  & = &
  \ln \eta_1 \ln \frac{2\eta_2}{1+\cos\psi} - \frac{1}{2} \ln^2 \eta_1
  \; .
\end{eqnarray}
We shall later see that this term contributes to the leading double
logarithms.

The second term in the decomposition (\ref{eq:decomp-E}) gives:
\begin{eqnarray}
  \int \frac{\dd\eta_1}{\eta_1} \;
  \ln \frac{E(\eta_1,x_2 r_1)}{\tilde{E}(\eta_1,0)}
  & = &
  \int \frac{\dd\eta_1}{\eta_1} \;
  \ln \frac{\chi \eta_1 + x_2 r_2}{\chi \eta_1}
  = \int \frac{\dd\eta_1}{\eta_1} \;
  \ln \left( 1 + \frac{x_2 r_2}{\chi \eta_1} \right)
  \nonumber \\
  & = &
  \dilog \left( - \frac{x_2 r_2}{\chi \eta_1} \right)
  \; .
\end{eqnarray}

The third pieces gives:
\begin{eqnarray}
  \int \frac{\dd\eta_1}{\eta_1} \;
  \ln \frac{\tilde{E}(\eta_1,\eta_2)}{\tilde{E}(0,\eta_2)}
  & = &
  \int \frac{\dd\eta_1}{\eta_1} \;
  \ln \frac{\eta_1 \cos\psi + \eta_2 + \sqrt{\tilde{D}}}%
	   {(1+\cos\psi)\eta_1}
  \nonumber \\
  & = &
  \Xi \left( \cos\psi, \frac{\eta_1}{\eta_2} \right)
  \; ,
\end{eqnarray}
where we introduced the auxiliary function
\begin{equation}
\label{def:Xi}
  \Xi(t;x) :=
  \int\limits_0^x \frac{\dd \xi}{\xi}
  \ln \frac{ \sqrt{1+2t\xi+\xi^2} + t\xi + 1 }{2}
  \; .
\end{equation}
Evaluating the integral for $-1 < t \leq 1$ and using various identities
for dilogarithms (see e.g., \cite{Lewin:1981}) yields:
\begin{eqnarray}
  \Xi(t;x)
  & = &
  \frac{1}{2} \ln^2 \left( \frac{ \sqrt{1+2tx+x^2} + tx + 1 }{2} \right)
  \nonumber
  \\ &+&
  \dilog\left( \frac{(1+t)x}{ \sqrt{1+2tx+x^2} + tx + 1 } \right)
  \\ &+&
  \dilog\left( - \frac{(1-t)x}{ \sqrt{1+2tx+x^2} + tx + 1 } \right)
  \; .
  \nonumber
\end{eqnarray}
The asymptotic behavior of this expression for $x \to +\infty$ is found to
be:
\begin{equation}
\label{eq:Xi-asympt}
  \Xi(t;x) =
    \frac{1}{2} \ln^2 \left( \frac{(1+t)x}{2} \right)
  + \frac{\pi^2}{6}
  + \dilog \left( - \frac{1-t}{1+t} \right)
  - \frac{1}{x}
  + \bigO{x^{-2}} \; .
\end{equation}

We are finally left with the fourth contribution,
\begin{equation}
  \int \frac{\dd\eta_1}{\eta_1} \;
  \ln \frac{E(\eta_1,\eta_2)}{\tilde{E}(\eta_1,\eta_2)}
  =
  \int \frac{\dd\eta_1}{\eta_1} \;
  \ln \frac{\eta_1 \cos\psi + \eta_2 + r_2(1-\cos\psi) + \sqrt{D}}%
	   {\eta_1 \cos\psi + \eta_2 + \sqrt{\tilde{D}}}
  \; ,
\end{equation}
which appears quite elaborate but can nevertheless be reduced to
dilogarithms.  Fortunately, this effort is not necessary as we need this
expression for large $\eta_2 = x_2 r_1(1+\zeta_0) \gg 1$.  As a consequence
the ratio $\xi=\eta_1/\eta_2$ does not become very large, since we
integrate symmetrically over the polar angles of the photons.  Substituting
$\eta_1=\xi\eta_2$ and expanding the integrand for large $\eta_2$, we
obtain:
\begin{equation}
  \int \frac{\dd\xi}{\xi} \;
  \left[ \frac{1}{\eta_2}
  \left( r_2(1-\cos\psi) + \bigO{\xi} \right)
    + \bigO{\frac{1}{\eta_2^2}}
  \right]
  \sim \frac{1}{\eta_2} \ln \frac{\eta_1}{\eta_2}
  \; .
\end{equation}
Therefore, this expression is $\bigO{1/\eta_2}$ and therefore always
negligible for symmetric integration.

Collecting the contributions from the first three terms of the
decomposition, inserting the integration limits and keeping only terms that
are not suppressed by factors $1/\zeta_0$, we find:
\begin{equation}
\label{eq:intlead2}
  \overline{ \left[ \frac{1}{z_1 \Delta} \right] }
  =
  \frac{1}{x_1 x_2 r_1}
  \Biggl[
  \frac{1}{2} L_0^2
  + L_0 \ln \frac{ x_2 r_1^2}{x_1 z}
  + \dilog \left( - \frac{x_2}{x_1 z} \right)
  + \Xi\left( \cos\psi; \frac{x_1 r_2}{x_2 r_1} \right)
  \Biggr]
  \; .
\end{equation}


\subsubsection{$\mathbf{m^2/(z_1^2 \Delta)}$:}

With the same steps as in the previous case we first find:
\begin{eqnarray}
  m^4 \int \dd\zeta_1 \, \dd\zeta_2
  \int\limits_{-\pi}^{\pi}
  \frac{\dd\phi_1}{2\pi} \, \frac{\dd\phi_2}{2\pi} \;
  \frac{m^2}{z_1^2 \Delta}
  & = &
  \frac{r_2}{x_1 x_2 r_1}
  \int \frac{\dd \eta_1}{\eta_1^2}
  \int \frac{\dd \eta_2}{ \sqrt{D(\eta_1,\eta_2)} }
  \nonumber \\
  & = &
  \frac{r_2}{x_1 x_2 r_1}
  \int \frac{\dd \eta_1}{\eta_1^2} \, \ln E(\eta_1,\eta_2) \; .
\end{eqnarray}
Again we apply the decomposition (\ref{eq:decomp-E}) to the logarithm in
the integrand.  Integrating the sum of the first and second term, we
obtain:
\begin{eqnarray}
  \int \frac{\dd\eta_1}{\eta_1^2} \;
  \ln \frac{\tilde{E}(0,\eta_2)}{E(\eta_1,x_2 r_1)}
  & = &
  \int \frac{\dd\eta_1}{\eta_1^2} \;
  \ln \frac{r_1 r_2\eta_2}{\chi\eta_1+x_2 r_2}
  \\
  & = &
  \frac{\chi}{x_2 r_2} \ln \left( \frac{\chi \eta_1+x_2 r_2}{\eta_1} \right)
  - \frac{1}{\eta_1} \ln \frac{r_1 r_2\eta_2}{\chi\eta_1+x_2 r_2}
  \; . \nonumber
\end{eqnarray}
The third term can be neglected for symmetric integration over the photon
polar angles.  This can be seen by substituting $\eta_1=\eta_2/\xi'$,
leading to
\begin{eqnarray}
  \frac{1}{\eta_2} \int \dd \xi' \;
  \ln \frac{\tilde{E}(\eta_2/\xi',\eta_2)}{\tilde{E}(0,\eta_2)}
  & = &
  \frac{1}{\eta_2} \int \dd \xi' \;
  \ln \frac{\xi' + \cos\psi + \sqrt{1+2\xi'\cos\psi+\xi'{}^2}}%
	   {1+\cos\psi}
  \nonumber \\
  & = & \bigO{ \frac{1}{\eta_2} }
  \; .
\end{eqnarray}
The contribution from the fourth term of decomposition (\ref{eq:decomp-E})
turns out to be negligible for the same reason.
We thus obtain:
\begin{eqnarray}
\label{eq:intlead3}
  \overline{ \left[ \frac{m^2}{z_1^2 \Delta} \right] }
  & = &
  \frac{1}{x_1^2 x_2 r_1} \left[
    L_0 +
    \ln \frac{ x_2 r_1 }{1-z}
  \right]
  -
  \frac{z}{x_1 x_2^2 r_1}
  \ln \frac{r_1 (1-z)}{ x_1 z } \; .
\end{eqnarray}


\subsubsection{$\mathbf{z_2/(z_1 \Delta^2)}$:}


After integration over azimuthal angles we have:
\begin{equation}
  m^4 \int \dd\zeta_1 \, \dd\zeta_2
  \int\limits_{-\pi}^{\pi}
  \frac{\dd\phi_1}{2\pi} \, \frac{\dd\phi_2}{2\pi} \;
  \frac{z_2}{z_1 \Delta^2}
  =
  \frac{1}{x_1 x_2 r_1^2}
  \int \dd \eta_1
  \int\dd \eta_2 \;
  \frac{\eta_2(\eta_1+\eta_2+2x_1x_2)}{\eta_1 \, [D(\eta_1,\eta_2)]^{3/2} }
  \; .
\end{equation}
It is interesting to note that the integral over $\eta_2$ can be split into
one piece that is precisely the same as occurred in the discussion of
$1/(z_1\Delta)$, plus another one which at this stage is free from
logarithms and involves only algebraic functions.
\begin{eqnarray}
  \int\dd \eta_2 \;
  \frac{\eta_2(\eta_1+\eta_2+2x_1x_2)}{\eta_1 \, [D(\eta_1,\eta_2)]^{3/2} }
  & = &
  \frac{1}{\eta_1}
  \int
  \frac{\dd \eta_2}{\sqrt{D(\eta_1,\eta_2)}}
  \nonumber \\
  &-&
  \frac{1}{1+\cos\psi}
  \frac{1}{\sqrt{D(\eta_1,\eta_2)}}
  \nonumber \\
  &-&
  (2-r_1)
  \frac{\eta_2}{\eta_1 \, \sqrt{D(\eta_1,\eta_2)}}
  \\
  &+&
  \left( \frac{1}{1+\cos\psi} - r_1 \right)
  \frac{\chi\eta_2+x_2(x_1-\chi)}{(\chi\eta_1+x_2 r_2)\sqrt{D(\eta_1,\eta_2)}}
  \; . \nonumber
\end{eqnarray}
The integration of the r.h.s.\ over $\eta_1$ now is quite simple.  The
first term leads to an expression that has been discussed further above.
The remaining antiderivatives read:
\begin{eqnarray}
  \int \frac{\dd \eta_1}{\sqrt{D(\eta_1,\eta_2)}}
  & = &
  \ln F(\eta_1,\eta_2)
  \; ,
  \nonumber \\
  \int \frac{\eta_2}{\eta_1 \, \sqrt{D(\eta_1,\eta_2)}} \, \dd \eta_1
  & = &
  G(\eta_1,\eta_2)
  \; ,
  \nonumber \\
  \int
  \frac{\chi\eta_2+x_2(x_1-\chi)}{(\chi\eta_1+x_2 r_2) \sqrt{D(\eta_1,\eta_2)}}
  \, \dd \eta_1
  & = &
  \ln K(\eta_1,\eta_2)
  \; ,
\end{eqnarray}
with
\begin{eqnarray}
\label{eq:FGK}
  F(\eta_1,\eta_2)
  & = & \sqrt{D} + \frac{1}{2} \frac{\df D}{\df \eta_1}
  = \eta_2 \cos\psi + \eta_1 + r_1(1-\cos\psi) + \sqrt{D}
  \; ,
  \nonumber \\
  G(\eta_1,\eta_2)
  & = &
  \sqrt{\frac{\eta_2}{\eta_2+2r_2(1-\cos\psi)}} \,
  \bigggl\{
  \ln \eta_1
  \\
  && \;\;{} -
    \ln \left[ \sqrt{D} +
    \frac{\eta_2^2+\eta_1\eta_2\cos\psi+(1-\cos\psi)(\eta_1 r_1 + 2\eta_2 r_2)}%
         {\sqrt{\eta_2[\eta_2+2r_2(1-\cos\psi)]}}
    \right]
  \bigggr\} ,
  \nonumber \\
  K(\eta_1,\eta_2)
  & = &
  \frac{\sqrt{D} - \eta_1\cos\psi - \eta_2 - r_2(1-\cos\psi)}%
       {\chi \eta_1 + x_2 r_2}
  +
  \frac{(1-\chi)(1-\cos\psi)}{\chi \eta_2 + x_2(x_1-\chi)}
  \; . \nonumber
\end{eqnarray}
Collecting the above expressions and inserting limits, we obtain for the
contribution in the doubly collinear region:
\begin{eqnarray}
\label{eq:intlead4}
  \overline{ \left[ \frac{z_2}{z_1 \Delta^2} \right] }
  & = &
  \frac{1}{x_1 x_2 r_1^2}
  \Biggl\{
  \frac{1}{2} L_0^2
  + L_0
    \left[ \ln \frac{ x_2 r_1^2}{x_1 \chi} - 1 + \frac{x_1x_2}{\chi} \right]
  \nonumber
  \\ && \qquad \qquad {}
   + \dilog \left( - \frac{x_2}{x_1 \chi} \right)
   + \Xi\left( \cos\psi; \frac{x_1 r_2}{x_2 r_1} \right)
  \\ && \qquad \qquad {}
  + \frac{r_1 r_2-2\chi}{\chi} \ln (2x_2)
  + \frac{r_1(r_2-2\chi)}{\chi} \ln \frac{x_1}{1-\chi}
  \nonumber
  \\ && \qquad \qquad {}
  + \frac{r_1(3r_2-4\chi)}{2\chi} \ln \chi
  - \frac{r_1r_2+4x_1\chi}{2\chi} \ln r_1
  - \frac{r_1 r_2}{2\chi}   \ln r_2
  \nonumber
  \\ && \qquad \qquad {}
  + \frac{4\chi-r_1r_2}{2\chi}
    \ln \left( \eta + x_2 r_1 + x_1 r_2 \cos\psi \right)
  \nonumber
  \\ && \qquad \qquad {}
  - \frac{r_1r_2}{2\chi}
    \ln \left( \eta + x_1 r_2 + x_2 r_1 \cos\psi \right)
  \Biggr\}
  \; ,
  \nonumber
\end{eqnarray}
with the abbreviation
\begin{equation}
  \eta =
  \sqrt{\tilde{D}(x_1 r_2,x_2 r_1)} =
  \sqrt{(x_1+x_2)(x_1+x_2-4x_1x_2)}
  \; .
\end{equation}
Comparing so far the above results with \cite{Mer88,Arbuzov:1997qb}, we
find that we reproduce the double and single logarithms in $L_0$.
Furthermore, our expressions also contain the `finite' terms for $\zeta_0
\gg 1$.

The double Compton tensor in the double collinear region
(\ref{eq:K-2-coll}) contains also further terms that contribute only
non-(logarithmically-)enhanced terms.  The corresponding integrals are
convergent if we remove the upper limit by taking $\zeta_0 \to \infty$.
Inspecting the expressions for these additional terms reveals that the
corresponding integrands contribute only when $\eta_i \lesssim \bigO{1}$,
i.e., the angle of both photons is of the order of $\vartheta_{1,2}
\lesssim \bigO{m/E_e}$.  Taking one photon at a large angle immediately
leads to a strong suppression.  We therefore take the upper limit of both
integration variables $\eta_i$ as infinite.

\subsubsection{$\mathbf{m^2/(z_1 z_2 \Delta)}$:}

\begin{equation}
  m^4 \int \dd\zeta_1 \, \dd\zeta_2
  \int\limits_{-\pi}^{\pi}
  \frac{\dd\phi_1}{2\pi} \, \frac{\dd\phi_2}{2\pi} \;
  \frac{m^2}{z_1 z_2 \Delta}
  =
  \frac{1}{x_1 x_2}
  \int
  \frac{\dd \eta_1 \, \dd \eta_2}{\eta_1 \, \eta_2 \, \sqrt{D(\eta_1,\eta_2)} }
\end{equation}
One integration, e.g., over $\eta_2$, is elementary:
\begin{equation}
\label{eq:auxint-1}
  \int
  \frac{\dd \eta_2}{\eta_2 \, \sqrt{D(\eta_1,\eta_2)} }
  =
  \frac{\ln \eta_2
      - \ln \left[ \sqrt{D} +
      \frac{\eta_1^2+\eta_1\eta_2\cos\psi+(1-\cos\psi)(2\eta_1 r_1 + \eta_2 r_2)}%
           {\sqrt{\eta_1[\eta_1+2r_1(1-\cos\psi)]}}
        \right]
  + c_1
       }{\sqrt{\eta_1[\eta_1+2r_1(1-\cos\psi)]}}
  \; .
\end{equation}
The integral is convergent for $\eta_2 \to \infty$, so we are free to
choose the constant $c_1$ (w.r.t.\ $\eta_2$) in such a way that the r.h.s.\
of (\ref{eq:auxint-1}) vanishes in that limit:
\begin{equation}
  c_1 =
  \ln \left[ 1 +
      \frac{ \eta_1\cos\psi+r_2(1-\cos\psi) }%
           { \sqrt{\eta_1[\eta_1+2r_1(1-\cos\psi)]} }
      \right]
  \; .
\end{equation}
We therefore obtain for the definite integral:
\begin{eqnarray}
\int\limits_{x_2 r_1}^\infty
  \frac{\dd \eta_2}{\eta_2 \, \sqrt{D(\eta_1,\eta_2)} }
  & = &
  \frac{
      \ln \left[ \eta_1+x_2(1+x_1) +
      \frac{\eta_1^2+r_1 \eta_1[x_2\cos\psi+2(1-\cos\psi)]+2x_1 x_2^2}%
           {\sqrt{\eta_1[\eta_1+2r_1(1-\cos\psi)]}}
        \right]
       }{\sqrt{\eta_1[\eta_1+2r_1(1-\cos\psi)]}}
  \nonumber \\
  &-& \frac{ \ln (x_2 r_1) + c_1
       }{\sqrt{\eta_1[\eta_1+2r_1(1-\cos\psi)]}}
  \; .
\end{eqnarray}
The integration over $\eta_1$ is attacked with the help of the substitution
\begin{equation}
\label{eq:subs-eta-rho}
  \eta_1 = \frac{2r_1(1-\cos\psi)}{\rho_1(\rho_1+2)}
  \; , \quad
  \frac{\dd\eta_1}{\eta_1^{3/2} \sqrt{\eta_1+2r_1(1-\cos\psi)}}
  = \frac{\dd \rho_1}{r_1 (1-\cos\psi)}
  \; ,
\end{equation}
leading to:
\begin{equation}
  \int\limits_{x_1 r_2}^\infty
  \frac{\dd \eta_1}{\eta_1}
  \int\limits_{x_2 r_1}^\infty
  \frac{\dd \eta_2}{\eta_2 \, \sqrt{D(\eta_1,\eta_2)} }
  =
  \frac{r_2}{2x_1x_2}
  \int\limits_0^{2x_2/r_2}
  \dd \rho_1 \;
  \ln \frac{(\rho_1+1)(\rho_1+2x_1/r_2)}{\rho_1(\rho_1+2\chi/r_2)}
  \; .
\end{equation}
Its evaluation yields:
\begin{eqnarray}
\label{eq:intsub1}
  \overline{ \left[ \frac{m^2}{z_1 z_2 \Delta} \right] }
  & = &
  \frac{1}{x_1^2 x_2^2}
  \Biggl[
   (1-\chi) \ln (1-\chi) + \chi \ln \chi
   - r_1 \ln r_1 - r_2 \ln r_2
  \nonumber \\ && \qquad \quad {}
   - x_1 \ln x_1 - x_2 \ln x_2
  \Biggr]
  \; .
\end{eqnarray}

\subsubsection{$\mathbf{m^2/(z_1 \Delta^2)}$:}

\begin{equation}
  m^4 \int \dd\zeta_1 \, \dd\zeta_2
  \int\limits_{-\pi}^{\pi}
  \frac{\dd\phi_1}{2\pi} \, \frac{\dd\phi_2}{2\pi} \;
  \frac{m^2}{z_1 \Delta^2}
  =
  \frac{1}{x_1 x_2 r_1}
  \int \dd \eta_1 \, \dd \eta_2 \;
  \frac{\eta_1+\eta_2+2x_1x_2}{\eta_1 \, [D(\eta_1,\eta_2)]^{3/2}}
  \; .
\end{equation}
Performing the integration over $\eta_2$, one finds for the r.h.s.:
\begin{equation}
\label{eq:subl2}
  \frac{1}{2 x_1 x_2^2}
  \int \frac{\dd \eta_1}{\sqrt{D(\eta_1,\eta_2)}}
  \left[
  \frac{\eta_2 - 2x_2}{\eta_1}
  - \frac{\chi \eta_2 + x_2 (x_1 - \chi)}{\chi \eta_1 + x_2 r_2}
  \right]
\end{equation}
Obviously, this can be expressed in terms of the already known expressions
$G$ and $\ln K$ given in (\ref{eq:FGK}).
%
%
However, this is not really necessary, as we only need the definite
integral which is absolutely convergent.  Evaluating (\ref{eq:subl2}) in
the limits $\eta_2=x_2r_1 \ldots \infty$, we arrive at a much simpler
expression:
\begin{equation}
  \frac{1}{x_1 x_2^2}
  \int \dd \eta_1 \; \frac{x_2 r_2}{\eta_1[\chi \eta_1 + x_2 r_2]}
  \; .
\end{equation}
This immediately leads to
\begin{equation}
\label{eq:intsub2}
  \overline{ \left[ \frac{m^2}{z_1 \Delta^2} \right] }
  =
  \frac{1}{x_1 x_2^2} \ln \frac{r_1(1-\chi)}{x_1 \chi}
  \; .
\end{equation}

\subsubsection{$\mathbf{m^4/(z_1^2 \Delta^2)}$:}

This integral can be dealt with in a similar way.  Starting from the
expression
\begin{equation}
  m^4 \int \dd\zeta_1 \, \dd\zeta_2
  \int\limits_{-\pi}^{\pi}
  \frac{\dd\phi_1}{2\pi} \, \frac{\dd\phi_2}{2\pi} \;
  \frac{m^4}{z_1^2 \Delta^2}
  =
  \frac{r_2}{x_1 x_2 r_1}
  \int \dd \eta_1 \, \dd \eta_2 \;
  \frac{\eta_1+\eta_2+2x_1x_2}{\eta_1^2 \, [D(\eta_1,\eta_2)]^{3/2}}
  \; ,
\end{equation}
and performing the integration over $\eta_2$, we arrive at the following
expression for the r.h.s.:
\begin{eqnarray}
  \frac{1}{2 x_1 x_2^3}
  \int \frac{\dd \eta_1}{\eta_1 \, \sqrt{D(\eta_1,\eta_2)}}
  \left[
  \frac{\eta_2 - 2x_2}{\eta_1}
  - \frac{\chi \eta_2 + x_2 (x_1 - \chi)}{\chi \eta_1 + x_2 r_2}
  \right] \; .
\end{eqnarray}
Again, we evaluate this in the limits $\eta_2=x_2r_1 \ldots \infty$, to
obtain:
\begin{equation}
  \frac{r_2}{x_1 x_2^2}
  \int \dd \eta_1 \; \frac{x_2 r_2}{\eta_1^2[\chi \eta_1 + x_2 r_2]}
  \; .
\end{equation}
The remaining integral is trivial, yielding:
\begin{equation}
\label{eq:intsub3}
  \overline{ \left[ \frac{m^4}{z_1^2 \Delta^2} \right] }
  =
  \frac{1}{x_1^2 x_2^2}
  \left[
    1 - \frac{x_1 \chi}{x_2} \ln \frac{r_1(1-\chi)}{x_1 \chi}
  \right]
  \; .
\end{equation}

\subsubsection{$\mathbf{m^4/(z_1 z_2 \Delta^2)}$:}

The last and most involved integral among the non-logarithmic ones that is
needed for the double collinear region is the following:
\begin{equation}
  m^4 \int \dd\zeta_1 \, \dd\zeta_2
  \int\limits_{-\pi}^{\pi}
  \frac{\dd\phi_1}{2\pi} \, \frac{\dd\phi_2}{2\pi} \;
  \frac{m^4}{z_1 z_2 \Delta^2}
  =
  \frac{1}{x_1 x_2}
  \int \dd \eta_1 \, \dd \eta_2 \;
  \frac{\eta_1+\eta_2+2x_1x_2}{\eta_1 \eta_2 \, [D(\eta_1,\eta_2)]^{3/2}}
  \; .
\end{equation}
We start with the integration over $\eta_2$ to obtain:
%
\begin{eqnarray}
\label{eq:subl4}
  &&
  \frac{1}{x_1 x_2}
  \int \frac{\dd \eta_1}{\sqrt{D(\eta_1,\eta_2)}}
  \Biggl[
    \frac{r_2(r_1 \eta_2 + 4 x_1 x_2)}{4 x_1 x_2 \eta_1^2}
  + \frac{r_1 \chi [\chi \eta_2 + x_2 (x_1 - \chi)]}{2 x_2^2 (x_1-\chi)[\chi \eta_1 + x_2 r_2]}
  \nonumber \\ && \qquad \qquad \quad
  + \frac{r_1^2 + x_1^2 - x_2^2}{4x_1 x_2 \eta_1}
  - \frac{r_1 \eta_2 [ \chi^2 + x_1(4\chi-x_1)]}{16 x_1^2 x_2^2 \eta_1}
  \\ && \qquad \qquad \quad
  - \frac{r_1 r_2^4 \eta_2}{16x_1^2 x_2^2 (x_1-\chi) [r_2 \eta_1 + 4x_1 x_2]}
  - \frac{r_2^2 (1+x_2)}{4x_1 x_2 [r_2 \eta_1 + 4x_1 x_2]}
  \Biggr]
  \nonumber \\ & + &
  \frac{1}{x_1 x_2}
  \int \dd \eta_1 \;
  \frac{\eta_1 + r_1 r_2 (1-\cos\psi)}
       {\eta_1^{5/2} \sqrt{\eta_1 + 2r_1(1-\cos\psi)}^3}
  \bigggl[
  \ln \eta_2
  \nonumber \\ && \qquad
    - \ln \left[ \sqrt{D(\eta_1,\eta_2)} +
        \frac{\eta_1^2+\eta_1\eta_2\cos\psi+(1-\cos\psi)(2\eta_1 r_1 + \eta_2 r_2)}%
             {\sqrt{\eta_1[\eta_1+2r_1(1-\cos\psi)]}}
          \right]
  \bigggr] . \nonumber
\end{eqnarray}
Again, these expressions confirm that the integrals are absolutely
convergent.  We therefore immediately insert the limits for $\eta_2$.  The
remaining $\eta_1$ integration is straightforward for the first part of
(\ref{eq:subl4}); the second is part is tackled with the help of the
substitution (\ref{eq:subs-eta-rho}):
\begin{eqnarray}
  &&
  \frac{1}{x_1 x_2}
  \Biggl[
  - \frac{1}{2 x_1 x_2}
  + \frac{r_1 \chi}{x_2^2 (\chi-x_1)}
    \ln \frac{r_1(1-\chi)}{x_1 \chi}
  + \frac{r_2^2 (1+x_2)}{4x_1 x_2^2(\chi-x_1)} \ln \frac{r_2}{1+x_2}
  \Biggr]
  \nonumber \\
  &+&
  \frac{r_2^2}{x_1^3 x_2^3}
  \int\limits_0^{2x_2/r_2} \dd \rho_1 \;
  \frac{\rho_1(2+\rho_1)[r_2(\rho_1+1)^2+2-r_2]}{16 (\rho_1+1)^2}
  \ln \frac{(\rho_1+2)(\rho_1+2x_1/r_2)}
           {\rho_1(\rho_1+2\chi/r_2)}
  \nonumber \\
  & = &
  \frac{1}{6 x_1^3 x_2^3}
  \Biggl[
      x_1^2 (3-2x_1) \ln \frac{(1-\chi)r_1}{x_1 \chi}
    + x_2^2 (3-2x_2) \ln \frac{(1-\chi)r_2}{x_2 \chi}
  \nonumber \\ && \qquad \quad
    + \ln \frac{\chi}{r_1 r_2} - 2 x_1 x_2
  \Biggr] \; .
\end{eqnarray}
Fortunately, an intermediate spurious singularity in $(\chi-x_1)$ cancels,
and we find:
\begin{eqnarray}
\label{eq:intsub4}
  \overline{ \left[ \frac{m^4}{z_1 z_2 \Delta^2} \right] }
  & = &
  \frac{1}{6 x_1^3 x_2^3}
  \Biggl[
      x_1^2 (3-2x_1) \ln \frac{(1-\chi)r_1}{x_1 \chi}
    + x_2^2 (3-2x_2) \ln \frac{(1-\chi)r_2}{x_2 \chi}
  \nonumber \\ && \qquad \quad {}
    + \ln \frac{\chi}{r_1 r_2} - 2 x_1 x_2
  \Biggr]
  \; .
\end{eqnarray}
The remaining integrals can be obtained from those given above by
exchanging the labels 1 and 2, i.e., $x_1 \leftrightarrow x_2$ and $r_1
\leftrightarrow r_2$.

Once again we stress that the coefficients of the double and single
logarithmic terms ($L_0^2$ and $L_0$) in eqs.\ (\ref{eq:intlead1}),
(\ref{eq:intlead2}), (\ref{eq:intlead3}) and (\ref{eq:intlead4}) agree with
refs.~\cite{Mer88,Arbuzov:1997qb}.  Furthermore we have calculated also the
non-logarithmic contributions.


\subsection{Integrals over relative photon energy}

Having performed the angular integration, we still need to integrate over
the relative photon energy, see (\ref{eq:2-coll-ps}).  As this integral
will be infrared-divergent, we introduce a soft-photon cutoff $\eps$ on the
minimum energy fraction of each photon, which is identical to the one used
in the soft-photon contribution.  We thus define:
\begin{equation}
  \left\langle
  \Bigl[
  \cdots
  \Bigr]
  \right\rangle
  :=
  \int\limits_\eps^1 x_1 \, \dd x_1 \int\limits_\eps^1 x_2 \, \dd x_2 \;
  \delta\left( (1-z) - (x_1+x_2) \right)
  \Bigl[
  \cdots
  \Bigr] \; .
\end{equation}
With the substitution $ x_1 \to (1-z) \cdot u $ and the abbreviation
$\tilde\eps=\eps/(1-z)$, we have, after elimination of the trivial
$\delta$-function:
\begin{eqnarray*}
  \left\langle
  \Bigl[
  \cdots
  \Bigr]
  \right\rangle
  & = &
  (1-z)
  \int\limits_{\tilde\eps}^{1-\tilde\eps} \dd u \;
  \, x_1 \, x_2
  \Bigl[
  \cdots
  \Bigr]_{x_1 = (1-z) u, x_2 = (1-z)(1-u), \ldots}
\end{eqnarray*}
For the logarithmically ($L_0$) enhanced leading terms, we find the familiar
result (\ref{eq:P2-log}).

The analytic calculation of the remaining, non-enhanced terms is quite
tedious, leading to lengthy expressions involving many dilogarithms and
trilogarithms (see e.g., \cite{Lewin:1981}).
As these terms also contain IR-divergent contributions, we shall pursue
here the following approach.  We analytically extract those terms in the
integrand that either contribute to the infrared-divergence as $\eps \to 0$
or survive in the limit $z \to 1$, before performing the integral over the
remaining expression numerically.  Besides, this separation improves the
stability of the numerical integration.

For the IR-divergent pieces of the non-enhanced terms we find
\begin{eqnarray}
\label{eq:P2-IR-div}
  P_\mathrm{nonlog}^\mathrm{IR-div}
  & = &
  \left\langle
    \frac{4z}{(x_1 x_2)^2}
  \right\rangle
  =
  \frac{4z}{1-z}
  \int\limits_{\tilde\eps}^{1-\tilde\eps} \dd u \;
  \frac{1}{u(1-u)}
  \simeq \frac{8z}{1-z} \ln \tilde\eps
  \; .
\end{eqnarray}
We then decompose the infrared-finite pieces as follows:
\begin{equation}
  P^{\mathrm{(2),IR-fin.}}_\mathrm{nonlog}(z)
  = P_\mathrm{nonlog}^{z\to 1} + P_\mathrm{nonlog}^\mathrm{rem}(z) \; ,
\end{equation}
so that $P_\mathrm{nonlog}^\mathrm{rem}(z=1)=0$.  The first term on the
r.h.s., obtained in the limit $z \to 1$ reads:
\begin{eqnarray}
  P_\mathrm{nonlog}^{z\to 1}
  & = &
  \Biggl\langle
    \frac{8}{3 x_1 x_2}
    \Biggl[
      \frac{x_1 + 3x_2}{x_2^2} \ln x_1
    + \frac{3x_1 + x_2}{x_1^2} \ln x_2
  \nonumber \\ && \qquad \qquad {}
    - \frac{(1-z)^3}{x_1^2 x_2^2} \ln (1-z)
    + \frac{1-z}{x_1 x_2}
    \Biggr]
  \Biggr\rangle
  \nonumber \\
  & = &
  \int\limits_{\tilde\eps}^{1-\tilde\eps} \dd u \;
  \frac{8}{3}
  \left[
    \frac{1}{u(1-u)}
    + \frac{3-2u}{(1-u)^2} \ln u
    + \frac{1+2u}{u^2} \ln (1-u)
  \right]
  \nonumber \\
  & = & - \frac{16}{9} \left( 3 + \pi^2 \right)
  \; ,
\end{eqnarray}
where we neglected terms of order $\eps$.

\begin{figure}
  \begin{center}
    \begin{picture}(120,70)
      \put(0,0){\includegraphics{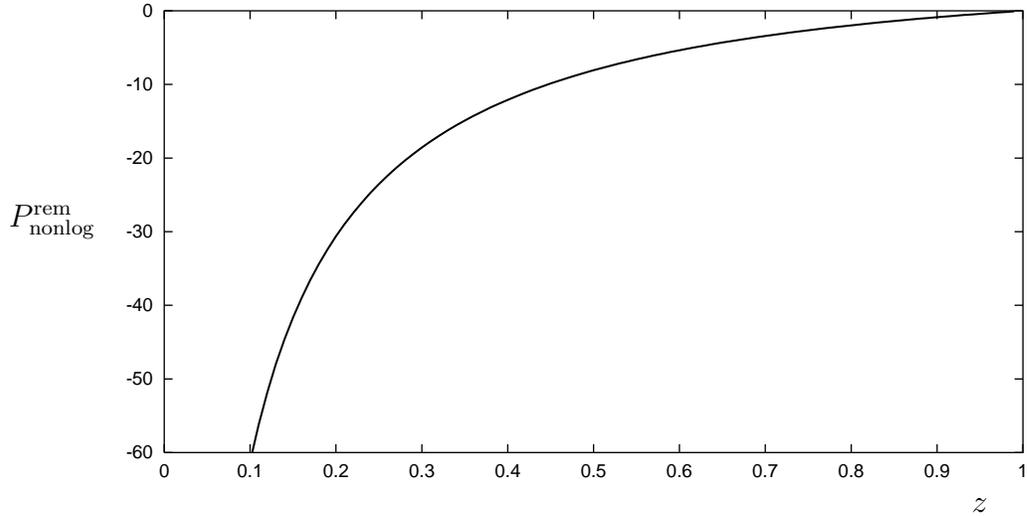}}
      \put(116,-3){$z$}
      \put(-12,35){$P_\mathrm{nonlog}^\mathrm{rem}$}
    \end{picture}
    \caption{The infrared-finite, `normalized' part of the double-collinear
      contribution, $P_\mathrm{nonlog}^\mathrm{rem}(z)$.}
    \label{fig:Pnonlog}
  \end{center}
\end{figure}

Finally, we plot in
figure~\ref{fig:Pnonlog}
the `remainder' $P_\mathrm{nonlog}^\mathrm{rem}(z)$, which we evaluated
numerically.


\chapter{The Semi-Collinear Contribution}
\label{sec:sc-calc}

In this appendix we shall describe the steps leading to the almost
factorization (\ref{eq:leptonic-semi-coll}) of the contributions from
double photon emission in the region of semi-collinear kinematics.

We assume that one photon, say photon 1, is almost collinear to the
incident electron and thus detected in the forward PD, i.e.,
$\vartheta_1<\vartheta_0$, while the other one, photon 2, is radiated at
larger angles, $\vartheta_2>\vartheta_0$.  For the computation of this
contribution we formally need to perform the phase space integration for
the two photons.  The solid angle part of this integral schematically
reads:%
\footnote{For the discussion below it is useful to temporarily consider the
  contraction of the Compton tensor with the hadron tensor.}
\begin{equation}
\label{eq:KdotH}
  \frac{E_e^4}{\pi^2}
  \int\limits_{\vartheta_1<\vartheta_0} \dd\Omega_1
  \int\limits_{\vartheta_2>\vartheta_0} \dd\Omega_2 \;
  K_{\mu\nu}^{\mathrm{semi-coll}}(\vartheta_1,\vartheta_2;\phi_1,\phi_2,\ldots)
  H^{\mu\nu}(\vartheta_1,\vartheta_2; \ldots)
  \; .
\end{equation}
The hadron tensor $H^{\mu\nu}$ depends on the photon angles only through
the transferred momentum $q=p-p'-k_1-k_2$.  It is thus natural to expect
that the tensor $H^{\mu\nu}(P,q)$ is a smooth function of the photon
angles, as long as the invariant mass of the final hadronic system,
$W^2=(P+q)^2$, stays well above the inelastic threshold of pion production,
$W^2 \gg (M+m_\pi)^2$.
%
%
Within the desired accuracy, which implies dropping terms of
$\bigO{\vartheta_0^2}$, we set $\vartheta_1$ to 0 in the argument of
$H^{\mu\nu}$.


In principle, given $K_{\mu\nu}^{\mathrm{semi-coll}}$, this allows us to
perform the integrations over $\phi_1$ and $\vartheta_1$ immediately.
Before doing this, however, we must first determine those terms of the full
expression for the double Compton tensor that lead to nonvanishing
contributions in the semi-collinear region.  Second, we have to find a
procedure to systematically expand the result in the limit of large
$\zeta_0$, which means to find the analogue of the logarithmic and
non-logarithmic terms in the double collinear region.  This last step is
actually non-trivial, which will become clear in the discussion below.

Let us first assume that photon 2 is sufficiently hard, $x_2 \sim
\bigO{x_1} \sim \bigO{1}$, and that it is emitted at a large angle,
$\vartheta_2 \gg \vartheta_0$, so that $z_2 \gg z_1$.  We find that under
these conditions the double Compton tensor simplifies considerably.
Approximating $k_1 \simeq x_1 p$, one has $1/\Delta \simeq 1/(r_1z_2)$,
with $r_1=1-x_1$, etc.  Terms of order $\bigO{(z_1)^0}$ can be dropped, as
they are expected not to contribute when integrated over the solid angle
of photon 1.  It is now straightforward to verify that the double Compton
tensor in this limit factorizes into a collinear radiation factor and the
Compton tensor for single photon emission, as in the quasi-real electron
approximation \cite{Baier:1973ms}:
\begin{eqnarray}
\label{eq:double-Compton-fact}
  \left.
  K_{\mu\nu}^{\mathrm{semi-coll}}(p,p',k_1,k_2)
  \right|_{z_2 \gg z_1}
  & \simeq &
  \frac{1}{r_1}
  \left[
    \frac{1+(1-x_1)^2}{x_1} \, \frac{1}{z_1}
    - 2(1-x_1) \frac{m^2}{z_1^2}
  \right]
  \nonumber \\
  & \times &
  K_{\mu\nu}^{1\gamma}(r_1 p,p',k_2)
  \; .
\end{eqnarray}
Strictly speaking, this expression is correct for emission at large angles
and for radiation collinear to the final state electron, while one should
drop those terms from the Compton tensor in the last line that are of order
$m^2$ and contribute only for small angles, i.e., terms like $m^2/z_2^2$.


However, if $\vartheta_2$ is small, eq.~(\ref{eq:double-Compton-fact}) not
only fails to reproduce the indicated terms that we dropped, it completely
misses the behavior of the angular distribution of photon 2 in that
region.  To understand this, let us consider the integral over the
azimuthal angle $\phi_1$.  Neglecting terms that are always suppressed by a
factor $m^2/E_e^2$ relative to the leading ones, we have:
\begin{eqnarray}
\label{eq:int-Delta}
  \int\limits_{-\pi}^{\pi} \frac{\dd \phi_1}{2\pi} \, \frac{1}{\Delta}
  & \simeq &
  \Biggl[
     (r_2 z_1)^2 + (r_1 z_2)^2 + 2r_1 r_2 z_1 z_2 \cos\psi
  \\ && \quad {}
     + 4 x_1 x_2 m^2 (z_1 + z_2))
     + \frac{ (z_1 + z_2)z_1 z_2}{E_e^2}
  \Biggr]^{-1/2} \; ,
  \nonumber
\end{eqnarray}
with $\cos\psi$ defined in (\ref{def:cospsi}).

Requiring one photon being almost collinear and the other one at a large
angle, we see that the terms in the last line of (\ref{eq:int-Delta}) can
be neglected in comparison to those of the first line.  Thus, for the
semi-collinear case we obtain:
\begin{equation}
\label{eq:Delta-large-theta2}
  \int\limits_{-\pi}^{\pi} \frac{\dd \phi_1}{2\pi} \, \frac{1}{\Delta}
  \simeq
  \frac{1}{\sqrt{(r_2 z_1)^2 + (r_1 z_2)^2 + 2r_1 r_2 z_1 z_2 \cos\psi}}
  \; .
\end{equation}
Under the same conditions, we find:
\begin{equation}
\label{eq:Delta2-large-theta2}
  \int\limits_{-\pi}^{\pi} \frac{\dd \phi_1}{2\pi} \, \frac{1}{\Delta^2}
  \simeq
  \frac{r_2 z_1+r_1 z_2}
       {[(r_2 z_1)^2 + (r_1 z_2)^2 + 2r_1 r_2 z_1 z_2 \cos\psi]^{3/2} }
  \; .
\end{equation}
It is easy to see that it is not sufficient to keep just the first terms of
the expansions for large $z_2$ when we are going to smaller angles for
photon 2.  Consider e.g., the r.h.s.\ of (\ref{eq:Delta-large-theta2}):
\begin{equation}
\label{eq:Delta-expansion}
  \frac{1}{\sqrt{(r_2 z_1)^2 + (r_1 z_2)^2 + 2r_1 r_2 z_1 z_2 \cos\psi}}
  =
  \frac{1}{r_1 z_2}
  - \frac{r_2 z_1 \cos\psi}{(r_1 z_2)^2}
  + \bigO{ \frac{z_1^2}{z_2^3} }
  \; .
\end{equation}
A similar expansion holds for $1/\Delta^2$.  Retaining only the first term
of this expansion in the expression for $K$ integrated over $\phi_1$ leads
to the factorized form (\ref{eq:double-Compton-fact}), as discussed above.
However, when integrating over the polar angle $\vartheta_1$, we find that
the higher terms in (\ref{eq:Delta-expansion}) lead to expressions of the
type:
\begin{equation}
  \frac{E_e^2 \vartheta_0^2}{z_2^2}
  \; , \quad
  \frac{(E_e^2 \vartheta_0^2)^2}{z_2^3}
  \; , \quad
  \ldots
\end{equation}
Although these expressions appear to be of formal order
$\bigO{\vartheta_0^2}$, they do contribute when integrating over the polar
angle of the second photon, $\vartheta_2$.  Since they fall off much faster
as a function of $\vartheta_2$ than the leading terms
(\ref{eq:double-Compton-fact}), their contribution is essentially
concentrated in a small region $\vartheta_2 \gtrsim \vartheta_0$.  We
express this by formally splitting the double Compton tensor as follows:
\begin{eqnarray}
\label{eq:double-Compton-split}
  K_{\mu\nu}^{\mathrm{semi-coll}}
  & \to &
  \frac{1}{r_1}
  \left[
    \frac{1+(1-x_1)^2}{x_1} \, \frac{1}{z_1}
    - 2(1-x_1) \frac{m^2}{z_1^2}
  \right]
  K_{\mu\nu}^{1\gamma}(r_1 p,p',k_2)
  \nonumber \\
  & + & R_{\mu\nu}(p,p',k_1,k_2)
  \; ,
\end{eqnarray}
thereby introducing the ``remainder'' $R_{\mu\nu}$ which falls off rapidly
as a function of $\vartheta_2$.

Inserting (\ref{eq:double-Compton-split}) into (\ref{eq:KdotH}) yields
\begin{eqnarray}
  &&
  \frac{1}{x_1 r_1} \, P(r_1,L_0)
  \cdot
  \frac{E_e^2}{\pi} \int\limits_{\vartheta_2>\vartheta_0} \dd\Omega_2 \;
  K_{\mu\nu}^{1\gamma}(r_1 p,p',k_2)
  \, H^{\mu\nu}(P,q) \Bigr|_{q=r_1p-p'-k_2}
  \nonumber \\
  &+&
  H^{\mu\nu}(P,q) \Bigr|_{q=(r_1-x_2)p-p'}
  \cdot
  \frac{E_e^4}{\pi^2}
  \int\limits_{\vartheta_1<\vartheta_0<\vartheta_2}
  \dd\Omega_1 \; \dd\Omega_2 \;
  R_{\mu\nu}(p,p',k_1,k_2)
  \; . \qquad \quad \;
\end{eqnarray}
In the second line we exploited again our assumption that the hadron tensor
is a smooth function of its arguments and the above finding that
$R_{\mu\nu}$ contributes to the integral only in a narrow region of the
$\vartheta_2$ integration.

We have not yet discussed the actual form of $R_{\mu\nu}$.  From current
conservation and the kinematic restrictions it is obvious that its tensor
structure should be as follows:
\begin{equation}
  R_{\mu\nu}(p,p',k_1,k_2)
  \simeq
  -\tilde{g}_{\mu\nu} R_g + \tilde{p}_\mu \tilde{p}_\nu R_{11}
  \; .
\end{equation}
Furthermore, we can actually guess the coefficients $R_g$ and $R_{11}$
without great effort just from the considerations above and from the
knowledge of the double Compton tensor in the double collinear region, for
the simple reason being that the ``problematic'' semi-collinear terms must
fall off at least like $1/z_2$, as well as contribute in the double
collinear case, too.  The ``remainder'' $R_{\mu\nu}$ is thus given by:
\begin{equation}
  R_{\mu\nu} \simeq K_{\mu\nu}^\mathrm{2-coll} -
  \left[ K_{\mu\nu}^\mathrm{2-coll} \right]_{\Delta \to r_1 z_2} \; .
\end{equation}
Looking at the expression for the Compton tensor in the double collinear
region (\ref{eq:K-2-coll}) and requiring that the resulting contribution
not be suppressed by a factor $1/\zeta_0$ after integration over the solid
angle of photon 1, we identify the following candidates with a denominator
$\Delta$ for closer investigation:
\begin{equation}
\label{eq:qc-cand}
  \frac{1}{z_1 \Delta}
  \; , \quad
  \frac{1}{z_2 \Delta}
  \; , \quad
  \frac{z_2}{z_1 \Delta^2}
  \; , \quad
  \frac{z_1}{z_2 \Delta^2}
  \; .
\end{equation}
All other terms, e.g., $m^2/(z_1^2 \Delta)$, will be suppressed and thus
harmless.

Analogous to the double collinear case it is convenient to define the
following abbreviation for the semi-collinear situation:
\begin{equation}
  \overline{ \biggl\{ \ldots \biggr\} }
  :=
  \frac{E_e^4}{\pi^2}
  \int \dd \Omega_1 \, \dd \Omega_2 \;
  \Theta(\vartheta_0 - \vartheta_1) \,
  \Theta(\vartheta_2 - \vartheta_0) \;
  \biggl\{ \ldots \biggr\} \; .
\end{equation}
We shall also use the substitutions described in
appendix~\ref{sec:double-coll-int} and exploit the fact that all integrals
below will be convergent even if the upper limit on the variable
$\vartheta_2$ is removed.

By straightforward calculation we obtain:
\begin{eqnarray}
  \overline{ \left\{
    \frac{1}{z_1 \Delta} - \frac{1}{z_1 (r_1 z_2)}
  \right\} }
  & \simeq &
  \frac{1}{x_1 x_2 r_1}
  \int\limits_{x_1 r_2}^{x_1 r_2 \zeta_0} \frac{\dd \eta_1}{\eta_1}
  \int\limits_{x_2 r_1 \zeta_0}^\infty \dd \eta_2 \;
  \left[
  \frac{1}{\sqrt{D(\eta_1,\eta_2)}} - \frac{1}{\eta_2}
  \right]
  \nonumber \\
  & = &
  \frac{1}{x_1 x_2 r_1}
  \left[
  \left[
   \Xi \left(\cos\psi, \frac{\eta_1}{\eta_2} \right)
  \right]_{\eta_1=x_1 r_2}^{x_1 r_2 \zeta_0}
  \right]_{\eta_2=x_2 r_1 \zeta_0}^{\infty}
  \nonumber \\
  & =: &
  \frac{1}{x_1 x_2 r_1} \,
  H_1(x_1,x_2)
  \; ,
\end{eqnarray}
with the function $\Xi$ being defined in (\ref{def:Xi}),
and the abbreviation:
\begin{equation}
  H_1(x_1,x_2)
  = - \Xi \left(\cos\psi, \frac{x_1 r_2}{x_2 r_1} \right)
  \; .
\end{equation}
The second candidate of (\ref{eq:qc-cand}) does not contribute to the naive
factorization (\ref{eq:double-Compton-fact}) at large angles.  Therefore no
`subtraction' is necessary, and we have:
\begin{eqnarray}
  \overline{ \left\{
    \frac{1}{z_2 \Delta}
  \right\} }
  & \simeq &
  \frac{1}{x_1 x_2 r_2}
  \int\limits_{x_2 r_1 \zeta_0}^\infty \frac{\dd \eta_2}{\eta_2}
  \int\limits_{x_1 r_2}^{x_1 r_2 \zeta_0} \eta_1
 \;
  \left[
  \frac{1}{\sqrt{D(\eta_1,\eta_2)}}
  \right]
  \nonumber \\
  & = &
  \frac{1}{x_1 x_2 r_2}
  \left[
  \left[
    \ln\eta_1 \ln\eta_2 +
    \Xi \left(\cos\psi, \frac{\eta_2}{\eta_1} \right)
  \right]_{\eta_1=x_1 r_2}^{x_1 r_2 \zeta_0}
  \right]_{\eta_2=x_2 r_1 \zeta_0}^{\infty}
  \nonumber \\
  & =: &
  \frac{1}{x_1 x_2 r_2} \,
  H_2(x_1,x_2)
    \; ,
\end{eqnarray}
with ($\chi = 1-x_1-x_2$):
\begin{eqnarray}
  H_2 & = &
  \frac{1}{2} \ln^2 \frac{x_1 r_2^2}{x_2 \chi}
  + \frac{\pi^2}{6}
  + \dilog\left( - \frac{x_1 x_2}{\chi} \right)
  - \Xi \left(\cos\psi, \frac{x_2r_1}{x_1r_2} \right)
  \, . \quad
\end{eqnarray}
Similarly, the third candidate yields:
\begin{eqnarray}
  \overline{ \left\{
    \frac{z_2}{z_1 \Delta^2} - \frac{z_2}{z_1 (r_1 z_2)^2}
  \right\} }
  & \simeq &
  \frac{1}{x_1 x_2 r_1^2}
  \int\limits_{x_1 r_2}^{x_1 r_2 \zeta_0} \frac{\dd \eta_1}{\eta_1}
  \int\limits_{x_2 r_1 \zeta_0}^\infty \dd \eta_2 \;
  \left[
    \frac{(\eta_1+\eta_2)\eta_2}{\sqrt{D(\eta_1,\eta_2)}^3}
    - \frac{1}{\eta_2}
  \right]
  \nonumber \\
  & = &
  \frac{1}{x_1 x_2 r_1^2}
  \left[
   \Xi \left(\cos\psi, \frac{\eta_1}{\eta_2} \right)
   + \Psi(\cos\psi;\eta_1,\eta_2)
  \right]_{\eta_1=x_1 r_2}^{x_1 r_2 \zeta_0}
  \nonumber \\
  & =: &
  \frac{1}{x_1 x_2 r_1^2} \, H_3 \; ,
\end{eqnarray}
where
\begin{eqnarray}
  \Psi(\cos\psi;\eta_1,\eta_2)
  & = &
  \frac{1+2\cos\psi}{1+\cos\psi}
  \ln \frac{\eta_2+\eta_1\cos\psi+\sqrt{D(\eta_1,\eta_2)}}{\eta_1}
  \nonumber \\
  &-&
  \frac{\ln[\eta_1+\eta_2\cos\psi+\sqrt{D(\eta_1,\eta_2)}]}{1+\cos\psi}
  \; .
\end{eqnarray}
One finds:
\begin{eqnarray}
  H_3 & = &
  H_1 +
  \frac{2\cos\psi}{1+\cos\psi}
  \ln (2 x_2 r_1)
  - \frac{1}{1+\cos\psi}
  \ln \frac{1+\cos\psi}{2}
  \nonumber \\
  & - &
  \frac{1+2\cos\psi}{1+\cos\psi} \ln ( \eta + x_2 r_1 + x_1 r_2 \cos\psi)
  \nonumber \\
  & + &
  \frac{1}{1+\cos\psi} \ln ( \eta + x_1 r_2 + x_2 r_1 \cos\psi)
  \; .
\end{eqnarray}
For the last candidate, the same arguments apply as to the second, and we
obtain:
\begin{eqnarray}
  \overline{ \left\{
    \frac{z_1}{z_2 \Delta^2}
  \right\} }
  & \simeq &
  \frac{1}{x_1 x_2 r_2^2}
  \int\limits_{x_2 r_1 \zeta_0}^\infty \frac{\dd \eta_2}{\eta_2}
  \int\limits_{x_1 r_2}^{x_1 r_2 \zeta_0} \dd \eta_1 \;
  \left[
    \frac{(\eta_1+\eta_2)\eta_1}{\sqrt{D(\eta_1,\eta_2)}^3}
  \right]
  \nonumber \\
  & = &
  \frac{1}{x_1 x_2 r_2^2}
  \left[
   \ln \eta_1 \ln \eta_2 +
   \Xi \left(\cos\psi, \frac{\eta_2}{\eta_1} \right)
   + \Psi(\cos\psi;\eta_2,\eta_1)
  \right]_{\eta_1=x_1 r_2}^{x_1 r_2 \zeta_0}
  \nonumber \\
  & =: &
  \frac{1}{x_1 x_2 r_2^2} \, H_4 \; ,
\end{eqnarray}
with
\begin{eqnarray}
  H_4 & = &
  H_2 +
  \frac{2\cos\psi}{1+\cos\psi} \ln [(1+\cos\psi) x_2 r_1 ]
  + \frac{1}{1+\cos\psi}       \ln \frac{1+\cos\psi}{2}
  \nonumber \\
  & + &
  \frac{1}{1+\cos\psi} \ln ( \eta + x_2 r_1 + x_1 r_2 \cos\psi)
  \nonumber \\
  & - &
  \frac{1+2\cos\psi}{1+\cos\psi} \ln ( \eta + x_1 r_2 + x_2 r_1 \cos\psi)
  \; .
\end{eqnarray}

As a representative of the harmless terms, we shall investigate whether the
contribution of $m^2/(z_1^2 \Delta)$ at large angles of the second photon
deviates from the naive approximation.  We find:
\begin{eqnarray}
  \overline{ \left\{
    \frac{m^2}{z_1^2 \Delta} - \frac{m^2}{z_1^2 (r_1 z_2)}
  \right\} }
  & \simeq &
  \frac{r_2}{x_1 x_2 r_1}
  \int\limits_{x_1 r_2}^{x_1 r_2 \zeta_0} \frac{\dd \eta_1}{\eta_1^2}
  \int\limits_{x_2 r_1 \zeta_0}^\infty \dd \eta_2 \;
  \left[
    \frac{1}{\sqrt{D(\eta_1,\eta_2)}}
    - \frac{1}{\eta_2}
  \right]
  \nonumber \\
  & = &
  \frac{r_2}{x_1 x_2 r_1}
  \int\limits_{x_2 r_1 \zeta_0}^\infty \frac{\dd \eta_2}{\eta_2^2} \,
  \bigggl[
    \frac{\eta_2-\sqrt{D(\eta_1,\eta_2)}}{\eta_1}
  \\ && \qquad {}
  - \cos\psi \ln \frac{\eta_2+\eta_1\cos\psi+\sqrt{D(\eta_1,\eta_2)}}{\eta_1}
  \bigggr]_{\eta_1=x_1 r_2}^{x_1 r_2 \zeta_0}
  \; .
  \nonumber
\end{eqnarray}
One can see even without explicit evaluation of the last integral that both
terms in the integrand actually contribute only at order $\bigO{1/\zeta_0}$
and are thus negligible.  We leave it as an exercise to the reader to
verify that the contributions of all other terms beyond the factorization
(\ref{eq:double-Compton-fact}) in the semi-collinear limit are suppressed
by a power of $1/\zeta_0$ for each power of $m^2$ in the numerator, except
to those given in (\ref{eq:qc-cand}) which after integration are leading to
$H_1 \ldots H_4$.

To summarize, the integral over the non-factorizing contribution is found
to be:
\begin{equation}
  \frac{E_e^4}{\pi^2}
  \int\limits_{\vartheta_1<\vartheta_0<\vartheta_2}
  \dd\Omega_1 \; \dd\Omega_2 \;
  R_{\mu\nu}
  =
  \left[
  - \tilde{g}_{\mu\nu} Q_l^2
  + 4 (r_1-x_2) \left( \tilde{p}_{\mu} \tilde{p}_{\nu} \right)
  \right] \cdot
  \frac{1}{x_1 x_2} \, H(x_1,x_2)
  \; ,
\end{equation}
with ($\chi = 1-x_1-x_2$):
\begin{equation}
\label{eq:function-H}
  H(x_1,x_2)
  =
  \frac{r_1^3+\chi r_2}{x_1 x_2 r_1} \, H_1
  + \frac{r_2^3+\chi r_1}{x_1 x_2 r_2} \, H_2
  - \chi \left(
  \frac{H_3}{r_1^2} + \frac{H_4}{r_2^2}
  \right)
  \; .
\end{equation}
This quasi-collinear contribution corresponding to one lost photon, i.e.,
being emitted outside the PD, is to be treated in collinear kinematics and
only depends on the energies of the photons.  Obviously, this can only be
true as long as the strong hierarchy $m/E_e \ll \vartheta_0 \ll \theta$
exists, so that the indicated approximations remain valid.

As a check on the quasi-collinear contribution, let us finally investigate
the leading behavior for $x_2 \to 0$.  We find:
\begin{eqnarray}
  H_1 & \simeq &
  - \left[
  \frac{1}{2} \ln^2 \frac{x_1}{x_2 r_1}
  + \frac{\pi^2}{6}
  \right] + \bigO{x_2} \; ,
  \nonumber \\
  H_2 & \simeq &
  \frac{1}{2} \ln^2 \frac{x_1}{x_2 r_1}
  + \frac{\pi^2}{6} + \bigO{x_2} \; ,
\end{eqnarray}
leading to a cancellation of the apparent $1/x_2$ singularity in
$H(x_1,x_2)$ and leaving an expression that is integrable for $x_2 \to 0$.
The entire soft behavior for photon 2 is thus contained in the factorizing
piece (\ref{eq:double-Compton-fact}).





\end{fmffile}



\end{document}